\documentclass[pstricks,siunitx,addfonts,theorem,font=palatino]{tumphthesis}

\newcommand{\nue}{\ensuremath{\nu_e}\xspace}
\newcommand{\numu}{\ensuremath{\nu_\mu}\xspace}
\newcommand{\nutau}{\ensuremath{\nu_\tau}\xspace}
\newcommand{\nux}{\ensuremath{\nu_x}\xspace}
\newcommand{\nuebar}{\ensuremath{\bar{\nu}_e}\xspace}
\newcommand{\numubar}{\ensuremath{\bar{\nu}_\mu}\xspace}

\newcommand{\nuxbar}{\ensuremath{\bar{\nu}_x}\xspace}

\newcommand{\mean}[1]{\ensuremath{\langle #1 \rangle}}
\DeclareSIUnit{\years}{years}
\DeclareSIUnit{\Msol}{\ensuremath{M_\odot}}
\DeclareSIUnit{\mwe}{m.\,w.\,e.} 

\renewcommand{\texttt}[1]{%
  \begingroup
  \ttfamily
  \begingroup\lccode`~=`/\lowercase{\endgroup\def~}{/\discretionary{}{}{}}%
  \begingroup\lccode`~=`[\lowercase{\endgroup\def~}{[\discretionary{}{}{}}%
  \begingroup\lccode`~=`.\lowercase{\endgroup\def~}{.\discretionary{}{}{}}%
  \catcode`/=\active\catcode`[=\active\catcode`.=\active
  \scantokens{#1\noexpand}%
  \endgroup
}

\usepackage{jm-epigraph}
\usepackage{pdfpages} 
\usepackage{wasysym} 
\usepackage{relsize} 

\subject{A thesis submitted in partial fulfilment of the requirements for the degree of Doctor of Philosophy}
\title{Supernova Model Discrimination\\with Hyper-Kamiokande}
\author{Jost Migenda}
\date{\today}
\renewcommand{\pdfauthor}{Jost Migenda}
\renewcommand{\pdftitle}{Supernova Model Discrimination with Hyper-Kamiokande}

\uppertitleback{However big you think supernovae are, they’re bigger than that.\\
			{\em ---\,Donald~Spector~\cite{Munroe2014}}}
\lowertitleback{\makebox[2.5cm]{Supervisors:} Dr.~Susan~L.~Cartwright\\
\makebox[2.5cm]{} Dr.~Matthew~S.~Malek
}

\begin{document}

\frontmatter
\maketitle
\chapter*{Declaration}\label{ch-declaration}

I, the author, confirm that the Thesis is my own work. I am aware of the University’s Guidance on the Use of Unfair Means\footnote{\href{https://www.sheffield.ac.uk/ssid/unfair-means}{\texttt{https://www.sheffield.ac.uk/ssid/unfair-means}}}.
This work has not been previously been presented for an award at this, or any other, university.

\begin{center}
  $\ast$~$\ast$~$\ast$
\end{center}

Due to the nature of a large experiment like Hyper-Kamiokande, this research is built on the contributions of hundreds of scientists.
In the following, I will describe the extent of my personal contributions.

Chapter~\ref{ch-intro} provides an overview over previous research on supernovae and neu\-trinos, summarising the work of many generations of scientists.
The detector design described in chapter~\ref{ch-hk} is the result of the combined work of the whole Hyper-Kamiokande proto-collaboration.
Both chapters do not constitute original research of this thesis.

Chapter~\ref{ch-sim} describes a software toolchain for generating, simulating and reconstructing supernova neutrino interactions in Hyper-Kamiokande.
While I have contributed to all parts of this toolchain, my main contributions are the event generator sntools and the energy reconstruction.
Both were largely developed by me, with some contributions by Liz Kneale and Owen Stone as part of undergraduate research projects under my co-supervision.

The data analysis performed in chapter~\ref{ch-ana} constitutes original research performed solely by me.

Finally, chapter~\ref{ch-conclusions} summarizes the results of this work and gives an outlook on future research.
\chapter*{Acknowledgements}

\setlength{\epigraphwidth}{.45\textwidth}
\epigraphhead[0]{\epigraph{If I have seen further it is by standing on the shoulders of giants.}{\textit{Isaac Newton}}}


First and foremost, I would like to thank my supervisors, Susan Cartwright and Matthew Malek.

The story of this PhD started in the spring of 2015, when I discovered the thesis topic “Supernova Neutrinos in TITUS and Hyper-K” being advertised on the University of Sheffield’s website.
The official application deadline had already passed, but since the topic matched my interests so perfectly, I wrote an email to Susan anyway. Without hesitation, she invited me to start a PhD in Sheffield.
In the following years, missing that application deadline has led to an \emph{unusual} funding situation.
However, Susan always went above and beyond to find the resources to allow me to attend numerous conferences, workshops, collaboration meetings and summer schools across three continents.
Time and again, her incredibly sharp mind and attention to detail have also helped me with issues ranging from physics to data analysis to proofreading this thesis.

Matthew is that legendary supervisor any PhD student can barely dare dream of---a brilliant physicist \emph{and} an equally wonderful human being, all in one person.
He has been my North Star during this research, always being available to provide guidance and direction while, at the same time, giving me the freedom to find my own way. His intimate knowledge of the science involved in this research and familiarity with the scientists working on it have been invaluable resources for me.
At the same time, Matthew has fostered my outreach efforts, hobbled next to me on the Big Walk, called me a gentleman%
\footnote{and apologized afterwards}
, invited me to a late-night cup of tea in the Austin Room and has always been available when I wanted to chat about important or not-so-important topics.
His encouragement has been genuine and his criticism fair and constructive (and more accurate than I wanted to admit at the time).

During the past four years, Marcus O’Flaherty has been almost like a brother to me. He helped me find my feet when I first arrived, shared in the joys and tribulations of being a PhD student, accompanied me to my first conference and summer school and, during my stay at Fermilab in August 2017, even let me stay in his living room for over two weeks.

If Marcus is like an older brother, Liz Kneale is probably most akin to a younger sister. When we first met in 2016, I was equal parts nervous and honoured to co-supervise her project work contributing to an early version of sntools. In the years since, I have seen her grow into an experienced neutrino physicist and spent many lunch breaks with her discussing physics and life in general, which has been incredibly enjoyable.

Completing the list of my academic siblings is Owen Stone. I started co-supervising him during a summer project in 2017 and was happy to see him return to work with Matthew and me for several other projects since. I’m really excited that he has now started a PhD in our group and can’t wait to learn what he will discover!

While I cannot list them all by name here, there are so many other members of the neutrino group, the HEP group and the whole department who have created a wonderful atmosphere that has truly allowed me to feel comfortable and do fantastic work here.
I’m also grateful to the staff in F10 (and elsewhere) without whom the group and the whole department couldn’t function. 

Finally, writing the early parts of this thesis would have been much slower and more frustrating if it hadn’t been for Connor Heapy’s help as part of the thesis mentoring programme.

\begin{center}
  $\ast$~$\ast$~$\ast$
\end{center}

Beyond Sheffield, I am grateful to the hundreds of members of the Hyper-Kamiokande proto-collaboration around the world, without whom none of this work could exist.
I particularly want to thank the leaders and fellow members of the astrophysics 
and software 
working groups for many interesting discussions and frequent feedback;
Yano-san for patiently answering countless questions about low-energy physics and the software tools;
Francesca Di Lodovico for giving me opportunities to contribute more than just my own research to the collaboration; 
and the group at Kobe University for their hospitality during my visit in February/March 2017.

Starting in mid-2017, I also spent a few months working as part of the DUNE collaboration.
While this did not produce any tangible results, I am grateful to the collaboration---particularly Kate Scholberg and the whole SNB/LE group---for welcoming me with open arms and teaching me so much about LAr TPCs.

For giving me access to the supernova models used in this thesis and patiently answering all my related questions, I am grateful to MacKenzie Warren, Nakazato-san, Totani-san, Adam Burrows, David Vartanyan and Irene Tamborra.


I would like to thank the organizers and participants of the various conferences, workshops and summer schools I have had the pleasure to attend.

Georg Raffelt introduced me to supernova neutrinos during my Master’s thesis, which led me to where I am today---even though neither of us could have seen this coming at the start. Thank you, Georg.

\begin{center}
  $\ast$~$\ast$~$\ast$
\end{center}

All of this work has been made so much easier by powerful tools that were often a joy to use:
online resources like InspireHEP, arXiv and NASA's Astrophysics Data System;
software like Python, the MacTeX distribution, Keynote and BBEdit;
and, of course, the operating system and trusty laptop on which all this software runs. 
Thanks to the people who created all these tools.

\begin{center}
  $\ast$~$\ast$~$\ast$
\end{center}


The past four years contained not only the work reflected in this thesis. They also contained copious amounts of joy and sorrow, infinite jest and depths of despair, love and heartbreak and beauty and wonder 
and one thousand green-tinted photos of a rapper.
One day, all these moments will be lost in time, like tears of a unicorn lost in the rain.
But today is not that day.
Today, as I am writing these lines, all these moments mean the world to me.

Thank you to my family:
Angela.
Thomas. 
Jakob.
Uwe.

To Alex and to Lisa:
We’ve known each other for over a decade now and I love you both dearly---in different ways yet equally much.

To Mel, for making me feel at home. 

To my “cuddly friends”, both current and former. 

To the CdE and all the wonderful people who make its events so special.

To Ashley, Izzy\,\&\,Jess, Ida, Abbie and Jenna for being incredibly generous with your time and advice and everything else.

To the whole LGBTSTEM community, particularly Beth and Alfredo and the many others who work tirelessly to build something wonderful.

To Héloïse
, Catherine
, Matt
, Katherine
, Morri%
, Burkhard\,\&\,John and all the many, many others who brightened up my days with open ears, deep conversations over a cup of tea, heartfelt messages and---literal or figurative---rainbows. 

To everyone who supported my outreach efforts---whether as organizers, as speakers or simply by attending and getting excited about the universe.


To all those people I only know from afar, yet who make my life and work beautiful with their writing or podcasts or music or videos.

Finally, to the many giants upon whose shoulders I stand. 

\chapter*{Abstract}\label{ch-abstract}

Supernovae are among the most magnificent events in the observable universe.
They produce many of the chemical elements necessary for life to exist and their remnants---neutron stars and black holes---are interesting astrophysical objects in their own right.
However, despite millennia of observations and almost a century of astrophysical study, the explosion mechanism of supernovae is not yet well understood.

Hyper-Kamiokande is a next-generation neutrino detector that will be able to observe the neutrino flux from the next galactic supernova in unprecedented detail.
In this thesis, I investigate how well such an observation would allow us to reconstruct the explosion mechanism.

I develop a high-precision supernova event generator and use a detailed detector simulation and event reconstruction to explore Hyper-Kamiokande’s response to five supernova models simulated by different groups around the world.
I show that 300 neutrino events in Hyper-Kamiokande---corresponding to a supernova at a distance of at least \SI{60}{kpc}---are sufficient to distinguish between these models with high accuracy.


These findings indicate that, once the next galactic supernova happens, Hyper-Kamiokande will be able to determine details of the supernova explosion mechanism.

\setcounter{tocdepth}{2}
\tableofcontents
\cleardoublepage
\listoffigures
\cleardoublepage
\listoftables

\mainmatter

\chapter{Introduction}\label{ch-intro}

\setlength{\epigraphwidth}{.37\textwidth}
\epigraphhead[0]{\epigraph{Remember when you were young\\You shone like the sun\\\emph{Shine on you crazy diamond}\\Now there’s a look in your eyes\\Like black holes in the sky\\\emph{Shine on you crazy diamond}}{\textit{Pink Floyd}}}


A long time ago\footnote{approximately \SI{160000}{years}}, in a galaxy far away\footnote{the Large Magellanic Cloud, a satellite galaxy of the Milky Way at a distance of approximately \SI{160000}{light years}} the blue supergiant Sanduleak~$-69\si{\degree}202$~\cite{Sanduleak1970} exploded, sending out $\sim$\num{e58} neutrinos. This was a common occurrence throughout the universe and would normally be unremarkable. This time, however, about two dozen of the resulting neutrinos were observed by humans---and that has made all the difference.\footnote{That difference, in this case, consists of more than 1600 papers written about those detected neutrinos. These not only improved our understanding of supernovae themselves; they also let us set new limits on properties of neutrinos as well as a wide range of hypothetical new elementary particles and test the theory of relativity to an accuracy not accessible to lab-based experiments.}

Supernovae, like SN1987A described above, are among the most energetic events in the universe---for a period of several days, a single star exploding in a supernova shines as bright as a galaxy consisting of billions of stars.
In a process called supernova nucleosynthesis, this explosion creates many of the chemical elements that are necessary for life as we know it to exist.
It then expels them in an outgoing shock wave that produces instabilities in the surrounding interstellar gas and can increase the local star formation rate, while also increasing the metallicity of that gas, which affects the evolution of those newly-forming stars.
The supernova remnant, meanwhile, forms a neutron star or a black hole, which are important subjects of astrophysical research in their own right.

Understanding exactly how supernovae explode is therefore an important goal of astrophysics.
However, in the electromagnetic spectrum we can only observe what happens after the supernova shock wave reaches the surface of the progenitor several minutes to hours after the start of the explosion.
This electromagnetic signal is largely decoupled from the processes that occur at the centre of the star and cannot help us understand the explosion mechanism.

Investigations of the precise explosion mechanism have thus far relied on computer simulations.
While these have progressed rapidly due to increases in available computing power as well as improvements to simulation codes, they still suffer from major limitations.
An ideal simulation would be fully three-dimensional and implement detailed flavour- and energy-dependent neutrino transport, all while including effects of general relativity.
However, these three features cannot be combined in a single simulation on current supercomputers, so current simulations often artificially impose rotational symmetry, effectively making them two-dimensional, and use various approximations which may include simplified neutrino transport schemes like the “ray-by-ray plus” method~\cite{Buras2006} and IDSA~\cite{Liebendorfer2009} or a modified gravitational potential that attempts to include relativistic effects in an otherwise Newtonian simulation~\cite{Marek2006}.
The effects of these simplifications are not yet completely understood and the resultant uncertainties often lead to physically meaningful differences that can even make the difference between a successful and a failed explosion.
Overall, “results of different groups are still too far apart to lend ultimate credibility to any one of them”~\cite{Skinner2015}.

The only way to settle this debate observationally is by detecting a high-statistics neutrino signal from the next galactic supernova. Approximately 99\,\% of the energy released in a supernova is in the form of neutrinos, which due to their weak interaction cross section are likely to travel through outer layers of the star unhindered, taking with them information about the processes happening at the centre of the supernova right in the moment of explosion.

In this thesis, I investigate how well the Hyper-Kamiokande detector will be able to distinguish different supernova simulations based solely on their respective neutrino signals.
Using the same methods, observing an actual supernova in the coming decades will let us determine which simulation most closely reproduces the explosion mechanism inside a real supernova.

The remainder of this chapter describes the history and fundamentals of neu\-trino physics and supernova observations, as well as summarizing previous work on supernova neutrinos. Chapter~\ref{ch-hk} describes Hyper-Kamiokande and its history, including the detector design and construction as well as calibration and sources of background that are relevant to supernova neutrino studies. The software toolchain for simulating and reconstructing supernova neutrino interactions in Hyper-Kamiokande is introduced in chapter~\ref{ch-sim}. Chapter~\ref{ch-ana} contains a detailed description and the results of my analysis. 
Finally, chapter~\ref{ch-conclusions} concludes by summarizing the main findings of this thesis and providing an outlook on possibilities of extending this work.

\section{Neutrinos}
\subsection{History}
\subsubsection{Theoretical Proposal}

In the 1920s, it was already known that some nuclei undergo beta decay by transforming into a nucleus of a different chemical element and expelling an electron:
\begin{equation*}
^A_Z X \rightarrow\ ^{\quad A}_{Z+1}X' + e^-\ \ (+ \nuebar )
\end{equation*}
As this was thought to be a two-body decay, physicists at the time expected the electrons to be monoenergetic due to conservation of energy and momentum. However, various measurements revealed a continuous spectrum whose upper endpoint was the theoretically expected energy~\cite{Ellis1927,Meitner1930}. Surprised by this, some physicists, including Bohr~\cite{Bohr1932}, considered the possibility that energy may not be conserved at a sub-atomic level---or that it may only be conserved “on average”. 

In 1930, in a letter to Meitner, Pauli proposed an alternative solution---that another, thus far unobserved particle was produced in beta decay, which transports away the missing energy and would thus lead to a continuous electron energy spectrum while maintaining conservation of energy~\cite{Pauli1930}.
Pauli originally called this particle “neutron”, but when Chadwick in 1932 discovered the particle that is nowadays known as the neutron~\cite{Chadwick1932}, Pauli’s particle was renamed “neutrino”.\footnote{This contraction of the Italian word “neutronino”---meaning “little neutron”---was first jokingly suggested by Amaldi and got popularized by Fermi~\cite{Amaldi1998,Bonolis2005}.}

While Pauli was originally hesitant to publish his proposal, word spread through discussions at various conferences and in 1934, Fermi published a first theoretical description of beta decay, including a discussion of how the shape of the electron spectrum near the end point depends on the neutrino mass~\cite{Fermi1934d}.
In the same year, Bethe and Peierls published a first theoretical estimate of the neutrino interaction cross section based on the known lifetime of beta decay nuclei, setting a limit of $\sigma < \SI{e-44}{cm^2}$ for \SI{2.3}{MeV} neutrinos~\cite{Bethe1934}. Since such a particle would traverse $\sim$\SI{e16}{km} of matter, they concluded “that there is no practically possible way of observing the neutrino.”

\subsubsection{Experimental Detection}

Over the following two decades, a wide range of experiments investigated beta decays with increasing precision and restricted alternative explanations for the continuous beta decay spectrum, without finding direct evidence for the existence of neutrinos~\cite{Leipunski1936,Crane1938,Jacobsen1945,Christy1947,Rodeback1952}.

After the second world war, in the wake of the Manhattan project, a group at Los Alamos Scientific Laboratory worked on nuclear weapons testing.
In 1951, Reines and Cowan developed a plan to use such a nuclear test to look for neutrinos, arguing that the intense yet brief burst would be advantageous in reducing possible backgrounds.
After presenting the planned experiment at an internal seminar the following year, they were encouraged to use a nuclear fission reactor as a neutrino source instead~\cite{LosAlamos1997}.
To reduce backgrounds, they relied on the spatial and temporal coincidence of the signals from the positron and neutron emitted in the inverse beta decay reaction ($\nuebar + p \rightarrow n + e^+$) in a liquid scintillator detector~\cite{Reines1953a,Cowan1953}.
While a first experiment in 1953 at the Hanford nuclear reactor showed some evidence for this process, results were not yet conclusive~\cite{Reines1953b}. In 1956, an improved experiment at the Savannah River nuclear reactor confirmed the previous observations, finally providing direct proof for the existence of neutrinos~\cite{Reines1956,Cowan1956}.

In the following two years, the parity and helicity of the neutrino were measured by Wu~\cite{Wu1957} and Goldhaber~\cite{Goldhaber1958} and their respective collaborators.

At that time, despite some hints from double beta decay experiments~\cite{Konopinski1953}, it was not yet clear whether neutrino and antineutrino were different particles.
Direct evidence was found by Davis, who exposed a tank containing carbon tetrachloride (CCl$_4$) to antineutrinos from the Savannah River nuclear reactor to look for the reaction
\begin{equation*}
^{37}\text{Cl} + \nuebar \rightarrow\ ^{37}\!\text{Ar} + e^-\!.
\end{equation*}
By 1959, Davis reached an upper limit on that cross section of $0.05$ times the cross section calculated for the equivalent neutrino reaction,
\begin{equation*}
^{37}\text{Cl} + \nue \rightarrow\ ^{37}\!\text{Ar} + e^-\!,
\end{equation*}
thus showing a difference in the behaviour of neutrinos and antineutrinos~\cite{Davis1955,Davis1959}.

Similarly, it was not known at that time whether electron neutrinos and muon neutrinos were different particles. This was experimentally determined in 1962 by producing a beam of muon neutrinos in pion decay, $\pi^\pm \rightarrow \mu^\pm + (\numu / \numubar)$, and showing that they produced muons, rather than electrons, in a nearby detector~\cite{Danby1962}.

Following the discovery of the $\tau$ lepton in 1975~\cite{Perl1975}, the existence of a corresponding neutrino, \nutau, was widely expected and experimental evidence was found in 2001 by the DONUT collaboration at Fermilab~\cite{Kodama2001}.
Measurements of the $Z$ boson’s decay width at LEP indicate that no other species of weakly interacting neutrinos with a mass $m < m_\text{Z}/2 = \SI{45.59}{GeV/c^2}$ exists~\cite{LEP2006,PDG2018}.

\subsubsection{The Solar Neutrino Problem}

For millennia, the Sun’s energy generation had been a mystery. Energy sources known before the twentieth century, like chemical or gravitational energy, would have given the Sun a life time of several thousand or several million years, which appeared irreconcilable with geological evidence that suggested an age of billions of years.\footnote{An account of the discussion between Kelvin and contemporary geologists is given in reference~\cite{England2007}.}

\begin{figure}[tbp]
	\centering
	\includegraphics[scale=0.77]{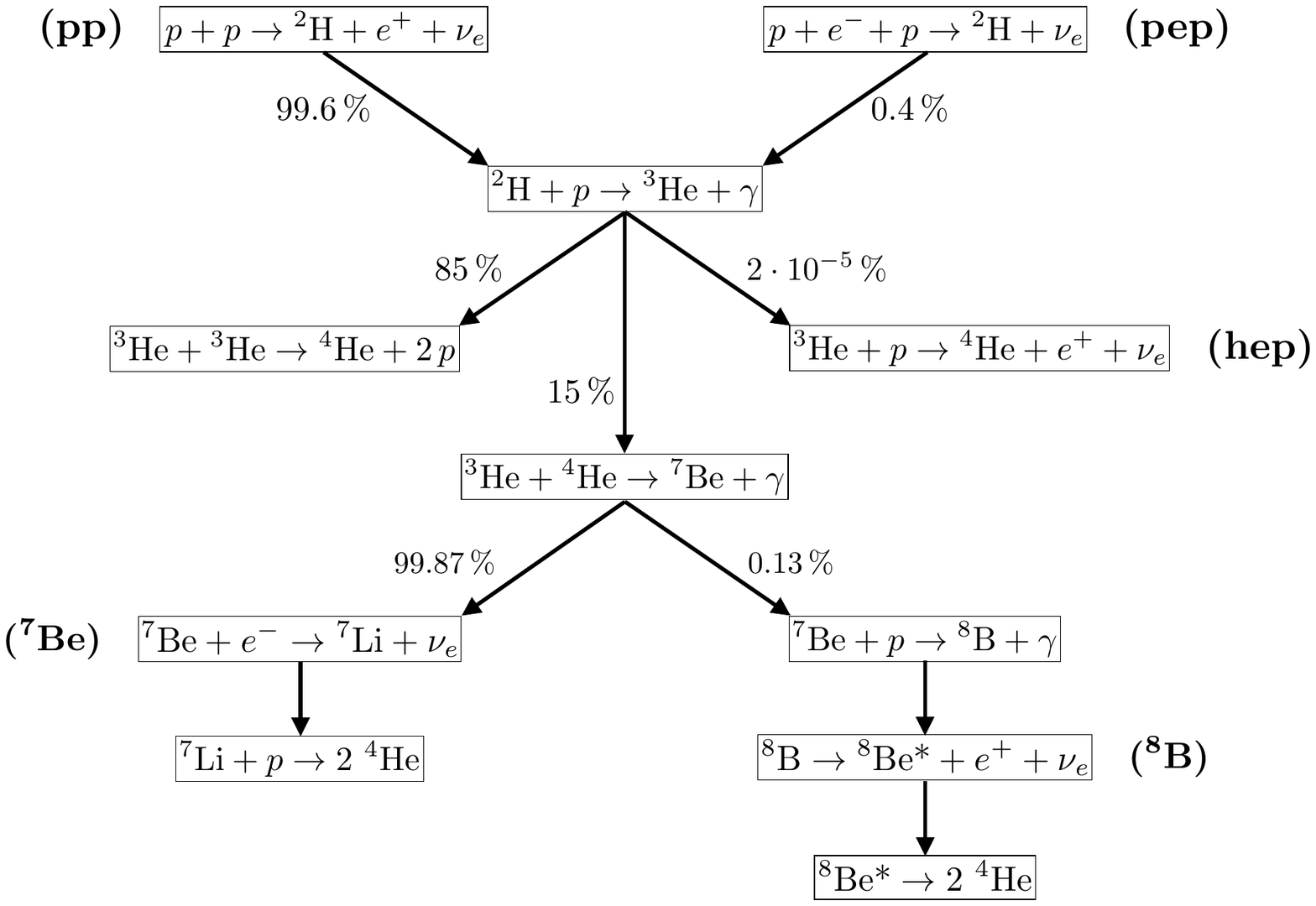}
	\caption[Solar neutrinos: The pp chain]{Reactions and branching ratios for the pp chain~\cite{Bilenky1999}.}
	\label{fig-intro-solar-pp}
\end{figure}

\begin{figure}[htb]
	\centering
	\includegraphics[scale=0.85]{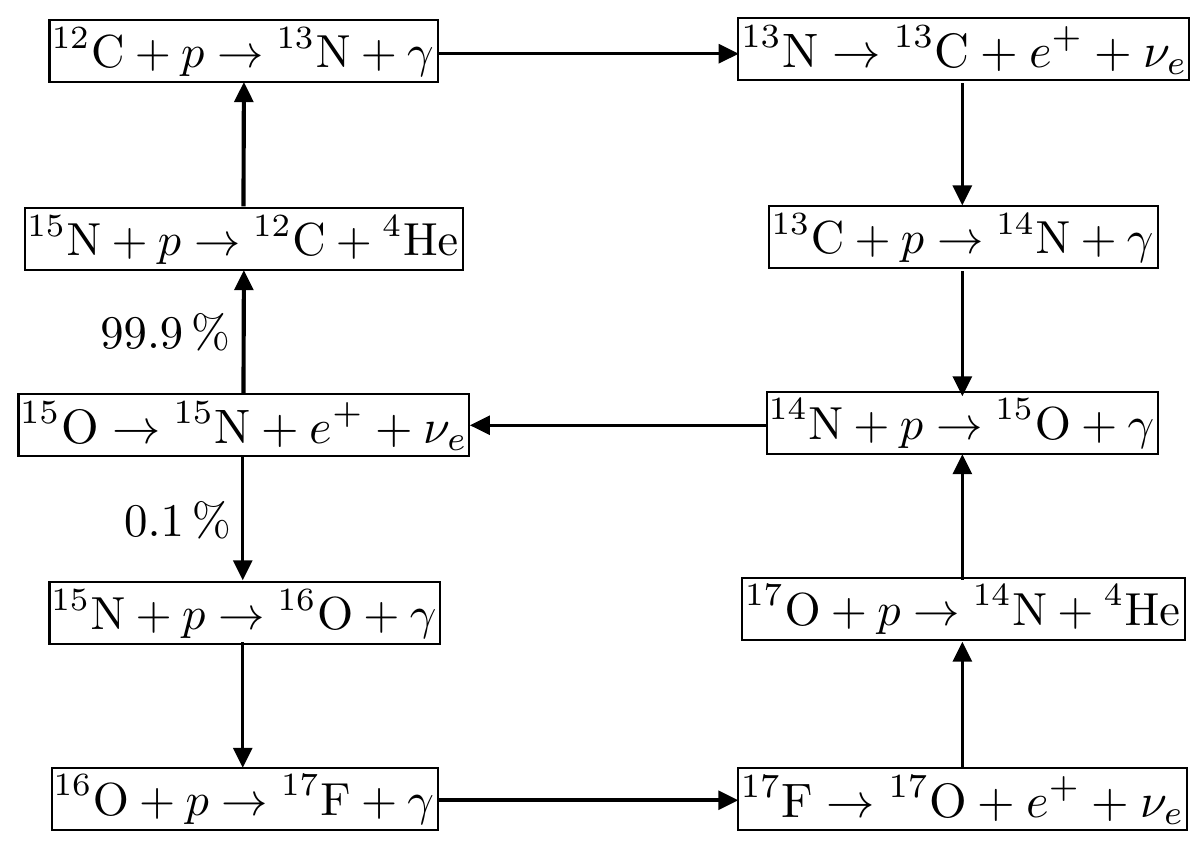}
	\caption[Solar neutrinos: The CNO cycle]{Reactions and branching ratios for the CNO cycle~\cite{Bilenky1999}.}
	\label{fig-intro-solar-cno}
\end{figure}

Only with the advent of nuclear physics in the early 1900s did it become possible to solve this question.
Eddington in 1920 was the first to suggest nuclear fusion as the source of energy production in stars~\cite{Eddington1920} and in 1939, Bethe expanded upon Eddington’s proposal by describing in detail the two participating reaction chains:
the pp chain (figure~\ref{fig-intro-solar-pp}), which fuses four hydrogen nuclei into a single $^4$He nucleus directly, and the CNO cycle (figure~\ref{fig-intro-solar-cno}), which uses a heavier nucleus as a catalyst for the same fusion process~\cite{Bethe1939}.
In both reaction chains, part of the mass difference between the four initial protons and the resulting $^4$He nucleus is emitted as several neutrinos with an energy at the \si{MeV} scale.



From an experimental point of view, the $^8$B neutrinos from the reaction chain
\begin{align}
^3\text{He} +\/ ^4\text{He} &\rightarrow\ ^7\text{Be} + \gamma \label{eq-intro-be7}\\
^7\text{Be} + p &\rightarrow\ ^8\text{B}\\
^8\text{B} &\rightarrow\ ^8\text{Be*} + e^+ + \nue
\end{align}
are a particularly interesting component of the solar neutrino flux due to their high energy of up to \SI{15}{MeV}.
While some other components of the solar neutrino flux are more abundant, their lower energy means that they are below the energy threshold of many common detector materials.
The flux of $^8$B neutrinos was first thought to be too small to be detected, since the cross section for $^7$Be production (reaction~\eqref{eq-intro-be7}) was underestimated.
After laboratory experiments found the cross section to be two orders of magnitude larger than expected~\cite{Holmgren1959}, Bahcall and Davis proposed to look for the reaction
\[
^{37}\text{Cl} + \nue \rightarrow\ ^{37}\!\text{Ar} + e^-
\]
using a detector filled with \SI{380000}{l} of perchlorethylene, C$_2$Cl$_4$~\cite{Bahcall1964,Davis1964}.
This proposal benefitted from an easily available detector material, the reaction’s low threshold energy of \SI{0.81}{MeV} and its enhanced cross section due to three excited states of $^{37}$\!Ar at energies of \SIrange{1.4}{5.1}{MeV}~\cite{Bahcall1964}.

The detector was built in the Homestake Gold Mine in South Dakota, with an overburden of \SI{1480}{m}---corresponding to \num{4400} metres of water equivalent (\si{\mwe})---to reduce the background from cosmic ray muons.
First results were published in 1968 and found no evidence for solar neutrinos, setting an upper limit on the product of flux and cross section of \SI{3}{SNU}\footnote{A “solar neutrino unit” (\si{SNU}) is defined as one interaction per \num{e36} target atoms per second.}~\cite{Davis1968}.
While this upper limit appeared to be in conflict with an updated theoretical prediction of \SI{7.5\pm3}{SNU}, it seemed likely at the time that this was caused by uncertainties in the solar model~\cite{Bahcall1968}.
Over the following decades, however, this conflict remained as theoretical improvements of the solar model led to an updated value of \SI[parse-numbers=false]{\left(7.6^{+1.3}_{-1.1}\right)}{SNU}~\cite{Bahcall2001}, while the value measured by the Homestake experiment was \SI{2.56\pm0.23}{SNU}~\cite{Cleveland1998}.

Starting in the late 1980s, Kamiokande and, since 1996, Super-Kamiokande (see section~\ref{ch-hk-history}) also searched for solar neutrinos. Due to their different detector technology, they were sensitive to the solar neutrino flux at higher energies only, yet by the year 2000 they had found a similar deficit, observing a $^8$B neu\-trino flux of \SI[parse-numbers=false]{\left(2.40^{+0.09}_{-0.08}\right)\cdot 10^6}{cm^{-2}s^{-1}}~\cite{Suzuki2001} compared to a theoretical prediction of \SI[parse-numbers=false]{\left(5.05^{+1.01}_{-0.81}\right)\cdot 10^6}{cm^{-2} s^{-1}}~\cite{Bahcall2001}.

In the 1990s, GALLEX~\cite{Hampel1999} and its successor GNO~\cite{Altmann2000} at Gran Sasso, as well as the Soviet-American collaboration SAGE~\cite{Gavrin2001} at Baksan began using the reaction
\[
^{71}\text{Ga} + \nue \rightarrow\ ^{71}\text{Ge} + e^-
\]
to look for solar neutrinos.
With an energy threshold of about \SI{0.2}{MeV}, much lower than that of chlorine- or water-based detectors, this reaction is sensitive to pp and $^7$Be neutrinos (see figure~\ref{fig-intro-solar-pp}), which are dominant components of the solar neutrino flux.
Combined, these gallium experiments measured a flux of \SI{74.7\pm5.0}{SNU} compared to a theoretical prediction of \SI[parse-numbers=false]{\left(128^{+9}_{-7}\right)}{SNU}, confirming the deficit found by other detectors~\cite{Bahcall2001}.

This decades-long conflict between theoretical predictions and experimental measurements of the solar neutrino flux became known as the “solar neutrino problem”.
When first results by the Homestake experiment hinted at a conflict, various theoretical explanations were soon proposed.
These included time-dependent variations in the solar fusion rate which may not be observable in other channels due to the large time scale required for radiation to diffuse from the centre of the Sun to its surface~\cite{Fowler1968}, neutrino decay~\cite{Bahcall1972}, rotation of the solar core~\cite{Demarque1973} or a central black hole inside the Sun~\cite{Clayton1975}. Neutrino oscillations were originally considered an unlikely explanation by many physicists~\cite{Davis2003}.

After Kamiokande and Super-Kamiokande found evidence for neutrino oscillations in atmospheric neutrinos~\cite{Hirata1988}, the solar neutrino problem was ultimately laid to rest by the Sudbury Neutrino Observatory (SNO) in 2002.
The SNO detector, located in one of the world’s deepest underground laboratories with an overburden of about \SI{2000}{m} of rock (\SI{6000}{\mwe}), contains \SI{1}{kt} of heavy water, D$_2$O, in a spherical acrylic vessel with a diameter of \SI{12}{m}.
This vessel is surrounded by a stainless steel support structure with a diameter of \SI{17.8}{m}, which carries over 9000 inward-looking photomultiplier tubes and acts as an active shield.
It is filled with light water, H$_2$O, and placed inside a barrel-shaped cavern filled with light water, which acts as a passive shield~\cite{Boger2000}.
Due to its unique detector material, SNO can detect solar neutrinos in three different interaction channels:
\begin{align}
\nu + e^-\!	&\rightarrow \nu + e^-	\tag{ES}\label{eq-intro-sno-es}\\
\nue + D	&\rightarrow p + p + e^-	\tag{CC}\label{eq-intro-sno-cc}\\
\nu + D	&\rightarrow p + n + \nu.	\tag{NC}\label{eq-intro-sno-nc}
\end{align}
The elastic scattering~\eqref{eq-intro-sno-es} channel, which was also used by Kamiokande and Super-Kamiokande to detect solar neutrinos, is sensitive to all neutrino flavours but has a lower sensitivity to \numu and \nutau.
The charged-current~\eqref{eq-intro-sno-cc} channel is sensitive only to \nue, while the neutral-current~\eqref{eq-intro-sno-nc} channel is sensitive to all neutrino flavours equally, allowing SNO to measure both the pure \nue flux and the total flux of all neutrino flavours independently.
While the measured ES event rate showed a deficit compared to theoretical predictions, in line with Super-Kamiokande, the total neutrino flux agreed with theoretical predictions, providing direct experimental evidence for neutrino oscillations~\cite{Ahmad2002}.

\subsection{Neutrino Oscillations}\label{ch-intro-oscillations}
\subsubsection{In Vacuum}

Mixing in the neutrino sector was first described in the context of neutrino-antineutrino oscillations in 1957 by Pontecorvo~\cite{Pontecorvo1958a,Pontecorvo1958b} in analogy with neutral kaon mixing described by Gell-Mann and Pais shortly before~\cite{Gell-Mann1955}.
Five years later\footnote{and within two weeks of the experimental observation that electron neutrinos and muon neutrinos behave differently~\cite{Danby1962}}, Maki, Nakagawa and Sakata described mixing between the flavour eigenstates \nue and \numu of weak interactions and the mass eigenstates $\nu_1$ and $\nu_2$~\cite{Maki1962}.\footnote{An obvious parallel exist to the quark sector, where eigenstates under strong and weak interaction are not identical, leading to mixing between generations of quarks. This mixing was first described in the case of two generations in 1963 by Cabibbo~\cite{Cabibbo1963} and CP violation in kaon decays was discovered experimentally shortly thereafter~\cite{Christenson1964}. In 1973, Kobayashi and Maskawa showed that this CP violation cannot be explained with two generations of quarks unless additional fields are introduced, but arises naturally with three generations~\cite{Kobayashi1973}; a prediction that was supported by the discovery of the bottom quark in 1976.

Unlike the quark sector, however, where mixing angles are very small, two of the mixing angles in the neutrino sector are large, with one being near-maximal. Whether this is coincidence or caused by some as yet unknown physics beyond the standard model is subject to speculation.}

Extending this model to three generations of neutrinos~\cite{Bilenky1978}, the relationship between flavour eigenstates and mass eigenstates is today written as
\begin{equation}
\begin{pmatrix}
	\nue\\
	\numu\\
	\nutau
\end{pmatrix}
= U_\text{PMNS}
\begin{pmatrix}
	\nu_1\\
	\nu_2\\
	\nu_3
\end{pmatrix},
\end{equation}
where $U_\text{PMNS}$ is called the Pontecorvo-Maki-Nakagawa-Sakata matrix or PMNS matrix. For Dirac neutrinos, it is commonly parametrized in terms of three mixing angles, $\theta_{ij}$, and a CP-violating phase $\delta$ in the form
\begin{align}
U_\text{PMNS}
&=
\begin{pmatrix}
	1 & 0 & 0\\
	0 & c_{23} & s_{23}\\
	0 & -s_{23} & c_{23}
\end{pmatrix}
\begin{pmatrix}
	c_{13} & 0 & s_{13} e^{- i \delta}\\
	0 & 1 & 0\\
	-s_{13} e^{i \delta} & 0 & c_{13}
\end{pmatrix}
\begin{pmatrix}
	c_{12} & s_{12} & 0\\
	-s_{12} & c_{12} & 0\\
	0 & 0 & 1
\end{pmatrix}
\\
&=
\begin{pmatrix}
	c_{12} c_{13} & s_{12} c_{13} & s_{13} e^{- i \delta}\\
	-s_{12} c_{23} - c_{12} s_{13} s_{23} e^{i \delta} & c_{12} c_{23} - s_{12} s_{13} s_{23} e^{i \delta} & c_{13} s_{23}\\
	s_{12} s_{23} - c_{12} s_{13} c_{23} e^{i \delta} & -c_{12} s_{23} - s_{12} s_{13} c_{23} e^{i \delta} & c_{13} c_{23}
\end{pmatrix},
\end{align}
where $s_{ij} = \sin{\theta_{ij}}$ and $c_{ij} = \cos{\theta_{ij}}$.
If neutrinos are Majorana particles~\cite{Majorana1937}, the PMNS matrix contains two additional phases,
\begin{equation}
U_\text{PMNS}^\text{Majorana} = U_\text{PMNS}
\begin{pmatrix}
	e^{i \alpha_1} & 0 & 0\\
	0 & e^{i \alpha_2} & 0\\
	0 & 0 & 1
\end{pmatrix},
\end{equation}
which do not affect flavour oscillations and are therefore usually omitted.

In this paradigm, neutrinos are created through the weak interaction in a flavour eigenstate $\nu_\alpha$, which is a superposition of three mass eigenstates.
While travelling, if the three mass eigenstates are independent (i.\,e. if at least two of them have different, non-zero masses), they can propagate independently such that at a later time the combination of mass eigenstates does not correspond to the original flavour eigenstate.
Instead, as it propagates in vacuum, its state over time is described by
\begin{align}
\ket{\nu_\alpha (0,0)} &= U^*_{\alpha 1} \ket{\nu_1} &+ U^*_{\alpha 2} &\ket{\nu_2}& + U^*_{\alpha 3} &\ket{\nu_3} &=& \ket{\nu_\alpha}\ \\
\ket{\nu_\alpha (t,x)} &= U^*_{\alpha 1} \ket{\nu_1(t,x)} &+ U^*_{\alpha 2} &\ket{\nu_2(t,x)}& + U^*_{\alpha 3} &\ket{\nu_3(t,x)} &\neq& \ket{\nu_\alpha}\!,
\end{align}
where $\ket{\nu_k(t,x)} = e^{-i (E_k t - p_k x)} \ket{\nu_k}$ and $E_k^2 = p^2 + m_k^2$.
When reaching the detector, the neutrino is detected in a flavour eigenstate $\nu_\beta$, so using $\ket{\nu_k} = \sum_{\beta = e,\mu,\tau} U_{\beta k} \ket{\nu_\beta}$ we find
\begin{align}
\ket{\nu_\alpha (t,x)} &= \sum_k U^*_{\alpha k} \ket{\nu_k(t,x)}\\
&= \sum_k U^*_{\alpha k} e^{-i (E_k t - p_k x)} \ket{\nu_k}\\
&= \sum_{\beta = e,\mu,\tau} \left( \sum_k U^*_{\alpha k} e^{-i (E_k t - p_k x)} U_{\beta k}\right) \ket{\nu_\beta}.
\end{align}
As a result, the probability of being detected in the flavour eigenstate $\nu_\beta$ is
\begin{equation}
P_{\alpha \rightarrow \beta} = \left| \sum_k U^*_{\alpha k} e^{-i (E_k t - p_k x)} U_{\beta k} \right|^2.
\end{equation}

Since neutrino energies considered here are $\mathcal{O}(\si{MeV})$ while neutrino masses are $\mathcal{O}(\si{eV})$, we can make the relativistic approximation $t \approx x = L$, where $L$ is the distance between neutrino source and detector.
Thus,
\begin{align}
E_k t - p_k x &\approx (E_k - p_k) L = \frac{E_k^2 - p_k^2}{E_k + p_k} L \approx \frac{m_k^2}{2 E} L,
\end{align}
and the transition probability becomes
\begin{align}
P_{\alpha \rightarrow \beta} &= \left| \sum_k U^*_{\alpha k} \exp\!\left(-i \frac{m_k^2 L}{2 E}\right) U_{\beta k} \right|^2\\
&= \sum_{k, j} U^*_{\alpha k} U_{\beta k} U_{\alpha j} U^*_{\beta j} \exp\!\left(-i \frac{\Delta m^2_{kj} L}{2 E}\right),
\end{align}
where $\Delta m^2_{kj} = m_k^2 - m_j^2$.
The transition probability for antineutrinos can be found by making the replacement $U \leftrightarrows U^*$.

Neutrino oscillations are thus described by three mixing angles, $\theta_{ij}$, two independent mass differences between the three mass eigenstates, $\Delta m^2_{ij}$, and a CP-violating phase $\delta$.

After early reactor and accelerator neutrino experiments showed no evidence of neutrino oscillations due to their short baseline, first evidence came from observations of atmospheric neutrinos in Kamiokande and Super-Kamiokande.
Atmospheric neutrinos are mainly produced when high-energy cosmic rays interacting with nuclei in the Earth’s atmosphere produce charged pions which decay via
\begin{align*}
\pi^\pm \rightarrow &\mu^\pm + \numu (\numubar)\\
&\mu^\pm \rightarrow e^\pm + \nue (\nuebar) + \numubar (\numu).
\end{align*}
This should result in a constant 2:1 ratio of muon neutrinos to electron neutrinos, almost independent of the cosmic ray flux model.
In the late 1980s, Kamiokande first found that the muon-to-electron ratio in atmospheric neutrino events was much lower than expected~\cite{Hirata1988}.
Over time, it became clear that this deficit in muon neutrinos varied with the azimuth angle, with downgoing events reproducing the expected ratio, while upgoing events, which had travelled through the Earth on their way to the detector, showed about half the expected number of muon neutrinos~\cite{Fukuda1994}.
Observations by Super-Kamiokande confirmed the Kamiokande data with much higher statistics, providing strong evidence for $\numu \rightarrow \nutau$ oscillations in atmospheric neutrinos~\cite{Fukuda1998}.

These measurements allowed a first determination of the mixing angle $\theta_{23}$ and the mass difference $\abs{\Delta m^2_{32}}$~\cite{Fukuda1998}, which are today usually determined from atmospheric neutrinos or in long baseline experiments~\cite{Aartsen2018,Abe2018,Acero2018}.
The second mixing angle, $\theta_{12}$, and the corresponding mass difference, $\Delta m^2_{21}$, were first measured using solar neutrinos by SNO~\cite{Ahmad2002a} and shortly thereafter using reactor antineutrinos by KamLAND~\cite{Eguchi2003}.
The third mixing angle, $\theta_{13}$, is much smaller than the others and early measurements remained compatible with zero. First hints of a non-zero value were found by T2K~\cite{Abe2011}, MINOS~\cite{Adamson2011} and Double Chooz~\cite{Abe2012} and within a few months, these hints were confirmed by Daya Bay~\cite{An2012} and RENO~\cite{Ahn2012}.

Today, the values of these mixing parameters have been measured to a high precision and are given by~\cite{PDG2018}
\begin{align*}
\sin^2{\theta_{12}} &= 0.307^{+0.013}_{-0.012}\\
\sin^2{\theta_{13}} &= 0.0212 \pm 0.0008\\
\sin^2{\theta_{23}} &=
	\begin{cases}
		0.417 ^{+ 0.025} _{- 0.028} &\text{(NO, if $\theta_{23}< \pi/4$)}\\
		0.597 ^{+ 0.024} _{- 0.030} &\text{(NO, if $\theta_{23}> \pi/4$)}\\
		0.421 ^{+ 0.033} _{- 0.025} &\text{(IO, if $\theta_{23}< \pi/4$)}\\
		0.592 ^{+ 0.023} _{- 0.030} &\text{(IO, if $\theta_{23}> \pi/4$)}
	\end{cases}\\
\Delta m^2_{21} &=\SI[parse-numbers=false]{\left(7.53 \pm 0.18\right)\cdot 10^{-5}}{eV^2}\\
\Delta m^2_{32} &=
	\begin{cases}
		\SI[parse-numbers=false]{\left(2.51 \pm 0.05\right)\cdot 10^{-3}}{eV^2} &\text{(NO)}\\
		\SI[parse-numbers=false]{\left(-2.56 \pm 0.04\right)\cdot 10^{-3}}{eV^2} &\text{(IO).}
	\end{cases}
\end{align*}
Here, “NO” stands for normal ordering, where the mass eigenstates obey the relation $m_1 < m_2 < m_3$, while “IO” stands for inverted ordering, i.\,e. $m_3 < m_1 < m_2$.
The value of the CP-violating phase $\delta$ is not yet known, though recent results by the long baseline experiments T2K and NOvA show a slight preference for a value near $\delta = 3 \pi / 2$, with T2K excluding the CP-conserving values 0 and $\pi$ at the $2\sigma$ level~\cite{Abe2018,Acero2018}.

\subsubsection{In Matter: The MSW Effect}\label{ch-intro-msw}
The equations in the previous section describe neutrino oscillations while propagating in vacuum. In 1978, Wolfenstein pointed out that the presence of matter can affect neutrino oscillations through coherent forward scattering.
While neutral-current scattering produces a similar phase shift for all neutrino flavours and thus has little practical effect, the presence of electrons in matter produces an additional potential for \nue through charged-current interactions, which is not present for other flavours.
Wolfenstein discussed the effects in a suggested long baseline experiment and on solar neutrinos, finding that this effect is not sufficient to solve the solar neutrino problem~\cite{Wolfenstein1978}.

Several years later, Mikheev and Smirnov showed that neutrinos propagating through a medium with a smoothly varying electron density can experience a resonance behaviour, where even a small mixing angle can lead to high transition probabilities~\cite{Mikheev1985}.
This effect is now called the Mikheev-Smirnov-Wolfenstein (MSW) effect.
While Mikheev and Smirnov already discussed implications of the MSW effect for solar and supernova neutrinos, Bethe in the following year expanded on their work by estimating neutrino oscillation parameters under the assumption that this resonance explained the results of the Homestake solar neutrino experiment~\cite{Bethe1986}.

In the following, I will briefly sketch out the derivation of this resonance effect.
For simplicity, I assume a two-flavour scenario and ignore neutral-current scattering as discussed above.

In vacuum, the Schrödinger equation for the neutrino mass eigenstates is
\begin{equation} \label{eq-intro-schrodinger}
i \tdiff{}{t} \begin{pmatrix} \nu_1\\ \nu_2 \end{pmatrix}
= \begin{pmatrix} E_1 & 0\\ 0 & E_2 \end{pmatrix} \begin{pmatrix} \nu_1\\ \nu_2 \end{pmatrix}
\sim \begin{pmatrix} m_1^2 & 0\\ 0 & m_2^2 \end{pmatrix} \begin{pmatrix} \nu_1\\ \nu_2 \end{pmatrix},
\end{equation}
where the mass matrix in the final step is the only contribution that differs between eigenstates.
Using $\ket{\nue} = \cos{\theta} \ket{\nu_1} + \sin{\theta} \ket{\nu_2}$ and $\ket{\numu} = -\sin{\theta} \ket{\nu_1} + \cos{\theta} \ket{\nu_2}$
as well as the relations $\sin{2\theta} = 2\sin{\theta}\cos{\theta}$ and $\cos{2\theta} = 1 - 2 \sin^2{\theta} = 2\cos^2{\theta} - 1$%
, the mass matrix for the flavour eigenstates can be written as
\begin{equation} \label{eq-intro-massmatrix}
M = \frac{m_1^2 + m_2^2}{2} \begin{pmatrix} 1 & 0\\ 0 & 1 \end{pmatrix}
+ \frac{m_2^2 - m_1^2}{2} \begin{pmatrix} -\cos{2\theta} & \sin{2\theta}\\ \sin{2\theta} & \cos{2\theta} \end{pmatrix},
\end{equation}
with off-diagonal elements, which were not present in equation~\eqref{eq-intro-schrodinger}, appearing due to neutrino mixing.

Charged-current interactions of \nue with electrons in the matter introduce an additional term in the Hamiltonian which is equivalent to an effective potential $V_\text{eff} = \sqrt{2} G_F n_e$, where $G_F$ is the Fermi constant of weak interaction and $n_e$ is the electron density.
Replacing $E^2$ in the relation $E^2 = m^2 + p^2$ with $(E - V_\text{eff})^2 \approx E^2 - 2 E V_\text{eff}$ shows that this new term changes the effective mass of \nue, adding a contribution of
\begin{equation*}
2 E V_\text{eff} \begin{pmatrix} 1 & 0\\ 0 & 0 \end{pmatrix}
\end{equation*}
to the mass matrix in equation~\eqref{eq-intro-massmatrix}.
Using $\Delta_m = m_2^2 - m_1^2$, the eigenstates of the modified mass matrix are given by
\begin{equation} \label{eq-intro-masseigenstates}
m_i^{*2} = \frac{1}{2} (m_1^2 + m_2^2 + 2 E V_\text{eff}) \pm \frac{1}{2} \sqrt{(\Delta_m \cos{2 \theta} - 2 E V_\text{eff})^2 + \Delta_m^2 \sin^2{2 \theta}}.
\end{equation}

\begin{figure}[tb]
	\centering
	\includegraphics[scale=0.53]{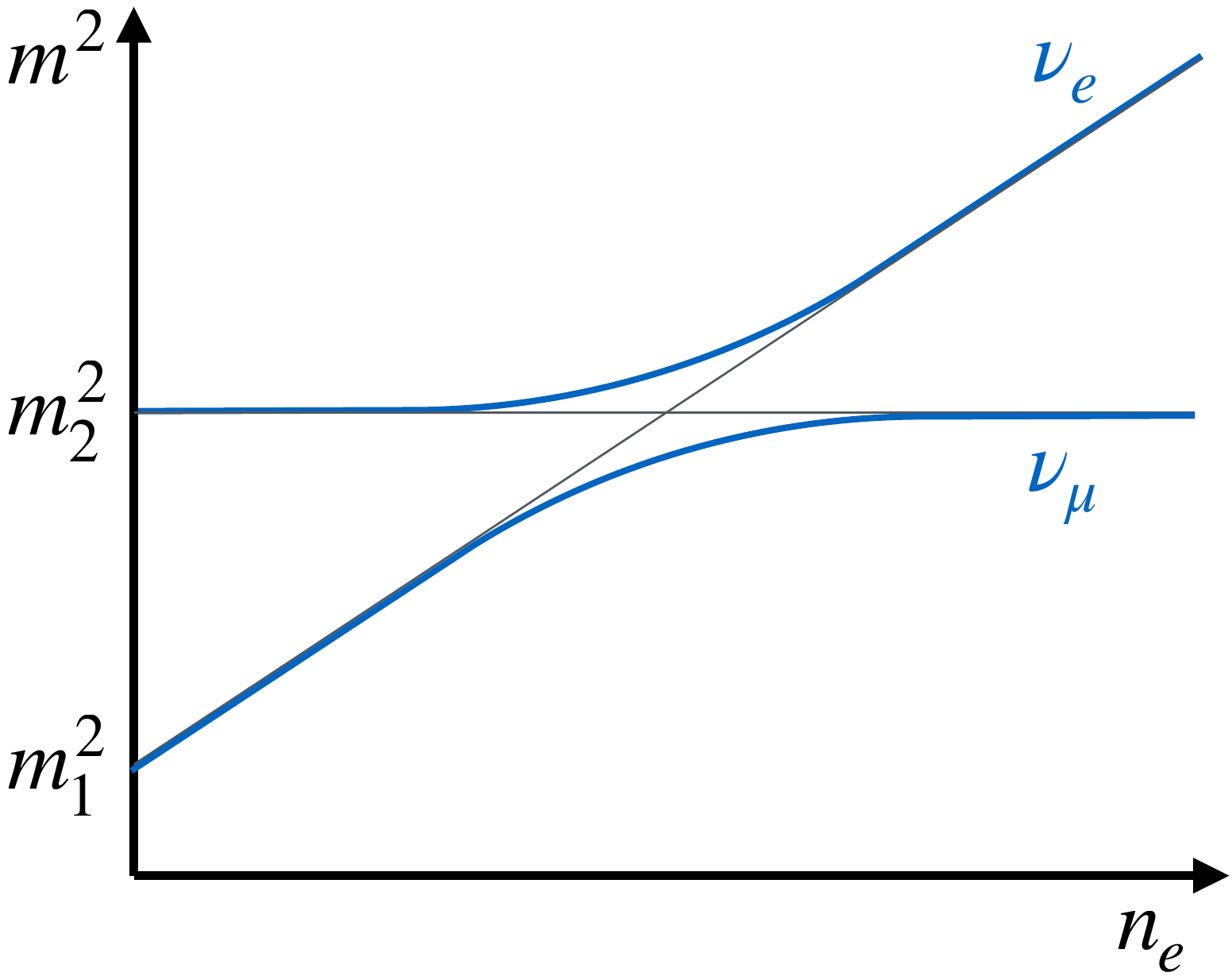}
	\caption[The MSW effect in a simplified two-flavour picture]{Impact of the presence of matter with electron density $n_e$ on the neutrino eigenstates in a simplified two-flavour picture. Thin black lines show the effective masses of the flavour eigenstates \nue (diagonal) and \numu (horizontal), while thick blue lines show the evolution of the propagation eigenstates given in equation~\eqref{eq-intro-masseigenstates}.}
	\label{fig-intro-msw}
\end{figure}

In figure~\ref{fig-intro-msw}, these mass eigenstates are plotted as a function of electron density.
When a \nue is produced in a region with a high electron density, like the interior of the Sun, it is in the mass eigenstate $m_2^*$.
Assuming a smoothly varying density, it stays in that eigenstate, propagating along the blue line and finally reaching the pure mass eigenstate $\nu_2$ as it exits the Sun, undergoing no vacuum oscillations as it travels to Earth.

Similar to this simplified example, neutrinos produced in the dense region near the centre of a supernova undergo this MSW effect as they travel through the outer layers of the star.
Antineutrinos experience an analogous effect, with the electron density being equivalent to a negative positron density.
The resulting fluxes $\Phi_i$, which can be detected on Earth, are linear combinations of the original fluxes produced inside the supernova, $\Phi_i^0$.
The resulting fluxes are described in section~\ref{ch-sim-sntools-ordering}.

\subsection{Detection}\label{ch-intro-detection}
Since the first detection of neutrinos, a wide range of different neutrino detection techniques have been developed that employ different target materials, interaction channels and readout techniques.
\enlargethispage{\baselineskip} 
As a result, an experimental collaboration can make trade-offs between various detector properties---like its cost, size, resolution, energy range and sensitivity to different neutrino flavours---to develop a detector design optimized for its specific physics goals.

In this section, I will summarize detection techniques used by current or planned detectors that are expected to play a major role in supernova neutrino physics in the foreseeable future. While a range of other detection techniques exist that may be of historic, niche or novelty interest, these are beyond the scope of this brief overview. 

\subsubsection{Water Cherenkov Detectors}
\begin{figure}[htbp]
	\centering
	\includegraphics[scale=1.2]{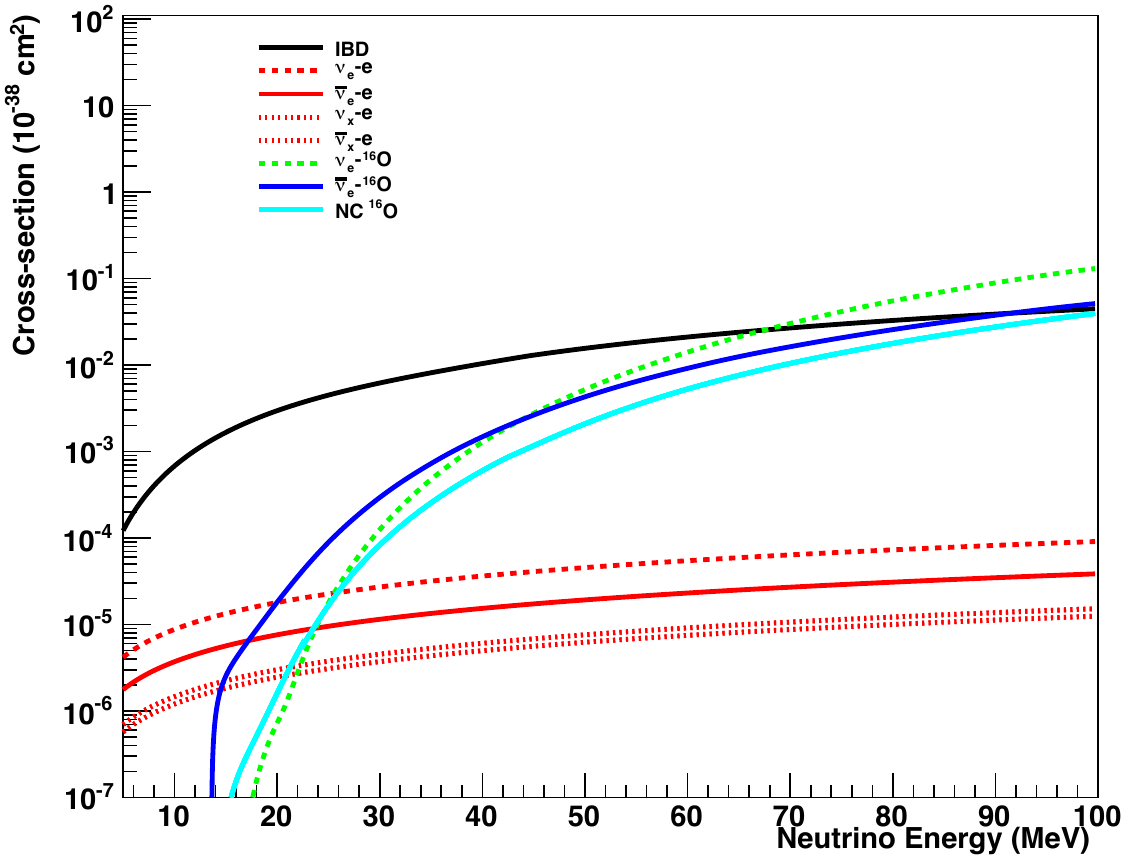}
	\caption[Cross sections of main interaction channels of supernova neutrinos in a water Cherenkov detector]{Cross sections of main interaction channels of supernova neutrinos in a water Cherenkov detector. The main interaction channel is inverse beta decay ($p + \nuebar \rightarrow n + e^+$; solid black), which due to its low threshold energy of \SI{1.8}{MeV} and high cross section makes up about 90\,\% of detected events. At low energies, a sizeable contribution comes from elastic scattering on electrons ($\nu + e^- \rightarrow \nu + e^-$; red, with solid, dashed and dotted lines for \nuebar, \nue and \nux, respectively), which provides the most precise information on the direction of incoming neutrinos. Charged-current interactions on oxygen-16 nuclei of \nue ($^{16}O + \nue \rightarrow X + e^-$; dashed green) and \nuebar ($^{16}O + \nuebar \rightarrow X + e^+$; solid blue) have a high threshold energy of about \SI{15}{MeV} and \SI{11}{MeV}, respectively, and mainly contribute to the high energy tail. Figure from SNOwGLoBES~\cite{Snowglobes}.}
	\label{fig-intro-detection-wch}
\end{figure}

Water Cherenkov detectors use pure water as a target material. Supernova neutrinos interacting through the processes displayed in figure~\ref{fig-intro-detection-wch} produce charged particles, which receive a large portion of the neutrino energy and move at relativistic speeds. If such a particle surpasses the speed of light in water, $c_{\text{H}_2\text{O}} \approx c_\text{vacuum} / 1.3$,
it sends out a cone of Cherenkov light which is then detected by photosensors in the detector.

Since water is an extremely cheap and common material that is easy to handle and has excellent optical properties, this detection technique can be used to build very large neutrino detectors with a relatively simple detector design.
However, reliance on the Cherenkov effect means that particles below the Cherenkov threshold cannot be detected and that the energy resolution is limited due to the low yield of Cherenkov photons.

Water Cherenkov detectors optimized for the energy range typical for supernova neutrinos, like Super- and Hyper-Kamiokande, use an enclosed design, consisting of a human-made water tank---typically located underground to reduce cosmic ray backgrounds---whose inside walls are covered with photosensors. Depending on the fraction of the inside walls covered with photosensors and the efficiency of these sensors, several Cherenkov photons per \si{MeV} energy will typically be detected.
Therefore, a simple trigger based on the number of hits above a threshold, which depends on the noise level of the photosensors, can be employed to identify individual neutrino interactions.

On the other hand, neutrino telescopes like IceCube~\cite{Abbasi2011} and KM3NeT~\cite{Adrian-Martinez2016} are optimized for high-energy astrophysical neutrinos and use an open design, where a naturally occurring body of water---like the antarctic ice or deep sea water---is instrumented with a sparse array of optical modules. With module separation ranging from a few metres to over \SI{100}{metres}, detector masses up to the \si{Gton} scale can be instrumented, making this design suitable for detecting neutrinos with energies up to several \si{PeV}, which have a much lower flux and produce particle showers that cannot be contained in enclosed detectors like Hyper-Kamiokande.
However, the sparse instrumentation means that neutrino telescopes are not sensitive to neutrinos below the \si{GeV} scale and will usually detect at most one photon from each supernova neutrino interaction in the instrumented volume. Any single neutrino interaction is therefore indistinguishable from noise and a supernova is only detectable as a temporary increase in the noise rate across the whole detector.

\subsubsection{Liquid Scintillator}
\begin{figure}[htbp]
	\centering
	\includegraphics[scale=1.2]{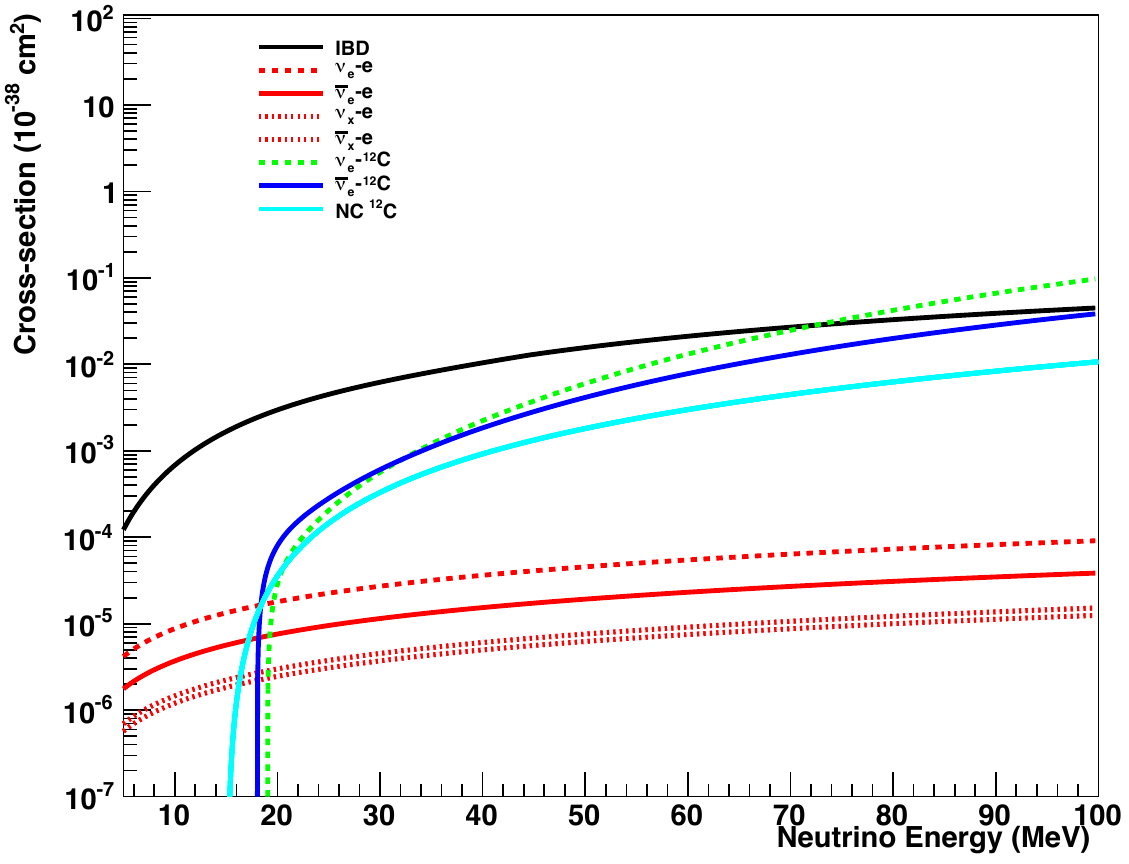}
	\caption[Cross sections of main interaction channels of supernova neutrinos in a liquid scintillator detector]{Cross sections of main interaction channels of supernova neutrinos in a liquid scintillator detector. The main interaction channel is inverse beta decay ($p + \nuebar \rightarrow n + e^+$; solid black) due to its low threshold energy of \SI{1.8}{MeV} and high cross section. At low energies, a sizeable contribution comes from elastic scattering on electrons ($\nu + e^- \rightarrow \nu + e^-$; red, with solid, dashed and dotted lines for \nuebar, \nue and \nux, respectively), while charged-current interactions on carbon nuclei of \nue ($^{12}C + \nue \rightarrow X + e^-$; dashed green) and \nuebar ($^{12}C + \nuebar \rightarrow X + e^+$; solid blue) have high threshold energies of nearly \SI{20}{MeV} and mainly contribute to the high energy tail. Elastic neutrino-proton scattering, which can be detected in liquid scintillator detectors with a sufficiently low energy threshold, is not included in this figure.
	Figure from SNOwGLoBES~\cite{Snowglobes}.}
	\label{fig-intro-detection-ls}
\end{figure}

Liquid scintillator detectors like KamLAND~\cite{Asakura2016}, SNO+~\cite{Andringa2016} or, in the near future, JUNO~\cite{An2016} use a design that is very similar to the enclosed water Cherenkov detectors described in the previous section. Instead of water, the target material inside the detector is a scintillating material---usually one of several hydrocarbons, like linear alkylbenzene---which may be doped with small amounts of wavelength shifting or stabilizing agents to achieve the desired properties. 
Supernova neutrinos interacting through the processes displayed in figure~\ref{fig-intro-detection-ls} produce charged particles that induce emission of scintillation light as they move through the detector.
The resulting light yield is about two orders of magnitude higher than that of Cherenkov light, depending on the composition of the medium. 

This scintillation light enables liquid scintillator detectors to detect particles below the Cherenkov threshold, which would remain undetected in water Cherenkov detectors. Furthermore, since the energy resolution at low energies is limited by Poisson fluctuations of the number of detected photons, the increased light yield of scintillation light improves the energy resolution.

On the other hand, since scintillation light is emitted nearly isotropically, it drowns out the directional information provided by Cherenkov light.\footnote{Scintillation light is slightly delayed compared to the Cherenkov light, so a separation based on emission time is in principle possible~\cite{Gruszko2019}. However, this requires sub-\si{ns} time resolution, which is beyond the capabilities of the photosensors that are currently available in sufficient quantities (and at sufficiently low prices) to equip a large neutrino detector.}
Furthermore, compared to water the higher cost and lower attenuation length of liquid scintillators limit the possible detector size.

In recent years, new water-based liquid scintillator (WbLS) materials have been developed which consist of pure water with a small admixture of a liquid scintillator~\cite{Yeh2011}.
While the resulting material has a lower light yield than pure liquid scintillators, it is still able to improve the energy resolution compared to a pure water Cherenkov detector and it enables detection of sub-Cherenkov threshold particles. At the same time, it largely eliminates the cost and light attenuation disadvantages of liquid scintillators.

A concept for a WbLS neutrino detector---the Advanced Scintillator Detector Concept, which has since been renamed THEIA---was presented in 2014~\cite{Alonso2014,Askins2019}.
A large worldwide R\&D programme consisting of several experimental collaborations including SNO+, WATCHMAN/\!AIT, ANNIE and EGADS is currently ongoing. Over the next years, these collaborations aim to develop and characterize the novel technologies and components expected to be used for THEIA.

\subsubsection{Liquid Argon Time Projection Chamber}
Liquid argon time projection chambers (LAr TPCs) employ a similar design to liquid xenon TPCs which are now common in dark matter direct detection experiments. 

Within a cryostat, a detector is filled with liquid argon as a detection material. Neutrino interactions produce energetic particles, which lose energy through ionization of argon atoms along their tracks.
Through an externally applied electric field, the resulting free electrons are drifted to readout planes on one side of the detector where they induce a signal on multiple layers of wires, enabling a 2D-reconstruction of each hit position along the track.
Detectors also contain some photosensors to detect Cherenkov light and determine the time of interaction, which can be combined with the time delay of the drifting electrons to reconstruct the interaction in 3D.

\begin{figure}[htbp]
	\centering
	\includegraphics[scale=0.7]{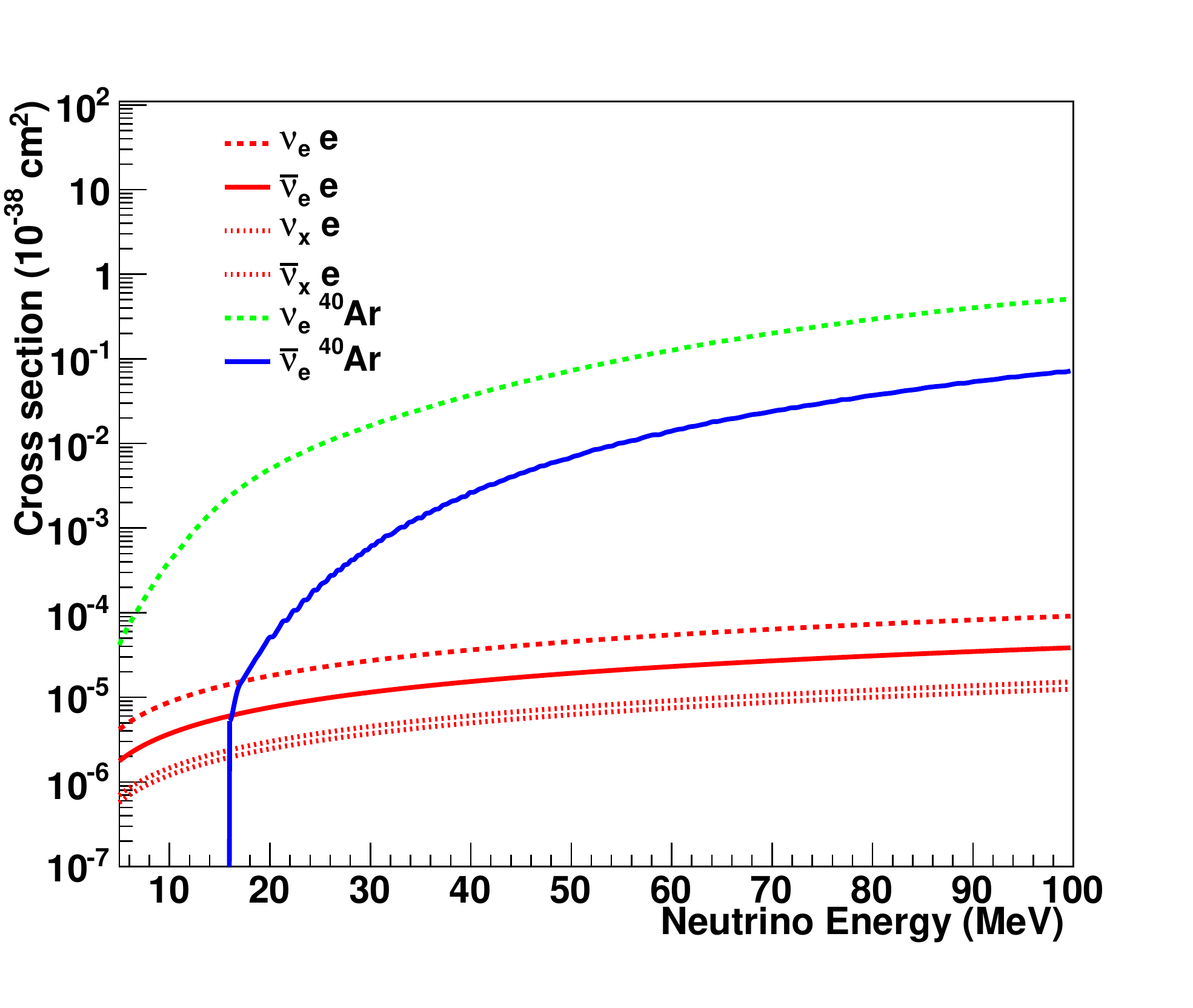}
	\caption[Cross sections of main interaction channels of supernova neutrinos in a LAr TPC]{Cross sections of main interaction channels of supernova neutrinos in a LAr TPC.
	The main interaction channel is the charged-current interaction of \nue on argon nuclei ($^{40}\!\text{Ar} + \nue \rightarrow X + e^-$; dashed green). At low energies, a sizeable contribution comes from elastic scattering on electrons ($\nu + e^- \rightarrow \nu + e^-$; red, with solid, dashed and dotted lines for \nuebar, \nue and \nux, respectively), while charged-current interactions of \nuebar on argon nuclei ($^{40}\!\text{Ar} + \nue \rightarrow X + e^-$; solid blue) have a high threshold energy of \SI{16}{MeV} and mainly contribute to the high energy tail. Figure from SNOwGLoBES~\cite{Snowglobes}.}
	\label{fig-intro-detection-lar}
\end{figure}

LAr TPCs like ICARUS~\cite{Amerio2004} or MicroBooNE~\cite{Acciarri2017} have been used for accelerator neutrinos in the $\sim$\si{GeV} energy range in recent years, since their active calorimetry enables a precise reconstruction of events.
However, this technique has not yet been used for the energy range required for supernova neutrino detection and a lot of work remains to precisely measure the interaction cross sections shown in figure~\ref{fig-intro-detection-lar}, optimize event reconstruction and characterize low-energy backgrounds including those from radioactive argon isotopes.

Currently operating LAr TPCs are small compared to water Cherenkov detectors and would only detect a small number of neutrinos from a galactic supernova.
The Deep Underground Neutrino Experiment (DUNE)~\cite{Acciarri2015}, which is currently under construction, plans to build four detector modules with a fiducial mass of \SI{10}{kton} each.
Due to the competitive size and the increased sensitivity to electron neutrinos, this will provide interesting complementarity to other detector types.

\section{Supernovae}
\subsection{History}

The history of supernova observations goes back almost two millennia. The first known observation of a supernova was reported in ancient China in the year 185~\cite{Zhao2006}, while other well-known supernovae were observed by Chinese and Japanese astronomers in 1054\footnote{This supernova is the progenitor of the Crab Nebula.}
, Tycho Brahe in 1572 and Johannes Kepler in 1604.

\enlargethispage{\baselineskip} 
The name \textit{supernova} refers to the fact that these were originally thought to be new stars (from Latin \textit{nova}, meaning \textit{new}).
Only in 1934 was it suggested by Baade and Zwicky that exactly the opposite might be true---that supernovae arise when massive stars reach the end of their lifetimes and explode~\cite{Baade1934a,Baade1934b,Baade1934}.
Around the same time, a first systematic survey and better telescopes led to a rapid increase in the number of observed supernovae~\cite{Zwicky1964}.
A classification of supernovae based on their spectral features and the time-dependence of their emission was first developed by Minkowski in 1941 and later extended by Zwicky~\cite{Minkowski1941,Zwicky1964}.

Today, supernovae are generally classified as type I if their spectra do not contain hydrogen lines, or type II if they do.
Type I is further subdivided into type Ia (if the spectrum contains silicon lines), type Ib (if it contains helium lines) or type Ic (if it contains neither).
Type II supernovae are subdivided based on the time evolution of the supernova brightness into type IIP (if the lightcurve shows a plateau of approximately constant brightness in the months after the explosion) or type IIL (if the brightness falls off linearly), as well as based on their spectra.
However, a number of supernovae have been discovered that do not neatly fit into this categorization scheme, e.\,g. due to their untypical luminosity or time-dependent changes to their spectra.
For a recent review of such so-called “peculiar” supernovae, see~\cite{Milisavljevic2018}.

This spectral classification, however, mainly reflects the properties of the progenitor star’s outer layers, like the size of its hydrogen envelope or its $^{56}$Ni content~\cite{Heger2003}.
Different spectral properties do not, generally, correspond to different explosion mechanisms, as shown in table~\ref{tab-intro-sn-classification}.

{\renewcommand{\arraystretch}{1.1}
\begin{table}[tbp]
\begin{center}
\begin{tabular}{l >{\centering}m{0.3\textwidth}  >{\centering\arraybackslash}m{0.32\textwidth}}
Spectral Type & Ia & Ib, Ic and II \\
\hline
Spectral Lines & Si & Ib: He \mbox{\hspace{1em}Ic: No H, He or Si\hspace{1em}} \mbox{II: H}\\ 
Explosion Mechanism & thermonuclear explosion & core collapse\\
Light Curve & reproducible & large variations \\
Neutrino Emission & minor & ca. 99\,\% of total energy \\
Remnant & none & neutron star or black hole
\end{tabular}
\end{center}
	\caption[Overview of spectral types, their main properties and the corresponding explosion mechanisms]{Overview of spectral types, their main properties and the corresponding explosion mechanisms.}
	\label{tab-intro-sn-classification}
\end{table}
}

An explosion mechanism responsible for type~Ia supernovae was suggested in 1960 by Hoyle and Fowler~\cite{Hoyle1960}, with Whelan and Iben proposing binary star systems as progenitors~\cite{Whelan1973}.
According to this model, thermonuclear supernovae originate in gravitationally bound systems of two stars, one of which is a white dwarf consisting primarily of carbon and oxygen, while the other is a low mass star that has entered its red giant phase.
If both stars orbit each other at a sufficiently close distance, the white dwarf will accrete matter from the envelope of its companion until it reaches the Chandrasekhar mass limit of about \SI{1.4}{\Msol}.
During accretion, the nuclear fuel in the white dwarf’s core heats up until a runaway fusion process starts, which releases \SI[parse-numbers=false]{\mathcal{O}(10^{51})}{erg} within seconds.
\enlargethispage{\baselineskip} 
Most of this energy is released in the form of kinetic energy, ripping the white dwarf apart and expelling its companion star.

While this mechanism is widely accepted, some recent supernova observations appear to point towards a different progenitor model being responsible for at least some type~Ia supernovae.
In that alternative model, two white dwarfs bound in a binary system merge to produce a supernova.
A recent review of the observational evidence can be found in reference~\cite{Maoz2014}, while computer simulations of these progenitor systems are reviewed in reference~\cite{Hillebrandt2013}.

Independent of the progenitor model, type~Ia supernovae create neutrinos mostly through electron capture on free protons or heavier nuclei, i.\,e.
\begin{equation*}
e^- + (A,Z) \rightarrow (A, Z-1) + \nue.
\end{equation*}
Neutrinos do not greatly influence the explosion and are responsible for only a small fraction of the total energy release. A type~Ia supernova at a distance of less than about \SI{1}{kpc} would be necessary to be able to detect neutrinos in current or next-generation neutrino detectors~\cite{Wright2016,Wright2017}.

Supernovae with the spectral types~Ib, Ic and II exhibit widely varying appearances, which are based on the properties of their progenitor. However, they are generally thought to all explode through the same core-collapse mechanism.
In contrast to thermonuclear (or type~Ia) supernovae, their progenitors are heavy stars with a mass of more than about \SI{8}{\Msol}.
Inside the core of such a star, temperatures and densities are sufficiently high to go through all stages of nuclear fusion and finally produce iron, which has the highest binding energy per nucleon such that further nuclear fusion is energetically disfavoured.
Once the mass of this iron core surpasses the Chandrasekhar mass limit of about \SI{1.4}{\Msol}, it starts to collapse and the core’s density rapidly increases until it surpasses nuclear density.
At this point, the equation of state of nuclear matter stiffens.
Infalling matter now bounces off the core and is reflected as an outgoing shock wave.
When this shock wave is reheated by neutrinos, the outer layers of the star are expelled, while the collapsed inner core leaves a neutron star behind.\footnote{If the progenitor has a very large mass, the neutrino emission from the core collapse may not be sufficient to reheat the shock wave and the star may instead collapse into a black hole without producing a strong signal in the electromagnetic spectrum. These so called “failed” or “dim” supernovae would produce a neutrino signal which first looks very similar to a regular supernova but has a characteristic sharp cut-off.}

Of the total explosion energy of such a core-collapse supernova, about 99\,\% or roughly \SI{3e53}{erg} is released in the form of neutrinos.
These are produced in or near the collapsing core of the star and traverse the outer layers of the star nearly unhindered.
Measuring properties of the neutrino flux therefore allows us to study the interior of the supernova and the processes at work during the explosion.

In the remainder of this thesis, I will focus on core-collapse supernovae.
The closest visible supernova of this type since the invention of the telescope happened in the Large Magellanic Cloud in 1987.
This supernova, called SN1987A, is also the only supernova whose neutrino emission was observed by humankind.
In the following section, I will describe the current understanding of the core-collapse explosion mechanism based on computer simulations and the observation of neutrinos from SN1987A.

\subsection{Explosion Mechanism}\label{ch-intro-explosion-mechanism}

Today, over \num{50000} supernovae or supernova candidates have been detected and detailed lightcurves and spectra are available for several thousand of them according to the Open Supernova Catalog~\cite{Guillochon2017,TOSC}.
Furthermore, these numbers are expected to grow rapidly in the near future, with the Large Synoptic Survey Telescope alone expected to observe \num{3e5} core-collapse supernovae per year when it starts operations in the early 2020s~\cite{Lien2009,Abell2009}.
However, measurements in the electromagnetic spectrum only allow observations of the surface layers of the supernova.
While these give detailed information on the composition of these outer layers and thus the properties of the progenitor, they give little information about the mechanism underlying the actual explosion, which takes place at the centre of the star several minutes to hours before the resulting shock wave reaches the stellar surface and the supernova becomes visible to telescopes.

Unlike electromagnetic radiation, neutrinos only interact weakly and are therefore unlikely to experience scattering or be absorbed in outer layers of star.
Neutrinos are thus the only known channel that allows us to directly observe the processes occurring near the centre of a star in the moment of explosion.
On the other hand, the small cross sections of weak interactions mean that most neutrinos pass through any detector unnoticed and only very large neutrino fluxes can be observed.
This limits the reach of current neutrino detectors to supernovae within the Milky Way or its immediate cosmic neighbourhood, where the rate of core-collapse supernovae is estimated to be about 2--3 per century~\cite{Tammann1994}.

Due to this severe scarcity of observational data, progress in understanding the core-collapse explosion mechanism has come mostly from computer simulations.
Pioneering contributions to numerical models of supernovae were made by Colgate, Grasberger\,\&\,White~\cite{Colgate1961,Colgate1966}, Arnett~\cite{Arnett1967} and Wilson~\cite{Wilson1971,Wilson1982}.
These early simulations often imposed spherical symmetry to reduce the required computing power, making them effectively one-dimensional.
Optical observations of SN1987A made it obvious that this assumption does not generally hold true in nature~\cite{Hillebrandt1989}, which increased the effort put into more complex simulations that only imposed rotational symmetry around one axis.
Today, these two-dimensional simulations are very common, while one-dimensional simulations are still used for parametric studies that compare a wide range of models.
Only in recent years has it become computationally feasible to simulate three-dimensional models that include detailed treatment of neutrino production and transport processes~\cite{Janka2012, Hanke2013}.

The difficulty in simulating supernova explosions stems both from the huge computing power required and the inherent complexity of the phenomenon itself:
Supernovae stand out from most other physical phenomena in that they involve all known fundamental forces---gravity as well as the strong and electroweak force---and operate at extreme conditions, which often cannot be reproduced in laboratory experiments.
Simulating them also requires solving difficult and non-linear hydrodynamical equations and taking into account relativistic effects. The latter are currently often treated as a modified potential in Newtonian gravity to simplify calculations, which might cause an error of tens of percent in some physical quantities and lead to qualitatively different outcomes~\cite{Muller2012}.

Computer simulations of supernovae have made remarkable progress in the last decades, in part due to a dramatic increase in available computing resources and in part due to an improved understanding of the neutrino physics and nuclear cross sections involved in the explosion of a core-collapse supernova.
Thus, while current simulations disagree on many points and “are still too far apart to lend ultimate credibility to any one of them”~\cite{Skinner2015}, they have reached widespread agreement on the basic explosion mechanism.

After Burbidge and others highlighted the gravitational instability of old massive stars~\cite{Burbidge1957}, Colgate and others proposed a solely hydrodynamical “bounce and shock” explosion mechanisms for core-collapse supernovae in 1961~\cite{Colgate1961}, where the equation of state of the collapsing core stiffens after it reaches nuclear density and infalling matter bounces off the now incompressible core resulting in an outgoing shock wave.
Several years later, they studied the role of the high neutrinos fluxes inside a supernova~\cite{Colgate1966}.

\begin{figure}[p]
	\centering
	\includegraphics[scale=0.74]{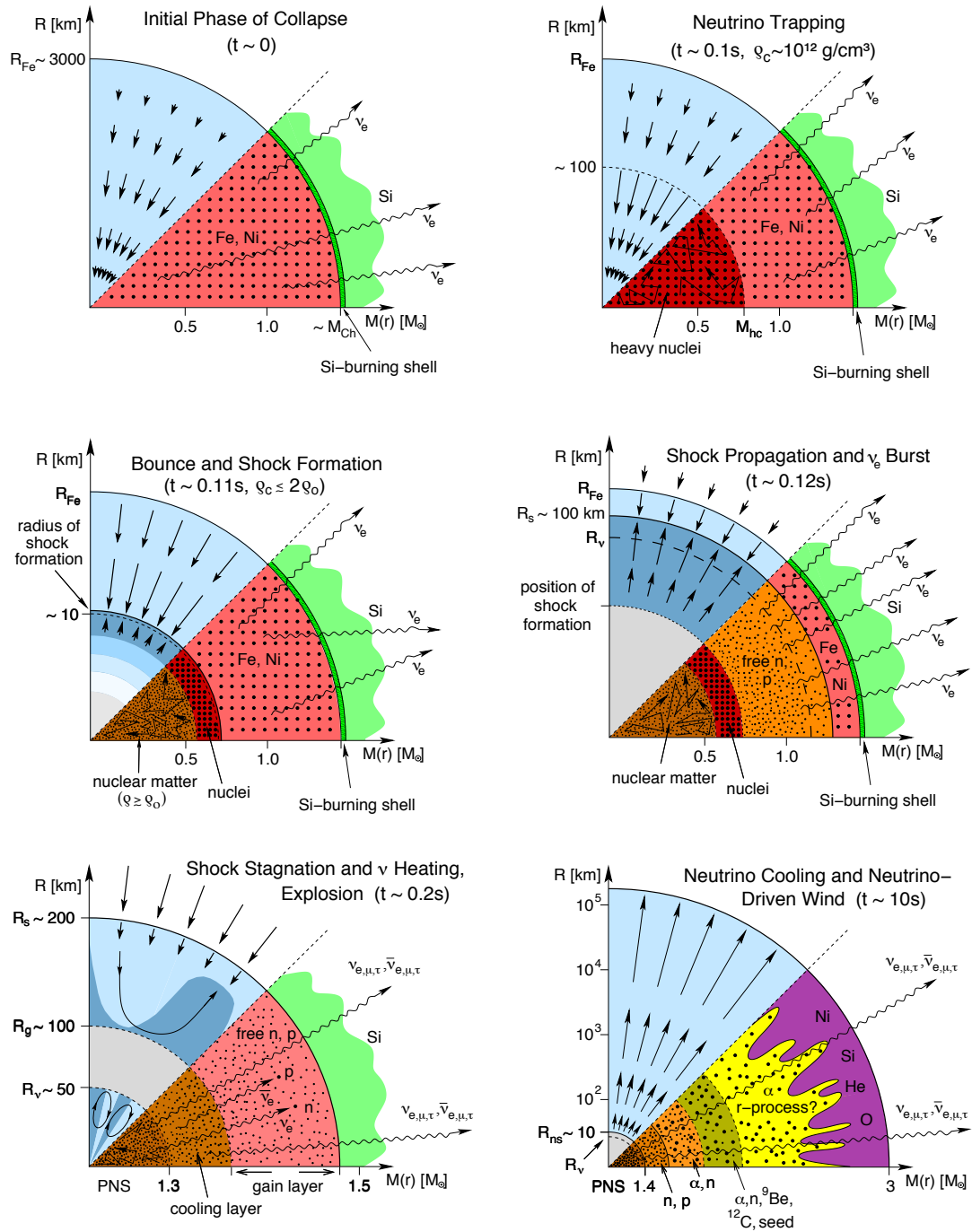}
	\caption[Six phases of the delayed explosion mechanism]{Sketch of the six phases of the delayed explosion mechanism as described in the text. In each panel, the upper section shows the dynamical processes, with arrows representing velocity vectors, while the lower section shows the nuclear composition of the star. Figure from reference~\cite{Janka2007}.}
	\label{fig-intro-explosion-mechanism}
\end{figure}

In the early 1980s, Wilson and Bethe~\cite{Wilson1982,Bethe1985} described the delayed neutrino-driven explosion mechanism generally accepted today.
This mechanism consists of six steps which are sketched in figure~\ref{fig-intro-explosion-mechanism}.

\begin{enumerate}
\item \emph{Initial phase:} The progenitor of a core-collapse supernova is a star with a mass of more than about \SI{8}{\Msol}, whose central region is sufficiently hot and dense to produce iron through nuclear fusion.
Since the iron cannot produce energy through further nuclear fusion steps it forms an inner core, which is held~up by electron degeneracy pressure whose density dependence is initially $P \propto \rho^\frac{5}{3}$.
At the same time, hydrostatic equilibrium requires $P \propto G M \rho R^{-1}$, which leads to a mass-radius relationship for the iron core of $R \propto M^{-\frac{1}{3}}$.
As silicon burning continues to produce iron that accretes onto the core, the core shrinks due to the increase in mass.
Electrons occupy increasingly higher energy states until they become relativistic and their equation of state changes to $P \propto \rho^\frac{4}{3}$.%
\footnote{Analogous to the mass limit for white dwarf stars, this happens when the iron core reaches the Chandrasekhar mass of about \SI{1.4}{\Msol}.}
During this transition to the relativistic regime, the mass-radius relationship becomes steeper and the shrinking of the core accelerates until the mass-radius relationship of the iron core breaks down, indicating that there is no stable configuration.
The core collapses.
During this phase, some electrons get captured by nuclei and the resulting neutrinos transport energy away from the core, thus reducing the degeneracy pressure which counteracts the collapse.

\item \emph{$\nu$ trapping:} After about \SI{100}{ms}, the inner core reaches a density of about $\SI{e12}{g/cm^3}$.
At this density, the mean free path of neutrinos becomes smaller than the radius of the inner core and they become trapped inside it.

\item \emph{Bounce and shock formation:} After about \SI{110}{ms}, the core has collapsed from a radius of about \SI{3000}{km} to just tens of kilometres, with infalling matter reaching about 10\,\% of the speed of light.
At this point, the density of the inner core surpasses nuclear density, reaching about \SI{3e14}{g/cm^3}, and its equation of state stiffens.
Infalling matter now hits a “wall” and is reflected, resulting in an outgoing shock wave.
Meanwhile, neutrinos are still trapped in the inner core due to its high density.

\item \emph{Shock propagation:} After about \SI{120}{ms}, the outgoing shock wave reaches the surface of the iron core at a radius of about \SI{100}{km}, dissociating the iron nuclei into free nucleons along the way.
Since the electron capture cross section on free protons ($e^- + p \rightarrow n + \nue$) is much higher than on the larger, neutron-rich nuclei, this leads to a sudden increase in the electron capture rate.
The matter density in the outer parts of the core is too low to trap the neutrinos, so a brief \nue burst is released.

\item \emph{Shock stagnation and $\nu$ heating:} After about \SI{200}{ms}, the shock wave stagnates at a radius of about \SIrange{100}{200}{km}, having used up most of its energy to dissociate heavy nuclei into their constituent nucleons.
Matter from outer layers infalling onto the almost stationary shock front creates an accretion shock, which powers neutrino emission. At this phase, convection sets in at the accretion shock layer.

Meanwhile, the neutrinos that were trapped inside the inner core are starting to diffuse out.
While some escape the supernova immediately, others deposit energy in the accretion shock layer mainly by neutrino capture on free nucleons, i.\,e. $\nuebar + p \rightarrow n + e^+$ and $\nue + n \rightarrow p + e^-$.
This heating increases the pressure in the region behind the shock front and reignites the shock wave.

\enlargethispage{\baselineskip} 
\item \emph{$\nu$ cooling:} During the following tens of seconds, the remnant of the core, a proto-neutron star (PNS), cools by diffusive neutrino transport.
The outgoing shock wave takes several minutes or hours to reach the surface of the star, where it will expel the matter in its outer shells and produce a signal that is visible in the electromagnetic spectrum.
\end{enumerate}

\begin{figure}[tb]
	\centering
	\includegraphics[scale=0.7]{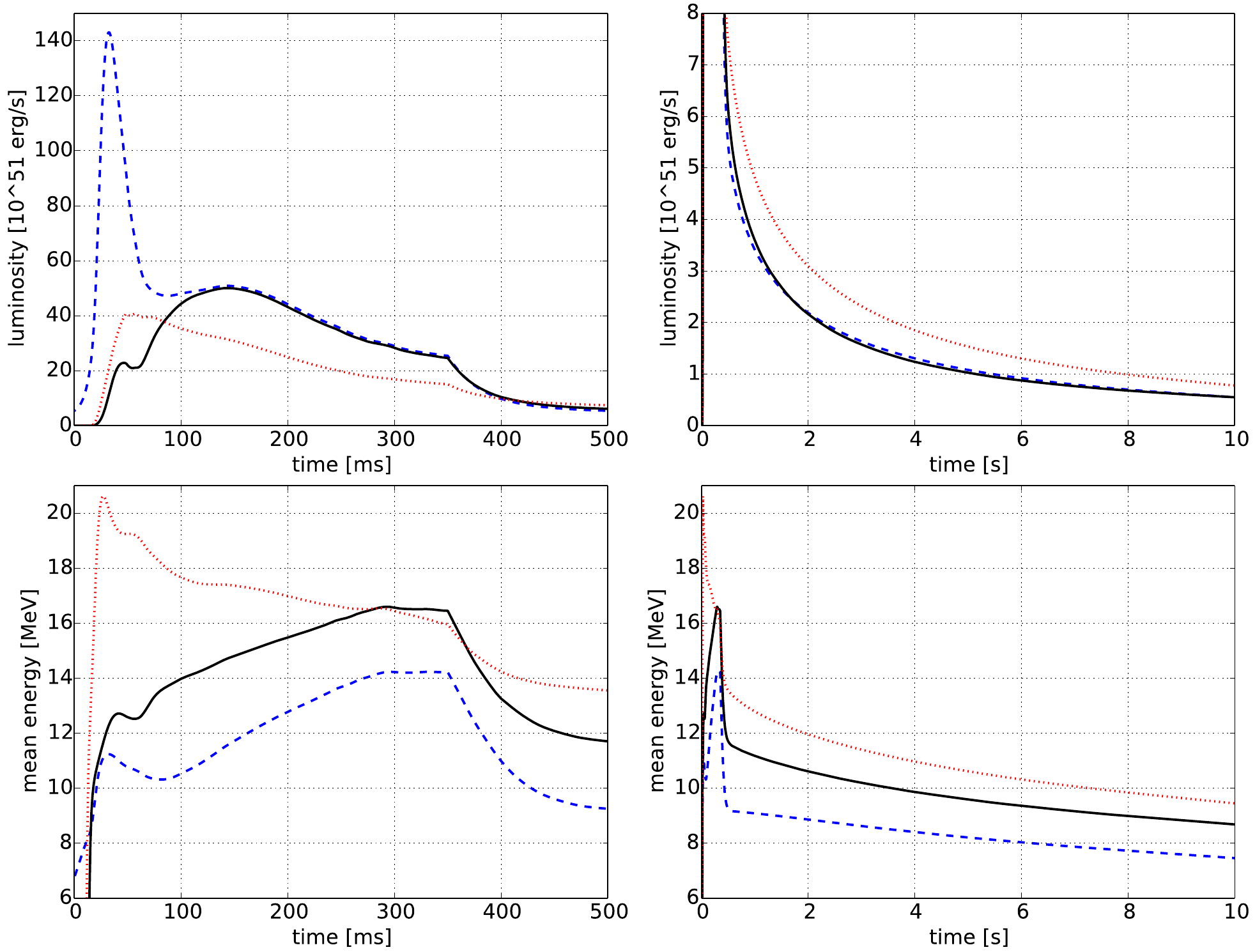}
	\caption[Luminosity and mean energy during three steps of neutrino emission]{Luminosity (top) and mean energy (bottom) of \nue (dashed blue line), \nuebar (solid black) and \nux (dotted red) for a \SI{20}{\Msol} model~\cite{Nakazato2013}. The left panels show the prompt \nue burst and the following phase of shock stagnation, while the right panels show the neutrino cooling phase.}
	\label{fig-intro-nu-emission}
\end{figure}

Throughout this process, neutrino emission occurs in three distinct steps, which are displayed in figure~\ref{fig-intro-nu-emission}.

The first step is a prompt \nue burst from electron capture during phase~4 above.
With a duration of roughly \SI{10}{ms}, this is expected to give a very sharp and unmistakeable feature that consists of almost pure \nue.
Since this signal originates in the iron core, which collapses at a well-defined set of physical conditions, independent of the properties of the outer shells of the star, it is very similar across a wide range of simulations with different progenitors~\cite{Kachelries2005}.

The second step has a duration of several \SI{100}{ms} and corresponds to the shock stagnation in phase~5 above.
Neutrino emission during this phase is powered by matter from outer layers accreting onto the shock front.
Radial movement of the shock front therefore leads to changes in the accretion rate, such that hydrodynamical features like the standing accretion shock instability could be observed as striking sinusoidal features in the neutrino event rate~\cite{Blondin2003,Lund2010,Tamborra2013,Migenda2015}.
In this phase, the luminosities $L_{\nue}$ and $L_{\nuebar}$ are roughly equal (and higher than $L_{\nux}$) while the average energies are unequal ($\mean{E_{\nue}} < \mean{E_{\nuebar}} \approx \mean{E_{\nux}}$), which contributes to the de-leptonization of the core.
This energy difference is caused by the higher cross section for \nue in electron-rich matter, which means that \nue leaving the star are, on average, emitted at larger radii and thus lower temperatures.

The final step of neutrino emission corresponds to the $\nu$ cooling (phase~6) and has a duration of some tens of seconds.
During this time, the supernova remnant cools through diffusive emission of neutrinos that became trapped in its core during the earlier phases of the explosion.
The composition of the neutrino flux at this stage is governed by a number of different physical processes, including nucleon-nucleon bremsstrahlung and neutrino-antineutrino pair annihilation~\cite{Hannestad1998,Thompson2000,Buras2003,Keil2003}, which lead to roughly equal luminosities of all neutrino species that fall off exponentially with time, while the average energies remain unequal.

\subsection{Open Questions}

On February 23$^\text{rd}$, 1987, three detectors around the world---Kamiokande~\cite{Hirata1987,Hirata1988a}, IMB~\cite{Bionta1987} and Baksan~\cite{Alekseev1987}---detected a total of two dozen neutrinos from the supernova SN1987A in the Large Magellanic Cloud.
In the decades since, these events have been analysed extremely closely and were used to set limits on neutrino masses, charges and life times, as well as on a wide range of hypothetical particles proposed by theorists\footnote{This is due to the fact that the neutrinos from SN1987A were observed over a duration of about \SI{10}{s}. Any weakly-interacting new particle that would cool the core by transporting away energy can therefore be produced only in limited quantities; otherwise, the core would cool down too quickly and there would not be sufficient energy left to power the observed neutrino emission at later times.}.
An overview over the information on particle physics that was gained from SN1987A can be found in reference~\cite{Raffelt1999}.
As the first and, for almost 30 years, only object outside of our solar system observed in a channel outside of the electromagnetic spectrum, SN1987A also enabled tests of relativity that were beyond the reach of lab-based experiments~\cite{Longo1987}. 

The neutrino events largely fell within the range of theoretical expectations, supporting the basics of the explosion mechanism described in the previous section~\cite{Arnett1989}.
The energy release in the gravitational collapse of a neutron star was expected to be on the order of \SI{2e53}{erg}~\cite{Cooperstein1988}, predominantly in the form of neutrinos.
This turned out to be in good agreement with the observation of SN1987A.
As expected, the detectors observed a brief period with a high rate of neutrino interactions at the start of the signal, followed by a period of much lower event rate with a duration of about \SI{10}{s}.

Some authors have pointed to irregularities in the neutrino signal from SN1987A and used these as evidence for nonstandard physical phenomena.
In addition to the \SI{7.3}{s} gap between the first 9 and the final 3 events observed in the Kamiokande detector, this includes differences in the reconstructed flux spectra and angular distributions between the Kamiokande and IMB detectors~\cite{Dadykin1989}.
Due to the low number of observed events, however, the alternative that these are simply artefacts caused by statistical fluctuations cannot be excluded.

Another puzzle is caused by a cluster of 5 events observed by the LSD experiment under Mt.\,Blanc almost \SI{5}{h} earlier~\cite{Aglietta1987}, at a time when none of the three other neutrino detectors observed an excess of events above background.
Some authors have, however, argued that there \emph{is} a coincidence between the cluster of events observed by LSD and individual events in the other neutrino detectors as well as possible events in two gravitational wave detectors~\cite{De-Rujula1987,Galeotti2016,Ryazhskaya2018}.
It has also been argued that the non-observation of the earlier burst in other detectors as well as LSD’s non-observation of the burst observed by the three other detectors could be explained by differences in the detector sizes and sensitivities as a function of energy~\cite{Aglietta1987a}.
To explain these observations, various multi-step supernova explosion mechanisms~\cite{De-Rujula1987,Berezinsky1988,Imshennik2004} or exotic particle physics models~\cite{Ehrlich2018} have been proposed; however, since these LSD-inspired models were invented \emph{ad hoc} to account for the observations, they are widely viewed sceptically.

Even putting aside these puzzles, SN1987A clearly demonstrated that previous computer simulations---most of which had imposed spherical symmetry---did not reflect realistic supernova explosions and that multidimensional effects needed to be taken into account.
Furthermore, the two dozen neutrinos detected were not sufficient to investigate details of the explosion mechanism.
Many open questions have therefore remained to this day.

First of all, the next observation of neutrinos from a galactic supernova would need to provide a higher-statistics signal that could provide incontrovertible evidence to identify the supernova explosion mechanism and clear up the irregularities in the SN1987A signal which are described above.
Furthermore, if the basic explosion mechanism described in section~\ref{ch-intro-explosion-mechanism} is confirmed, a high-statistics neutrino signal could be used to investigate details of the explosion mechanism where current simulations give conflicting results.

One recent example for this is a hydrodynamic feature of the shock front called the “Standing Accretion Shock Instability” (SASI), which occurs during the shock stagnation phase shown in the fifth panel of figure~\ref{fig-intro-explosion-mechanism}.
While this self-sustained oscillation of the shock front was observed in a number of two-dimensional simulations and contributed to explosions there, it did not play a role in early three-dimensional simulations, leading some authors to conclude that it had to be an artefact of the rotational symmetry imposed in two-dimensional simulations~\cite{Burrows2012}.
While other authors have since found SASI oscillations in three-dimensional simulations~\cite{Hanke2013}, there is still no consensus.
Some recent work suggests that the presence and importance of SASI may depend on the equation of state of nuclear material~\cite{Kuroda2016,Kuroda2017}, the shock radius~\cite{OConnor2018} or rotation~\cite{Summa2018, Walk2018}, while other authors find that SASI is either completely absent or negligible compared to the role of neutrino-driven convection in a layer interior to the accretion shock~\cite{Burrows2012,Vartanyan2019}.

More generally, progenitors in computer simulations often do not explode on their own but need to be triggered artificially.
For example, the first successful neutrino-driven explosion in a three-dimensional simulation was only achieved in 2015 by the Garching group~\cite{Melson2015}.
While this lack of explosions may simply be the effect of imperfect modelling of the known physical processes, it could also be the result of new physics---non-standard interactions of neutrinos or completely new particles like axions or a dark sector~\cite{Turner1988,Janka1996,Dreiner2014,Kazanas2015}---which are not included in current models.


\section{Summary of Previous Work}

After describing neutrino detection and supernova simulations in the earlier parts of this chapter, in this section I will broadly summarize ongoing work in these fields and how it contributes to using supernova neutrinos to determine details of the explosion mechanism.
Astronomers and astrophysicists also perform extensive research of supernovae that improves our understanding of stellar evolution and enables us to create increasingly detailed models of supernova progenitors.
While this work is extremely important, it is only indirectly connected to the neutrino signal from a supernova and therefore beyond the scope of this overview.


Due to the severe lack of supernova neutrino observations, ongoing research largely depends on computer simulations.
A number of different groups around the world perform computer simulations of supernovae with progenitors that have a wide range of different masses, compositions, rotational velocities and other properties.
Some of these simulations are very complex and aim to reproduce a realistic supernova in the greatest possible amount of detail, which requires huge amounts of computational resources and thus only allows the simulation of individual or very few progenitors.
On the other end of the spectrum are very simplified studies that usually impose rotational or even spherical symmetry and use various approximations in neutrino transport and other physical parameters to reduce the computational needs.
While these simulations are less realistic, they allow groups to perform a large number of simulations and study the effect of varying parameters of the progenitor (like its mass or metallicity), numerical resolution or individual approximations.

Various current and next-generation neutrino detection collaborations have determined the ability of their respective detectors to observe supernova neutrinos; see e.\,g. references~\cite{An2016,Alberini1986,Suzuki1987,Abbasi2011,Acciarri2015,HKDR2018}.
These collaborations commonly use a small number of supernova models as benchmarks to showcase the event counts \mbox{expected} in their detector, its sensitivity to different interaction channels and ability to distinguish supernova neutrinos from the dominant sources of background.
In the absence of any observations of supernova neutrinos, long-running experiments can also use those analyses to set limits on the local supernova rate~\cite{Ikeda2007,Vigorito2018}.

Depending on the sensitivity of their detector, collaborations may also produce more advanced analyses, such as the ability of a detector to detect pre-supernova neutrinos~\cite{Asakura2016} or to determine the direction of a supernova~\cite{Abe2016}.
These analyses often contain a very detailed treatment of detector efficiencies, reconstruction uncertainties and backgrounds.
However, since they are restricted to a small number of benchmark models, they are unable to investigate in detail the wide range of features displayed in modern computer simulations.

There are a number of publications, particularly by simulations groups or theorists, that focus on a specific aspect of the neutrino signal and try to determine how well it could be detected in current or future detectors based on individual computer simulations; for recent examples see e.\,g. references~\cite{Lund2010,Tamborra2013,Migenda2015,Dighe2003a,Kato2017,Capozzi2018,Scholberg2018}.
Partly for simplicity and partly due to a lack of access to or experience using collaboration-internal tools and data, these authors often employ a simplified treatment of detector effects, like backgrounds and reconstruction uncertainties, or of interaction channels. 

Of fundamental importance for comparing supernova neutrino observations to computer simulations is the ability to reconstruct the flux of emitted neutrinos from the detected events.
The spectrum of emitted neutrinos in each flavour can be described by a Gamma distribution~\cite{Keil2003,Tamborra2012},
\begin{equation}
f (E_\nu) = \frac{E_\nu^\alpha}{\Gamma (\alpha + 1)} \left( \frac{\alpha + 1}{\mean{E_\nu}} \right)^{\alpha + 1} \exp \left[ - \frac{(\alpha + 1) E_\nu}{\mean{E_\nu}} \right],
\end{equation}
where $\mean{E_\nu}$ is the mean energy of neutrinos and $\alpha$ is a shape parameter, with $\alpha = 2$ corresponding to a Maxwell-Boltzmann distribution and $\alpha > 2$ corresponding to a “pinched” spectrum.
The neutrino flux at Earth is thus described by three parameters: $\mean{E_\nu}$, $\alpha$ and an overall normalization, which depends on the luminosity of the supernova and its distance.

Different groups recently showed that Hyper-Kamiokande will be able to reconstruct the time-integrated fluxes of \nuebar and \nux to few percent precision, while the time-integrated flux of \nue can be determined to similar precision by exploiting the complementarity between Hyper-Kamiokande and DUNE~\cite{Gallo-Rosso2018,Nikrant2018}.
Since the flux is time-dependent, however, it is important to determine how these flux parameters change with time.
This can in principle be done by simply splitting data into multiple time bins and reconstructing the spectrum in each time bin separately\footnote{with an associated loss in precision due to the lower number of events in each time bin}, though a more advanced approach using a likelihood function can be employed---especially if a relatively low number of events is observed---to reduce the information loss inherent in binning~\cite{Totani1998}.

A recent review of supernova neutrino detection and the lessons we can learn about supernovae can be found in reference~\cite{Horiuchi2018}.

Despite this wide range of ongoing research, however, there is thus far no analysis showing how well we can discriminate between different computer models of supernovae in a realistic detector.
This thesis is the first such analysis and includes precision cross sections, subdominant interaction channels, a detailed detector simulation including reconstruction uncertainties and an unbinned likelihood analysis that makes optimal use of the available time and energy information for all reconstructed events.

\chapter{The Hyper-Kamiokande Detector}\label{ch-hk}

\setlength{\epigraphwidth}{.46\textwidth}
\epigraphhead[0]{\epigraph{Things are easy when you’re big in Japan.}{\textit{Alphaville}}}

Hyper-Kamiokande~\cite{HKDR2018} is a next-generation ring imaging water Cherenkov detector that is currently in the planning stages and expected to start taking data in 2027. 
It is designed to be a general purpose detector contributing to various different fields of particle and astrophysics.

In searches for proton decay, the large detector mass will enable it to explore proton lifetimes of up to $\sim$\SI{e35}{years} in its most sensitive channel ($p \rightarrow e^+ + \pi^0$), and $\sim$\SI{e34}{years} in a range of other channels.
Both represent roughly an order of magnitude improvement over previous limits set by Super-Kamiokande and would cover large parts of the parameter space predicted by current theoretical models.

Hyper-Kamiokande will act as the far detector for a long baseline neutrino experiment, referred to as T2HK. Building upon the currently ongoing work of the T2K collaboration, this successor experiment will combine upgrades to the neutrino beam and existing near detector that are already planned by the T2K collaboration with a new intermediate water Cherenkov detector and a larger far detector, with Hyper-Kamiokande replacing Super-Kamiokande. As a result, higher statistics and reduced systematic uncertainties are expected to enable a determination of the neutrino mass ordering and a first measurement of $\delta_{CP}$, as well as precision measurements of other oscillation parameters.
Hyper-Kamiokande will also perform detailed measurements of atmospheric neutrinos, which are the product of cosmic rays interacting with nuclei in the Earth’s atmosphere. In addition to being sensitive to the mass ordering and the octant of $\theta_{23}$, this data set will help eliminate parameter degeneracies in a joint analysis together with accelerator neutrinos.

\enlargethispage{\baselineskip} 
For solar neutrinos, higher statistics will allow Hyper-Kamiokande to perform a precise measurement of the day-night asymmetry caused by neutrinos traversing the Earth’s matter potential before reaching Hyper-Kamiokande during the night.
While Super-Kamiokande has already measured this effect at $3\sigma$, that measurement resulted in an unexpectedly small value of $\Delta m^2_{21}$, producing a tension with KamLAND measurements of reactor antineutrinos which Hyper-Kamiokande is expected to help clear up.
Hyper-Kamiokande will also try to observe neutrinos from the $^3\text{He} + p$ fusion reaction inside the Sun, which have not yet been observed due to the very low branching ratio of this reaction. Compared to other solar neutrino species, these \textit{hep} neutrinos have a higher energy and are expected to be produced at larger distances from the centre of the Sun, making them an excellent tool to test the standard solar model.
Finally, Hyper-Kamiokande will try to detect the transition in neutrino oscillation probability between the region below $\sim$\SI{1}{MeV}, where vacuum oscillations on the way to Earth dominate, and the region above $\sim$\SI{10}{MeV}, where MSW oscillations in the Sun dominate.\footnote{See section~\ref{ch-intro-oscillations} for an explanation of the MSW effect.}
Various exotic models like non-standard interactions~\cite{Friedland2004}, mass-varying neutrinos~\cite{Barger2005} or sterile neutrinos~\cite{Holanda2004} predict modifications to the shape of this spectral upturn, which could be tested by Hyper-Kamiokande.

If a supernova within our Milky Way happens during Hyper-Kamiokande’s lifetime, we would observe \numrange{e4}{e6} neutrino interactions and measure the precise arrival time and energy of each event. Such a data set would not only enable a detailed investigation of the supernova explosion mechanism as described in this thesis, but also provide valuable knowledge to many related areas of particle and astrophysics.
Even in the absence of a galactic supernova, Hyper-Kamiokande could perform a high statistics measurement of supernova relic neutrinos (SRN), with $\sim$\num{e2} events expected within 10 years.
This first determination of the SRN spectrum would constrain parameters of theoretical models, such as the star formation rate in the universe as a function of redshift or the fraction of “failed” supernovae that collapse to a black hole without an optically visible explosion.

Finally, Hyper-Kamiokande will explore a wide range of rare or more speculative topics including searches for an excess of \si{GeV}-scale neutrinos from the Sun or the centre of the Milky Way, which might originate from annihilation of dark matter, or high-energy neutrinos associated with astrophysical sources like solar flares, gamma ray burst jets, new-born pulsars or colliding binary systems containing at least one neutron star similar to the neutron star merger observed in gravitational waves and various electromagnetic wavelengths on 17 August 2017~\cite{Abbott2017}. 

This chapter begins by describing the history leading up to Hyper-Kamiokande, including a brief discussion of its predecessor experiments, Kamiokande and Super-Kamiokande.
Section~\ref{ch-hk-detector} describes the detector design and its components in detail.
Finally, sections~\ref{ch-hk-calibration} and~\ref{ch-hk-background} discuss detector calibration and sources of background that are relevant to the investigation of supernova neutrinos.
By necessity, discussion of calibration and backgrounds will largely remain qualitative, since reliable quantitative data will not be available until the detector is actually running.

Throughout, I will focus on the single tank planned to start construction in April 2020 in Japan. 
A proposed second Hyper-Kamiokande tank in South Korea~\cite{Abe2018a} would likely be very similar to the one described here, though plans have not yet progressed far enough to discuss them in detail. Other elements of the experiment, such as the beamline or near and intermediate detectors, are not related to the investigation of supernova neutrinos and will therefore not be discussed here.

\section{History}\label{ch-hk-history}
\subsection{The Past: Kamiokande}
The 1960s were a period of major changes across particle physics.
In the lepton sector, Glashow, Weinberg and Salam provided a unified description of electromagnetism and weak interaction~\cite{Glashow1961,Weinberg1967,Salam1968}.
Around the same time, Gell-Mann and Zweig proposed the quark model~\cite{Gell-Mann1964,Zweig1964} to categorize the rapidly growing “hadron zoo” found in experiments.
By the early 1970s, the theory of strong interactions (known as quantum chromodynamics) was formulated~\cite{Gross1973,Politzer1973,Fritzsch1973}, completing the Standard Model (SM) of particle physics in its modern form.

While experimental particle physicists spent the next four decades detecting the matter particles and bosons predicted by the SM, theoretical physicists soon began work on so-called “Grand Unified Theories” (GUTs) that unified electroweak and strong interaction, the first one being Georgi and Glashow’s SU(5) model~\cite{Georgi1974}.
In many of these GUTs, instead of the baryon number, B, and the lepton number, L, only their combination B--L was conserved.
This allowed the proton---which, in the Standard Model, was predicted to be stable---to decay via channels such as $p \rightarrow e^+ + \pi^0$ and with predicted life times as low as $10^{30}$ years in some models, close to lower limits from contemporary experiments~\cite{Langacker1981}.

Encouraged by these predictions, several groups of physicists began work on experiments to search for proton decay.
These experiments consisted of a large tank of pure water---containing a large number of hydrogen nuclei, i.\,e. free protons---whose inside walls were equipped with photosensors to detect the flash of light expected from a decaying proton.
One such experiment---known as IMB---was built in the U.S. by groups from Irvine, Michigan and Brookhaven, while a group of Japanese physicists built a detector in the Kamioka mine in Japan’s Gifu region, which they called the Kamioka Nucleon Decay Experiment or Kamiokande.

The Kamiokande detector~\cite{Nakamura1989} was a cylindrical tank with a diameter of \SI{15.5}{m} and a height of \SI{16}{m}, containing \SI{3}{kt} of water.
It started operations in 1983 with a focus on looking for proton decay.
Soon afterwards, it was pointed out that Kamiokande would be able to observe solar $^8$B neutrinos if the low-energy backgrounds were reduced sufficiently and beginning in 1985, the collaboration added an outer detector and improved the electronics and water purification systems~\cite{Suzuki1988}. 
Following completion of this upgrade, in February 1987 Kamiokande-II observed a burst of 12 events caused by neutrinos from the supernova SN1987A~\cite{Hirata1987,Hirata1988a}.
Together with the 13 events observed by the IMB~\cite{Bionta1987} and Baksan~\cite{Alekseev1987} detectors, these marked the beginning of extra-solar neutrino astronomy and remain the only supernova neutrinos observed to this day.
In 2002, part of the Nobel Prize for Physics was jointly awarded to Ray Davis~Jr. and to the Kamiokande collaboration’s Masatoshi Koshiba for their “pioneering contributions to astrophysics, in particular for the detection of cosmic neutrinos”~\cite{NobelPrize2002}.

Kamiokande also observed solar neutrinos, confirming the deficit observed by the Homestake experiment~\cite{Davis1968}, as well as atmospheric neutrinos.
Notably, the ratio of muon-like to electron-like events in atmospheric neutrino interactions in Kamiokande deviated from theoretical expectations, hinting at neutrino oscillations\footnote{At the time, these observations were not yet fully accepted as evidence for neutrinos oscillations. Instead, systematic errors due to misidentification of the observed particles were suggested as an alternative explanation. To investigate this, an experiment with a scaled model of the Kamiokande detector was performed in a charged particle beam at KEK, which confirmed the accuracy of Kamiokande’s particle identification~\cite{Kasuga1996}.}~\cite{Fukuda1994}.

\subsection{The Present: Super-Kamiokande}

\begin{figure}[htbp]
	\centering
	\includegraphics[scale=1.0]{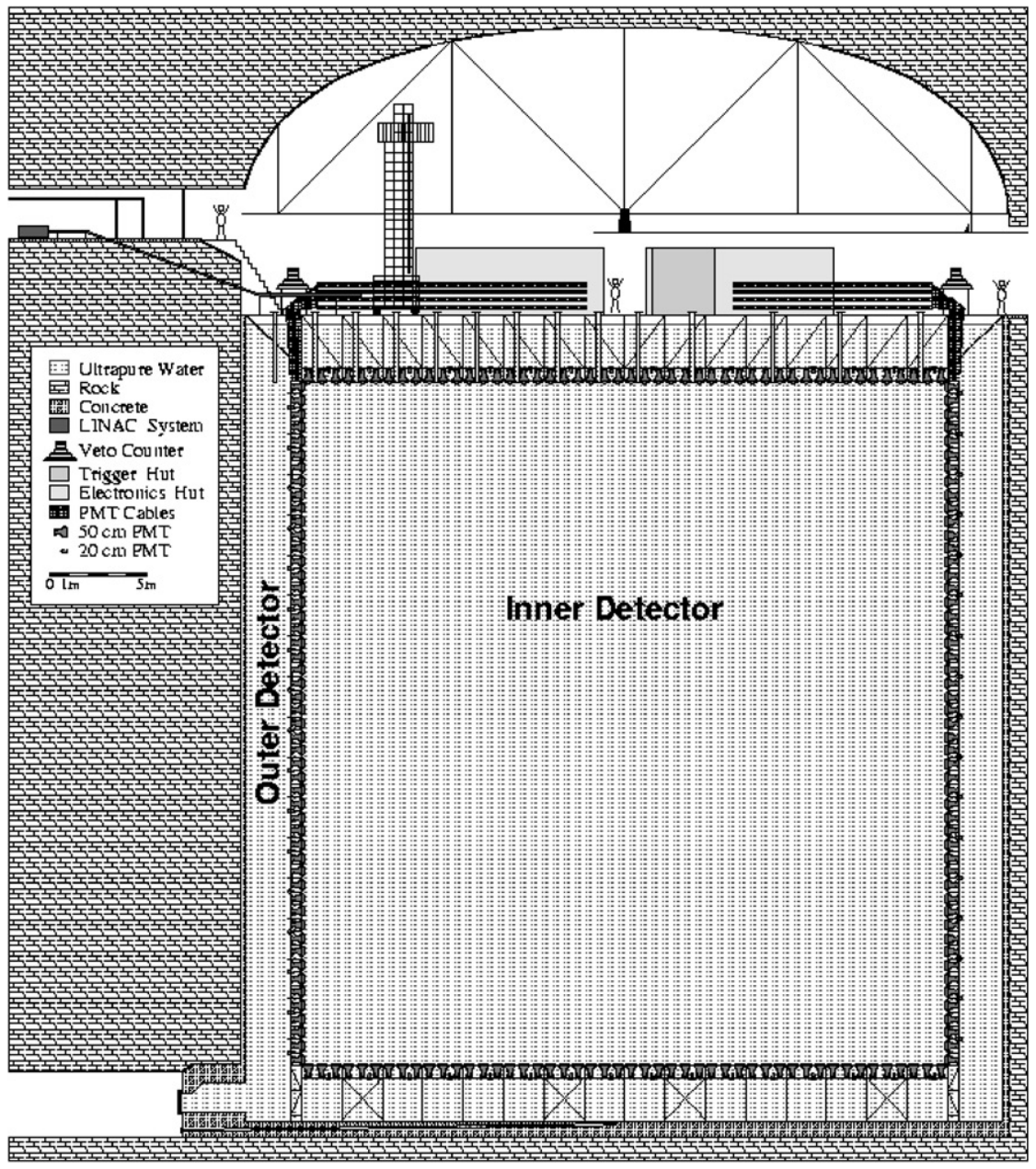}
	\caption[Schematic view of Super-Kamiokande]{Schematic view of Super-Kamiokande. Figure from reference~\cite{Fukuda2003}.}
	\label{fig-hk-superk}
\end{figure}

By the end of the 1980s it became clear that Kamiokande, while very successful, would soon be limited by its size and plans for a successor experiment called Super-Kamiokande were made.
Located in the same mine as its predecessor below \SI{1000}{m} of rock shielding, corresponding to \num{2700}\,m.\,w.\,e., Super-Kamiokande~\cite{Fukuda2003} started data-taking in April 1996 after \SI{4.5}{\years} of construction and commissioning.
Similar in shape to Kamiokande, it is a cylindrical tank with a diameter of \SI{39}{m} and a height of \SI{42}{m}, containing \SI{50}{kt} of water.
This total volume is divided by a stainless-steel support structure into two optically separated regions: an inner detector (ID) region with a diameter of \SI{33.8}{m} and a height of \SI{36.2}{m} and an outer detector (OD) region with a width of approximately \SI{2}{m}.
See figure~\ref{fig-hk-superk} for an overview.

The OD is sparsely instrumented by 1885 photomultiplier tubes (PMTs) with \SI{20}{cm} diameter and wavelength shifting plates to increase light collection.
It acts as an active veto against incoming particles such as cosmic-ray muons and as a passive shield against radioactivity from the surrounding rock.

The ID contains \SI{32}{kt} of water, which is divided up for analysis purposes into a central, fiducial volume of approximately \SI{22.5}{kt} of water\footnote{Depending on the background levels, a larger or smaller fiducial volume is used for some analyses.} surrounded by an outer shell of \SI{2}{m} width. 
That shell acts as a passive shield against radioactive backgrounds, which mainly come from the PMT glass and enclosure or the stainless-steel support structure.

The ID is densely instrumented with PMTs with \SI{50}{cm} diameter, which are mounted on the support structure.
Between April 1996 and June 2001, a period referred to as Super-Kamiokande-I, the ID was instrumented by \num{11146} PMTs, effectively covering 40\,\% of its inner surface.
When refilling the detector with water in November 2001 after being shut down for maintenance, an imploding PMT caused a chain reaction that destroyed over half the PMTs in the detector.
After fitting the PMTs with pressure-resistant covers to avoid a reoccurrence, data-taking resumed with a 20\,\% photocoverage between December 2002 and October 2005, a period referred to as Super-Kamiokande-II.
Newly manufactured PMTs which restored the photocoverage to 40\,\% were added for the Super-Kamiokande-III period which started in October 2006.
For the Super-Kamiokande-IV period starting in September 2008, new front-end electronics and a new data acquisition system were installed, while the photocoverage stayed constant.
Together with improvements to the water system, calibration and analysis, this has allowed Super-Kamiokande-IV to observe electrons with kinetic energies as low as \SI{3.5}{MeV}~\cite{Abe2016a}.
Most recently, Super-Kamiokande was shut down for maintenance in June 2018 to fix some water leaks and prepare for the upcoming addition of gadolinium, which would strongly enhance its ability to detect neutron captures and thus reduce backgrounds in a variety of analyses.

Super-Kamiokande confirmed the deficit of muon-like atmospheric neutrino events observed in Kamiokande and was able to measure its dependence on the azimuth angle, providing strong evidence for neutrino oscillations.
In 2015, the Nobel Prize for Physics was jointly awarded to Takaaki Kajita (Super-Kamiokande) and Arthur McDonald (SNO) for discovering neutrino oscillations~\cite{NobelPrize2015}.
Since 2010, Super-Kamiokande acts as the far detector of the T2K experiment~\cite{Abe2011a}, which produces a beam of neutrinos at J-PARC in the city Tokai on the east coast of Japan and sends it over a distance of \SI{295}{km} to Super-Kamiokande to measure neutrino oscillation parameters.

Super-Kamiokande also performs precision measurements of the solar $^8$B neutrino flux~\cite{Abe2016a}, searches for nucleon decay channels, with lower limits on the proton lifetime now surpassing \SI{e34}{\years} in the most sensitive channel~\cite{Abe2017}, and a wide range of other analyses.

\subsection{The Future: Hyper-Kamiokande}

Just a few years after Super-Kamiokande started taking data, the prospects of a megaton-scale successor were first being explored~\cite{Suzuki2000,Shiozawa2000}.
A letter of intent to build Hyper-Kamiokande was presented in 2011 which proposed to build two horizontally segmented tanks with an egg-shaped cross section~\cite{Abe2011b}.
After several years of ongoing R\&D, a design report was published in 2016~\cite{HKDR2016} and most recently updated in 2018~\cite{HKDR2018}.
The design report presented an optimized design that returned to the cylindrical shape of its predecessors.

The collaboration has received seed funding from the Japanese Ministry of Education, Culture, Sports, Science and Technology in 2019 and expects to get fully approved by 2020~\cite{Gonokami2018}.
Construction is scheduled to start in April 2020, with commissioning and data taking expected by 2027.

{\renewcommand{\arraystretch}{1.1}
\begin{table}[htbp]
\begin{center}
\begin{tabular}{m{0.24\textwidth} >{\centering}m{0.21\textwidth} >{\centering}m{0.22\textwidth}  >{\centering\arraybackslash}m{0.22\textwidth}}
& Kamiokande & Super-Kamiokande & Hyper-Kamiokande\\
\hline
Depth & \SI{1000}{m} & \SI{1000}{m} & \SI{650}{m}\\
\quad \emph{water equivalent} & \SI{2700}{\mwe} & \SI{2700}{\mwe} & \SI{1750}{\mwe}\\
Height & \SI{16}{m} & \SI{42}{m} & \SI{60}{m}\\
Diameter & \SI{15.6}{m} & \SI{39}{m} & \SI{74}{m}\\
Volume &  &  & \\
\quad \emph{total} & \SI{4.5}{kt} & \SI{50}{kt} & \SI{258}{kt}\\
\quad \emph{inner} & \SI{3}{kt} & \SI{32.5}{kt} & \SI{216}{kt}\\
\quad \emph{fiducial} & \SI{0.68}{kt} & \SI{22.5}{kt} & \SI{187}{kt}\\
ID Photocoverage & 20\,\% & 40\,\% & 40\,\%\\
ID PMTs & 948 (\SI{50}{cm} \diameter) & \num{11129} (\SI{50}{cm} \diameter) & \num{40000} (\SI{50}{cm} \diameter)\\
OD PMTs & 123 (\SI{50}{cm} \diameter) & \num{1885} (\SI{20}{cm} \diameter) & \num{6700} (\SI{20}{cm} \diameter)\\
Single-photon detection efficiency & \emph{unknown} & 12\,\% & 24\,\%\\
Single-photon \mbox{timing} resolution & \SI{4}{ns} & \SIrange{2}{3}{ns} & \SI{1}{ns}\\
\end{tabular}
\end{center}
\caption[Comparison of Kamiokande, Super-Kamiokande and Hyper-Kamiokande]{Comparison of Kamiokande, Super-Kamiokande and Hyper-Kamiokande. Note that this reflects Kamiokande-II and Super-Kamiokande-IV; some parameters have been different during earlier phases.}
\label{tab-hk-comparison}
\end{table}
}

Table~\ref{tab-hk-comparison} shows a comparison of Hyper-Kamiokande and its predecessors.
The design of Hyper-Kamiokande will be described in more detail in the following section.

\section{Detector Design and Construction}\label{ch-hk-detector}
Hyper-Kamiokande is a ring imaging water Cherenkov detector and relies on the basic detection principle described in section~\ref{ch-intro-detection}.
The detector will be built inside a mountain to shield it against muons from cosmic ray interactions in the Earth’s atmosphere. It consists of a cavern filled with ultra-pure water that is split by a stainless steel structure into the outer detector, which is used as an active veto region, and the inner detector. This stainless steel structure also holds photodetectors for both the outer and inner detector, which will detect light from particle interactions in the detector.
Signals from the photodetectors will be collected and digitized by front-end electronics and then transferred to a data acquisition system, which combines information from all photosensors to identify and reconstruct events.
Finally, reconstructed events will be stored by a multi-tiered computing infrastructure and used for physics analyses.

This section will describe these components of the detector in detail. It is based on the November 2018 update to the Hyper-Kamiokande Design Report~\cite{HKDR2018} and reflects the status of the detector design at that time.
At that time, R\&D on many components of the detector was still ongoing and final design decisions on these components had not yet been made. Throughout this section, where applicable, I will therefore describe the baseline design of that component and briefly give an overview over alternatives. While major changes are unlikely at this stage, the final detector design may diverge in some aspects from the descriptions on the following pages. 

\subsection{Location and Cavern}
Hyper-Kamiokande will be built in the Tochibora mine of the Kamioka Mining and Smelting Company, near the town Kamioka in Japan’s Gifu Prefecture.
The proposed location at geographic coordinates N\,\ang{36;21.330;} E\,\ang{137;18.820;} (using the WGS 84 standard) is approximately \SI{8}{km} south of Super-Kamiokande but lies at same \ang{2.5} off-axis angle from the J-PARC neutrino beamline.
Located below the peak of Mount Nijugo at an altitude of \SI{514}{m} above sea level, Hyper-Kamiokande will have an overburden of \SI{650}{m} of rock (\SI{1750}{\mwe}).

\begin{figure}[htbp]
	\centering
	\includegraphics[scale=0.59]{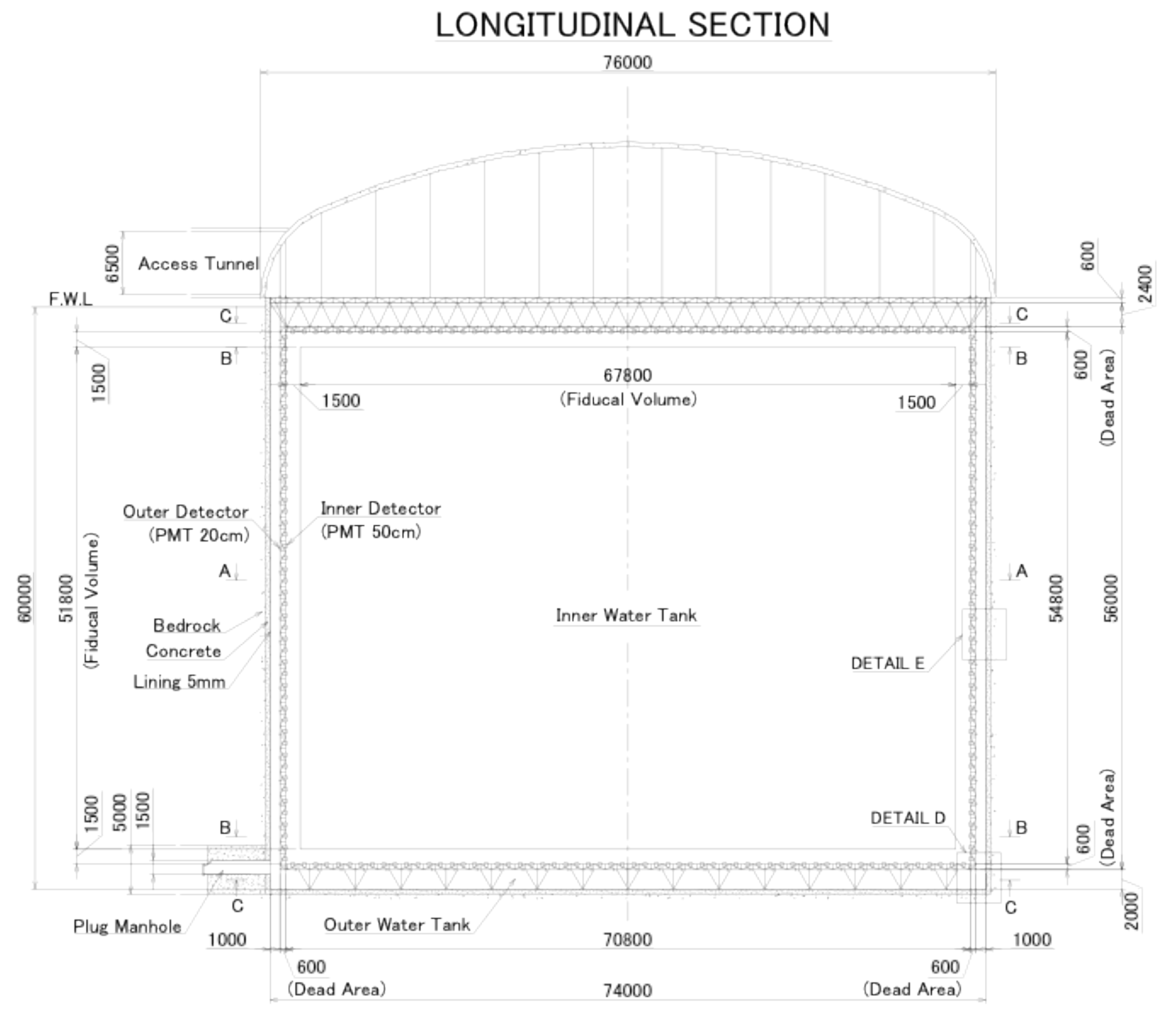}
	\caption[Schematic view of Hyper-Kamiokande]{Schematic view of the Hyper-Kamiokande detector. Figure from reference~\cite{HKDR2018}.}
	\label{fig-hk-tank_schematic}
\end{figure}

The cavern design consists of a cylindrical (or “barrel”) section with a height of \SI{62}{m} and a diameter of \SI{76}{m}, which will house the detector volume\footnote{This excavated volume is slightly larger than the detector volume to account for the water containment system described below.}, and a \SI{16}{m} high “dome” section on top of the cylinder, which will house detector infrastructure such as the data acquisition and calibration systems.
A schematic drawing of the cavern and detector is shown in figure~\ref{fig-hk-tank_schematic}.
The excavation work will also include smaller caverns for the water circulation and purification systems as well as access tunnels.
Optimization of the cavern dimensions and the layout of the access tunnels is still ongoing.%
\footnote{In September 2019, the Hyper-Kamiokande proto-collaboration announced an updated detector design with a diameter of \SI{68}{m} and a water depth of \SI{71}{m}. The total and fiducial volumes both changed by less than 1\,\%; the effect on the results described in this thesis, which still used the previous design, will therefore be negligible.}

To avoid fault lines and determine the rock quality, the collaboration has performed a geological survey of the mountain using pre-existing tunnels, some newly drilled boreholes and seismic prospecting with acoustic waves.
Using these results, the location of Hyper-Kamiokande has been narrowed down to a candidate region of approximately $\SI{200}{m} \times \SI{150}{m}$ which exhibits the best and most even rock quality.
A detailed survey of the geological conditions in this region is planned, which will culminate in a precise cavern design including a pattern of pre-stressed anchors necessary to support the rock and ensure structural stability.
Several possible rock quality distributions have been simulated and in all cases the cavern construction is feasible with existing techniques.

The excavation will start by constructing “access tunnels”, which lead from the mine entrance to the vicinity of the detector, and “approach tunnels”, which connect the access tunnels to the water rooms and various levels of the main cavern.
Once the tunnels are constructed, the cavern will be excavated starting with the dome and concluding with the bottommost part of the barrel region.

An intermediate deposition site for excavated rock 
and a final disposal site were identified and the geological stability of the final disposal site was confirmed through a boring survey and several computer simulations.
Both sites can be accessed via existing roads, though some widening or re-routing of roads will be necessary to allow a large number of dump trucks to use them.

\subsection{Water Tank}
\begin{figure}[htbp]
	\centering
	\includegraphics[scale=0.7]{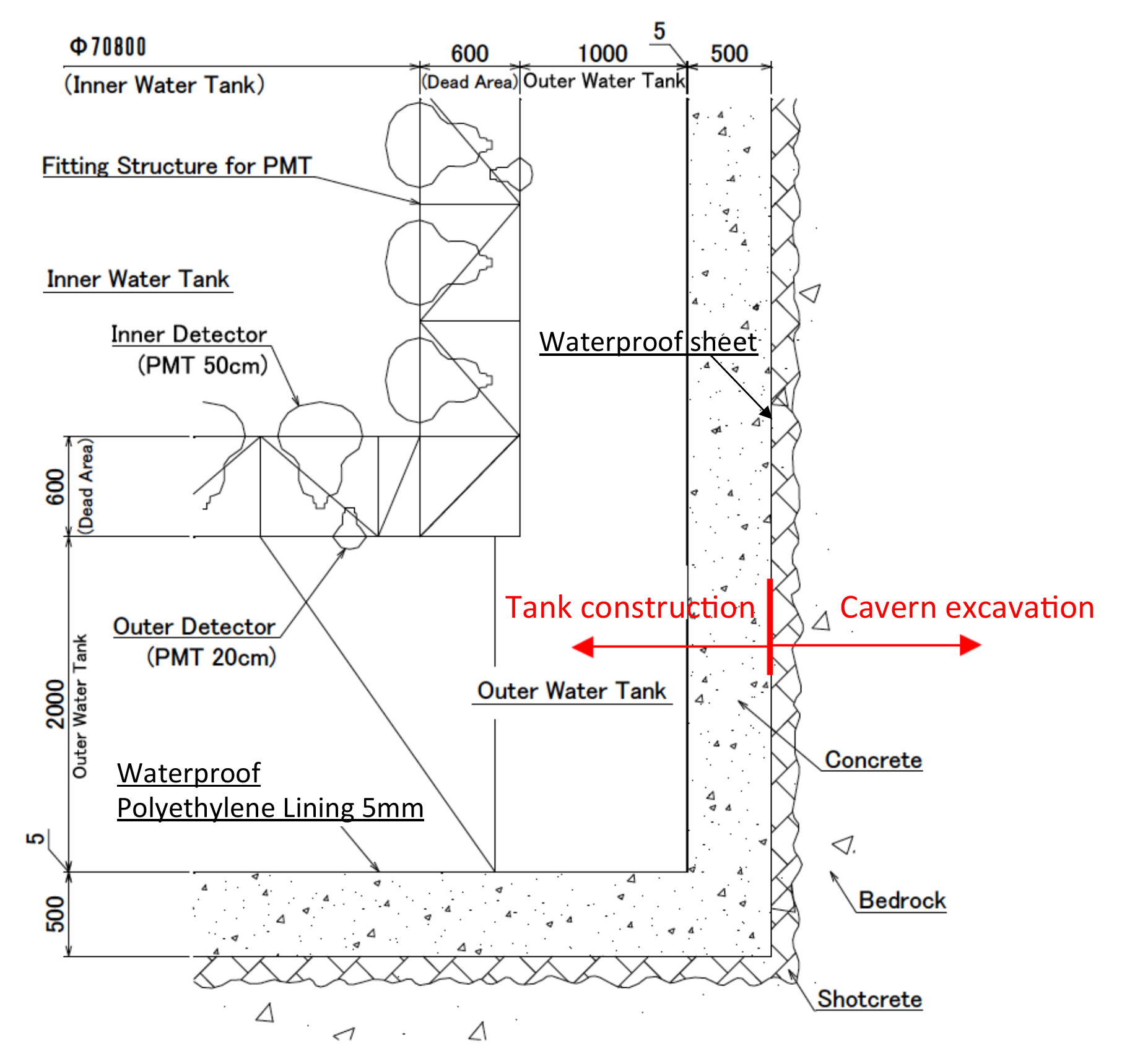}
	\caption[Schematic view of the outer edge of the water tank]{Schematic view of the outer edge of the water tank. Figure from reference~\cite{HKDR2018}.}
	\label{fig-hk-cavern_tank_boundary}
\end{figure}

Once the cavern has been fully excavated, its surface will be treated with a layer of shotcrete, which will then be covered by a waterproof sheet to stop ultra-pure water from leaking out and to stop external sump water from entering the detector.
To ensure the stability of the cavern, a \SI{50}{cm} thick layer of concrete is then added which is reinforced with steel rods and lined on the inside with a \SI{5}{mm} thick waterproof high-density polyethylene (HDPE) sheet.
HDPE was chosen for its flexibility, which allows it to cover cracks in the concrete wall without breaking, its low water permeability and its high chemical resistance to both ultra-purified water and a gadolinium sulfate solution.
A schematic drawing of the outer edge of the water tank is shown in figure~\ref{fig-hk-cavern_tank_boundary}.

Magnetic fields perpendicular to a PMT’s direction reduce the collection efficiency of large PMTs.
For the B\&L PMT described in section~\ref{ch-hk-id-blpmt}, the reduction was measured to be 1\,\% at \SI{100}{mG}, growing to 3\,\% at \SI{180}{mG}.
To counteract the effects of the geomagnetic field, which at the detector site is $(B_x, B_y, B_z) = (-303, 0, -366)$\,\si{mG}, a combination of rectangular coils (in the y-z plane) and circular coils (in the x-y plane) will be embedded in the concrete layer.
A preliminary design found that the residual perpendicular magnetic field can be reduced to $B_\perp < \SI{100}{mG}$ for 97.8\,\% of the ID PMTs, with only PMTs at the top and bottom edge of the detector experiencing a higher residual $B_\perp$.
Optimizations of the coil design are still ongoing.

Inside the water tank, a stainless-steel structure will separate the inner and outer detector regions and support the PMTs and their covers (see sections~\ref{ch-hk-id} and~\ref{ch-hk-od}) as well as front-end electronics (see section~\ref{ch-hk-electronics}).
In Super-Kamiokande, the bottom and barrel sections are a self-supported framework standing on the bottom of the tank.
In Hyper-Kamiokande, due to the larger size of the detector, only the bottom section will be free-standing on the bottom of the tank while the barrel and top sections will be suspended from the ceiling of the cavern.
This allows the frame to be thinner and thus lighter and cheaper, since it is not at risk of buckling under its own weight.

The strength of the frame is determined by the weight of the PMTs, covers and electronics it has to support while the tank is empty.
When determining the configuration of photosensors and covers, care must be taken to ensure that the buoyancy of the PMTs in water does not overcompensate for the weight of the structure.
Otherwise, additional weights would need to be added to stop the structure from floating, which would require a stronger frame to support the additional weights when the tank is empty and thus increase the cost of the structure.

The inner and outer detector regions will be optically separated to allow the OD to act as an active veto; however, the details of this are not yet fixed.
In the simplest case, as in Super-Kamiokande, this could be achieved with two layers of light-proof sheets covering the inner and outer surfaces of the PMT support structure.
However, if Hyper-Kamiokande is to be loaded with gadolinium at some point, a hermetic separation of the ID and OD may be desirable to ensure that the gadolinium remains contained within the ID, which would simplify the water system and reduce costs.

\subsection{Water Purification and Circulation System}\label{ch-hk-water}

All materials used in detector construction are extensively tested to avoid introducing sources of radioactivity or water-soluble impurities into the detector.
Despite these efforts however, radon emanating from materials in the tank, particularly the glass of photosensors and the support structure, and from the rock surrounding the detector is a major source of low-energy backgrounds.
In addition, light scattering and absorption due to impurities in the water are a major source of uncertainty in event reconstruction.
To reduce these effects, the water in the detector is constantly recirculated and purified.

In Super-Kamiokande, after continuous improvements the water system is now able to purify the water inside the detector to reach a water transparency of over \SI{100}{m} and a radon concentration in the ID of less than \SI{1}{mBq / m^3}.
In Hyper-Kamiokande, where the diagonal size of the detector increases to nearly \SI{100}{m}, a similar or better water quality will be required.
To achieve this goal, the design of the water system will be similar to that employed in Super-Kamiokande but scaled up to account for the larger detector mass.

\begin{figure}[tbp]
	\centering
	\includegraphics[scale=0.53]{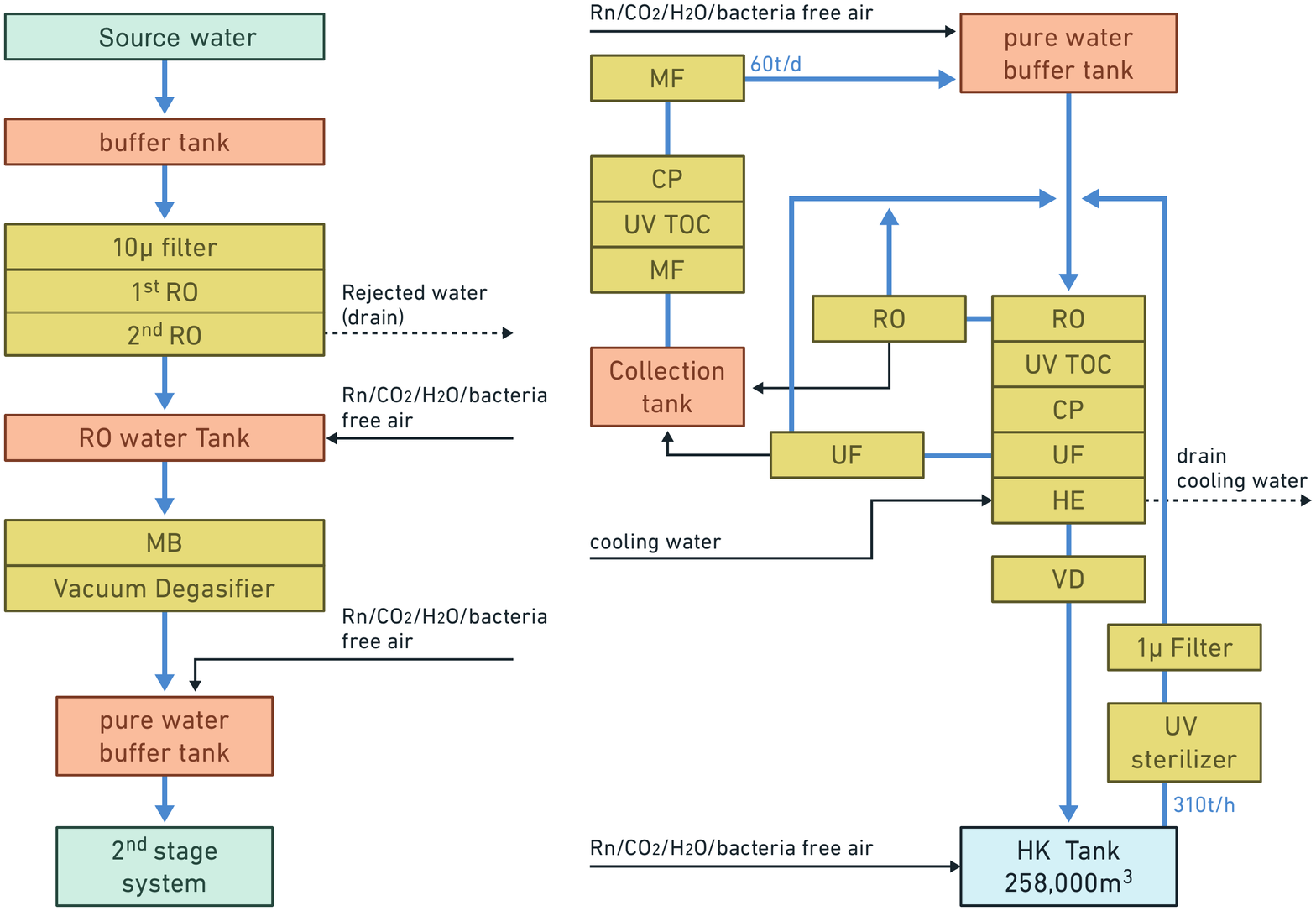}
	\caption[Water system design for Hyper-Kamiokande]{Water system design for Hyper-Kamiokande. See text for an explanation of the individual steps. Figure from reference~\cite{HKDR2018}.}
	\label{fig-hk-1stand2nd}
\end{figure}

The system consists of two separate stages (see figure~\ref{fig-hk-1stand2nd}), one for initial filling of the tank and one for ongoing recirculation of the water during operations.
The water to fill the detector will come from the storage well of the snow-melting system of the nearby Kamioka town.
During filling, \SI{105}{t/h} of source water will be needed to fill the detector with purified water at a rate of \SI{78}{t/h}, with approximately half a year needed to completely fill the detector.
The water will be recirculated at a rate of \SI{310}{t/h}, such that the total detector mass is recirculated approximately once per month, at the same rate as in Super-Kamiokande.

In the first stage, the raw water is passed through a \SI{10}{\micro m} filter to eliminate dust and larger particles from the raw water, before going through reverse osmosis (RO) and additional filters (MB) to remove smaller particulates.
A vacuum degasifier (VD) removes dissolved oxygen (which encourages growth of bacteria in the water) and radon from the water before the pre-cleaned water is fed into the second stage of the system.

In the second stage, water coming from the Hyper-Kamiokande tank is sterilized with UV light and filtered before going through a multi-stage process including RO, further UV irradiation (UV TOC), a cartridge polisher (CP) that removes heavy ions and more advanced filtration (MF, UF).
The purified water is then cooled down in a heat exchanger (HE) to remove heat produced by photosensors and electronics in the water as well as the water system itself.
Finally, the cooled water is degasified in a VD and supplied back into the tank.

Radon-free air with a concentration of less than \SI{1}{mBq / m^3} is used as a cover gas for the Hyper-Kamiokande tank as well as for buffer tanks which are part of the water system.
To produce enough radon-free air, the system employed in Super-Kamiokande will be scaled up to the larger detector size.

\begin{figure}[tbp]
	\centering
	\includegraphics[scale=0.41]{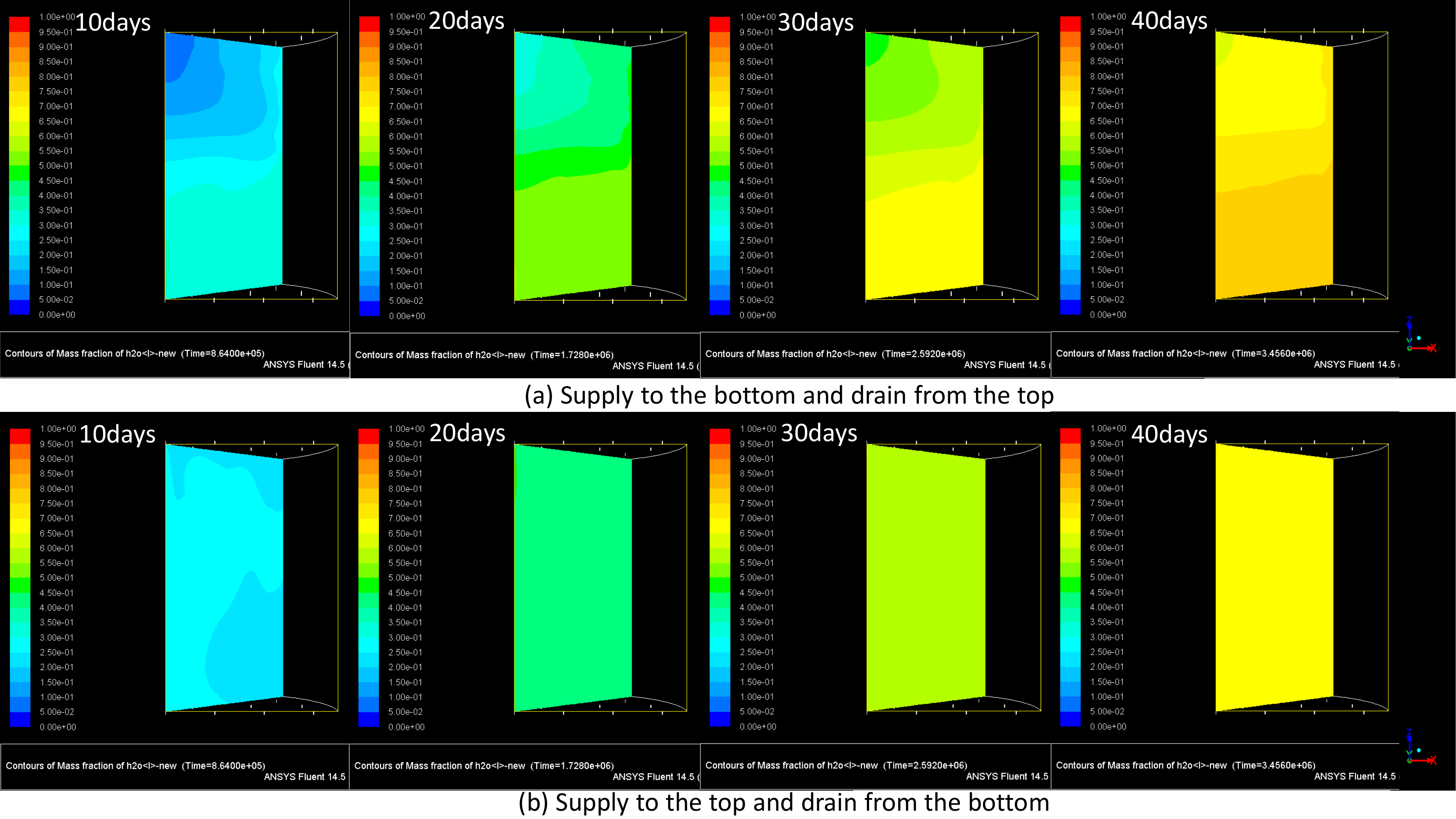}
	\caption[Simulation of water replacement efficiency in Hyper-Kamiokande]{Simulation of water replacement efficiency in Hyper-Kamiokande. The tank is filled with old water (blue) at the start of the simulation and fresh water (red) is then supplied: (a) at the bottom of the tank while draining from the top; (b) at the top of the tank while draining from the bottom. Supplying fresh water from the bottom leads to a higher replacement efficiency, displayed as a more reddish colour, while supplying it from the top leads to large-scale convection in the tank and a more uniform water quality. Figure from reference~\cite{HKDR2018}.}
	\label{fig-hk-FlowSim}
\end{figure}

To avoid radon from the surrounding rock entering the inner detector, there is no water exchange between the inner and outer detector.
Radioactive impurities in the inner detector mostly originate from the photomultipliers and their support structure.
Controlling the water flow in the detector is essential to limit the spread of these impurities in the inner detector and reduce their impact on the physics performance of the detector.
Computer simulations of the water flow (see figure~\ref{fig-hk-FlowSim}) show that supplying cold water at the bottom of the tank and draining water at the top leads to laminar flow and ensures effective water replacement.
Supplying cold water at the top of the tank and draining water from the bottom would instead lead to convection in the tank, which leads to uniform water quality throughout the tank and decreases the efficiency of water replacement.
These simulations agree with observations in Super-Kamiokande.

Adding gadolinium to a water Cherenkov detector to detect neutron captures and thus better identify events was originally suggested by Beacom and Vagins in 2003~\cite{Beacom2004}.
After extensive testing, gadolinium will be added to Super-Kamiokande in the near future~\cite{Marti-Magro2018} and is being explored as an option for Hyper-Kamiokande.
This would require changes to the water system to remove the gadolinium from drained water using molecular bandpass filters.
Both components would then be cleaned separately and recombined before supplying the water back into the tank.
The necessary technologies have been developed for Super-Kamiokande and use a modular design, ensuring they can be scaled up for use in Hyper-Kamiokande.
Since gadolinium loading is not part of the Hyper-Kamiokande baseline design and may instead be added in a later upgrade, I am not considering its benefits in this thesis.

\subsection{Inner Detector}\label{ch-hk-id}

The inner surface of the structure separating the inner and outer detector is divided into \num{40000} “cells” with a size of $\SI{70}{cm} \times \SI{70}{cm}$, each of which can house either a single \SI{50}{cm} diameter PMT or a so-called multi-PMT module (mPMT) consisting of an array of smaller PMTs with \SI{8}{cm} diameter.

In the reference design, every cell houses one \SI{50}{cm} PMT, resulting in 40\,\% of the inner detector surface being light-sensitive.
The newly developed PMT model and two alternative designs that are currently under development are described in sections~\ref{ch-hk-id-blpmt} and~\ref{ch-hk-id-alt50cm}, while covers to increase pressure-resistance of these PMTs are described in section~\ref{ch-hk-id-cover}.

Alternative configurations, which employ half the number of \SI{50}{cm} PMTs and augment them with mPMTs or other modifications that increase light collection, are briefly described in sections~\ref{ch-hk-id-mpmt} and~\ref{ch-hk-id-lc}.
Studies to determine the impact of these alternative designs on the performance of Hyper-Kamiokande are currently ongoing.

\subsubsection{B\&L PMTs}\label{ch-hk-id-blpmt}

A new \SI{50}{cm} PMT model with a higher quantum efficiency and a box-and-line dynode (Hamamatsu R12860-HQE~\cite{R12860}, hereafter referred to as the B\&L PMT) is being developed for Hyper-Kamiokande.
\begin{figure}[htbp]
	\centering
	\includegraphics[scale=1.19]{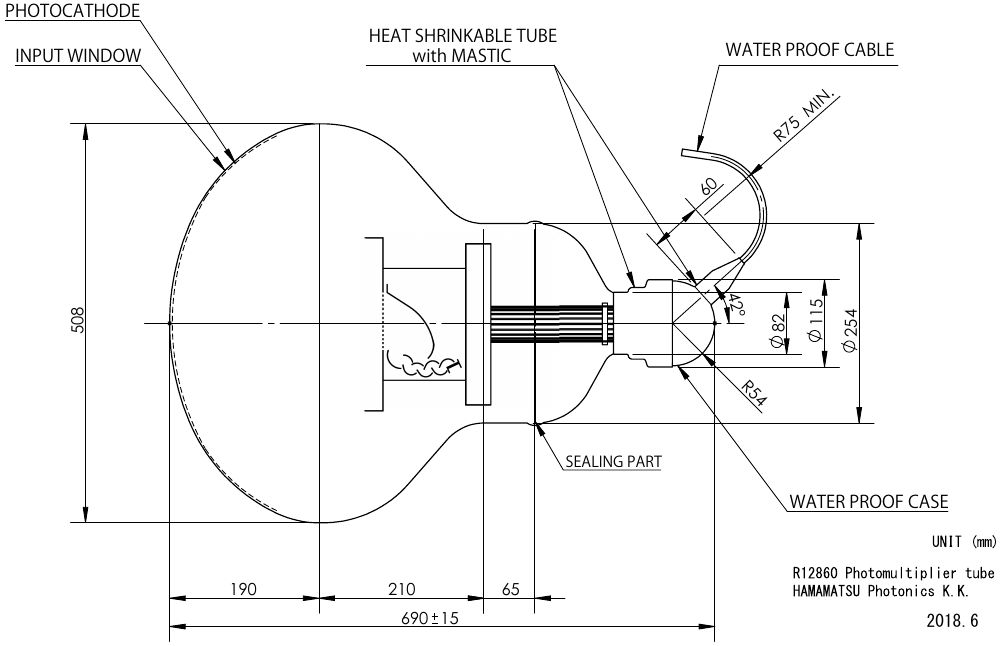}
	\caption[Side view of the B\&L PMT]{Side view of the B\&L PMT. Figure from reference~\cite{HKTR2018}.}
	\label{fig-hk-box-and-line}
\end{figure}
Figure~\ref{fig-hk-box-and-line} shows the B\&L PMT in a side view.
Compared to Hamamatsu’s R3600 model used in Super-Kamiokande, it offers improvements to both detection performance and mechanical stability.

\begin{figure}[tbp]
	\hspace{-0.4pc}\begin{minipage}{15pc}
	\includegraphics[width=15pc]{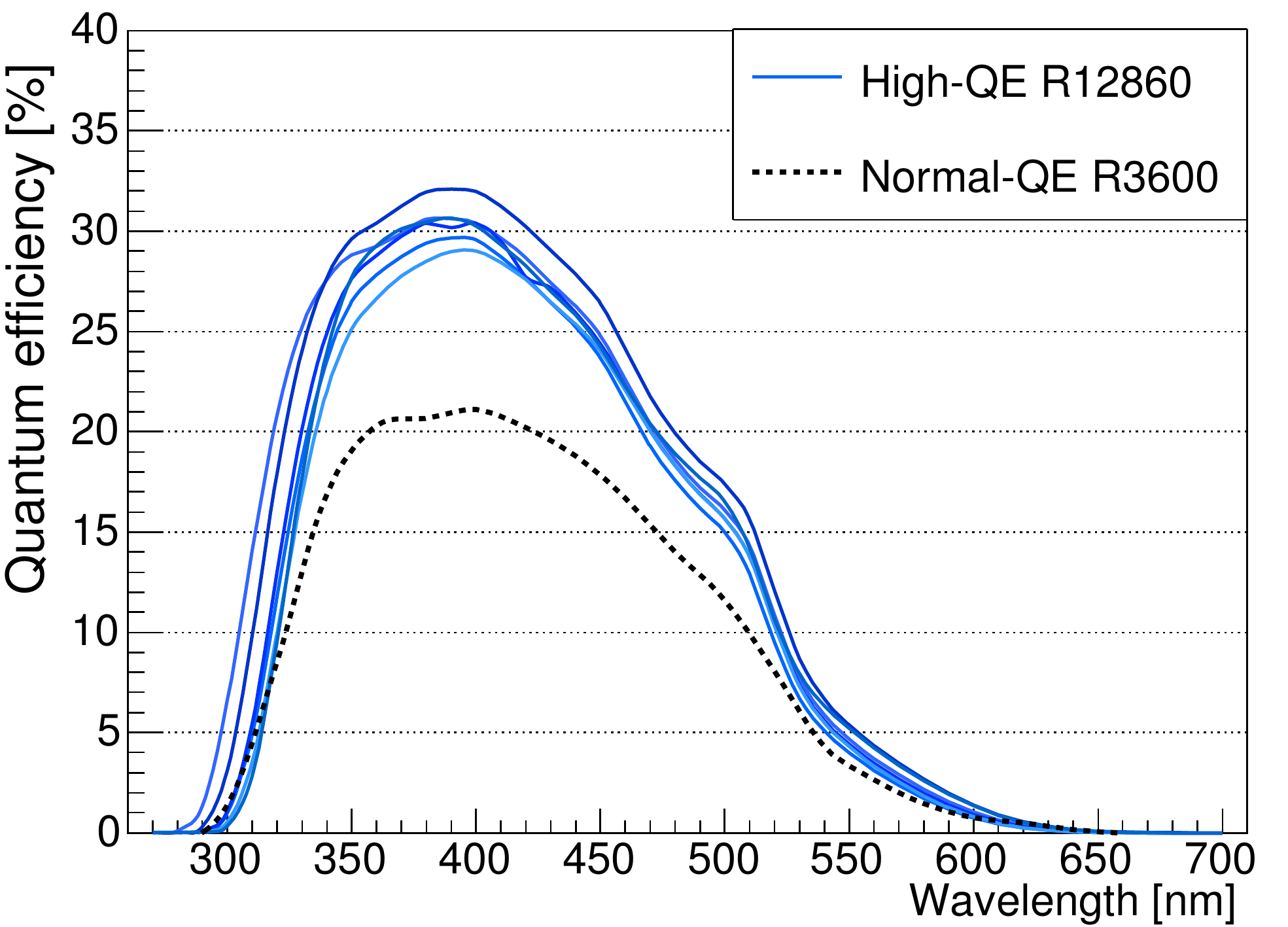}
	\end{minipage}
	\begin{minipage}{18.5pc}
	\includegraphics[width=18.5pc]{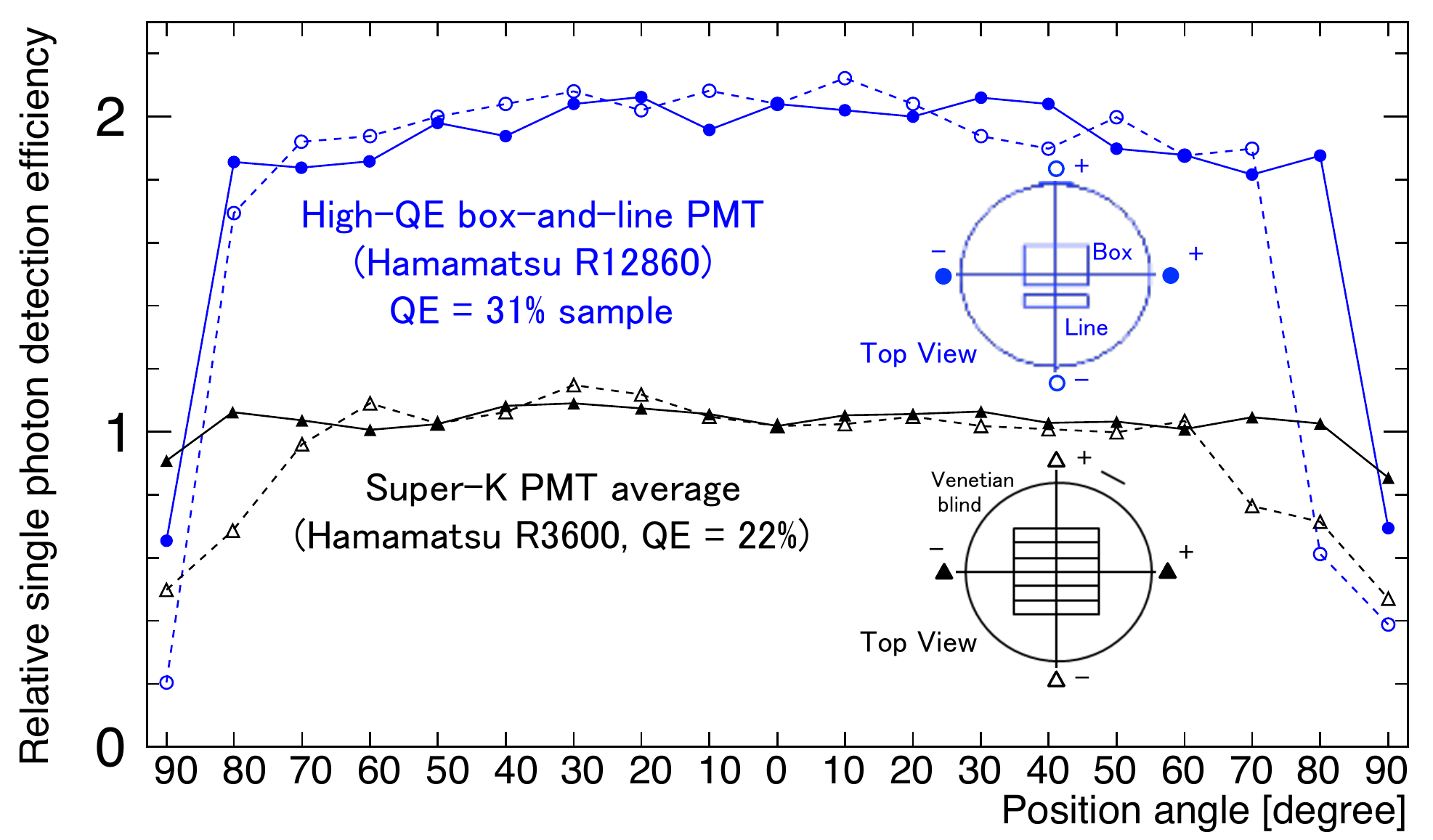}
	\end{minipage}
	\caption[Quantum efficiency and single photon detection efficiency of B\&L PMT and Super-Kamiokande PMT]{Left: Quantum efficiency as a function of wavelength for six different B\&L PMTs (blue, solid) and a Super-Kamiokande PMT (black, dotted). Right: Single photon detection efficiency as a function of photocathode position for a B\&L PMT (blue) and a Super-Kamiokande PMT (black, normalized to 1). Figure from reference~\cite{HKDR2018}.}
	\label{fig-hk-HQEspectra}
\end{figure}

The total detection efficiency for a single photon was increased by a factor of two (see figure~\ref{fig-hk-HQEspectra}) by combining improvements to the quantum efficiency, which now reaches 30\,\% at a wavelength of \SI{390}{nm}, and the capture efficiency, which was increased from 73\,\% to 95\,\% in the central \SI{46}{cm} diameter and is more uniform near the edges of the detection area.
The increased capture efficiency was reached through changes to the glass curvature and the focusing electrode, as well as by using the box-and-line dynode.

\begin{figure}[htbp]
	\hspace{-0.5pc}\begin{minipage}{17pc}
	\includegraphics[width=17pc]{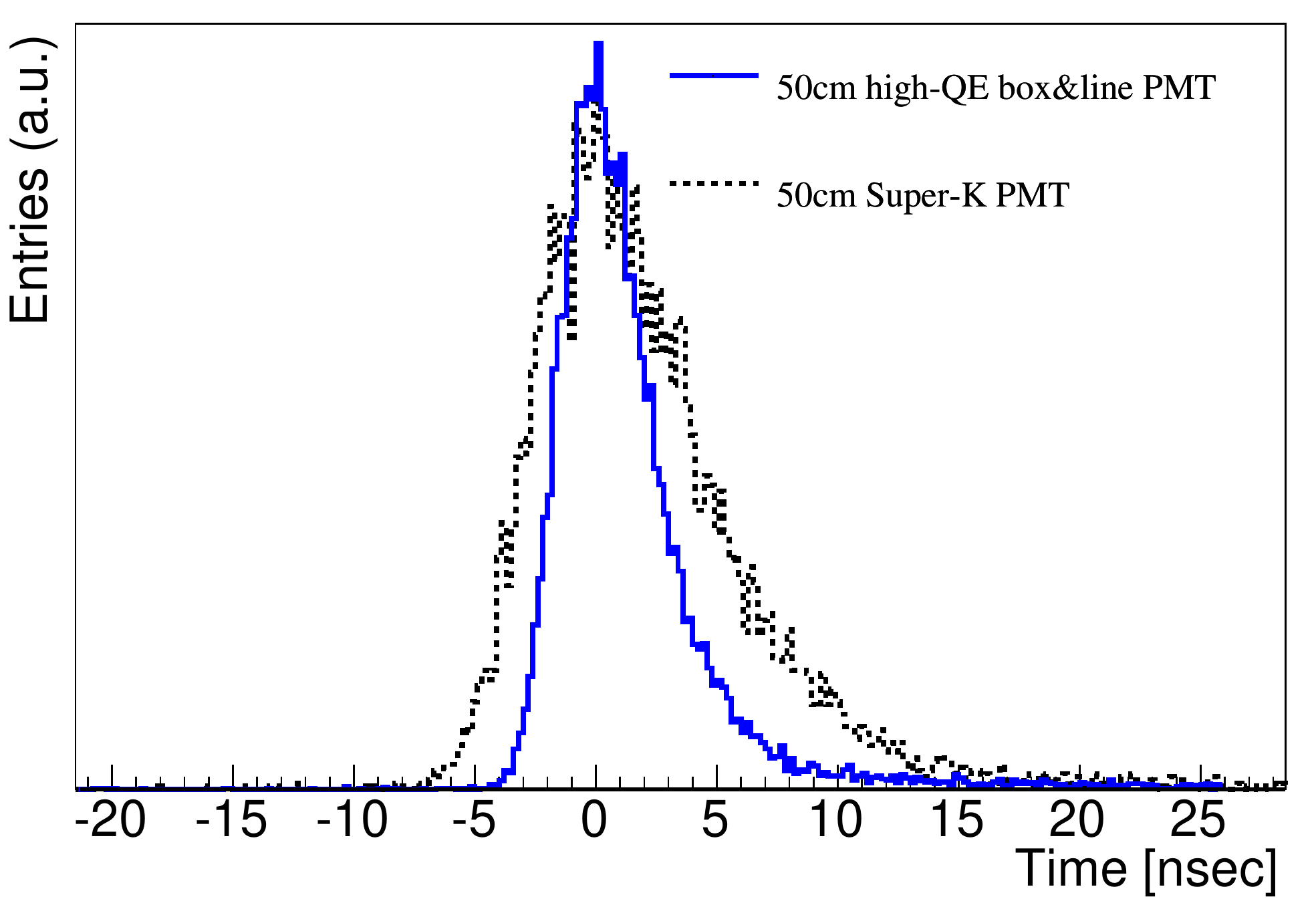}
	\end{minipage}
	\begin{minipage}{17pc}
	\includegraphics[width=17pc]{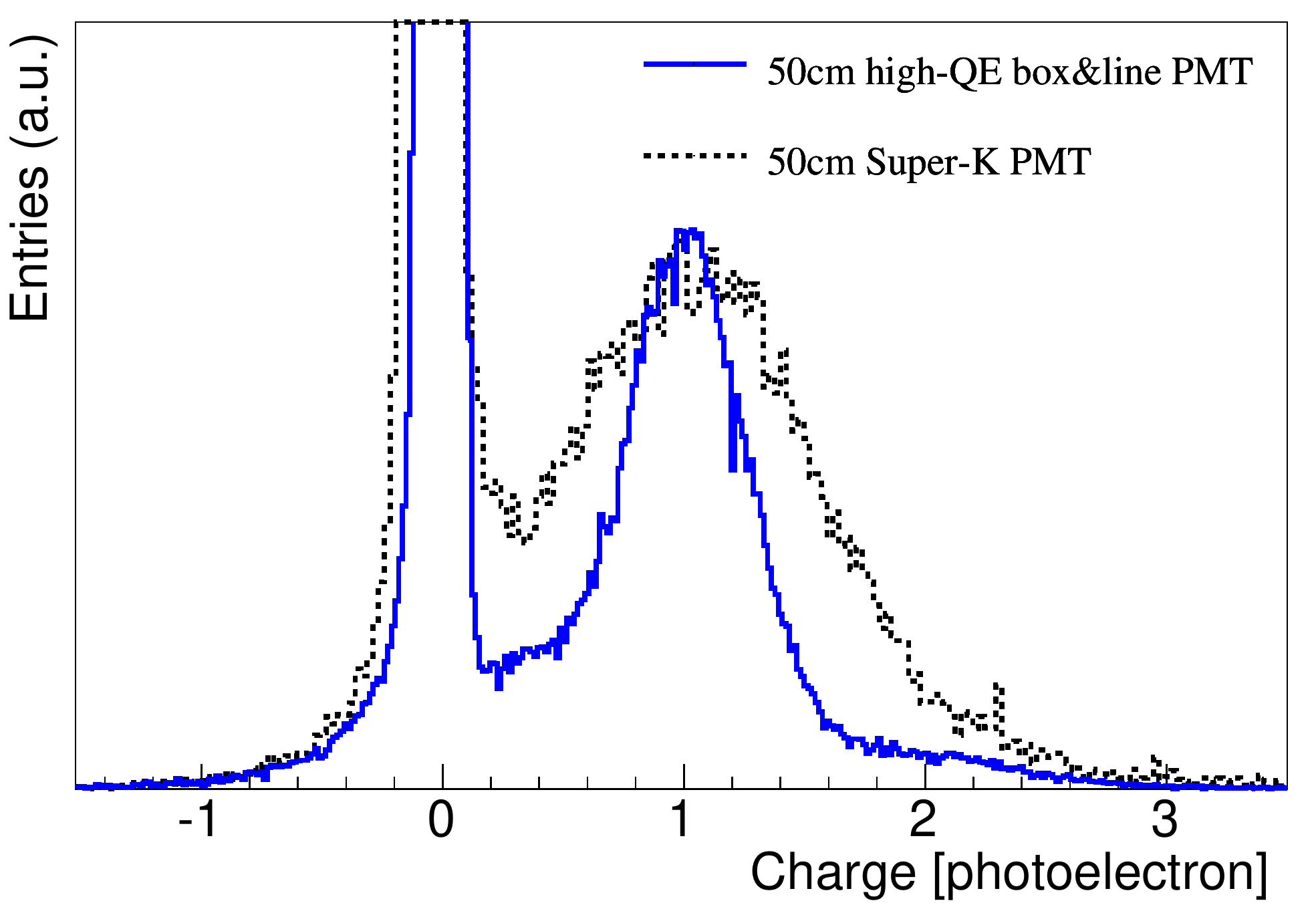}
	\end{minipage}
	\caption[Transit time and charge for a single photoelectron in the B\&L PMT and Super-Kamiokande PMT]{Distribution of the single photoelectron transit time (left) and charge (right) in the B\&L PMT (blue, solid) and the Super-Kamiokande PMT (black, dotted). Figure from reference~\cite{HKDR2018}.}
	\label{fig-hk-photosensor_1petts}
\end{figure}

In the Venetian blind dynode used in the Super-Kamiokande PMTs, some photo\-electrons would miss the first stage of the dynode, while the larger first stage box-and-line has a much higher acceptance.
In addition to the capture efficiency, this also improves the time resolution to \SI{4.1}{ns} at FWHM---approximately half the value of Super-Kamiokande PMTs---and the charge resolution to 35\,\% compared to 50\,\% in Super-Kamiokande PMTs (see figure~\ref{fig-hk-photosensor_1petts}).

The nominal gain of the B\&L PMTs is \num{1e7} at \SI{2000}{V}, though it can be adjusted by changing the bias voltage in a range of \SIrange{1500}{2200}{V}.

The charge response was measured to be linear within 5\,\% at up to 470 photoelectrons and the saturation threshold beyond which the nonlinear response cannot be corrected for is higher than \SI{2000}{PE}.
This high dynamic range is sufficient to reconstruct events across Hyper-Kamiokande’s physics areas, covering the \si{MeV} to multi-\si{GeV} range.

Along with the increased efficiency, the B\&L PMTs have an increased dark noise rate compared to the \SI{4.2}{kHz} in Super-Kamiokande PMTs.
For an earlier batch of B\&L PMTs, a dark noise rate of \SI{8.3}{kHz} was measured at a temperature of \SI{15}{\celsius} after a month-long stabilization period.
By early 2019, improvements in the manufacturing process had reduced the dark rate to \SIrange{6}{7}{kHz}.
Further work to lower the dark rate by eliminating radioactive impurities from the glass manufacturing process is currently ongoing.

As part of the maintenance work in 2018, 136 B\&L PMTs were added to Super-Kamiokande to test them under real conditions and explore the long-term stability of gain and dark rate.

In addition to these sensitivity changes, the B\&L PMTs includes mechanical improvements.
The shape and thickness of the glass bulb were optimized for pressure resistance to ensure survival at the bottom of Hyper-Kamiokande below approximately \SI{60}{m} of water, corresponding to \SI{0.6}{MPa} of pressure. 
Fifty sample B\&L PMTs were tested in water at pressures of up to \SI{1.25}{MPa} and no damage was found.
However, due to the large number of photosensors in Hyper-Kamiokande, it is difficult to ensure that there is no glass failure and additional covers described in section~\ref{ch-hk-id-cover} will be employed to improve pressure resistance.

\subsubsection{Alternative 50\,cm Photosensors}\label{ch-hk-id-alt50cm}

In addition to the B\&L PMTs, two alternative photosensors with a \SI{50}{cm} diameter are currently under consideration for the inner detector.

The first alternative are micro-channel plate (MCP) PMTs, which were developed for the JUNO experiment and are produced by North Night Vision Technology.
Their detection efficiency is comparable with that of the B\&L PMTs and while the time resolution of the GDB-6201 model used by JUNO was insufficient for Hyper-Kamiokande, a time resolution of \SI{5.5}{ns} was achieved in a newly developed model.
At the moment, however, their high dark noise rate is an issue and further tests for reducing it are ongoing.

Another alternative are the Hamamatsu R12850-HQE hybrid photodetectors (HPD), which use a bulb and photocathode that are almost identical to those of the B\&L PMT.
Instead of multiplying photoelectrons on a metal dynode, however, they are accelerated and focussed on an avalanche diode by a high voltage of \SI{8}{kV}.
The HPD offers an improved charge resolution of $\sigma = 15\,\%$ for \SI{1}{PE}, while its other characteristics are similar to the B\&L PMT.
At the moment, however, no capacities for mass production of HPDs exist.

\subsubsection{PMT Covers}\label{ch-hk-id-cover}

Even with a range of pressure tests before installation in the tank, the risk of a PMT glass failure cannot be fully eliminated.
In November 2001 during refilling of Super-Kamiokande, the pressure wave from an imploding PMT led to a chain reaction that destroyed almost \num{7000} PMTs.
To avoid a reoccurrence, Super-Kamiokande afterwards added protective covers to all \SI{50}{cm} PMTs, which consist of a transparent front section allowing Cherenkov light from the inner detector to enter and a light-tight base section.
Both components are assembled around the photosensor and connected before being installed in the photosensor support structure inside the tank.
For Hyper-Kamiokande, these covers are being redesigned to account for the increased water depth in the detector.

The front section of the cover will be manufactured from acrylic and has a hemispherical shape with a flat flange around the edge to attach it securely to the base.
In the centre, its maximum height is \SI{19}{cm} above the flange.
The front section contains small holes to allow water to flow past the photosensitive surface and to reduce buildup of radioactive material or biofilms.
The optical properties of the acrylic are tested to ensure a transparency of at least 50\,\% for photons with \SI{300}{nm} wavelength and 90\,\% at \SIrange{400}{800}{nm}.
A low reflectivity in water is also required to reduce reflected photons that effectively increase the dark noise rate in other PMTs.

\begin{figure}[tbp]
	\centering
	\includegraphics[scale=0.51]{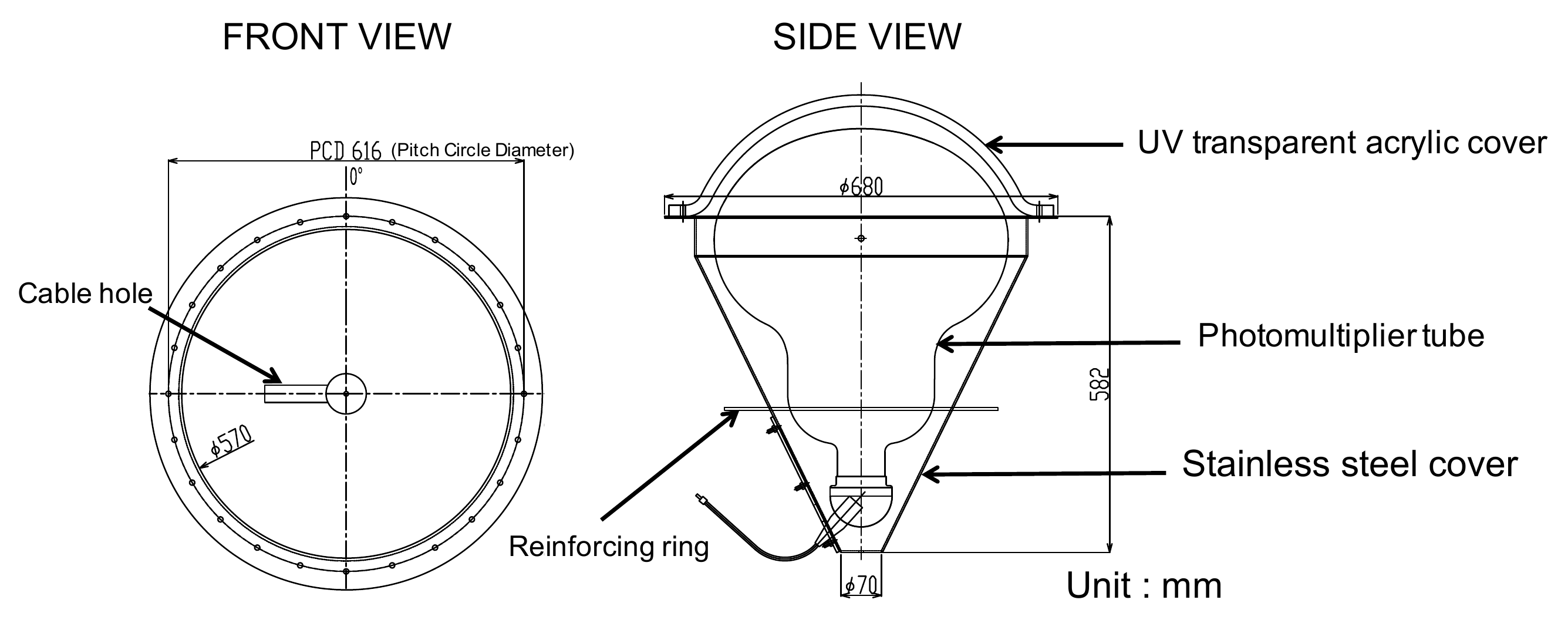}
	\caption[Front and side view of baseline design for protective cover]{Front and side view of the baseline design for the protective cover. Figure from reference~\cite{HKDR2018}.}
	\label{fig-hk-protective_cover_shape}
\end{figure}

The base section is intended to be water- and light-tight to separate the inner detector from the dead region between inner and outer detector and will also be used to mount the photosensor assembly on the support structure.
As a baseline, a conical design made from stainless steel has been developed, which is shown in figure~\ref{fig-hk-protective_cover_shape}. 

To ensure that the cover is able to withstand the implosion of a PMT, hydrostatic pressure tests at up to \SI{1.5}{MPa} and a set of shock wave tests were performed.
In these shock wave tests, a $3 \times 3$ array of covered PMTs is placed under hydrostatic pressure corresponding to \SI{60}{m} or \SI{80}{m} of water and the centre PMT is artificially imploded, while pressure sensors and high-speed cameras observe the other PMTs.
The cover incurred no damage while reducing the pressure experienced by neighbouring PMTs to less than \SI{0.05}{MPa}, sufficient to prevent a chain reaction.
During maintenance work in 2018, eight of these covers were installed in Super-Kamiokande to test them under real conditions.

Two alternative designs for the base section are currently studied.
The first uses a polyphenylene sulfide resin mixed with carbon fibre.
While this would be lighter and cheaper to manufacture, an early version of this design failed hydrostatic pressure tests at \SI{0.6}{MPa}.
Another design uses a stainless steel cover in a tubular shape instead of the conical shape of the baseline design.
While this design has passed hydrostatic pressure tests at \SI{0.8}{MPa} and would be easier and cheaper to manufacture, it is currently significantly heavier than the baseline design.
Improvements to both alternative designs are currently under development.
For the final detector design, a mix of covers could be used at different depths to optimize the cost and weight while ensuring pressure resistance.

\subsubsection{Multi-PMT Optical Module}\label{ch-hk-id-mpmt}
While the baseline design relies on \num{40000} large-area photosensors with a \SI{50}{cm} diameter as described above, alternative designs are considered that employ \num{20000} of these large photosensors and supplement them with optical modules containing multiple smaller \SI{7.7}{cm} PMTs.
These multi-PMT (mPMT) modules are based on a design by the KM3NeT collaboration~\cite{Adrian-Martinez2014}, who are using modules comprising 31 smaller PMTs in a spherical glass pressure vessel for a neutrino detector in the Mediterranean Sea.

\begin{figure}[tbp]
	\centering
	\includegraphics[scale=0.6]{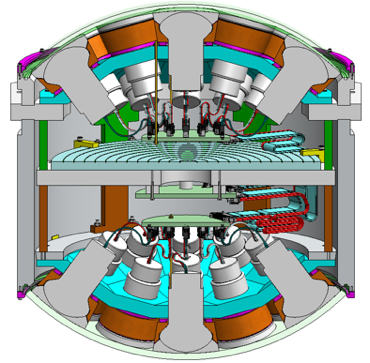}
	\caption[Drawing of the two-sided mPMT module]{Drawing of the two-sided mPMT module described in the text. Figure from reference~\cite{HKDR2018}.}
	\label{fig-hk-mPMT_NuPRISM}
\end{figure}

For Hyper-Kamiokande, this design was initially adapted into a cylindrical module that combines photosensors for the inner and outer detector into one pressure module by placing a hemispherical section on each end (see figure~\ref{fig-hk-mPMT_NuPRISM}).
A single-sided module containing only PMTs for the inner detector is currently under development.
While the PMT arrangement and thus the physics sensitivities are unchanged compared to the two-sided module, the resulting design will be lighter and simpler to produce and install, thus reducing the cost.

The mPMT module consists of a cylindrical structure, which contains electronics and a support structure holding the PMTs in place,
The photosensitive surface of each PMT is surrounded by a conical reflector that increases the light collection efficiency by 20\,\%.
An acrylic window is placed in front of the PMTs and coupled to them with optical gel.

Multiple \SI{7.7}{cm} PMT models from different manufacturers are available that fulfil the technical requirements for use in Hyper-Kamiokande.
Prototype mPMT modules have been constructed; however, the design of the electronics and the detailed procedures for final assembly are still under development.

The main advantage of mPMTs over \SI{50}{cm} photosensors is their higher granularity, which may be particularly useful for reconstructing events close to the edge of the inner detector 
or for separating overlapping Cherenkov rings in proton decay candidate events.
Due to the hemispherical shape of the mPMT module, different PMTs have different fields of view.
This effect could be used to identify dark noise in PMTs facing away from the reconstructed event vertex, which would improve reconstruction particularly at low energies.
The smaller PMTs may also be less sensitive to external magnetic fields and offer a better timing resolution.
However, the total photosensitive area of an mPMT module is significantly smaller than that of a \SI{50}{cm} photosensor. 

\subsubsection{Light Collection}\label{ch-hk-id-lc}

The photon detection efficiency of the \SI{50}{cm} photosensors could be increased by collecting photons that did not hit the photosensitive area directly.

Such a light collection system should have a high angular acceptance, which ensures that events near the edge of the inner detector can still be reconstructed accurately, and it should delay arrival times of the collected photons by no more than \SI{5}{ns} in order to not degrade the timing resolution of the photosensors.
The system should also minimize reflection of photons back into the inner detector and avoid introducing radioactive backgrounds or other impurities that decrease water transparency.

Multiple designs for this light collection system have been suggested, including reflective cones similar to those used in the mPMT modules, wavelength shifting plates as used in the outer detector and Fresnel lenses mounted in front of the photosensors.
All these options are currently under investigation.

\subsection{Outer Detector}\label{ch-hk-od}

The main function of the outer detector is distinguishing neutrino interactions in the detector from external backgrounds.
Incident cosmic ray muons, which are a major background in Hyper-Kamiokande, can be identified due to spatial and temporal coincidence of energy deposition in the outer and inner detector.
The outer detector also provides passive shielding against gamma rays or neutrons from natural radioactivity in the surrounding rock.

To serve as an active veto, the outer detector needs to detect the presence or absence of a signal but does not need to provide precise reconstruction capabilities.
It is therefore equipped only with a sparse array of photodetectors.

In Super-Kamiokande~\cite{Fukuda2003}, the outer detector is approximately \SI{2}{m} wide and equipped with 1885 \SI{20}{cm} PMTs that provide a photocoverage of 1\,\%.
The light collection is increased by approximately 50\,\% using wavelength shifting plates attached to the PMTs which absorb UV light and re-emit photons in the visible spectrum, better matching the spectral sensitivity of the PMTs.
Inside the support structure separating the inner and outer detector, a dead zone of about \SI{0.6}{m} provides additional passive shielding.

In Hyper-Kamiokande, the width of the outer detector is reduced to \SI{1}{m} in the barrel region, while remaining at \SI{2}{m} at the top and bottom.
The dead zone between inner and outer detector remains at \SI{0.6}{m} and is limited by the depth of the covers for the \SI{50}{cm} photosensors.
While this narrower outer detector increases the size of the fiducial volume, the background reduction capabilities must be equivalent to those of Super-Kamiokande to enable the physics goals.

An outer detector design analogous to that of Super-Kamiokande has been considered, which would require 6700 \SI{20}{cm} PMTs  to provide a 1\,\% photocoverage.
Due to the narrower outer detector in the barrel region, however, it is advantageous to reduce the distance between PMTs and provide a more uniform coverage.
The baseline design for the Hyper-Kamiokande outer detector therefore uses \numrange{10000}{20000} \SI{7.7}{cm} PMTs for a photocoverage of \numrange{0.21}{0.42}\,\% as well as wavelength shifting plates to increase light collection.
Performance studies to investigate the veto efficiency as a function of the number of PMTs are currently ongoing.

\subsubsection{PMTs}
The ET9302KB~\cite{ET9302} and ET9320KFLB~\cite{ET9320KFLB} models manufactured by Electron Tubes have been studied for use in the outer detector.
Both are \SI{7.7}{cm} PMTs with a peak quantum efficiency of about 30\,\% and a dark rate which, at \SI{400}{Hz}, is about ten times lower than for typical \SI{20}{cm} PMTs.
The gain of the ET9302KB (ET9320KFLB) is \num{3e6} at \SI{950}{V} (\SI{800}{V}) and the charge response is linear within a few percent up to a light intensity of \num{1500} photoelectrons.
The ET9302KB (ET9320KFLB) has a rise time of \SI{7.5}{ns} (\SI{2.5}{ns}), both providing a timing resolution that is much better than required for the active veto.

Compared to the \SI{50}{cm} photosensors in the inner detector, the volume of outer detector PMTs is much smaller while the distance between them is larger.
In the case of an implosion, the resulting pressure wave would thus be much smaller and there is no risk of a chain reaction.
As in Super-Kamiokande, no PMT covers are therefore planned for the outer detector.

\subsubsection{Light Collection}
\begin{figure}[tbp]
	\centering
	\includegraphics[scale=1.1]{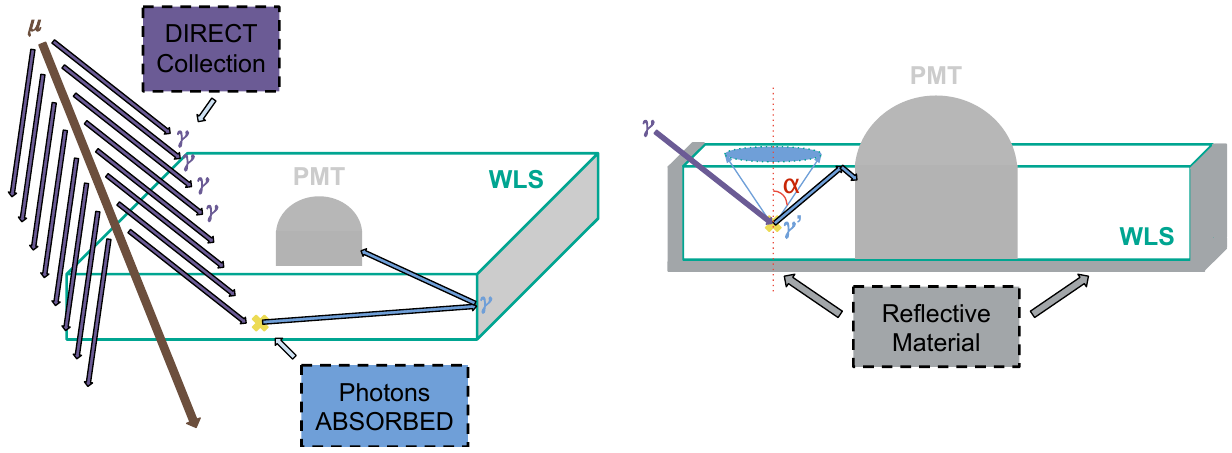}
	\caption[Schematic drawing of the wavelength shifting plate]{Schematic drawing of the wavelength shifting plate. Figure from reference~\cite{HKTR2018}.}
	\label{fig-hk-wlsplate}
\end{figure}

To increase the photosensitive surface, wavelength shifting plates (model Eljen EJ-286) will be mounted around each PMT with light coupling achieved through close contact between the plate and the side of the PMT’s hemispherical photocathode.
The plate absorbs UV light at \SIrange{280}{400}{nm} and re-emits approximately one secondary photon for every absorbed photon.
Figure~\ref{fig-hk-wlsplate} shows a conceptual drawing of this process.

The secondary photon is emitted with a wavelength of \SIrange{410}{460}{nm}, where the quantum efficiency of the PMT is near its peak, and at a random angle.
The refractive index of the WLS plate is 1.58 and the critical angle at a surface with water is \ang{57}, meaning that approximately 54\,\% of light is trapped inside the plate due to total internal reflection.
Very close to the PMT, almost half the trapped light is emitted in the direction of the PMT.
At larger distances $r$ from the PMT, this geometric factor is inversely proportional to $r$, though adding reflectors to the sides of the WLS plate can slow down this decrease.

Based on computer simulations and laboratory tests, adding a WLS plate with dimensions $\SI{60}{cm} \times \SI{60}{cm}$ around each PMT would increase light collection in the outer detector by a factor of more than 3, leading to an effective photocoverage of about 1\,\%.

While the primary goal of the outer detector is to identify muons entering the detector, tagging outgoing muons is desirable in the analysis of events from atmospheric or beam neutrinos.
The Cherenkov light from outgoing muons passing through the outer detector is largely emitted towards the outside wall of the tank, away from the photosensors.
By covering the outer wall with a sheet of highly reflective Tyvek, these photons can be collected by the outer detector PMTs or WLS plates after a delay of about \SIrange{5}{10}{ns}, which is shorter than the timing resolution required for the active veto.

\subsection{Electronics}\label{ch-hk-electronics}

The main tasks of the front-end electronics are to provide PMTs inside the detector with the high voltage necessary for regular operations and to collect and digitize time and charge data for all detected hits.
Where necessary, front-end electronics will then buffer this data, before finally delivering it to the data acquisition system described in section~\ref{ch-hk-daq}, which performs triggering and reconstruction.

The electronics are designed to provide the \si{ns}-level timing precision that is necessary to reconstruct events and to deal with very high data rates from a combination of the dark noise in each PMT, radioactive backgrounds in the detector and the possibility of extremely high peak event rates in the case of a nearby SN.%
\footnote{If Hyper-Kamiokande were to observe the supernova explosion of Betelgeuse, a red supergiant with a distance of about \SI{0.2}{kpc} from Earth, it is expected to observe a peak event rate of approximately \SI{e8}{Hz}.}

Since the planned lifetime of the detector is at least \SI{20}{\years} and many components cannot be replaced while the detector is running, we also require a high reliability of components and have added redundancy in some key areas.

In Super-Kamiokande~\cite{Fukuda2003}, every PMT is connected to a single cable which provides it with the high voltage necessary to operate and transports away the analog output signal.
All cables lead to the dome on top of the detector, where four "electronics huts" (one per quadrant of the detector) contain high voltage power supplies and electronics racks that digitize the signals from all PMTs.
A “central hut” in the dome combines the digitized signals from all quadrants and contains a trigger system and control electronics.
The readout electronics and trigger system have changed multiple times during the operation of Super-Kamiokande and are beyond the scope of this thesis.

In Hyper-Kamiokande, such a design is not practical any more due to the increased size of the detector and number of PMTs.
The total weight of the cables would reach several hundred tons, which would require a stronger and more expensive support frame, while the length of cables would need to be larger than \SI{100}{m} for some PMTs, which would degrade the quality of the analog signal.

Instead, front-end electronics modules mounted on the PMT support structure will each have a power and data connection to the electronics huts located in the dome on top of the detector.
Every module then provides high voltage power to 24 nearby PMTs and receives and digitizes the signal from those same PMTs before transferring it to the data acquisition system located in the dome.

However, placing the electronics modules inside the water tank creates a new set of challenges.
Modules need to be contained in a watertight and pressure-resistant case and require watertight connectors for incoming and outgoing connections to transfer power and data.
In the current, preliminary design, the case will be constructed mainly from stainless steel to ensure efficient heat exchange.
Heating the water in the inner detector too much would lead to convection, which decreases the efficiency of the water purification system as discussed in section~\ref{ch-hk-water}.
This effectively limits the power consumption of the front-end electronics module and needs to be taken into account during development.

In this section, I will first describe the requirements and preliminary designs of the electronics modules and its individual components for the baseline design of the inner detector, which uses exclusively \SI{50}{cm} photosensors.
For the outer detector and for alternative inner detector designs that include mPMT modules, possible modifications are described below.
Development of the components and design of the interfaces and communication protocols between them is still ongoing and many details are not yet determined.
In some cases, multiple designs for a given component exist; I will summarize those briefly.

\subsubsection{Inner Detector}

\begin{figure}[tbp]
	\centering
	\includegraphics[scale=0.79]{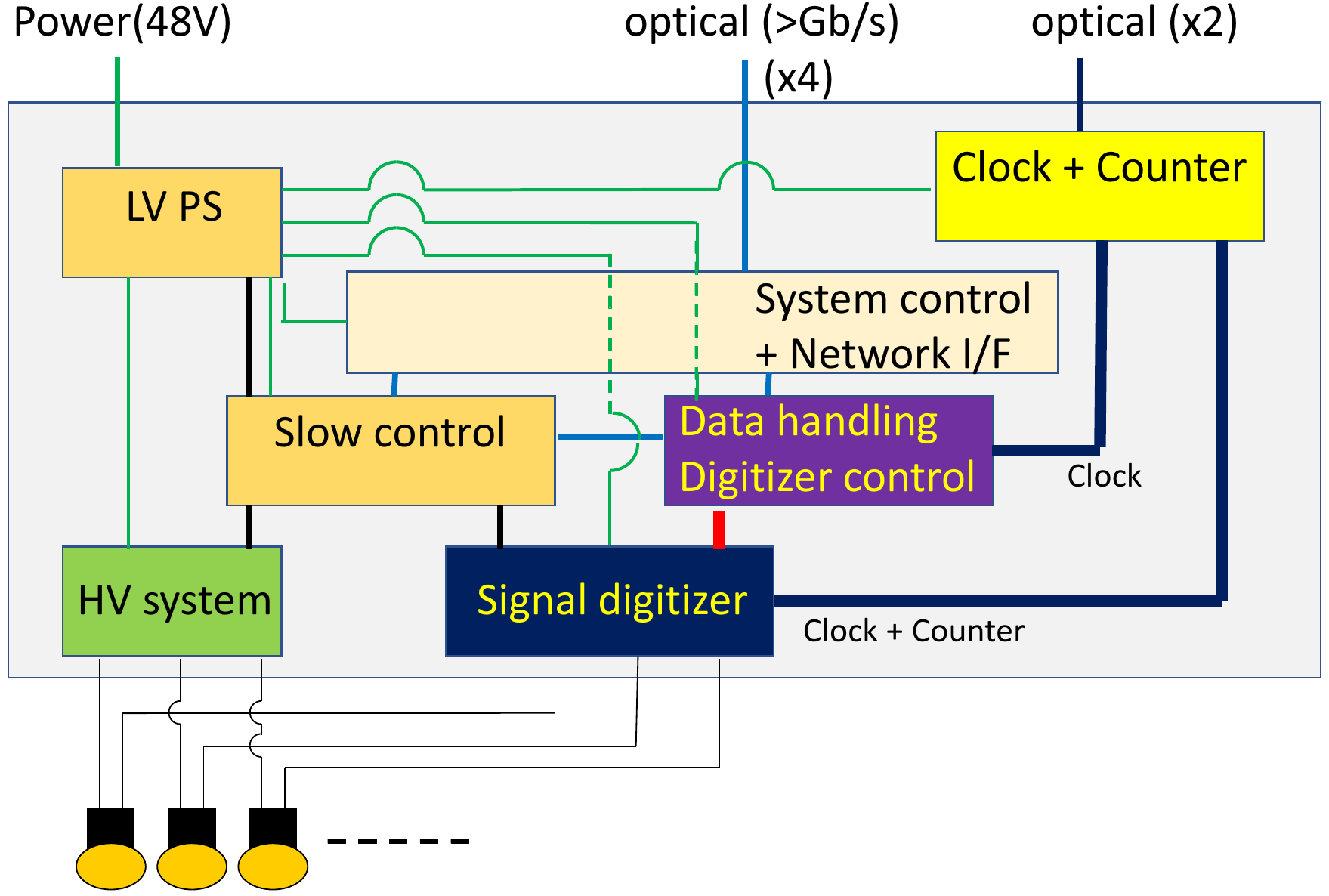}
	\caption[Block diagram of the front-end electronics module for Hyper-Kamiokande]{Block diagram of the front-end electronics module for Hyper-Kamiokande. See text for description of individual components. Figure from reference~\cite{HKTR2018}.}
	\label{fig-hk-fee-blockdiagram}
\end{figure}

Figure~\ref{fig-hk-fee-blockdiagram} gives an overview over the components of the electronics module for the inner detector, which are described in this section.

\begin{description}
\item[Power Supply] \hfill \\
DC power is produced in the dome on top of the detector and provided to each electronics module via a single cable.
A voltage of \SI{48}{V} is assumed here, though this could be changed to \SI{24}{V} or \SI{12}{V} if a lower voltage is advantageous. 
A low voltage power supply block (LV PS) provides power to other blocks in the electronics module, while a separate high voltage (HV) system transforms this to the stable and low noise voltage required by the  photosensors, which is \SIrange{1.5}{2.5}{kV} for the \SI{50}{cm} B\&L PMTs and MCP PMTs or \SI{8}{kV} for HPDs.
Power consumption for the HV system should be below \SI{1}{W} per channel or \SI{24}{W} per module.

\item[Signal Digitizer] \hfill \\
The signal digitizer digitizes the charge information from 24 PMTs and needs to be able to handle a maximum data rate of \SI{1}{MHz} per PMT.
There are currently two different designs for this component under development.

The first design combines a charge-to-time converter (QTC) with a time-to-digital converter (TDC).
Whenever the signal pulse height from the PMT exceeds the threshold of \SI{0.25}{PE}, the QTC integrates the charge in a predefined time window and outputs a single square shape pulse whose start time indicates the signal time and whose length is proportional to the integrated input charge.
The TDC, which is synchronized to an external reference clock described below, then digitizes the time of the rising and falling edge.
Since the output signal of the QTC has a finite width, this system introduces a dead time, which can be kept below \SI{1}{\micro\second} and therefore satisfies the requirement of a \SI{1}{MHz} peak event rate.
This design is similar to that used in Super-Kamiokande since 2008.
Based on over ten years of experience, this design has been shown to be satisfy the requirements of Hyper-Kamiokande. 

An alternative design would digitize the waveform directly and thus be deadtime-free.
This could be done either by using a Flash ADC (analog-to-digital converter) with a sampling rate of at least \SI{100}{MHz} or by a capacitor array, which could reach extremely fast sampling rates of up to a few \si{GHz}.

Work on all these approaches is currently ongoing.
Total power consumption for the digitizer should be below \SI{1}{W} per channel or \SI{24}{W} per module.

\item[Clock and Counter System] \hfill \\
The goal of the clock system is to provide a detector-wide uniform timing information with a long-term stability better than \num{2e-11} and a phase change across different parts of the detector, e.\,g. due to jitter or a reset of individual components, of less than \SI{100}{ps}.

A master clock generator located in the dome above the water tank uses an atomic clock and GPS signals as an external reference and provides a reference clock, counter and control signal.
These signals are sent via optical fibres to distributor modules located in the dome and each distributor then distributes this signal to 48 electronic modules in the water via individual optical fibres.

In the electronics module in the water tank, a clock and counter block with two optical interfaces for increased fault tolerance receives these signals and distributes them to the signal digitizer and data handling blocks.
It also receives status information and number of hits from the digitizer and sends them back to the distributor.

\item[Data Handling and Digitizer Control] \hfill \\
The data handling and digitizer control block receives data from the signal digitizer and buffers it before sending it to the readout system.
This requires at least \SI{8}{GB} of dedicated memory per module, to buffer all hits for up to several minutes.
The block also checks for errors in received data and could include data compression to reduce memory usage.

Based on commands it receives from the readout system, it can initiate pedestal data taking or other calibration data taking.
It also receives commands to control other blocks, including slow control and high voltage, which it processes and then forwards accordingly.

This block may be integrated with the system control and network interface into one module.

\item[Slow Control] \hfill \\
The slow control block monitors environmental conditions inside the electronics module using temperature and humidity sensors.

It also monitors voltages and currents of the low voltage and high voltage systems and can control the current and voltage provided to each channel by the HV system following commands from the digitizer control block.

\item[System Control and Network Interface] \hfill \\
The system control and network interface block provides communication between other blocks in the module and the main electronics outside of the water tank via multiple optical interfaces to ensure redundancy and fault tolerance.
Data transfers are expected to use TCP/IP and support error detection and correction, while control commands and monitoring are expected to use UDP/IP.

\end{description}

\subsubsection{Outer Detector}
For the outer detector, front-end electronics are likely to be largely similar to those for the inner detector described above.

Since the \SI{7.7}{cm} PMTs in the outer detector require a lower voltage of \SIrange{600}{1100}{V} and their use as a veto necessitates less precise charge and timing information, the signal digitizer and HV system may be replaced with simpler and cheaper designs.
However, cost savings from mass production of uniform electronics for both inner and outer detector may outweigh the advantages of simpler components.

\subsubsection{Multi-PMT Modules}
\begin{figure}[tbp]
	\centering
	\includegraphics[scale=0.8]{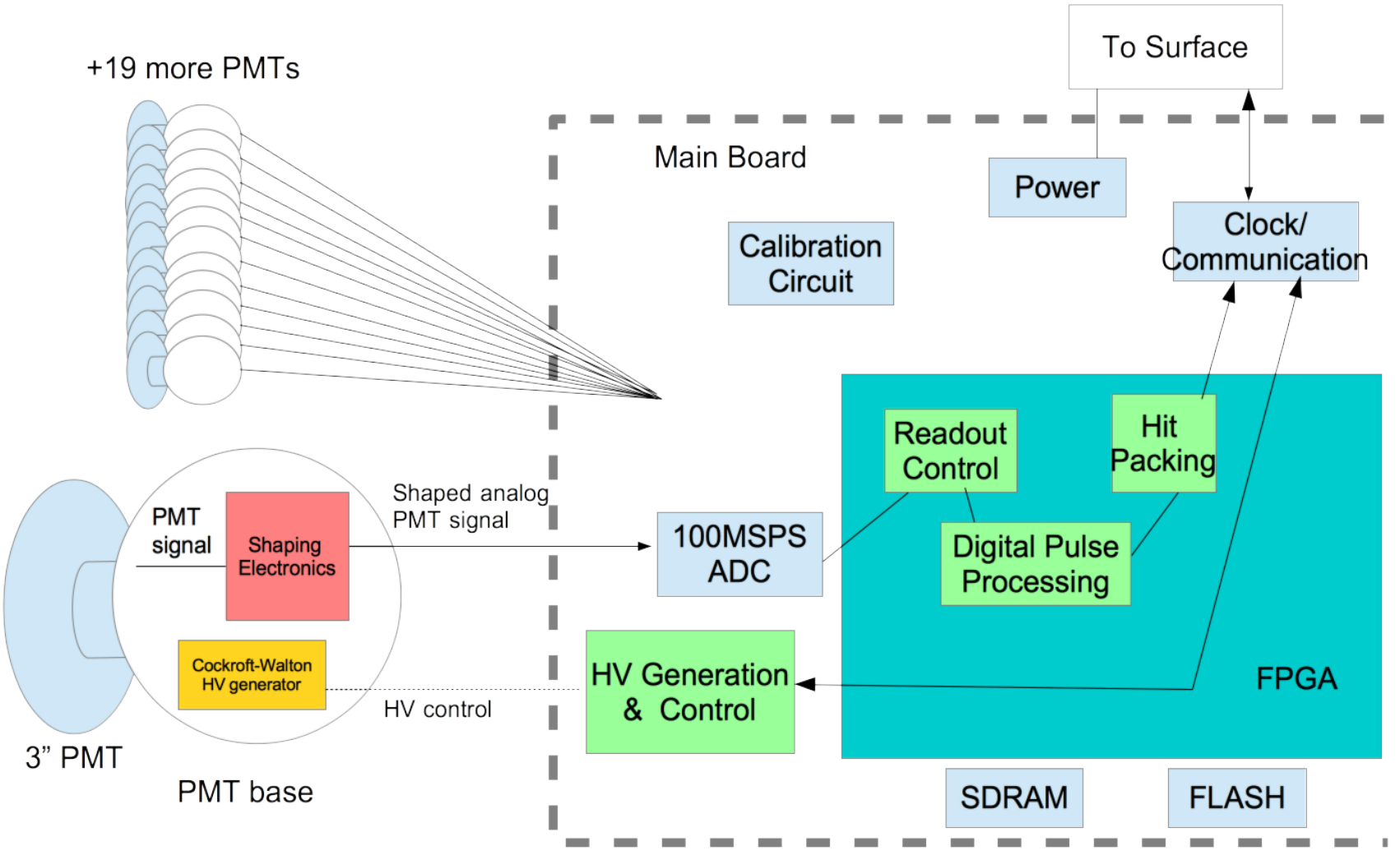}
	\caption[Block diagram of the front-end electronics inside an mPMT module]{Block diagram of the front-end electronics inside an mPMT module. Figure from reference~\cite{HKTR2018}.}
	\label{fig-hk-mPMT-electronics}
\end{figure}

The mPMT modules contain individual front-end boards for every PMT and one main board per module, as shown in figure~\ref{fig-hk-mPMT-electronics}.
The main board largely fulfils the same role as the inner detector electronics modules for the inner detector described above, combining power supply, digitization, clock and communication.
Front-end boards located at the base of each individual PMT contain a Cockcroft-Walton voltage multiplier that generates the high voltage required by the PMT.
Designs for the individual blocks are still being worked on.

\subsection{DAQ and Computing}\label{ch-hk-daq}

After the front-end electronics located in the water read out and digitize data from individual photosensors, this data is sent to the data acquisition (DAQ) system in the dome above the water tank.
The goal of the DAQ system is to combine data from all PMTs, identify and reconstruct events in the detector and write them to storage for later analysis.
This section will first discuss the basic design of the DAQ system and its operations, before describing the separate operations mode designed for supernova bursts.
Finally, the multi-tiered computing system used for data analysis will be described briefly.

\subsubsection{Design of the DAQ System}
\begin{figure}[tbp]
	\centering
	\includegraphics[scale=0.8]{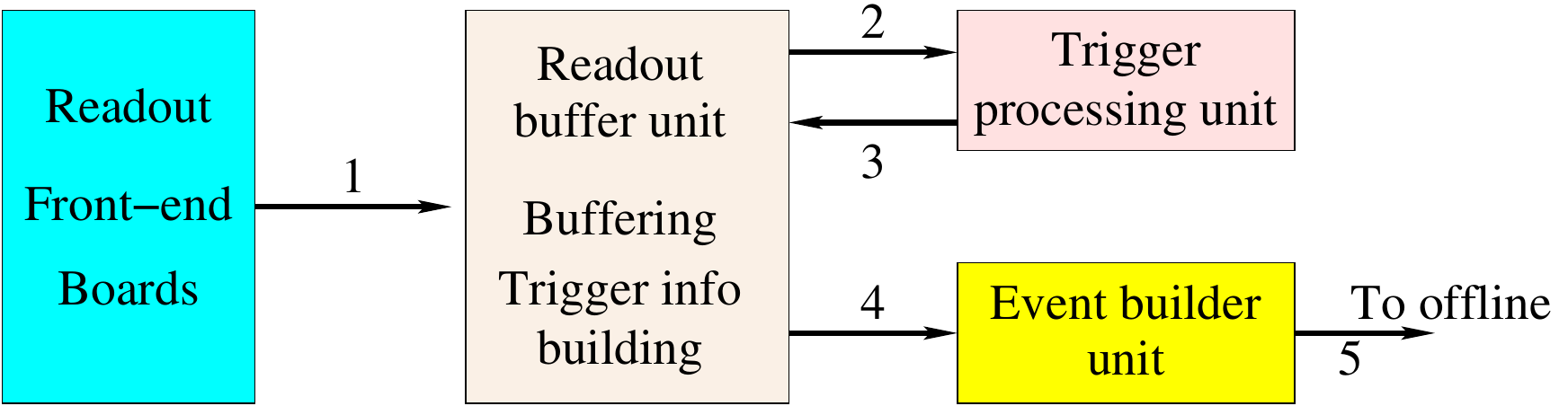}
	\caption[Simplified block diagram of the DAQ system for Hyper-Kamiokande]{Simplified block diagram of the DAQ system for Hyper-Kamiokande. Figure from reference~\cite{HKDR2018}.}
	\label{fig-hk-daq}
\end{figure}

The basic design of the DAQ system is shown in figure~\ref{fig-hk-daq}.
The system will be implemented using the ToolDAQ framework~\cite{Richards2018}, and consist of four main components which are built using off-the-shelf server hardware. 

Due to the high expected data rates, the system is designed to be modular and highly parallelizable.
The ToolDAQ framework uses messaging protocols, redundant connections and buffers along with dynamic service discovery to increase fault tolerance by detecting and dynamically replacing unresponsive computing nodes.

\begin{description}
\item[Readout Buffer Units (RBU)] \hfill \\
The DAQ system will contain approximately 70 RBUs that are connected to the front-end electronics modules in the water via a gigabit network switch, allowing data to be rerouted to other RBUs if a failure occurs.
Each RBU is responsible for reading out the digitized signals from about 30 front-end electronics modules in the water.
It then buffers all data in active memory for about 100 seconds and temporarily saves older data to hard drives for about one hour.

During normal operations, RBUs additionally reduce data by eliminating all PMT hits where the signal is below a threshold of 0.25\,photoelectrons.
This reduced data in a given time window is then provided to trigger processing and event building units upon request.

\item[Trigger Processor Units (TPU)] \hfill \\
Once all data is read out by the RBUs, the TPUs will analyse every time window in the data for possible signals.
To do this, the TPU requests the reduced data for this time window from RBUs and then applies various trigger algorithms to look for events.

The simplest trigger is the “NDigits” or “Simple Majority” trigger, which applies a sliding time window to the data and triggers if the number of events in that window surpasses a given threshold.
This trigger will be able to identify high-energy events but is not viable for low-energy events due to the high dark-noise rates of the PMTs.
Instead, time windows that fail this NDigits trigger get passed on to a more sophisticated trigger optimized for low-energy events.

This “Vertex Reconstruction” trigger relies on the fact that a low-energy lepton travels only a few centimetres in water before its energy falls below the Cherenkov threshold, so it is well approximated as a point source that emits all Cherenkov photons at the same time and position.
The trigger uses a uniformly spaced three-dimensional grid of test vertices in the detector.
For each vertex, it corrects the recorded hit times in all PMTs by the time-of-flight of a Cherenkov photon originating at that vertex and then applies an NDigits trigger to a shorter sliding time window of \SI{20}{ns}.
If the test vertex is close to the true vertex of an event, this time-of-flight correction leads to a narrow peak in the corrected arrival times, while dark noise hits remain randomly distributed.

Furthermore, the TPUs will take into account external calibration triggers as well as GPS timestamps sent from the J-PARC accelerator to help identify events during beam spills.

Since all triggers are software-based, they can be updated if algorithmic improvements are available or increases in computing power enable a lower threshold.

\item[Event Builder Units (EBU)] \hfill \\
Once a TPU has identified an event, the timestamps of that event are sent to an EBU, which requests the data in that time window from the RBUs.
The EBU then identifies hits within the trigger time window that are associated with the event and writes them to disk for permanent archival.

\item[Brokers] \hfill \\
A central broker is tasked with coordinating operations of the DAQ system.
To increase fault tolerance, two identical machines act as broker.
The primary one handles all tasks during normal operations, while the secondary keeps track of all decisions of the primary and is ready to replace it at any time in case of a failure.

The broker distributes tasks to TPUs and EBUs.
To reduce load, it does not transfer the data itself; instead, it tells a TPU or EBU which time window to analyse and the TPU or EBU will then request the data for that time window from the RBUs directly.

The broker handles failures of individual TPUs or EBUs by redistributing jobs to other available units, while failures of individual RBUs are handled by reassigning front-end electronics modules to other RBUs.

\end{description}

\subsubsection{Supernova Mode}
The DAQ design also contains two dedicated supernova trigger machines that examine the event rate in each \SI{1}{ms} time slice as well as in a sliding \SI{20}{s} time window.

If a significant increase in a \SI{20}{s} window is detected, which could be the signal of a distant supernova, all data from that time period is saved to long-term storage.
In the case of a galactic supernova, an increased event rate should be visible in a \SI{1}{ms} time slice shortly after the start of the burst.
In that case, an alert is sent out to all machines in the DAQ system to switch into a dedicated supernova operations mode.
While in this mode, RBUs will temporarily stop the processor-intensive data reduction and stream all data to the EBUs for permanent storage as fast as the network connections allow.

Buffer capacities and bandwidths throughout the DAQ system are designed to be able to handle a nearby supernova at a distance of \SI{0.2}{kpc} with a peak event rate of about \SI{e8}{Hz}, corresponding to peak data rates of about \SI{100}{GB/s}.

Like Super-Kamiokande, Hyper-Kamiokande is likely to participate in the SuperNova Early Warning System (SNEWS~\cite{Scholberg2000}) and would send out an alert to that system in the case of a supernova trigger.


\subsubsection{Computing}
As common in modern high-energy physics experiments, Hyper-Kamiokande will adopt a multi-tiered computing system based on the Worldwide LHC Computing GRID.
Kamioka, which hosts the Hyper-Kamiokande detector, as well as KEK, which hosts the neutrino beamline and near detector, will be Tier-0 sites storing all raw event data.
Several Tier-1 sites hosted by major research facilities distributed around the world will store all reduced data, while individual institutions participating in the experiment will typically host Tier-2 sites that store subsets of the data as required.

The Hyper-Kamiokande software will be made available via the Cern Virtual Machine File System (CVMFS), a read-only file system optimized for software distribution, to ensure that all users have access to the most recent versions.

\section{Calibration}\label{ch-hk-calibration}

In a sensitive detector like Hyper-Kamiokande, precise understanding of all components of the detector is essential to be able to properly identify and reconstruct events.
Gaining that understanding requires extensive commissioning before the start of operations, which includes comprehensive calibrations.
Furthermore, variations in water quality, defects, ageing of photosensors or a number of other effects lead to changes of the detector response over time which requires regular re-calibration throughout the lifetime of the detector.

Super-Kamiokande has been running successfully for over \SI{20}{\years} and the collaboration has developed expertise in calibrating a large water Cherenkov detector.
The design of the Hyper-Kamiokande calibration system will therefore be based on the techniques used in Super-Kamiokande, while adding improvements in crucial areas.
In particular, the increased detector volume compared to Super-Kamiokande means that calibration needs to be performed in more locations inside the detector, which requires increased automation.

In this section, I will first discuss initial calibration of photosensors before installation in the detector.
I will then discuss the various planned calibrations that will be performed regularly during detector operations and that are essential for reconstructing low-energy events.
At high energies, additional natural particle sources like cosmic ray muons or $\pi_0$ events can be used for calibration; however, these are beyond the scope of this thesis. 
Finally, I will discuss the deployment infrastructure for radioactive sources. 
Since results of calibration and comparison of data with MC simulations will only be available once the detector is actually running, in this thesis I will, by necessity, only give a qualitative overview over the calibration strategy.

\subsection{Pre-Calibration of Photosensors}

Before installation in the inner detector, all \SI{50}{cm} photosensors will undergo brief, automated tests to reject any defects and to confirm that properties like the gain and dark noise rate are within the expected ranges.

About 2\,\% of photosensors will undergo more extensive tests to measure the quantum efficiency and characterize the gain as function of high voltage.
These tests will take place in a dark room surrounded by coils that completely compensate the geomagnetic fields.
Photosensors calibrated like this will be installed uniformly throughout the detector and used as a reference for calibrating the high voltage settings of other sensors to ensure equal gain throughout the detector.

In addition, about 0.5\,\% of photosensors will undergo extensive tests to determine the gain, quantum efficiency and timing performance as a function of the location and angle of incident light, as well as the residual magnetic field.
The performance of the \SI{50}{cm} B\&L PMTs is known to depend significantly on these factors and this detailed characterization is essential to account for photosensor performance in event reconstruction.

All \SI{7.7}{cm} PMTs used in the outer detector will undergo similar automated tests to reject defects and check basic properties.
In addition, once they are attached to the wavelength-shifting plates, automated tests will check whether photon hits on the plates are registered by the PMTs.

Since the outer detector is only used as an active veto and does not need to reconstruct Cherenkov rings, calibration requirements for it are less strict.
Only about 0.1\,\% of PMTs will undergo extensive tests which include measurements of the gain and dark noise rate as a function of high voltage, of the charge and timing resolution as well as of the spatial dependence of the quantum efficiency.
The smaller PMTs have a much lower sensitivity to residual magnetic fields, which therefore does not need to be measured.

If mPMT modules are used in the inner detector, they will undergo a separate calibration procedure starting with tests of individuals PMTs similar to those described above.
After those tests, fully assembled modules will undergo additional tests to characterize the photon detection efficiency as a function of hit position. These include effects of the reflectors around each PMT and changes to the reflectivity due to the pressure vessel.

\subsection{Light Sources}

The Hyper-Kamiokande calibration system will use a commercially available Xenon lamp located in the dome above the tank, which is connected via an optical fibre to an acrylic ball containing \num{2000}\,ppm of MgO as a diffuser to ensure uniform light emission in all directions.
This diffuser ball is then lowered into the centre of detector to produce uniform illumination of all photosensors in the inner detector.
Using the 2\,\% of photosensors that underwent more extensive pre-calibration as a reference, this will be used during initial detector commissioning to tune the high voltage settings for every individual photosensor and equalize the detector response.
During data taking, the Xenon lamp can be used to monitor uniformity of response throughout the detector and correct for long-term drift of individual PMTs.

Hyper-Kamiokande will include an integrated light injection system that will be used to regularly monitor the optical properties of the water like scattering and absorption as well as their spatial dependence in both the inner and outer detector.
It will also be used to monitor photosensor timing, drift of gain over time and multi-photon response.

The light injection system uses light sources in the dome on top of the tank, which could be LEDs or similar sources capable of producing light pulses of approximately \SI{1}{ns} length in several different wavelengths in the range from \SIrange{320}{500}{nm}.
This light source is connected via optical fibres to a number of permanently installed light injection points which are located on the support structure in gaps between photosensors for the inner detector and in the detector wall for the outer detector.
Light could be injected into the detector by a collimator, which produces a narrow-angle beam illuminating only a small number of PMTs, by a naked fibre, which produces a light cone with approximately a \ang{12} opening angle, or by a diffusor which produces a wide-angle beam to illuminate a high number of PMTs.

For the inner detector, a system combining all three injection approaches has been designed and was deployed for tests in Super-Kamiokande in the summer of 2018.
The combination of all three approaches creates a versatile system that allows a wide range of calibration schemes and cross checks.

The outer detector has a width of \SIrange{1}{2}{m}, so a narrow beam would only reach individual PMTs.
To limit the number of injection points required, the calibration strategy for the outer detector will therefore rely solely on the diffuser.
This is sufficient to fulfil the less stringent calibration requirements for an active veto that does not need to precisely reconstruct events.

\subsection{Radioactive Sources}

Three different radioactive sources are planned to be used for calibration in Hyper-Kamiokande.

A nickel-californium source consists of a Cf source surrounded by a sphere of high-density polyethylene containing NiO$_2$ powder.
$^{252}$Cf produces neutrons which are thermalized and then captured on $^{58}$Ni.
When the resulting excited state of $^{59}$Ni decays, it releases gamma rays with a total energy of approximately \SI{9}{MeV}.
This can act as a uniform, stable source of low-intensity Cherenkov light to measure the gain and single photoelectron charge distribution of photosensors.
Since the signal contains a background component coming mainly from neutron capture on hydrogen, this source cannot be used to calibrate the absolute energy scale.

An americium-beryllium source uses $\alpha$ decays of $^{241}$\!Am to induce an ($\alpha$, n) capture reaction on $^9$Be.
The resulting excited state of $^{12}$C then decays emitting a \SI{4.44}{MeV} $\gamma$, while the free neutron gets captured on hydrogen nuclei on a timescale of several \si{\micro\second}, resulting in a \SI{2.2}{MeV} $\gamma$ emission.
The coincidence of the $^{12}$C deexcitation signal and the neutron capture signal will be used to determine the neutron capture detection efficiency of Hyper-Kamiokande.
If gadolinium is added to the detector, about half the neutrons would get captured on gadolinium, producing several photon with a total energy of about \SI{8}{MeV}.
In that case, this calibration source could be used to determine the relative capture rates on Gd and H as well as the combined neutron detection efficiency.

The third calibration source uses a commercial deuterium-tritium generator to produce \SI{14.2}{MeV} neutrons in the water.
This is above the \SI{11}{MeV} threshold to produce $^{16}$N from $^{16}$O nuclei in the water via a (n, p) reaction.
$^{16}$N decays with a half life of \SI{7.13}{s} producing both an electron with $\beta$ endpoint of \SI{4.3}{MeV} and a \SI{6.1}{MeV} $\gamma$ in the dominant decay branch.
This calibration will be repeated at multiple vertices within detector to calibrate the energy scale as a function of position and direction.
If a LINAC cannot be used to calibrate Hyper-Kamiokande, this source will also be used to calibrate the collection efficiency of photosensors and to fix the absolute energy scale.

\subsection{LINAC}
The Super-Kamiokande calibration system contains a LINAC that delivers a low-intensity beam of single electrons with a well-defined energy of around \SI{6}{MeV} or \SI{13}{MeV} to calibrate the absolute energy scale at low energies.
The beam is transported into the dome and deployed into the water using a guide tube that allows injecting electrons into the tank at varying depth.
For Hyper-Kamiokande, to achieve the physics goals in full, it is required to reduce the energy scale uncertainty to below 0.3\,\% and the energy resolution uncertainty to 2\,\% across the whole fiducial volume, which can only be achieved with a LINAC.

While the design of the LINAC system for Hyper-Kamiokande is not yet completed, two main improvements are currently under consideration.
A commercially available LINAC model could be used that produces electrons with a wider range of energies of \SIrange{4}{20}{MeV}.
At the bottom end of the guide tube, a magnet system could be installed to vary the electron direction in the tank and probe the dependence of energy scale on the particle direction, which is one of the major uncertainties in Super-Kamiokande.

\subsection{Deployment System}

Due to the large detector volume, calibrating Hyper-Kamiokande requires deploying sources in a large number of different positions throughout the detector to understand the position and direction dependence of reconstruction.
To reduce the manual labour required and the detector downtime associated with calibration, a source deployment system needs to be designed.

For Hyper-Kamiokande, the deployment system is designed to allow vertical deployment through calibration ports located along two perpendicular axes across the detector, exploiting cylindrical symmetry to achieve a full three-dimensional calibration.

Different types of calibration ports will be provided.
Regular ports will have a diameter of approximately \SI{22}{cm} and be located in gaps between neighbouring photo\-sensors.
They will typically be located at a distance of \SI{3}{m} from each other, which is reduced to \SI{0.5}{m} near the edges of the inner detector to achieve precise calibration near the edge of the fiducial volume and reduce systematic uncertainties.
Regular size ports will be used both for deployment of sources and for the LINAC, though the LINAC may have higher loading requirements due to the more complicated transport and deployment system.
These ports will also be available in the outer detector.
Near the centre of the detector, one oversized port with a diameter of at least \SI{75}{cm} will be provided for deployment of larger sources, which may require removal of one photosensor.

A prototype of the automatic source deployment system for radioactive sources designed for Hyper-Kamiokande was installed in Super-Kamiokande during 2018 for testing.

\section{Backgrounds}\label{ch-hk-background}

Supernova burst neutrinos are observed in the energy range of \SI{5}{MeV} to almost \SI{100}{MeV}.
At the upper end, this overlaps with the low end of the energy range of atmospheric neutrinos, 
while at intermediate energies of about \SIrange{20}{40}{MeV}, supernova relic neutrinos are another main signal that Hyper-Kamiokande will try to observe.
Below about \SI{20}{MeV}, Hyper-Kamiokande will observe solar neutrinos, as well as a small number of reactor antineutrinos near its lower energy threshold.

However, the event rate of each of these conflicting signals is between less than one and up to about one hundred events per day, with zero or at most one event expected during the \SIrange{10}{20}{s} duration of a supernova burst.
These are therefore negligible as background sources.

The main sources of background in the energy range relevant for supernova burst neutrinos are radioactive decays in the detector with energies below \SI{5}{MeV} and spallation events induced by cosmic ray muons, which can reach energies of up to \SI{20}{MeV}.
Since these backgrounds cover the full energy range of solar neutrinos, which are a major area of physics in both Super- and Hyper-Kamiokande, they have been investigated in great detail and are well understood.
The Super-Kamiokande collaboration has been able to reduce these backgrounds significantly, which has enabled them to routinely detect and perform precision measurements with solar neutrinos down to a kinetic energy of \SI{3.49}{MeV} in recent analyses~\cite{Abe2016a}.
In this section, I will discuss these backgrounds in Hyper-Kamiokande and approaches to eliminate them as far as possible.

Since the flux of supernova burst neutrinos is several orders of magnitude higher than that of solar neutrinos, with about \numrange{e4}{e6} events expected in a \SIrange{10}{20}{s} time window, applying the background reduction techniques developed for solar neutrino analyses would guarantee an effectively background-free signal.
As discussed in section~\ref{ch-hk-daq}, the DAQ system for Hyper-Kamiokande is designed to save a complete, unreduced data set to permanent storage if a supernova burst is detected.
After a burst, it will therefore be possible to developed optimized triggering and background reduction algorithms and apply them to the data.
It will likely be possible to relax some of the cuts used in solar neutrino analyses and, for example, to increase the size of the fiducial volume, which has the potential to significantly increase the number of observed events.

Due to the advanced triggering algorithms described in section~\ref{ch-hk-daq}, dark noise of photosensors does not produce background events above \SI{5}{MeV} and will therefore not be considered as a background source here.
Its effects on energy reconstruction will be discussed in section~\ref{ch-sim-reco-e}.

\subsection{Radioactive Decays}
In Super-Kamiokande, radioactive decays in the detector are the dominant background at the lowest energies and effectively preclude Super-Kamiokande from detecting solar neutrino events with an energy below \SI{3.49}{MeV}~\cite{Abe2016a}.

Measurements in Super-Kamiokande found that the decay chain of $^{222}$Rn is the largest component of this background, while decay products of $^{220}$Rn give only a minor contribution.
Many daughter isotopes of $^{222}$Rn have an energy that is too low to produce Cherenkov light and the biggest exception is beta decay of $^{214}$Bi, which has a Q-value of \SI{3.27}{MeV}.

There are three main sources introducing radon into the Hyper-Kamiokande detector.

The first source is air inside the mine, which contains radon emanating from the surrounding rock.
Measurements with a new radon monitoring system~\cite{Pronost2018} showed that the radon level in the Tochibora mine, where Hyper-Kamiokande will be built, is about \SI{1200}{Bq/m^3}, which is comparable to that in the Mozumi mine near the Super-Kamiokande detector.
Like in Super-Kamiokande, a fresh air system will be installed for Hyper-Kamiokande, which pumps outside air into the experimental area to keep the radon level in the dome below about \SI{100}{Bq/m^3}.

Another source of radon is the rock surrounding the detector itself.
Traces of radon will be able to penetrate the surface of the tank and enter the detector.
A first estimate of the radon levels that will be introduced in this way indicates that the concentration in the outer detector will be \SI[parse-numbers=false]{\mathcal{O}(10)}{mBq/m^3}, similar to that observed in the Super-Kamiokande outer detector.
This is achieved to a large part 
by lining the cavern with a \SI{5}{mm} thick layer of HDPE which has a low radon permeability.
To achieve a more reliable estimate of the radon levels that can be achieved, the Hyper-Kamiokande collaboration is testing the radon permeability of HDPE sheets in water.
First results found the permeation to be less than \SI{5e-7}{cm^2 /s}, while high-sensitivity measurements are still ongoing.

From experience in Super-Kamiokande, we know that the radon concentration in the outer detector has little influence on the concentration in the inner detector as long as there is no water exchange between both.
In Hyper-Kamiokande, the inner detector will be completely separated from the outer detector. 

\begin{table}[tbp]
\begin{center}
\begin{tabular}{ll}
Radioactive Source & Requirement\\
\hline
U decay chain & $\leq \SI{3}{Bq}$ per PMT\\
Th decay chain & $\leq \SI{1}{Bq}$ per PMT\\
$^{40}$K & $\leq \SI{10}{Bq}$ per PMT\\
Rn emanation & $\leq \SI{3}{mBq/m^3}$
\end{tabular}
\end{center}
\caption[Upper limits on radioactivity for photosensors in the Hyper-Kamiokande inner detector]{Upper limits on radioactivity for photosensors in the Hyper-Kamiokande inner detector. Values from reference~\cite{HKTR2018}.}
\label{tab-hk-radioactivity}
\end{table}

The main source of radon in the inner detector is therefore expected to be impuri\-ties in the glass used in photosensors.
Requirements defined by the collaboration for \SI{50}{cm} PMTs are listed in table~\ref{tab-hk-radioactivity}.
We are currently in the process of determining the origin of radioactive impurities in prototype PMTs and aim to improve the production process to reduce these impurities.

In addition to photosensors, all other materials that are planned to be used inside the detector---including photosensor covers, the support structure, sheets separating the inner and outer detector, cables, electronics modules and components of mPMT modules---will undergo screening in an effort to reduce radioactive backgrounds through appropriate choice of raw materials or changes to production processes.

\begin{figure}[tbp]
	\centering
	\includegraphics[scale=0.42]{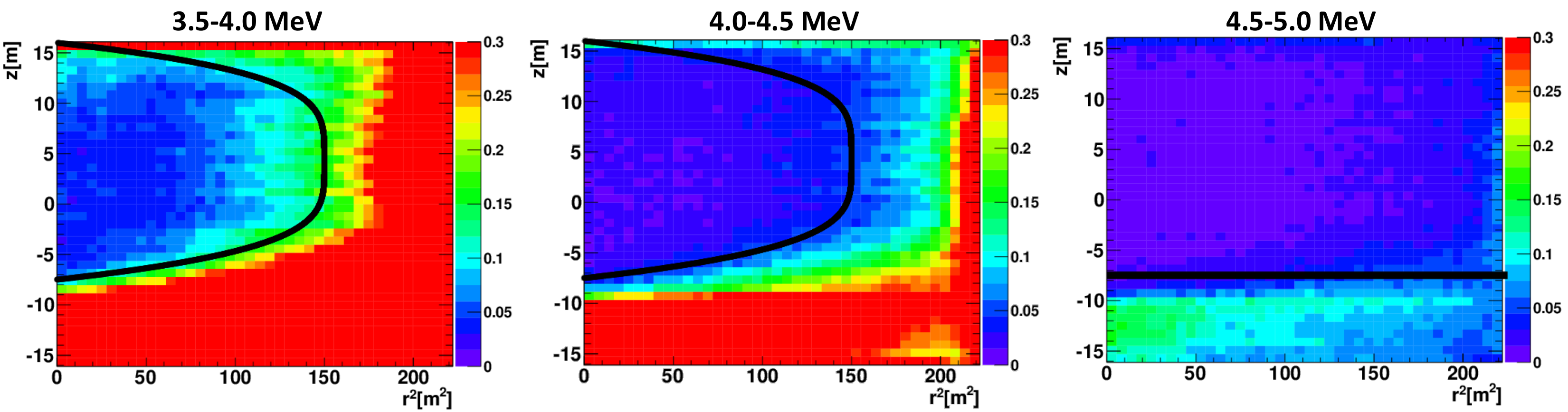}
	\caption[Distribution of low-energy events in Super-Kamiokande during phase IV]{Distribution of low-energy events in the Super-Kamiokande detector during phase IV. Black lines indicate the reduced fiducial volume at the respective energies, while the fiducial volume above \SI{5}{MeV} is the whole area, corresponding to \SI{22.5}{kt}. Colours show the event rate per day and bin, while r and z correspond to the horizontal and vertical axis of the detector, respectively. Figure from reference~\cite{HKDR2018}.}
	\label{fig-hk-sk4-vertex}
\end{figure}

As long as water flow in the detector is well-controlled, radioactive elements emanating from photosensors or other detector components will remain near the edges of the inner detector so that any background events are concentrated in that region.
Removing all events that are reconstructed to be outside of the fiducial volume, i.\,e. less than \SI{1.5}{m} away from the walls of the ID, therefore dramatically reduces these backgrounds.
For solar neutrino analyses in Super-Kamiokande, in the energy range below \SI{5}{MeV} the collaboration uses a much stricter fiducial volume cut shown in figure~\ref{fig-hk-sk4-vertex}.
That stricter cut eliminates all events with a distance of less than about \SI{5}{m} from the wall of the inner detector or less than about \SI{10}{m} from the bottom of the inner detector, since radioactive backgrounds are largest in those parts of the detector.

In the current Super-Kamiokande solar neutrino analysis, a number of additional cuts are used, which are based on the spatial and temporal distribution of observed hits~\cite{Abe2016a}.
While changes in detector geometry would likely require modifications to these cuts, they could, in principle, be used for background reduction in Hyper-Kamiokande as well.

\subsection{Muon-Induced Spallation}
Cosmic ray muons entering the detector can produce unstable isotopes via the interaction $\mu + ^{16}\text{O} \rightarrow \mu + \text{X}$ or via capture on $^{16}$O to produce $^{16}$N.
The resulting nuclei will then decay by emitting beta or gamma particles with an energy of up to about \SI{20}{MeV}, making them an important background for neutrino detection at these energies.

\begin{figure}[tbp]
	\centering
	\includegraphics[scale=0.6, angle=270]{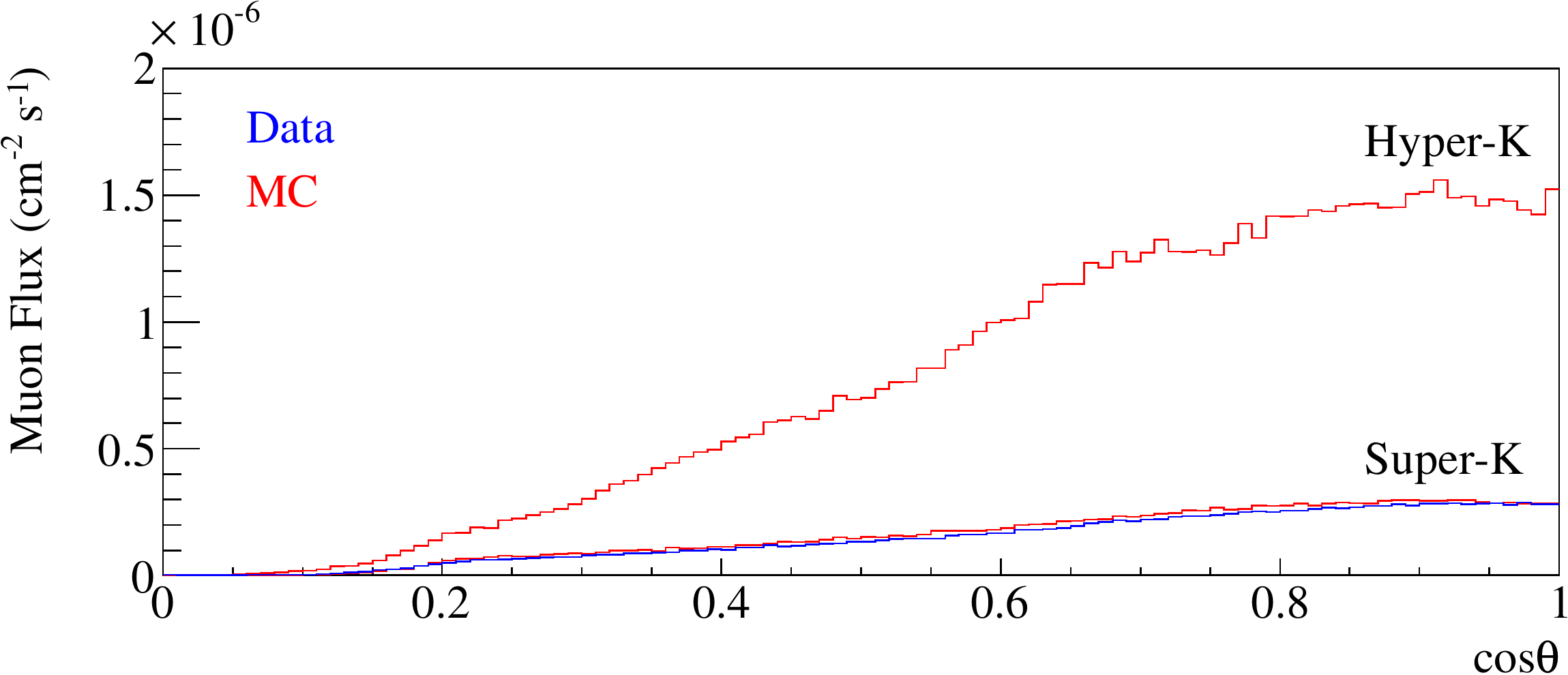}
	\includegraphics[scale=0.6, angle=270]{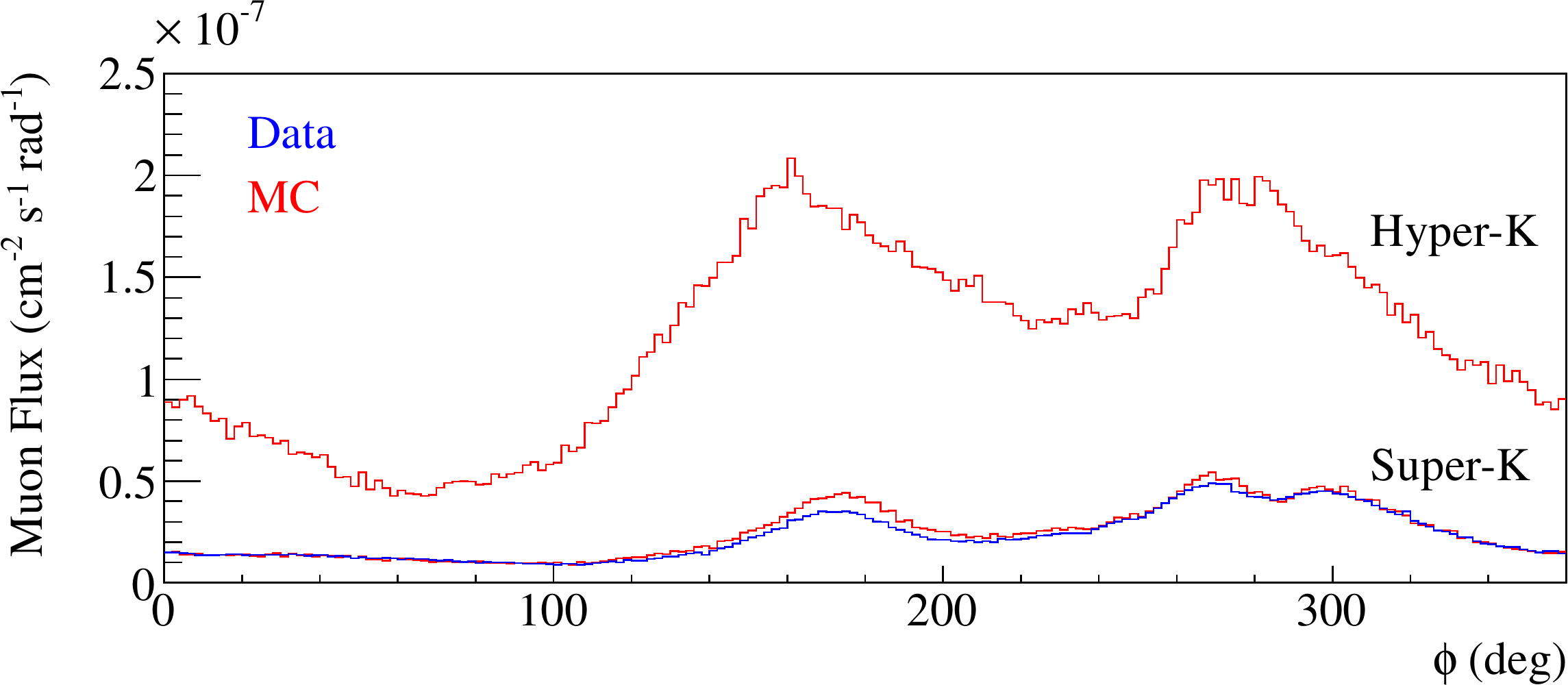}
	\caption[Muon flux as a function of zenith and azimuth angle in Super- and Hyper-Kamiokande]{Muon flux as a function of zenith angle $\theta$ (top) and azimuth angle $\phi$ (bottom). Red lines show the result of simulation with MUSIC in Super- and Hyper-Kamiokande, while blue lines show the measured fluxes in Super-Kamiokande. Figure from reference~\cite{HKDR2018}.}
	\label{fig-hk-MuonFlux}
\end{figure}

The flux and average energy of cosmic ray muons in the Tochibora mine, where Hyper-Kamiokande will be located, was simulated with the muon simulation code MUSIC~\cite{Antonioli1997} using a topological map with \SI{5}{m} mesh resolution and assuming a rock density of about \SI{2.7}{g/cm^3}.
Figure~\ref{fig-hk-MuonFlux} shows the directional dependence of the simulated muon flux in Super- and Hyper-Kamiokande as well as the measured muon flux in Super-Kamiokande, which shows good agreement with the simulation.

The expected muon flux in Hyper-Kamiokande is $J_\mu = \SI{7.55e-7}{cm^{-2} s^{-1}}$ with an average energy of $\mean{E_\mu} = \SI{203}{GeV}$, compared to $J_\mu = \SI{1.54e-7}{cm^{-2} s^{-1}}$ and $\mean{E_\mu} = \SI{258}{GeV}$ in Super-Kamiokande.
The higher overburden of Super-Kamiokande leads to a lower total flux than in Hyper-Kamiokande and has a relatively stronger shielding effect for low-energy muons, which leads to an increase in the average energy of observed muons.
Thus, while the muon flux in Hyper-Kamiokande is increased by a factor of five, the average spallation yield per muon calculated with the FLUKA code~\cite{Ferrari2005} is reduced by about 20\,\% due to the lower muon energy so that the total rate of spallation events per unit volume is about four times as high as in Super-Kamiokande.

The rate of downward-going muons is about \SI{2}{Hz} in Super-Kamiokande~\cite{Hosaka2006} and will increase to about \SI{50}{Hz} in Hyper-Kamiokande due to the lower overburden and larger detector size.
In Super-Kamiokande, the yield of unstable isotopes with decay energies of more than \SI{3.5}{MeV} for a single muon was calculated to be \SI{5e-6}{g^{-1} cm^2}~\cite{Li2014}.
Assuming the density of water in the detector to be \SI{1}{g / cm^3} and a track length of \SI{32.2}{m} (corresponding to the height of the fiducial volume), this gives about 0.016 spallation events per muon.
In Hyper-Kamiokande, a 20\,\% lower spallation yield per unit length combined with a 1.6 times taller fiducial volume yields about 0.02 spallation events per muon,
As a result, the rate of spallation events in Hyper-Kamiokande is expected to be approximately \SI{1}{Hz}.

Since the half-life of spallation products is between several milliseconds and a few seconds, these events can in principle be identified through spatial and temporal coincidence with cosmic ray muons that pass through the detector.
The first modern search for supernova relic neutrinos with Super-Kamiokande employed a two-step process to reduce spallation backgrounds~\cite{Malek2003,Malek2003a}, starting with a time correlation cut which removed all events within \SI{0.15}{s} after a muon event.
Remaining events were then subject to a likelihood function cut.
In addition to the time delay after the muon event, this took into account the distance between the reconstructed event vertex and the preceding muon track as well as the residual charge $Q_\text{res}$, which was defined as the detected charge (measured in photoelectrons) that is above the typical ionization loss of \SI{2300}{PE} per metre track length.
A large and positive value of $Q_\text{res}$ indicates energy loss through showers, which could produce spallation products.
This likelihood cut reduces the spallation background by an order of magnitude while introducing a detector dead time of about 20\,\%.

A later analysis discovered that spallation events are correlated with peaks in the energy loss rate dE/dx along the muon track, which can be used for improved background rejection~\cite{Bays2012}.
In a range of papers over the following years, Li and Beacom provided a theoretical description of spallation processes, which put this empirical observation onto a theoretical foundation, and suggested a number of significant improvements to muon reconstruction and spallation cuts~\cite{Li2014,Li2015,Li2015a}.

While the lower overburden and increased rate of downward-going muons in Hyper-Kamiokande will require some modifications to these cuts, it is clear that the spallation event rate can be reduced to much less than \SI{1}{Hz}, which enables an effectively background-free observation of supernova burst neutrinos.

\chapter{A Software Toolchain for Supernova Neutrino Events in Hyper-Kamiokande}\label{ch-sim}

\setlength{\epigraphwidth}{.435\textwidth}
\epigraphhead[0]{\epigraph{True heroism is minutes, hours, weeks, year upon year of the quiet, precise, judicious exercise of probity and care—with no one there to see or cheer.}{\textit{David Foster Wallace}}}

\begin{figure}[tbp]
	\centering
	\includegraphics[scale=0.51]{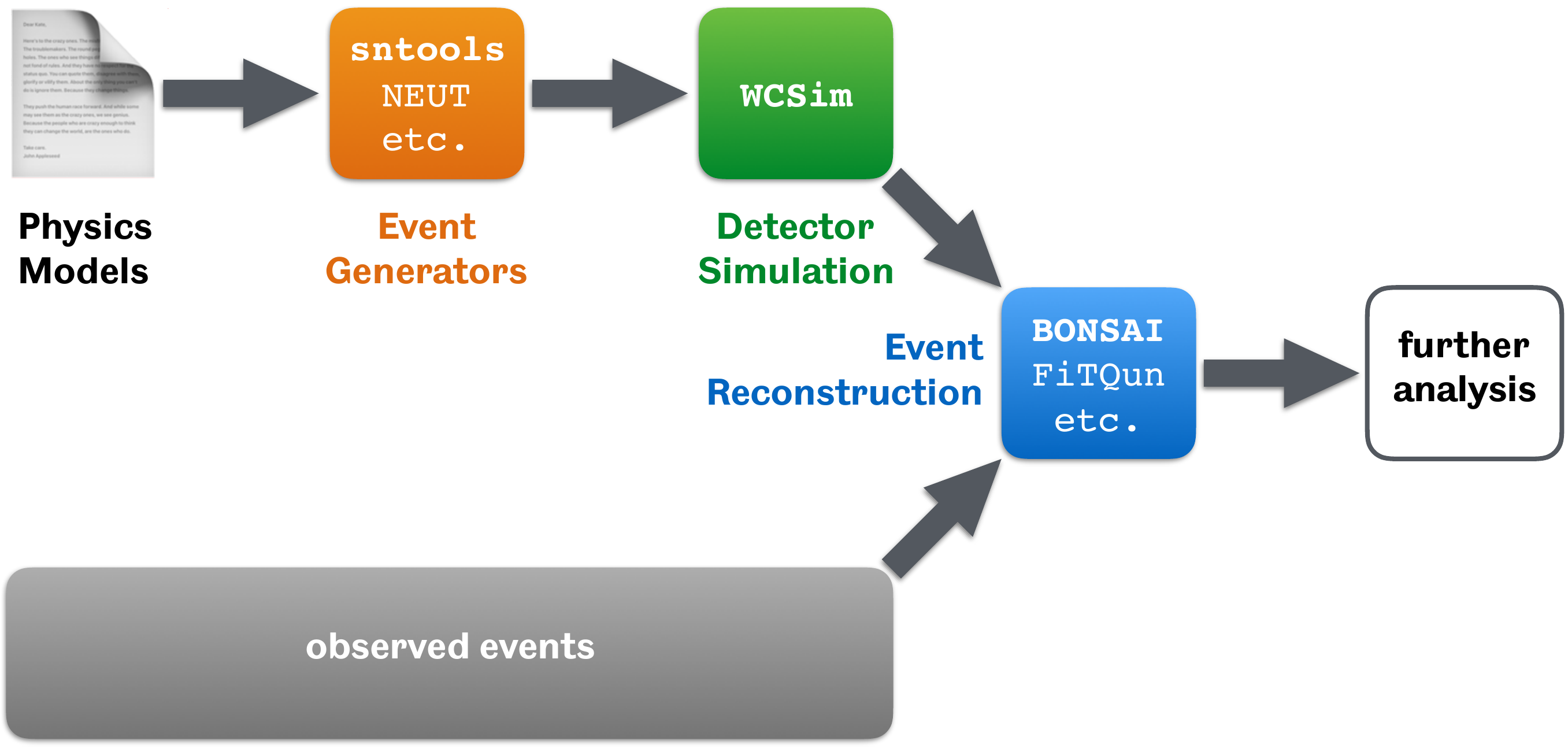}
	\caption[Overview over the software toolchain for Hyper-Kamiokande]{Overview over the software toolchain for Hyper-Kamiokande. Software used for supernova burst neutrinos is highlighted in bold.}
	\label{fig-sim-toolchain}
\end{figure}

In addition to the detector described in the previous chapter, the Hyper-Kamiokande collaboration is developing a software toolchain for generating and simulating events as well as reconstructing simulated or observed events to enable physics analyses.
A schematic overview of this toolchain is given in figure~\ref{fig-sim-toolchain}.

In this chapter, I will describe the toolchain insofar as it applies to supernova burst neutrinos.
It starts with neutrino fluxes as a function of neutrino flavour, time and energy that are produced by computer simulations of supernovae.
Section~\ref{ch-sim-models} describes the supernova models I use in this thesis.
I then use a custom software called sntools, introduced in section~\ref{ch-sim-sntools}, to generate neutrino interactions in the detector volume from these neutrino fluxes.
Section~\ref{ch-sim-wcsim} describes the detector simulation software, WCSim, which simulates propagation of particles and Cherenkov light in the detector and applies detector effects, including digitization and triggering.
The output of this detector simulation should be equivalent to the output of the DAQ system once Hyper-Kamiokande is operational and observes actual neutrino interactions. 
Finally, reconstruction of the vertex, direction and energy of each simulated event is described in section~\ref{ch-sim-reco}.
Further analysis of the simulated data sets, which builds on the results of the event reconstruction, is described in chapter~\ref{ch-ana}.

\section{Supernova Models}\label{ch-sim-models}

In this thesis, I use five supernova models that are developed and run by external groups not affiliated with the Hyper-Kamiokande proto-collaboration.
These models are simulated with custom software suites that typically combine a hydrodynamics solver and a neutrino interaction code.
The hydrodynamics code simulates self-interactions of matter by solving the differential equations for conservation of mass, momentum and energy in the system.
The neutrino interaction code implements transport of neutrinos and their interactions with the background matter.

In these simulations, it is typically assumed that temperatures inside the supernova are too low to produce $\mu$ and $\tau$ leptons.%
\footnote{A recent study indicated that this assumption may not be realistic and that $\mu$ production near the centre of the supernova can facilitate explosions in some models~\cite{Bollig2017}. However, such effects are not yet taken into account in most recent simulations, including the ones used in this thesis.}
As a result the emitted fluxes of \nue and \nuebar differ significantly from each other and from those of heavy lepton flavour neutrinos, while the differences between \numu, \nutau and their respective antineutrinos are assumed to be negligible.
It is therefore common to label the different neutrino species as \nue, \nuebar and \nux.%
\footnote{Throughout this thesis, I will use \nux to refer to any one of the four heavy lepton flavours. Note that this is not handled uniformly in literature and some authors use \nux to refer to the sum of all four flavours instead.}

These simulations are then performed on supercomputers using a numerical grid with model-dependent resolution.
After post-processing, the output of these models is generally provided in the form of text files describing the neutrino flux at the outer edge of the simulation volume---i.\,e. before traversing the outer layers of the star---as a function of neutrino species, time and energy.


To truly understand the explosion mechanism, high-fidelity three-dimensional models are required that solve the magnetohydrodynamical equations in high spatial resolution and full general relativity, while taking into account complex microphysics of the equation of state for nuclear matter and accurate energy- and flavour-dependent neutrino transport.
Current simulations are still severely limited by the available computing power and thus unable to simultaneously include all these effects to the highest possible accuracy.
Furthermore, even the most sophisticated simulations available today require millions of node-hours on modern supercomputers and may take weeks or months to run, so that only a small number of models can be simulated at the highest possible levels of precision.

It is thus impossible to perform such sophisticated simulations for a large number of possible progenitors, which span a wide range of masses, metallicities, rotational velocities and magnetic fields, all of which may have a significant quantitative and qualitative impact on the outcome of the simulation (for recent studies of these parameters, see e.\,g. references~\cite{Nakazato2013,Sukhbold2016,Couch2019,Ibeling2013,Takiwaki2016,Andresen2019,Obergaulinger2018}).
Additionally, in binary systems, interactions with a companion star may affect stellar evolution and thus the fate of a star~\cite{Podsiadlowski2004}.

One-dimensional (i.\,e. spherically symmetric), computationally much less expensive models are therefore necessary to explore this wide range of progenitors and study the population of supernovae.
However, they face a major problem: The core-collapse supernova explosion mechanism is fundamentally multi-dimensional and imposing spherical symmetry suppresses explosions in many simulations.
Several different methods may be used to artificially initiate explosions.
Two early approaches, which neglected neutrino contributions to the explosion mechanism and were mainly used for studying nucleosynthesis yields in the ejecta, are the “piston”, where an outgoing shock wave was artificially started within the stellar core, or the “thermal bomb”, where matter in the stellar core is artificially heated to cause an expanding shock wave~\cite{Aufderheide1991}.
More recently introduced approaches for triggering explosions (see e.\,g. references~\cite{Ugliano2012,Perego2015,Couch2019}) include more realistic neutrino physics but are still sensitive to the choice of free parameters, which may be chosen such that the outcome reproduces observations or results of more sophisticated simulations.

A comparison of six different modern simulation codes that simulated an identical \SI{20}{\Msol} progenitor with largely identical input physics showed good qualitative agreement in spherical symmetry.
Differing treatments of hydrodynamics or neutrino transport caused \SI[parse-numbers=false]{\mathcal{O}(10)}{\percent} differences in the neutrino luminosity and mean energy, as well as the event rate expected in a neutrino detector such as Super-Kamiokande~\cite{OConnor2018a}.
When comparing simulations between different publications, however, there may be significantly stronger disagreement since simulations may also differ in aspects of input physics---such as the equation of state for nuclear matter or the set of included neutrino interactions---that were intentionally kept identical in the above comparison.

In this thesis, I use five different models: a one-dimensional model that is primari\-ly of historic interest, two one-dimensional models from recent parametric studies and two more complex multi-dimensional models.
These simulations were performed by different groups using different simulation codes that implement different approximations.
They are intended to represent the much wider range of available models.

\subsection{Totani}
This model~\cite{Totani1998}, which is also referred to as the “Livermore model” or “Wilson model” in literature, was published in 1997 and is one of a small number of models that include the late-time evolution of the neutrino emission.
While it is now dated and has been surpassed by more accurate models, it is still used for comparisons with literature in some recent publications, for example by the Super-Kamiokande~\cite{Abe2016} and DUNE~\cite{Acciarri2015} collaborations.
I include it here since it is used as a baseline model in the Hyper-Kamiokande Design Report~\cite{HKDR2018}.

It uses a \SI{20}{\Msol} progenitor, which was modelled to resemble the progenitor of SN1987A, 
and a simulation code developed by Wilson and Mayle~\cite{Wilson1986,Mayle1987}.
Neu\-tri\-no transport is modelled by the flux-limited diffusion approximation with 20 logarithmically spaced energy groups up to \SI{322.5}{MeV}.
The simulation is one-dimensional and was performed from start of collapse to \SI{18}{s} after the core bounce.

Neutrino data for this model was provided in the Totani format described below in section~\ref{ch-sim-sntools-formats}.
Spacing of time steps was \SI{0.2}{ms} during the neutronization burst (for \nue only), \SIrange{10}{25}{ms} during the first \SI{250}{ms} and \SIrange{100}{1000}{ms} at later times.

\subsection{Nakazato}\label{ch-sim-models-nakazato}
This family of models~\cite{Nakazato2013} offers a modern successor to the Totani model described above.
It was used by the Super-Kamiokande collaboration for designing their real-time supernova burst monitor~\cite{Abe2016} as well as by the Hyper-Kamiokande collaboration in their design report~\cite{HKDR2018}.

It contains simulations of multiple progenitors with a range of initial masses ($M_\text{init} = \SI{13}{\Msol}$, \SI{20}{\Msol}, \SI{30}{\Msol} and \SI{50}{\Msol}), each with solar metallicity ($Z = 0.02$) and a lower metallicity of $Z =0.004$, which is typical for supernova progenitors in the Small Magellanic Cloud.\footnote{The metallicity of the Large Magellanic Cloud, where the progenitor of SN1987A was located, is about half the solar metallicity~\cite{Geha1998,Mottini2006}.}
In this work, I focus on the \SI{20}{\Msol} progenitor with solar metallicity.
Additionally, I will use several other progenitors for comparison in section~\ref{ch-ana-siblings}.

The one-dimensional simulation was performed from start of collapse to \SI{20}{s} after core bounce in two stages.
Both stages used the Shen equation of state~\cite{Shen1998}, the same set of neutrino interactions~\cite{Sumiyoshi2005} and 20 variably spaced energy groups up to \SI{300}{MeV}.

The first part of the simulation starts with the collapse of the stellar core 
and ends at \SI{550}{ms}, while the outer layers of the core are accreting onto a standing shock front.
This part is simulated with a general relativistic neutrino-radiation-hydrodynamics ($\nu$RHD) code that solves the differential equations for hydrodynamics and neutrino transport simultaneously~\cite{Sumiyoshi2005}.

The second stage of the simulation starts at an arbitrarily chosen shock revival time of $t_\text{revive} = \SI{100}{ms}$, \SI{200}{ms} or \SI{300}{ms} and ends at \SI{20}{s} after the core bounce.
Using the result of the $\nu$RHD simulation as an initial condition, the proto-neutron star cooling (PNSC) is then simulated in general relativity by solving the hydrostatic equations and using a multi-group flux-limited diffusion scheme to model neutrino transport.
Since the amount of matter falling back onto the proto-neutron star after the revival of the shock-wave is small, effects of accretion are not simulated.

Combining these two simulation stages allows performing a more detailed and computationally expensive simulation for the accretion phase, where dynamics of the shock front are essential, while still determining the evolution of the supernova out to late times using a simpler simulation.
However, the physical differences between both stages lead to a discontinuity in the resulting neutrino flux.

During the $\nu$RHD stage, neutrino emission is dominated by the accretion shock resulting from outer layers of the stellar core falling onto the outgoing shock wave.
Most one-dimensional simulations, including the one described here, do not successfully produce a supernova explosion except when artificially triggered, since the lack of multi-dimensional effects 
leads to an unrealistically high mass accretion rate.
Therefore, the neutrino emission during this phase is likely overestimated.
On the other hand, during the PNSC stage, neutrino emission from the accretion onto the revived shock front and the proto-neutron star is completely ignored, which underestimates the emission particularly shortly after the shock revival.

The authors recommend interpolating between these stages using an exponential function, such that the flux $F_{\nu_i}$ of each neutrino species is given by
\begin{equation}
F_{\nu_{i}}(E, t) = f(t) F_{\nu_i}^{\nu\mathrm{RHD}} (E, t) + (1-f(t)) F_{\nu_i}^{\mathrm{PNSC}} (E, t),
\end{equation}
where
\begin{equation}
f(t) = \left\{
	\begin{array}{ll}
		1, & t \leq t_\text{revive} + t_\text{shift} \\
		\exp \left(-\frac{t - (t_\text{revive} + t_\text{shift})}{\tau_\text{decay}}\right), & t_\text{revive} + t_\text{shift} < t
	\end{array}\right.
\end{equation}
is an interpolating function, $t_\text{shift} = \SI{50}{ms}$ is a time shift necessary to avoid discontinuous effects during the switch from the $\nu$RHD to the PSNC stage and $\tau_\text{decay} = \SI{30}{ms}$ is a decay time scale.

Unfortunately, while this approach eliminates the discontinuity it introduces an exponential drop-off in the luminosity and mean energy across all neutrino species at $t_\text{revive} + t_\text{shift}$, which would make it easy to identify this model.
Since this thesis focusses on the early part of the neutrino signal, I will use the fluxes from the $\nu$RHD phase only and not use this interpolation. 
Insofar as this overestimates the neutrino luminosity by a constant factor, that will not affect the results of the analysis in chapter~\ref{ch-ana} since that analysis assumes an unknown supernova distance and normalizes the differential event rates to a fixed total number of events.

If the overestimation of the neutrino luminosity is time-dependent, the size of this effect depends on the shock revival time, which is not known.
For the purposes of this thesis, I assume that this effect is negligible.
This is similar to assuming a shock revival time larger than about \SI{500}{ms}, which does not appear improbable.\footnote{While the paper by Nakazato \emph{et al.} only considered revival times up to \SI{300}{ms}~\cite{Nakazato2013}, different simulations of the same progenitor often produce large variations in explosion time; see e.\,g. table~1 of~\cite{Couch2019}.}
For the progenitor comparison in section~\ref{ch-ana-siblings}, this effect will largely cancel out under the conservative assumption that shock revival times are similar across progenitors.
If the shock revival time varies significantly between progenitors, that would increase the difference in their neutrino emission and thus increase the identification accuracy above the results shown in this thesis.

Neutrino data for this model was provided in the Nakazato format described below in section~\ref{ch-sim-sntools-formats} with sub-\si{ms} spacing of time steps. 

\subsection{Couch}
This family of models~\cite{Couch2019,Warren2019} introduces a new approach for including effects of convection and turbulence in a one-dimensional simulation, which the authors call STIR (Supernova Turbulence In Reduced-dimensionality).
In this approach, the effective strength of convection depends on one tuneable parameter, $\alpha_\Lambda$.
The authors find that a value of \numrange{0.8}{0.9} best reproduces the results from a three-dimensional simulation of the same progenitor~\cite{OConnor2018}.
In the following, I therefore use results from the simulation with $\alpha_\Lambda = 0.8$.

The model family contains 138 solar-metallicity progenitors with masses from \SIrange{9}{120}{\Msol}.
Here, I use results from the simulation of a \SI{20}{\Msol} progenitor.
This progenitor was originally described in reference~\cite{Sukhbold2014}.

The simulation was implemented in the FLASH simulation framework~\cite{Fryxell2000,Dubey2009} using a newly-implemented hydrodynamics solver 
with a modified effective potential to approximate effects of general relativity~\cite{Marek2006,OConnor2018b} and the SFHo equation of state~\cite{Steiner2013}.

Neutrino transport is simulated using a so-called “M1” transport scheme~\cite{OConnor2015,OConnor2018b} with 12 logarithmically spaced energy groups up to \SI{250}{MeV}.
Starting at \SI{5}{ms} post-bounce, effects of inelastic neutrino-electron scattering are turned off to reduce the computational power required.
While this has little impact on the supernova dynamics, it may result in an increased mean energy for \nux~\cite{OConnor2018b}. 

Neutrino data for this model was provided in the Gamma format described below in section~\ref{ch-sim-sntools-formats} in time steps of approximately \SI{0.5}{ms}.

\subsection{Tamborra}
This model~\cite{Hanke2013,Tamborra2014} is a pioneering three-dimensional supernova simulation with sophisticated neutrino transport.
It reported significant effects of the standing accretion shock instability (SASI)~\cite{Blondin2003}, large-scale sloshing motions of the shock front that may compete with neutrino-driven convection in facilitating explosions in multi-dimensional simulations and are fundamentally multi-dimensional and thus not observable in one-dimensional simulations of identical progenitors.

I use results from simulations of a \SI{27}{\Msol} progenitor from reference~\cite{Woosley2002}.
To seed the growth of hydrodynamic instabilities, random density perturbations of 0.1\,\% were manually introduced at the start of the simulation.

The simulation was performed using the \textsc{Prometheus-Vertex} code consisting of the \textsc{Prometheus}~\cite{Fryxell1991} code, a hydrodynamics solver which implements the piecewise-parabolic method~\cite{Colella1984}, and the neutrino transport code \textsc{Vertex}~\cite{Rampp2002}, which uses the “ray-by-ray-plus” approach for velocity- and energy-dependent neutrino transport~\cite{Buras2006}.
In this approximation, the neutrino moments equations for different angular bins (“radial rays”) decouple and can be solved independently.
It assumes that neutrino fluxes are symmetric around the radial direction, while including non-radial neutrino advection and pressure terms.

The simulation employs a sophisticated set of neutrino interaction rates described in reference~\cite{Muller2012}. 
It uses the Lattimer and Swesty equation of state with compressi\-bility $K = \SI{220}{MeV}$~\cite{Lattimer1991} and an effective potential to account for general relativistic corrections to Newtonian gravity~\cite{Marek2006}.

In multi-dimensional simulations, the neutrino signal inherently depends on the direction of the observer relative to the progenitor and extensive post-processing is necessary to determine the directionality dependence of the neutrino signal.
Here, I use the fluxes in the “violet” observer direction identified in reference~\cite{Tamborra2014}, which exhibits a particularly large amplitude of the SASI oscillations in the luminosity and mean energy of neutrinos.

Neutrino data for this model was provided in the Gamma format described below in section~\ref{ch-sim-sntools-formats} in time steps of approximately \SI{0.5}{ms}.

\subsection{Vartanyan}
This model is a very recent two-dimensional simulation with high-precision neutrino radiation hydrodynamics.
It is similar to the simulations presented in references~\cite{Radice2017, Seadrow2018} but used a different equation of state and grid resolution, which caused some physical and numerical differences.
As a result, while the luminosity and mean energy are qualitatively very similar to those described in reference~\cite{Seadrow2018}, exact values may differ by several percent.

I use results of a simulation of a \SI{9}{\Msol} progenitor with solar metallicity from reference~\cite{Sukhbold2016}.

This two-dimensional simulation was performed using the neutrino-radiation-hydrodynamics code \textsc{Fornax}~\cite{Skinner2019}, which combines a radiation hydrodynamics solver using a generalized variant of the piecewise-parabolic method~\cite{Colella1984} with neutrino transport using the “M1” scheme~\cite{Shibata2011,Murchikova2017}, similar to that used in the Couch model.
It used 20 logarithmically spaced energy groups with energies up to \SI{300}{MeV} for \nue and up to \SI{100}{MeV} for \nuebar and \nux.
The detailed set of neutrino-matter interactions employed are summarized in reference~\cite{Burrows2006}. 
The simulation used the SFHo equation of state~\cite{Steiner2013} and an effective potential to account for general relativistic corrections to Newtonian gravity~\cite{Marek2006}.

Neutrino data for this model was provided in the Princeton format described below in section~\ref{ch-sim-sntools-formats} in time steps of \SI{1}{ms}.

\section{sntools: A Supernova Event Generator}\label{ch-sim-sntools}

sntools~\cite{sntools} is a Monte Carlo (MC) event generator for supernova neutrino interactions in water Cherenkov detectors.
Based on detailed time- and energy-dependent neutrino fluxes provided by the supernova models described in the previous section, it generates interactions within the detector for the dominant and the most important subdominant interaction channels, before writing them to event files that can be used as an input for the full detector simulation described in section~\ref{ch-sim-wcsim}.

While a similar event generator was already developed by the Super-Kamiokande collaboration~\cite{Abe2016}, that code is proprietary and not currently available to non-members.
Furthermore, it would have likely required extensive modifications for compatibility with the new detector simulation for Hyper-Kamiokande and to use more recent cross-section calculations and different input formats.

The SNOwGLoBES software~\cite{Snowglobes} is widely used to compute event rates and energy distributions for supernova burst neutrinos in various different detectors.
While it is an excellent tool for preliminary studies or quick comparisons of different detector configurations, it uses simplified approximations for detector effects like energy resolution or threshold and does not take into account time-dependence of input fluxes.
It is an event rate calculator---not an event generator---and is by its own admission “not intended to replace full detector simulations”~\cite{Snowglobes}.

Finally, a number of other event generators like \textsc{neut}~\cite{Hayato2009} or \textsc{genie}~\cite{Andreopoulos2015} exist.
However, these are focussed on atmospheric or accelerator neutrinos which typically have much higher energies than supernova burst neutrinos and may therefore be incomplete or inaccurate at these low energies.
They would require extensive modifications to be used for here.

Since none of these existing codes are suitable for the purposes of this thesis, I have developed a completely new event generator called sntools.
In this section, I will start by describing its overall design.
I will then discuss the supported input formats, its treatment of neutrino flavour conversion and the implemented interaction channels.

\subsection{Design}
\begin{figure}[htbp]
	\centering
	\includegraphics[scale=0.52]{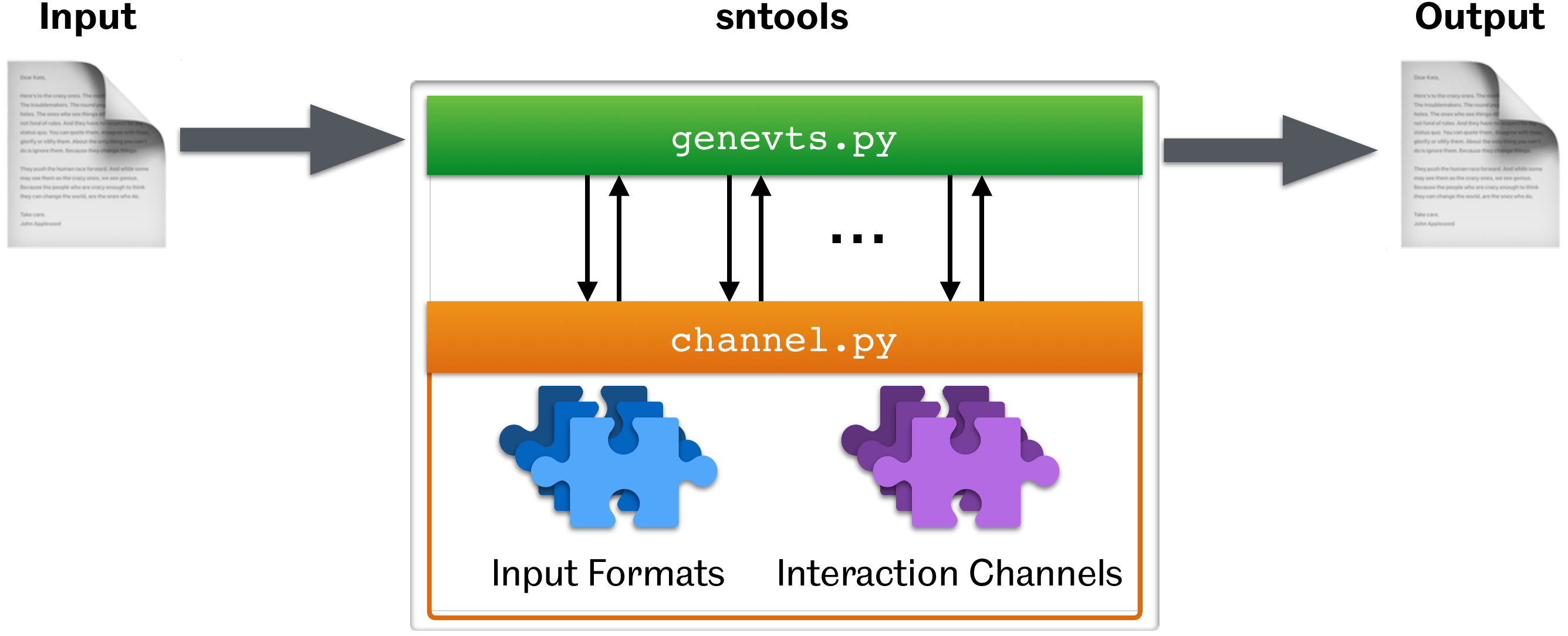}
	\caption[Overview over the structure of sntools]{Overview over the structure of sntools. See text for a detailed description.}
	\label{fig-sim-sntools-overview}
\end{figure}

An overview over the structure of sntools is given in figure~\ref{fig-sim-sntools-overview}.
sntools is written in Python, making it easy to read and extend the code, and designed to be extensible and accurate.
It makes use of the scipy and numpy libraries~\cite{Virtanen2019,Walt2011}, which implement many numerical calculations in Fortran for performance reasons, and I have tuned some performance-critical parts of sntools for increased performance.
On a current desktop computer, sntools needs \SI[parse-numbers=false]{\mathcal{O}(10)}{min} to generate events for a supernova at the fiducial distance of \SI{10}{kpc} in Hyper-Kamiokande.

The main user interface is provided by the file \texttt{genevts.py}.
It requires an input file containing neutrino fluxes from the supernova, while other, optional arguments include the format of the input file (see section~\ref{ch-sim-sntools-formats}), the neutrino mass ordering (see section~\ref{ch-sim-sntools-ordering}), the distance to the supernova\footnote{which can be selected using the \texttt{--distance <value>} command line argument, where the value is in \si{kpc}} and the interaction channels to consider (see section~\ref{ch-sim-sntools-channels}).
A full list of possible arguments can be found by executing \texttt{python genevts.py -h}. 

After parsing the arguments, events are generated separately for each combination of interaction channel and input species by calling code in the file \texttt{channel.py}.
The code first calculates the total number of events expected in each \SI{1}{ms} bin, which is given by
\begin{equation}
N(t) = \mathop{\mathlarger{\mathlarger{\iint}}} \tdiff{\Phi (t, E_\nu)}{E_\nu} \tdiff{\sigma (E_\nu, E_e)}{E_e}\, \d E_e\,\d E_\nu,
\end{equation}
where $\Phi (t, E_\nu)$ is the neutrino flux and $\sigma (E_\nu, E_e)$ is the cross section of the current interaction channel.
It then picks the actual number of events to generate within that time bin from a Poisson distribution with expectation value $N(t)$.
Finally, it generates events by rejection sampling from the energy spectrum of neutrino interactions at that time and the distribution of outgoing particle directions.

The event generation code relies on a plug-in architecture to support various different input formats and interaction channels.
Input format plug-ins provide functions that read in the data from an input file and return the number luminosity as a function of time and energy.
Interaction channel plug-ins specify properties of the interaction channel, like the number of targets per water molecule or the neutrino species that undergo this interaction, and provide functions to calculate quantities like the differential cross section $\tdiffx{\sigma (E_\nu, E_e)}{E_e}$ or the kinematically allowed energy range.
This modular design makes sntools easily extensible, with roughly 100 lines of code required to add a new input format or interaction channel.

Finally, \texttt{genevts.py} collects the events generated in all interaction channels and writes them to a text file, which can be used as an input file for a full detector simulation.

\subsection{Input Formats} \label{ch-sim-sntools-formats}
sntools supports multiple different input formats for the neutrino fluxes from a simulation, which can be selected using the \texttt{--format <value>} command line argument.
This section will briefly describe each format and the processing steps necessary to calculate the spectral number luminosity.

All formats contain separate information on the three species \nue, \nuebar and \nux.
I will, however, omit the reference to each species in the following for simplicity.

\subsubsection{Totani Format}
The files provided by Totani contain, for each time step $t_n$, the total number of neutrinos emitted until that time, which makes it possible to calculate the number $N_n$ of neutrinos emitted since the previous time step.

For 20 energy bins $E_k$ per time step, a quantity $X_k$ is provided, which is proportional to the number of neutrinos emitted during that time step and in that energy bin.
I divide this by the width of each energy bin to get
\begin{equation}
X_k^\text{spec} = \frac{X_k}{E_{k+1} - E_k},
\end{equation}
which is proportional to the spectral number emission during that time step.

Integrating $X_k^\text{spec}$ over all energy bins and dividing it by that integral gives the spectral number emission during that time step normalized to 1, $X_k^\text{norm}$.
The spectral number luminosity at time $t_n$ and energy $E_k$ is given by
\begin{equation}
\tdiff{\text{NL} (t_n, E_k)}{E} = \frac{N_n}{t_n - t_{n-1}} \cdot X_k^\text{norm}.
\end{equation}
Finally, a linear interpolation in time and log cubic spline interpolation in energy are used to determine the spectral number luminosity at an arbitrary time and energy.

This approach closely follows that used in code provided by Totani.
I have confirmed that there is excellent agreement of the calculated fluxes between Totani’s code and sntools, with differences of at most a few per mille due to slight differences in the numerical interpolation algorithms used.

\subsubsection{Nakazato Format}
Files in this format contain, for 20 energy bins $E_k$ during each time step $t_n$, the quantities $\Delta N_k (t_n) / \Delta E_k$ and $\Delta L_k (t_n) / \Delta E_k$, which reflect the number luminosity and luminosity at those energies, respectively.
For each energy bin, I calculate the mean energy within that bin, which is given by
\begin{equation}
\mean{E_k} = \frac{\quad  \frac{\Delta L_k (t_n)}{\Delta E_k}  \quad}{  \frac{\Delta N_k (t_n)}{\Delta E_k}  },
\end{equation}
and set the differential neutrino number flux at that energy to $\Delta N_k (t_n) / \Delta E_k$.
Finally, a linear interpolation in time and cubic spline interpolation in energy are used to determine the spectral number luminosity at an arbitrary time and energy, as recommended by Nakazato\footnote{private communication, April 7, 2018}. 

\subsubsection{Gamma Format}
Files in this format contain, for each time step $t_n$, the luminosity $L$, mean energy $\mean{E_\nu}$ and mean squared energy $\mean{E_\nu^2}$ of neutrinos.
To reconstruct the spectrum from this, I assume that the neutrino spectrum is described by a normalized Gamma distribution~\cite{Keil2003,Tamborra2012} given by
\begin{equation}
f (E_\nu) = \frac{E_\nu^\alpha}{\Gamma (\alpha + 1)} \left( \frac{\alpha + 1}{A} \right)^{\alpha + 1} \exp \left[ - \frac{(\alpha + 1) E_\nu}{A} \right].
\end{equation}
In this formula, $A$ is an energy scale, while $\alpha$ determines the shape of the distribution: $\alpha = 2$ corresponds to a Maxwell-Boltzmann distribution, while $\alpha > 2$ corresponds to a “pinched” spectrum, which is more typical for neutrino spectra from supernovae.

The first two energy moments of the distribution are
\begin{align}
\mean{E_\nu}	&= \int_0^\infty \d E_\nu\, E_\nu f(E_\nu) = A\\
\mean{E_\nu^2}&= \int_0^\infty \d E_\nu\, E_\nu^2 f(E_\nu) = \frac{\alpha + 2}{\alpha + 1} A^2,
\end{align}
and therefore,
\begin{equation}
\alpha = \frac{\mean{E_\nu^2} - 2 \mean{E_\nu}^2}{\mean{E_\nu}^2 - \mean{E_\nu^2}}.
\end{equation}
Thus, the shape of the spectral number luminosity is uniquely determined by the mean energy \mean{E_\nu} and the mean squared energy \mean{E_\nu^2}, while the normalization is provided by $L / \mean{E_\nu}$.
To determine the spectral number luminosity at arbitrary times, each of the three parameters is interpolated separately before calculating the spectral number luminosity using the interpolated values.

\subsubsection{Princeton Format}
Files in this format contain, for each time step $t_n$, the spectral luminosity $\tdiffx{L}{E}$ for 20 logarithmically spaced energy bins $E_k$.
I divide this by the central energy $\sqrt{E_k E_{k+1}}$ of the respective bin to get the spectral number luminosity at that energy.
Finally, a linear interpolation in time and cubic interpolation in energy are used to determine the spectral number luminosity at an arbitrary time and energy.

This follows the procedure described in reference~\cite{Seadrow2018}.
It is similar to that used for the Nakazato format described above, though with a different definition of the bin energy.

\subsection{Treatment of Neutrino Flavour Conversion} \label{ch-sim-sntools-ordering}

sntools implements three different mass ordering scenarios that can be selected by using the \texttt{--ordering <value>} command line argument.%
\footnote{The alias \texttt{--hierarchy <value>} also exists.}

The first scenario, \texttt{noosc}, assumes that neutrino oscillations do not take place such that the flux of a neutrino species $\nu_i$ observed by a detector on Earth, $\Phi_{\nu_i}$, is identical to the fluxes originating within the supernova, $\Phi_{\nu_i}^0$.\footnote{For simplicity, throughout this section I omit the geometrical factor $\frac{1}{4 \pi d^2}$ which depends on the distance $d$ of the supernova.}

The other two scenarios, \texttt{normal} and \texttt{inverted}, assume that adiabatic flavour conversion happens via the MSW effect since neutrinos traverse a smoothly varying density profile while exiting the star.
The resulting observed fluxes are linear combinations of the initial fluxes, which, for normal mass ordering, are given by~\cite{Dighe2000}
\begin{align}
\begin{split}
\Phi_{\nue} &= \sin^2 \theta_{13} \cdot \Phi_{\nue}^0 + \cos^2 \theta_{13} \cdot \Phi_{\nux}^0\\
\Phi_{\nuebar} &= \cos^2 \theta_{12} \cos^2 \theta_{13} \cdot \Phi^0_{\nuebar} + (1 - \cos^2 \theta_{12} \cos^2 \theta_{13}) \cdot \Phi^0_{\nuxbar} \\
2 \Phi_{\nux} &= \cos^2 \theta_{13} \cdot \Phi^0_{\nue} + (1 + \sin^2 \theta_{13}) \cdot \Phi^0_{\nux} \\
2 \Phi_{\nuxbar} &= (1 - \cos^2 \theta_{12} \cos^2 \theta_{13}) \cdot \Phi^0_{\nuebar} + (1 + \cos^2 \theta_{12} \cos^2 \theta_{13}) \cdot \Phi^0_{\nuxbar},
\end{split}
\end{align}
while for inverted ordering, they are
\begin{align}
\begin{split}
\Phi_{\nue} &= \sin^2 \theta_{12} \cos^2 \theta_{13} \cdot \Phi_{\nue}^0 + (1 - \sin^2 \theta_{12} \cos^2 \theta_{13}) \cdot \Phi_{\nux}^0\\
\Phi_{\nuebar} &= \sin^2 \theta_{13} \cdot \Phi_{\nuebar}^0 + \cos^2 \theta_{13} \cdot \Phi_{\nuxbar}^0\\
2 \Phi_{\nux} &= (1 - \sin^2 \theta_{12} \cos^2 \theta_{13}) \cdot \Phi_{\nue}^0 + (1 + \sin^2 \theta_{12} \cos^2 \theta_{13}) \cdot \Phi_{\nux}^0\\
2 \Phi_{\nuxbar} &= \cos^2 \theta_{13} \cdot \Phi_{\nuebar}^0 + (1 + \sin^2 \theta_{13}) \cdot \Phi^0_{\nuxbar}.
\end{split}
\end{align}


In both cases, the factor of 2 in the last two equations accounts for the fact that I combine the fluxes of \numu and \nutau, as well as those of the corresponding antineutrinos, into \nux as well as \nuxbar, respectively.
These equations assume purely adiabatic transition (corresponding to $P_H = 0$ in~\cite{Dighe2000,Fogli2005}) as explained below.

In cases where the detected flux is a mixture of original fluxes of different species, sntools generates events for each original species separately with the appropriate weighting factor applied.
For example, when generating inverse beta decay events in the normal ordering, sntools will generate events first using the input flux $\Phi_{\nuebar} = \cos^2 \theta_{12} \cos^2 \theta_{13} \cdot \Phi^0_{\nuebar}$ and then using the input flux $\Phi_{\nuebar} = (1 - \cos^2 \theta_{12} \cos^2 \theta_{13}) \cdot \Phi^0_{\nuxbar}$, before finally combining both sets of events into one output file.

Several other effects may induce additional time- and energy-dependent flavour conversion.
These effects, together with a brief explanation of why they are not currently implemented in sntools, are discussed in the following.

After the accretion phase, the revived shock front travels outwards and passes through the layer within the star where the adiabatic flavour conversion described above takes place.
This causes a sudden change in the matter and electron density and can severely impact the flavour conversion processes~\cite{Schirato2002}.
Since this occurs after the shock wave is revived, it mainly affects the late-time part of the supernova neutrino signal\footnote{For the Totani model, this effect is expected to become relevant more than \SI{1}{s} after core-bounce~\cite{Fogli2005}.} and would almost certainly not impact the analysis in chapter~\ref{ch-ana}.
Furthermore, this effect is highly dependent on the the density structure of each individual progenitor and cannot be taken into account by sntools.
Instead, where appropriate, groups performing supernova simulations will need to include this effect in their codes and publish neutrino flux data that takes it into account.

Near the centre of the supernova, the high neutrino density could induce a matter effect that causes self-induced flavour conversion~\cite{Duan2006a,Duan2006}.
These collective effects---and their observable consequences, which could include energy-dependent flavour conversion (so called “spectral splits”)---are the subject of intense theoretical study~\cite{Raffelt2007,Dasgupta2009,Friedland2010,Duan2011,Izaguirre2017}, though no clear picture has yet emerged of how these effects will manifest in a given supernova.
See reference~\cite{Chakraborty2016} for a recent review.
Flavour conversion may also be suppressed in dense matter~\cite{Esteban-Pretel2008,Zaizen2018}.
As a result, the consequences in a realistic supernova are currently not well understood and may depend on the progenitor~\cite{Chakraborty2014}. 
They can therefore not be taken into account by sntools.
Once theoretical understanding has improved, it may be more appropriate to include these effects in individual supernova simulations and publish neutrino flux data that takes this into account.

Finally, depending on the location of the supernova relative to the detector, the neutrino detector may be “shadowed” by the Earth.
For a detector in Kamioka, the shadowing probability for a galactic supernova is 56\,\%~\cite{Mirizzi2006}.
As neutrinos traverse the Earth’s matter potential before detection, they can undergo energy-dependent flavour transitions which are, in principle, detectable in Hyper-Kamiokande~\cite{Dighe2003}.
Since the presence and amplitude of this shadowing effect depends on the location of the progenitor within the Milky Way as well as, due to the Earth’s rotation, on the time of day that the neutrinos arrive on Earth, I will not consider it in this thesis.
However, Hyper-Kamiokande will be able to measure the position of a galactic supernova to within a few degrees uncertainty.
That would allow us to theoretically calculate the influence of this shadowing effect, if present, which could then be compared to the observed energy-dependent variations in the signal to set limits on non-standard interactions and reduce the uncertainty introduced by the shadowing effect.

\subsection{Interaction Channels} \label{ch-sim-sntools-channels}
sntools supports multiple different interaction channels described in this section.
\enlargethispage{\baselineskip} 
By default, it will generate events across all supported channels, but it can be restricted to a single channel by using the \texttt{--channel <value>} command line argument, where \texttt{<value>} can be one of \texttt{ibd}, \texttt{es}, \texttt{o16e} or \texttt{o16eb}.

In water Cherenkov detectors like Hyper-Kamiokande, the dominant interaction channel for supernova neutrinos is inverse beta decay, which makes up about 90\,\% of events.
Another important interaction channel is elastic scattering on electrons, which makes up only a few per cent of events but provides precise information on the direction of the supernova.
The cross sections for both interactions have been calculated to a high level of precision.

Another important subdominant channel are charged-current interactions of \nue and \nuebar on $^{16}$O nuclei.
While this channel suffers from large theoretical uncertainties, it is very sensitive to the high-energy tail of supernova neutrino fluxes, so that the number of events in this channel can vary greatly between models.

Consistent with recent work on the Super-Kamiokande supernova burst monitor~\cite{Abe2016} and with the Hyper-Kamiokande Design Report~\cite{HKDR2018}, I have not included additional subdominant interaction channels like neutral-current interactions on $^{16}$O nuclei or neutral- and charged-current interactions on heavier oxygen isotopes.
Due to their low event rates, these channels would have a relatively small influence on the observed event spectra, while introducing additional uncertainties since their cross sections are not well known.
Furthermore, their contributions would be mainly at low or medium energies, where inverse beta decay dominates, while their contributions at high energies would be much smaller than that of the charged-current $^{16}$O channel.
As a result, these channels have a very minor dependence on the supernova model and would contribute little to the analysis presented in this thesis.

Neutral-current scattering on free protons, which may be an important detection channel in scintillator detectors, cannot be detected by Hyper-Kamiokande since the outgoing proton is below the Cherenkov threshold in water~\cite{Beacom2002}.

\subsubsection{Inverse Beta Decay}
In Hyper-Kamiokande, inverse beta decay (IBD; $\nuebar + p \rightarrow n + e^+$) is the dominant interaction channel for supernova neutrinos due to its relatively high cross section and low energy threshold of $E_\nu^\text{thr} \approx \SI{1.8}{MeV}$, as well as the large number of free protons in the detector.
The observed energy of IBD events is closely related to the neutrino energy, making this an excellent channel to reconstruct the \nuebar spectrum.

In sntools, I have implemented IBD using the full tree-level cross section calculated in reference~\cite{Strumia2003} and including radiative corrections based on the approximation from reference~\cite{Kurylov2003}.\footnote{That calculation uses the limit $m_e \rightarrow 0$. This approximation is accurate to better than 0.1\,\% above $E_e = \SI{1}{MeV}$ and the effect in Hyper-Kamiokande, whose energy threshold is much higher than \SI{1}{MeV}, is completely negligible.}
The calculation of the cross section is summarized in appendix~\ref{apx-xs}.
Due to uncertainties in experimental measurements of input parameters, the overall uncertainty is estimated to be 0.4\,\% at low energies and reaches about 2\,\% at $E_\nu = \SI{100}{MeV}$, which is the upper end of the energy range of supernova neutrinos.

\subsubsection{Neutrino-Electron Scattering}
In Hyper-Kamiokande, elastic neutrino-electron scattering  ($\nu + e^- \rightarrow \nu + e^-$) is a subdominant interaction channel due to its low cross section, which is only partially compensated by the large number of electrons in the detector which is 5 (10) times bigger than the number of free protons ($^{16}$O nuclei).
Elastic scattering events make up only a few per cent of all events but their angular distribution is strongly peaked into a forward direction, pointing away from the supernova.
They can therefore be used to determine the direction of a supernova at the fiducial distance of \SI{10}{kpc} with an accuracy of \SIrange{3}{5}{\degree} in Super-Kamiokande~\cite{Abe2016} or about \SI{1}{\degree} in Hyper-Kamiokande~\cite{HKDR2018}.

Elastic scattering is the only interaction channel considered here which is sensitive to all neutrino flavours.
However, the cross section of \nue and \nuebar, which can interact through both neutral and charged currents, is higher than that of \nux and \nuxbar, which can interact only through neutral currents.

In sntools, I have implemented elastic scattering using the tree-level cross sections from standard electroweak theory calculated by ’t~Hooft~\cite{t-Hooft1971} and including one-loop electroweak and QCD corrections as well as QED radiative corrections as calculated in reference~\cite{Bahcall1995}.
The calculation of the cross section is summarized in appendix~\ref{apx-xs}.
Similar to the calculation of the IBD cross section above, the uncertainty due to experimental measurements of some input parameters is on a per mille level.

\subsubsection{Charged-Current Interactions on $^{16}$O}
In Hyper-Kamiokande, charged-current interactions of \nue and \nuebar on $^{16}$O nuclei,
\begin{align}
\nue + ^{16}\text{O} &\rightarrow e^- + X \\
\nuebar + ^{16}\text{O} &\rightarrow e^+ + X,
\end{align}
are a subdominant interaction channel.\footnote{Charged-current interactions of other neutrino species do not occur, since the energy of supernova neutrinos is too small to produce muons or $\tau$ leptons.}
Due to the high energy threshold of both interactions of approximately \SI{15}{MeV} and \SI{11}{MeV}, respectively, as well as the steep energy-dependence of the cross sections, the number of events in each channel is a very sensitive probe of the high-energy tail of the supernova neutrino flux.
It may vary by more than two orders of magnitude depending on the supernova models and oscillation scenario, making up anywhere from $< 1\,\%$ to over 10\,\% of all events observed in Hyper-Kamiokande.

In sntools, I have implemented a recent approximation of the cross section for both interaction channels~\cite{Nakazato2018}, which is based on a new shell model calculation~\cite{Suzuki2018}.

An earlier calculation used the continuum random phase approximation (CRPA) and tabulated the resulting total cross sections, instead of partial cross sections for each excitation energy~\cite{Kolbe2002}.
A fit based on those results therefore only considered a simplified scenario where all final nuclear states shared the energy of the ground state~\cite{Tomas2003}.
This significantly overestimated the energy spectrum of the emitted $e^\pm$.

The modern calculation selected 42 different nuclear states and calculated their respective partial cross sections.
To simplify the evaluation of the cross section, the authors of reference~\cite{Nakazato2018} divide these states into four groups such that the total cross section is
\begin{equation}
\sigma (E_\nu) = \sum_{g=1}^{4} \sigma_\text{g} (E_\nu),
\end{equation}
where the partial cross sections for each group, $\sigma_\text{g}$, are given by the expression
\begin{align}
\log_{10} \left( \frac{\sigma_\text{g} (E_\nu)}{\si{cm^2}} \right) &= a_\text{g} + b_\text{g} \Lambda (E_\nu) + c_\text{g} \left[ \Lambda (E_\nu) \right]^2 \\
\Lambda (E_\nu) &= \log_{10} \left[ \left( \frac{E_\nu}{\si{MeV}} \right)^{1/4} - \left( \frac{ E_\text{g}}{\si{MeV}} \right)^{1/4} \right].
\end{align}
The excitation energies $E_\text{g}$ and fit parameters $a_\text{g}$, $b_\text{g}$ and $c_\text{g}$ of each group of states are shown in table~\ref{tab-sim-o16fit}.

\begin{table}[tbp]
\begin{center}
\begin{tabular}{lccccc}
reaction & group $g$ & $E_\text{g}$ (\si{MeV}) & $a_\text{g}$ & $b_\text{g}$ & $c_\text{g}$ \\
\hline
\multirow{4}{*}{$^{16}\mathrm{O}\left(\nue, e^-\right) \mathrm{X}$} & 1 & 15.21 & -40.008 & 4.918 & 1.036 \\
& {2} & {22.47} & {-39.305} & {4.343} & {0.961} \\
& {3} & {25.51} & {-39.655} & {5.263} & {1.236} \\
& {4} & {29.35} & {-39.166} & {3.947} & {0.901} \\
\hline
\multirow{4}{*}{$^{16}\mathrm{O}\left( \nuebar, e^{+}\right) \mathrm{X}$} & {1} & {11.23} & {-40.656} & {4.528} & {0.887} \\
& {2} & {18.50} & {-40.026} & {4.177} & {0.895} \\
& {3} & {21.54} & {-40.060} & {3.743} & {0.565} \\
& {4} & {25.38} & {-39.862} & {3.636} & {0.846}
\end{tabular}
\end{center}
\caption[Excitation energies and fitting parameters for charged-current interactions on $^{16}$O]{Excitation energies and fitting parameters for charged-current interactions on $^{16}$O. Values from reference~\cite{Nakazato2018}.}
\label{tab-sim-o16fit}
\end{table}

This fit matches the cross sections calculated from the full set of nuclear states to within a few per cent at neutrino energies of up to \SI{100}{MeV}.
For a typical supernova neutrino flux, the difference in the resulting event spectra when using the four groups instead of all 42 nuclear states is also very small.

\section{Detector Simulation}\label{ch-sim-wcsim}
To simulate events in Hyper-Kamiokande, we use WCSim~\cite{WCSim}, a package for simulating water Cherenkov detectors that is based on the physics simulation framework \textsc{Geant4}~\cite{Agostinelli2003} and the data analysis framework \textsc{root}~\cite{Brun1997}.

It was originally developed by the Long Baseline Neutrino Experiment collaboration\footnote{The LBNE collaboration initially considered several different designs for their far detector, including one based on water Cherenkov technology~\cite{Goon2012}. They have since decided to use a different design based on a liquid argon time projection chamber and became the Deep Underground Neutrino Experiment (DUNE) collaboration. See also section~\ref{ch-conclusions-dune}.} 
and is now used primarily by the Hyper-Kamiokande collaboration, where it will replace the \textsc{Geant3}-based \textsc{SKDetSim} software which is currently used for Super-Kamiokande. 

As an input, WCSim takes text files in the \textsc{Nuance} format (see references~\cite{Casper2002,nuance-format}), which contain the particle species ($e^+$ or $e^-$) as well as the initial energy, position and direction for each event.
Based on physics lists included in \textsc{Geant4}, WCSim then simulates the propagation of each primary particle through the detector, including production of secondary particles like Cherenkov photons and their respective propagation.
WCSim registers when photons hit photosensors around the edge of the detector and applies the quantum and collection efficiency.
It can also simulate dark noise or after-pulsing in the photosensors and digitization of the signal.

Currently, WCSim also includes simple triggers that decide which data is written to the output file.
In the future, triggering will be performed by a specialized framework as part of the DAQ system described in section~\ref{ch-hk-daq}.
However, this framework was still in the very early stages of development when the analysis in this thesis was performed.

\section{Event Reconstruction}\label{ch-sim-reco}

After simulating the detector response to supernova neutrino events, this section describes how the digitized photosensor signals are used to reconstruct the event.
In contrast to the previous parts of this chapter, which are concerned only with simulated events, the reconstruction methods described in this section apply equally to simulated events and---once Hyper-Kamiokande is operating---actual observed events.
As far as possible, the same code will be used for both simulated and observed events in order to avoid duplication of effort and reduce discrepancies and systematic errors between MC and observations.

Despite this, there are some inherent differences between simulations and actual data taking.
These include, for example, the occurrence of technical issues (like defects in individual photosensors or electronics modules), variations in the noise rate of individual photosensors or time-dependent changes to the water quality.
These often require additional steps in data processing that are not necessary for simulated data. 
An extensive programme of calibrations discussed in section~\ref{ch-hk-calibration} is under development to understand and---wherever possible---correct for these detector effects.
Throughout this section, I will discuss these issues where appropriate and describe how they are handled in the simulations.

Due to the wide energy range of events in Hyper-Kamiokande, several different reconstruction algorithms are used which are optimized for specific energy ranges or event topologies. 
For low-energy events up to about \SI{100}{MeV}, the Super-Kamiokande collaboration has developed the \textsc{BONSAI} (Branch Optimization Navigating Successive Annealing Interactions) code for phase II of the experiment, which was later optimized for the higher density of PMTs in Super-Kamiokande-III~\cite{Smy2007}.
For Hyper-Kamiokande, we will continue using \textsc{BONSAI} with minor modifications to work with the new simulation software (WCSim) and re-tuning for the new photosensors and detector size.

\textsc{BONSAI}’s method for reconstructing the vertex of an event is described in section~\ref{ch-sim-reco-vertex} and the method for reconstructing the direction of the outgoing particle is described in section~\ref{ch-sim-reco-direction}.
A separate script then uses these results to reconstruct the particle’s energy as described in section~\ref{ch-sim-reco-e}.

\subsection{Vertex Reconstruction}\label{ch-sim-reco-vertex}
Supernova neutrino interactions typically produce electrons and positrons with energies of a few tens of \si{MeV}.
At these energies, the typical track length of $e^\pm$ in water is at most a few \si{cm}.
Since they are required to be above the Cherenkov threshold, i.\,e. $v > c_\text{vacuum} / n_\text{water} \approx \SI{22}{cm / ns}$ to be detected in Hyper-Kamiokande, this track length corresponds to a travel time of less than \SI{1}{ns}.
This travel time is smaller than the time resolution of the photosensors and small compared to the total detector size. 
We therefore ignore the track length and assume a point-like interaction for the purposes of reconstructing the event vertex.

In Super-Kamiokande-I, vertex reconstruction relied on a grid of test vertices spaced at about \SI{4}{m} distance throughout the detector~\cite{Hosaka2006}.
The reconstruction software would calculate an ad-hoc goodness for each test vertex and then repeat the procedure with a finer grid spacing around the vertex from the previous test grid that gave the highest goodness.

Due to the lower photocoverage of Super-Kamiokande-II, the reconstruction performance of the previous approach deteriorated significantly at low energies and a new vertex reconstruction code named \textsc{BONSAI} was developed~\cite{Cravens2008}.

\begin{figure}[tbp]
	\centering
	\includegraphics[scale=0.25]{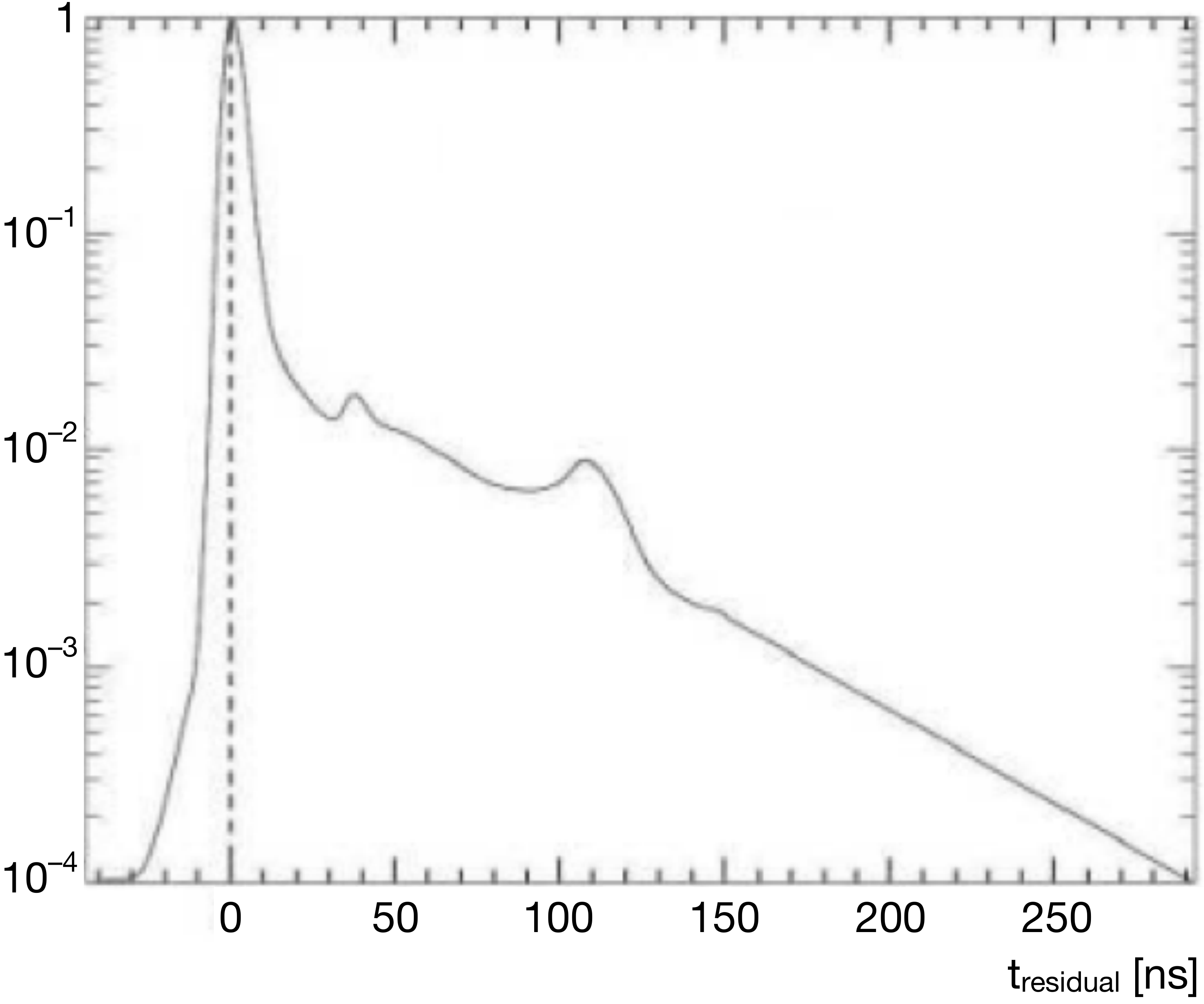}
	\caption[Probability density $P(t_\text{residual})$ used for vertex reconstruction in Super-Kamiokande]{Probability density $P(t_\text{residual})$ used for vertex reconstruction in Super-Kamiokande. The width of the peak around $t_\text{residual} = 0$ depends on the timing resolution of PMTs, while the additional peaks at \SI{30}{ns} and \SI{100}{ns} are caused by after-pulsing in the PMTs. Figure from reference~\cite{Nakano2016}.}
	\label{fig-sim-reco-timing}
\end{figure}

Instead of using a fixed grid, \textsc{BONSAI} generates test vertices by selecting tuples of four PMT hits and calculating the event vertex which would reduce the timing residuals of all four hits to zero.
This improves reconstruction performance and simultaneously reduces the risk of getting stuck in a local maximum that is at a large distance from the global maximum.
After identifying these test vertices, \textsc{BONSAI} performs a maximum likelihood fit over all $N_\text{hit}$ PMT hits in the event, which is defined as
\begin{equation}
\mathcal{L} \left(\vec{x_0}, t_{0}\right) = \sum_{i=1}^{N_\text{hit}} \log P (t_\text{residual}),
\end{equation}
where $(\vec{x_0}, t_0)$ is the test vertex and $P(t_\text{residual})$ is a probability density function which is determined through calibration.
The function used in Super-Kamiokande-IV is shown in figure~\ref{fig-sim-reco-timing}.
It depends on the time-of-flight-corrected timing residual, which is defined as $t_\text{residual} = t_i - t_0 - t_\text{tof}$, where $t_i$ is the time of the $i$-th PMT hit and $t_\text{tof} = |\vec{x_i} - \vec{x_0}| / c_\text{water}$ is the light travel time between the test vertex and the PMT location $\vec{x_i}$.

\begin{figure}[htbp]
	\centering
	\includegraphics[scale=0.35]{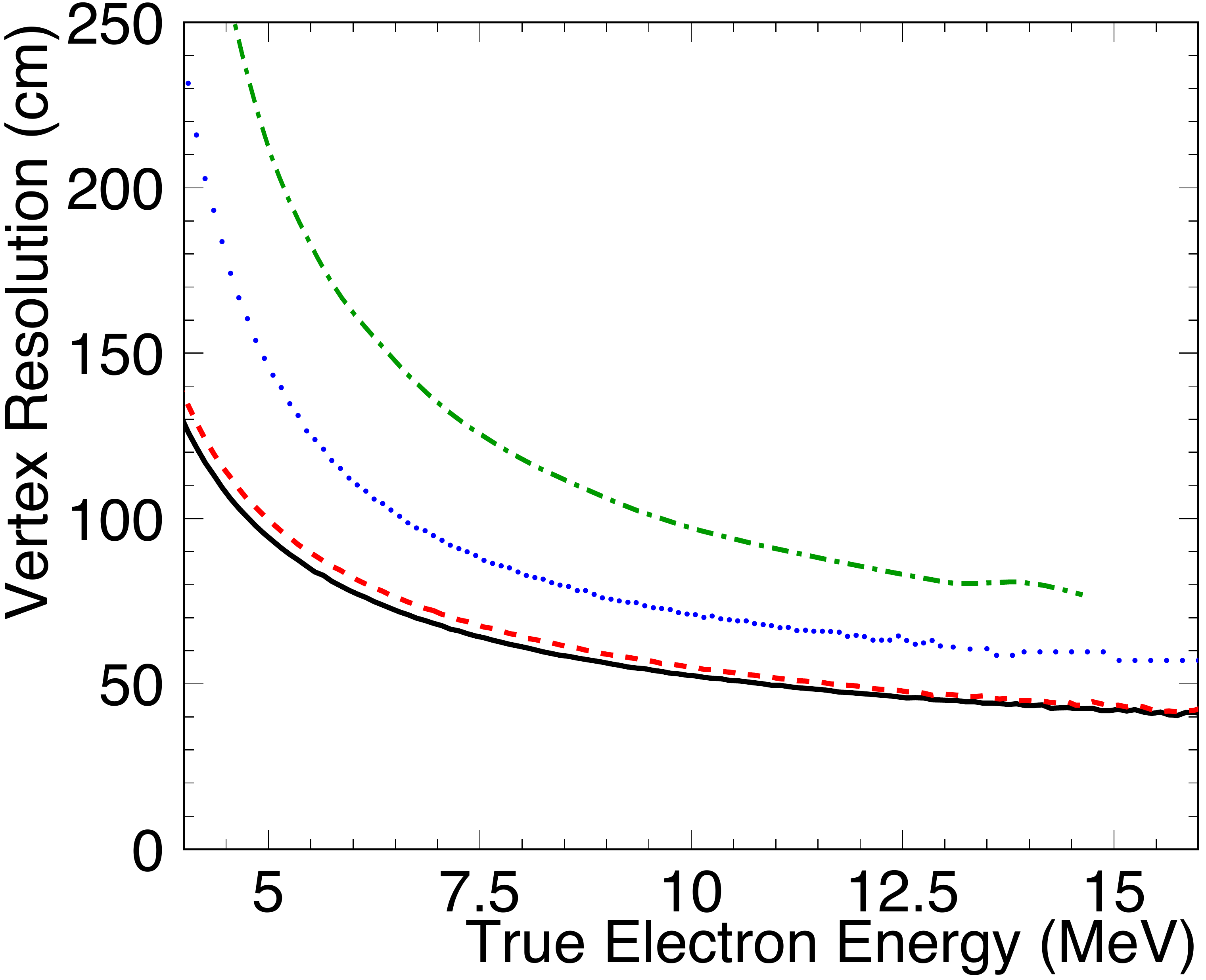}
	\caption[Vertex resolution in Super-Kamiokande]{Vertex resolution in Super-Kamiokande-I (dotted blue line), -II (dash-dotted green), -III (dashed red) and -IV (solid black) as a function of the true electron energy. While the resolution in phase II is worse due to the lower photocoverage, the improvement in phases III and IV compared to phase I is due to the improved vertex reconstruction described in the text. Figure from reference~\cite{Abe2016a}.}
	\label{fig-sim-reco-vertex}
\end{figure}

The resulting vertex resolution achieved across different phases of Super-Kamiokande is shown in figure~\ref{fig-sim-reco-vertex}.
At \SI{22}{cm / ns}, this corresponds to a few \si{ns}, which is comparable in magnitude to the time resolution of PMTs.
Overall, the event time can be reconstructed with an uncertainty of a few \si{ns}, which is much smaller than the \si{ms}-scale time resolution of neutrino fluxes provided by most supernova simulations (see section~\ref{ch-sim-models}).%
\footnote{Due to their finite mass, neutrinos travel slightly slower than the speed of light in vacuum, with the exact speed depending on the neutrino energy.
The resulting time-of-flight difference between high- and low-energy neutrinos from a galactic supernova is~\cite{Lund2010}
\begin{equation}
\Delta t=0.57 \mathrm{ms}\left(\frac{m_{v}}{\mathrm{eV}}\right)^{2}\left(\frac{30 \mathrm{MeV}}{E}\right)^{2}\left(\frac{D}{10 \mathrm{kpc}}\right),
\end{equation}
implying that sub-\si{ms} features in the neutrino flux may get washed out on the way to Earth.
}
For all practical purposes, Hyper-Kamiokande’s event time reconstruction is thus perfect.

\subsection{Direction Reconstruction}\label{ch-sim-reco-direction}
Electrons and positrons travelling through water emit Cherenkov light in a cone with an angle
\begin{equation}
\cos \theta_\text{Ch} = \frac{1}{n \beta}
\end{equation}
relative to the particle’s direction.
Since the index of refraction in water is $n \approx 1.34$ and $\beta = v/c \approx 1$ for electrons and positrons with an energy of $E_e \gg m_e$, that angle is $\theta_\text{Ch} \approx \SI{42}{\degree}$.

By combining this knowledge with the reconstructed vertex, \textsc{BONSAI} is able to reconstruct the direction of individual particles.
To reduce the contribution of dark noise and scattered photons, it considers only the $N_{20}$ hits whose time-of-flight-corrected hit time is within \SI{20}{ns} of the reconstructed event time.
It then maximizes the likelihood function~\cite{Nakano2016}
\begin{equation}
\mathcal{L}(\vec{d}) = \sum_{i}^{N_{20}} \log \left( f(\cos \theta_i, E_e) \right) \times \frac{\cos \theta^\text{inc}_i}{a (\theta^\text{inc}_i)}.
\end{equation}

\begin{figure}[htbp]
	\centering
	\includegraphics[scale=1.2]{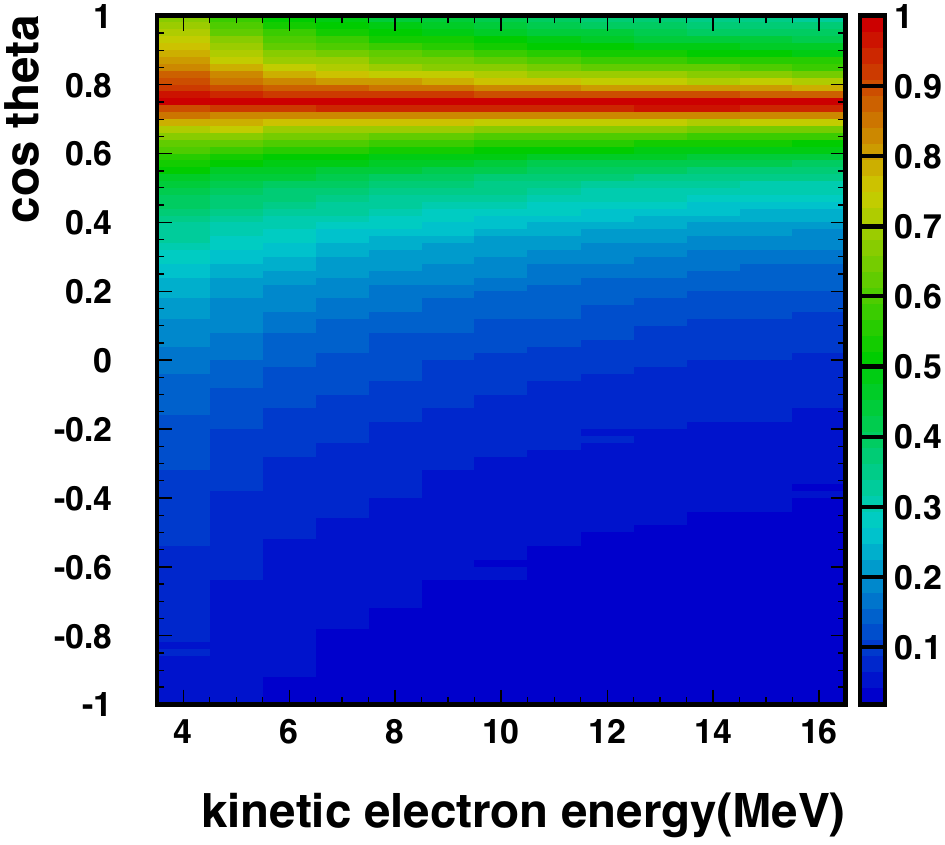}
	\caption[Distribution of angle between particle direction and the vector from event vertex to PMT location as a function of energy]{Distribution of angle between particle direction and the vector from event vertex to PMT location as a function of energy. At lower energies, multiple Coulomb scattering deflects the particle more strongly, which leads to a wider angular spread. Figure from reference~\cite{Abe2011c}.}
	\label{fig-sim-reco-openingangle}
\end{figure}

Here, $f (\cos \theta_i, E_e)$, is the expected distribution of the angle between the direction $\vec{d}$ of the particle and the vector pointing from the reconstructed vertex to the location of the hit PMT.
While the maximum of this distribution is approximately at $\cos \SI{42}{\degree} \approx 0.75$ as discussed above, an energy-dependent spread in the opening angle occurs due to multiple Coulomb scattering of the particle in water. 
The exact shape of this distribution is determined from MC simulations; the one used in Super-Kamiokande-IV is shown in figure~\ref{fig-sim-reco-openingangle}.

The second term in that likelihood function depends on the angle of incidence onto the PMT, $\theta^\text{inc}_i$, which is defined in the same way as $\theta$ in figure~\ref{fig-sim-reco-photocoverage}.
$a(\theta^\text{inc}_i)$ is determined by the shape of the PMT and the acrylic cover.
In Super-Kamiokande, the expression used since phase II is
\begin{equation}
a(\theta^\text{inc}_i) = 0.205 + 0.524 \cos \theta^\text{inc}_i + 0.390 \cos^{2} \theta^\text{inc}_i - 0.132 \cos^{3} \theta^\text{inc}_i.
\end{equation}

\subsection{Energy Reconstruction}\label{ch-sim-reco-e}
Based on the reconstructed event vertex, it is now possible to reconstruct the energy of the detected particle from the amount of Cherenkov light it produced.
To do this, I have developed a \textsc{root} script~\cite{energetic-bonsai} that implements the algorithm used in Super-Kamiokande~\cite{Abe2011c}.
The algorithm itself and changes due to the different detector are described in this section.

We first calculate the light travel time from the reconstructed event vertex to each hit photosensor to determine the time-of-flight-corrected times of each hit in the event.
Assuming a sufficiently accurate position reconstruction, Cherenkov photons should now form a narrow peak around the event time, while the dark noise hits will remain randomly distributed.
We then search for the \SI{50}{ns} interval with the highest number of time-of-flight-corrected hit times and refer to the number of hits as $N_{50}$.

Broadly speaking, a higher $N_{50}$ indicates a higher-energy particle; however, at a given particle energy $N_{50}$ has a significant position dependence due to geometric and detector effects.
To correct for this, we estimate the effective number of Cherenkov photons sent out by the detected particle, which is given by
\begin{equation}\label{eq-sim-reco-energy}
N_\text{eff} = \sum_{i}^{N_{50}} \left[ (X_i + \varepsilon_\text{tail} - \varepsilon_\text{dark}) \times \frac{N_\text{PMT}}{N_\text{alive}} \times \frac{1}{S (\theta_i, \phi_i)} \times \exp \left(\frac{r_i}{\lambda_\text{eff}}\right) \times \frac{1}{\text{QE}_i}\right].
\end{equation}
Each term in equation~\ref{eq-sim-reco-energy} is discussed in a separate subsection below.

$N_\text{eff}$ is a position-independent quantity that can be used to estimate the energy of the detected particle.
The relation between both quantities is described in the final subsection below.

\subsubsection{Occupancy Correction}\label{ch-sim-reco-e-occupancy}
For the purposes of this section, we assume that each photosensor was hit by at most one photon.%
\footnote{This reflects the poor charge resolution of the PMTs used in Super-Kamiokande (see figure~\ref{fig-hk-photosensor_1petts}), which meant that their \SI{1}{PE} and \SI{2}{PE} peaks are not well separated.
The B\&L PMTs developed for Hyper-Kamiokande (see section~\ref{ch-hk-id-blpmt}) have an improved charge resolution which may in the future be used instead of or in addition to this occupancy correction.
However, since the B\&L PMTs are still under development, trying to include this within this thesis would be premature.}
That assumption may not be true for high-energy events or events that are close to and pointing towards the wall such that the Cherenkov cone did not have the opportunity to spread out.

To correct for this, we take advantage of the fact that if a photosensor detected more than one photon, it is very likely that its neighbouring photosensors have also detected photons.
We define $x_i$ as the fraction of photosensors within a $3 \times 3$ grid around the original photosensor that registered a hit within that same time-of-flight-corrected \SI{50}{ns} window.%
\footnote{Photosensors at the top or bottom edge of the barrel have five neighbours instead of eight, which I take into account.
For photosensors located at the edge of the top or bottom plane of the detector, there may be between three and seven neighbours within the plane.
Since this depends on the exact layout of photosensors, it is not yet determined and I will omit this correction here.
Due to the small number of affected photosensors, the effect of this is negligible.}
The corrected number of hits in the $i$-th photosensor is then given by
\begin{equation}
X_i =
	\begin{cases}
		1 & \text{if } x_i = 0\\
		\frac{- \ln(1-x_i)}{x_i} & \text{if } x_i < 1\\
		3 & \text{if } x_i = 1.
	\end{cases}
\end{equation}

This correction term is derived from a Poisson distribution by assuming that the $3 \times 3$ photosensors receive the same expected number of photons.
For the detector configuration with 20\,\% photocoverage, the distance between neighbouring photosensors is increased and this assumption starts to break down---particularly for high-energy events that occur close to the wall.
In that configuration, I therefore use a modified occupancy correction given by $X_i^\text{mod} = X_i^{1.4}$, where the value $1.4$ in the exponent was calibrated using Monte Carlo simulations to reproduce an approximately linear relation between true particle energy and reconstructed $N_\text{eff}$.%
\footnote{An alternative approach would be to consider only the four closest neighbouring photosensors instead of all eight. While this would salvage the assumption of equal illumination, it would lead to a lower dynamic range and decrease reconstruction accuracy in that way.}

\subsubsection{Late Hit Correction}
Scattering in the water means that the time-of-flight correction underestimates the path length travelled by some Cherenkov photons before arriving at the photosensor.
As a result, their corrected arrival time may be outside of the \SI{50}{ns} window.

To account for these scattered photons, we define $N_{100}$ analogous to $N_{50}$ and determine the fraction of delayed hits by calculating
\begin{equation}
\varepsilon_\text{tail} = \frac{N_{100} - N_{50} - N_\text{alive} \cdot R_\text{dark} \cdot \SI{50}{ns}}{N_{50}},
\end{equation}
where $N_\text{alive}$ is the number of working photosensors and $R_\text{dark}$ is their dark noise rate.

\subsubsection{Dark Noise Correction}
To substract dark noise events, we determine the fraction of dark noise events expected within that \SI{50}{ns} window by calculating
\begin{equation}
\varepsilon_\text{dark} = \frac{N_\text{alive} \cdot R_\text{dark} \cdot \SI{50}{ns}}{N_{50}}.
\end{equation}

\subsubsection{Dead Photosensor Correction}
Despite testing before installation, some photosensors will break during the lifetime of the detector.
Out of the total number of photosensors, $N_\text{PMT}$, only a smaller number ($N_\text{alive}$) of photosensors may be working correctly at any point in time.
To correct for this, the formula for $N_\text{eff}$ contains the factor $N_\text{PMT} / N_\text{alive}$.

However, such defects do not affect the simulations performed here and as a result, I assume $N_\text{alive} = N_\text{PMT}$ here.

\subsubsection{Photocoverage Correction}
The effective photocoverage $S (\theta_i, \phi_i)$ depends on the angle at which the incoming photon hits the detector wall.
For small $\theta_i$, the effective photocoverage is equal to the nominal photocoverage, i.\,e. either 40\,\% or 20\,\%.
At large $\theta_i$, the protruding PMTs shadow parts of the detector wall so that the effective photocoverage increases.

\begin{figure}[htbp]
	\centering
	\includegraphics[scale=0.82]{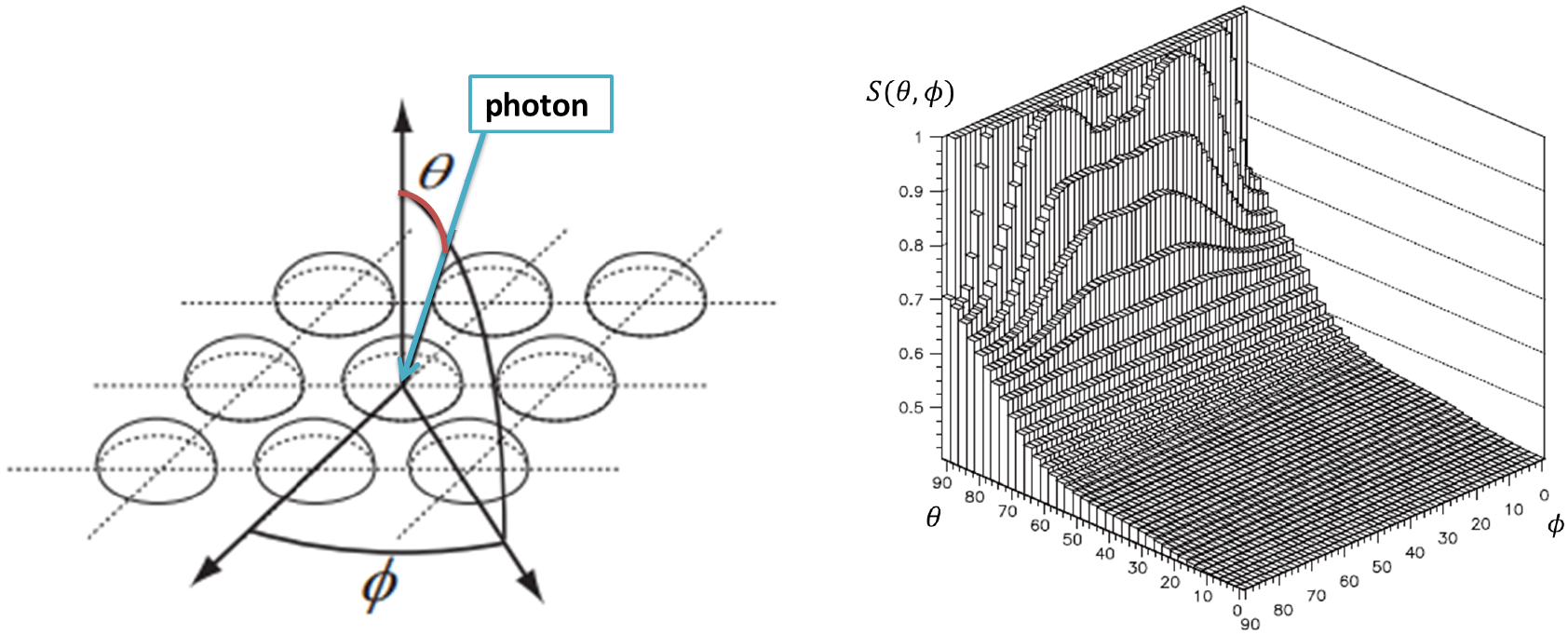}
	\caption[Angular dependence of photocoverage correction in Super-Kamiokande]{Angular dependence of photocoverage correction. Left: Sketch of inner wall of detector, showing the definition of $\theta_i$ and $\phi_i$. Right: Shape of $S (\theta_i, \phi_i)$ used by the Super-Kamiokande collaboration. Figures from references~\cite{Ikeda2009,Nakano2016}.}
	\label{fig-sim-reco-photocoverage}
\end{figure}

The exact shape of $S (\theta_i, \phi_i)$ is determined via Monte Carlo simulation of the detector and depends on the exact arrangement of photosensors as well as the shape of the photosensors and acrylic covers.
In Super-Kamiokande IV, the shape shown in figure~\ref{fig-sim-reco-photocoverage} is used.
Since the final design of those components for Hyper-Kamiokande is not yet fixed, in this thesis I use a simplified angular dependence with ten bins in $\theta_i$ and I omit the dependence on $\phi_i$.
This qualitatively reproduces the more precise angular bins used in Super-Kamiokande and changes $N_\text{eff}$ by $\mathcal{O}(1\,\%)$.

\subsubsection{Water Transparency Correction}
Absorption losses in the water scale as
\begin{equation}
\exp \left(\frac{r_i}{\lambda_\text{eff}}\right),
\end{equation}
where $\lambda_\text{eff}$ is the water transparency and $r_i$ is the distance between the event vertex and the position of the $i$-th photosensor.
In Super-Kamiokande, the water transparency varies over time but is typically above \SI{100}{m}~\cite{Abe2016a}.
In this thesis, I assume a fixed water transparency of \SI{100}{m} for simulations in WCSim as well as for energy reconstruction.

\subsubsection{Quantum Efficiency Correction}
The quantum efficiency $\text{QE}_i$ of photosensors in the detector can differ due to individual variations or if different types of photosensors are used, e.\,g. due to defective photosensors being replaced with modern ones during upgrades of Super-Kamiokande. 
This variation is accounted for by dividing by the quantum efficiency of each individual photosensor.

In the simulations performed here, the quantum efficiency of all PMTs is identical.
As a result, $\text{QE}_i$ becomes a constant that I absorb into the relation between $N_\text{eff}$ and the reconstructed energy described below.

\subsubsection{Determination of Particle Energy}
\begin{figure}[b]
	\begin{minipage}{17pc}
	\includegraphics[width=16pc]{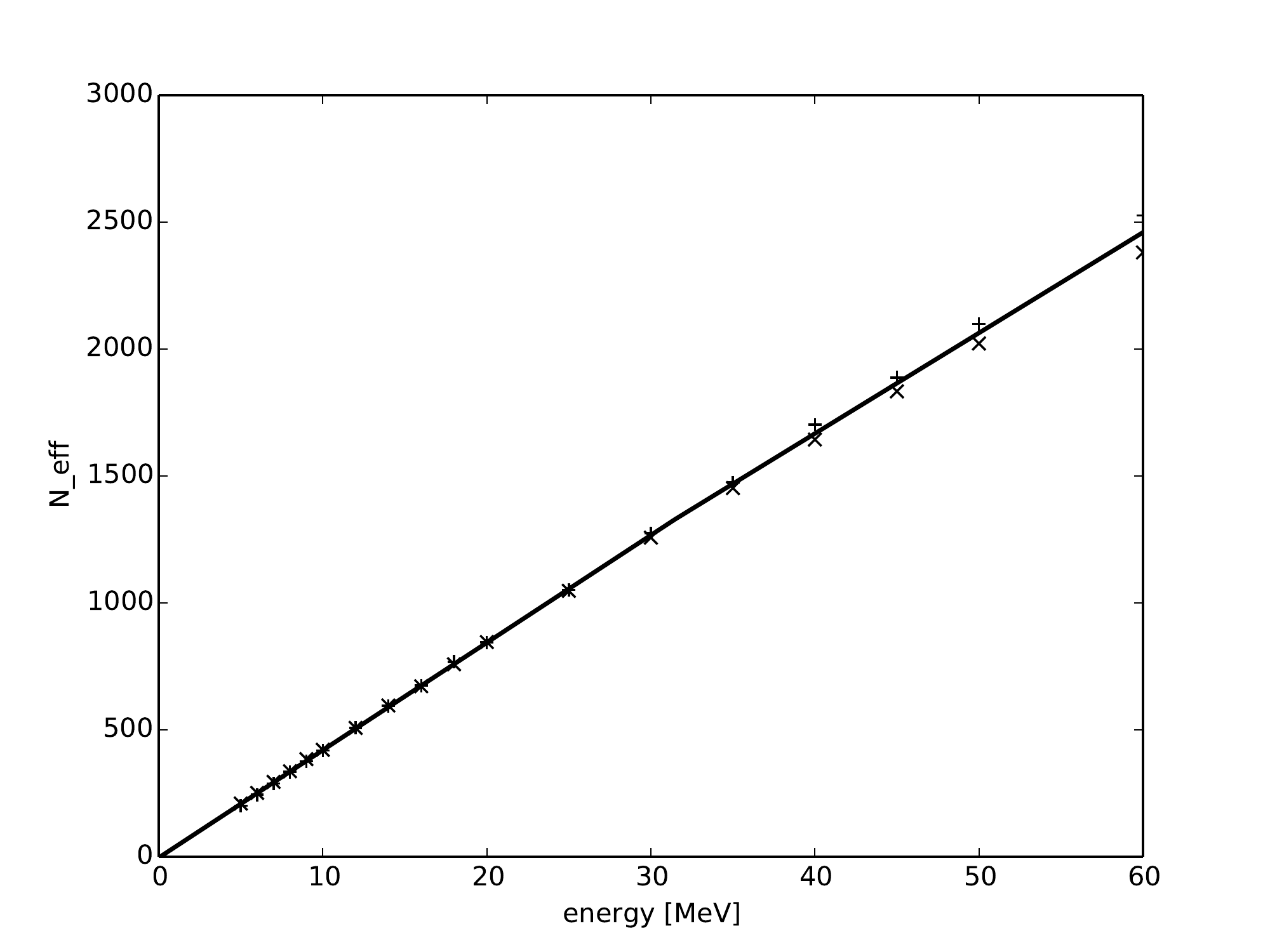}
	\end{minipage}
	\begin{minipage}{17pc}
	\includegraphics[width=16pc]{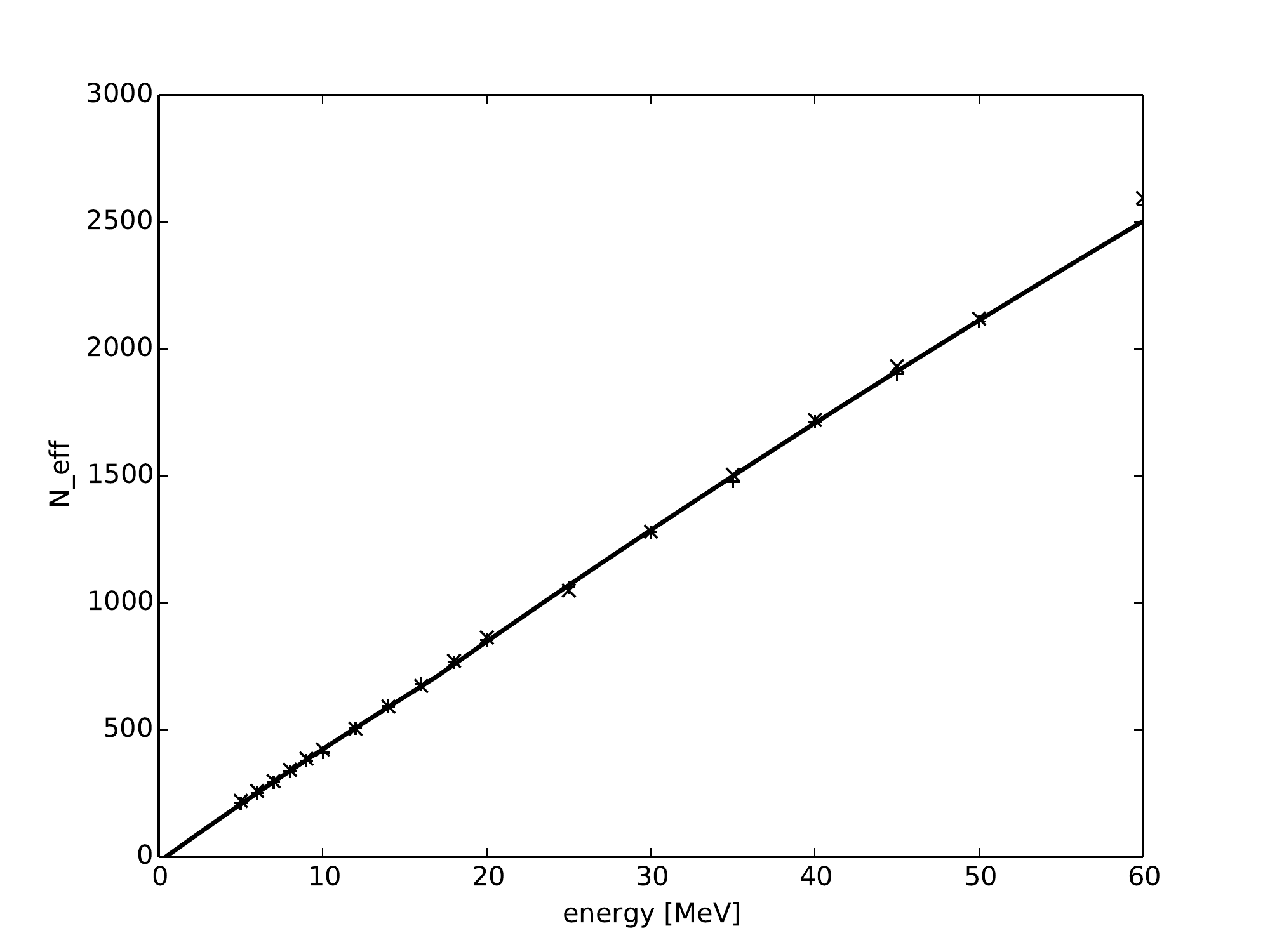}
	\end{minipage}
	\caption[Relation between $N_\text{eff}$ and reconstructed energy]{Relation between $N_\text{eff}$ and $E_\text{reco}$ for 40\,\% photocoverage (left) and 20\,\% photocoverage (right). Marks give MC results while the solid line is the fit described in the text.}
	\label{fig-sim-reco-neff2energy}
\end{figure}

After including all corrections described above to calculate $N_\text{eff}$, I use MC simulations of mono-energetic electrons 
to determine the relation between $N_\text{eff}$ and the particle energy, as shown in figure~\ref{fig-sim-reco-neff2energy}.

While this relation is expected to be approximately linear, the Super-Kamiokande collaboration has found a slightly non-linear relation at low energies and uses a fourth-order polynomial fit in that region~\cite{Abe2011c}.
In my calibration, using a higher order polynomial has not produced a noticeably improved fit for the 40\,\% photocoverage configuration.
For the 20\,\% photocoverage configuration, a quadratic function has produced an improved fit. In addition, as discussed earlier I use the modified occupancy correction $X_i^{1.4}$ for that configuration, which affects the relation between energy and $N_\text{eff}$ at high energies.

As a result, in this thesis I use the relations
\begin{equation}
E_\text{reco} =
	\begin{cases}
		0.02360 N_\text{eff} + 0.082 &\text{if } N_\text{eff} < 1320\\
		0.02524 N_\text{eff} - 2.081 &\text{if }  N_\text{eff} \ge 1320
	\end{cases}
\end{equation}
for the detector configuration with 40\,\% photocoverage and
\begin{equation}
E_\text{reco} =
	\begin{cases}
		\num{2.55e-6} N_\text{eff}^2 + 0.0215 N_\text{eff} + 0.429 &\text{if } N_\text{eff} < 701\\
		\num{1.148e-6} N_\text{eff}^2 + 0.02032 N_\text{eff} + 1.94 &\text{if }  N_\text{eff} \ge 701
	\end{cases}
\end{equation}
for the detector configuration with 20\,\% photocoverage.
However, given the various approximations discussed throughout section~\ref{ch-sim-reco-e}, this relation needs to be recalibrated once the detector design has been finalized.

\chapter{Supernova Model Discrimination}\label{ch-ana}

\setlength{\epigraphwidth}{.45\textwidth}
\epigraphhead[0]{\epigraph{All supernova models are different,\\ but some are more different than others.}{\textit{George Orwell}}}

While the fundamental explosion mechanism of core-collapse supernovae is believed to be understood thanks to a combination of computer simulations and observations of the neutrino burst from SN1987A, details of the explosion mechanism still remain unclear.
Even for identical progenitors, different simulations---which use various different approximations due to computing power limitations---give quantitatively~\cite{OConnor2018a} and, in some cases, qualitatively~\cite{OConnor2018b} different results.
It is therefore essential to use the next galactic supernova---which may well be a once-in-a-lifetime opportunity---to compare model predictions with observations and figure out which model best represents reality. 

Among current and planned neutrino detectors, Hyper-Kamiokande is unique in its ability to detect both a high number of neutrinos from a galactic supernova---about an order of magnitude higher than detectors like Super-Kamiokande, DUNE or JUNO---and provide precise energy information for every single event, whereas IceCube would only be able to determine the average neutrino energy.
The goal of this thesis is to develop a method of distinguishing between different supernova models that makes optimal use of Hyper-Kamiokande’s capabilities.

After laying the groundwork in previous chapters by introducing Hyper-Kamiokande and the software toolchain for simulating and reconstructing supernova neutrino events, this chapter describes my analysis and its results.
I start in section~\ref{ch-ana-gen} by describing how the data sets used for this analysis were generated, simulated and reconstructed.
Section~\ref{ch-ana-cuts} discusses the cuts applied to reconstructed events.
In section~\ref{ch-ana-ll}, I derive the log-likelihood function used, before showing results of a comparison of five different supernova models in section~\ref{ch-ana-modelfamilies}.
I give one example of a more targeted study by showing how this same method can be used to distinguish between closely related simulations in section~\ref{ch-ana-siblings}.
Finally, section~\ref{ch-ana-bayes} discusses how the analysis of an observed supernova neutrino burst will differ from the one presented here.

\enlargethispage{\baselineskip} 
Throughout this chapter I will assume that the distance to the supernova---and thus the normalization of the neutrino flux---is completely unknown.
If additional distance information is available---e.\,g. because an optical counterpart is identified---this could in principle be used to further distinguish between different supernova models as described in section~\ref{ch-conclusions-mma}.

\section{Generating Data Sets}\label{ch-ana-gen}
I have used the event generator sntools (described in section~\ref{ch-sim-sntools}) to generate data sets of 100 and 300 events from the five supernova models described in section~\ref{ch-sim-models} and for both normal and inverted mass ordering.
For every combination of these parameters, I have generated 1000 data sets in order to estimate how well Hyper-Kamiokande is able to identify the true model despite the random fluctuations in the observed events.

Since the main goal of this thesis was a broad model comparison demonstrating the reach of Hyper-Kamiokande, the two sizes of data sets were chosen to represent supernovae at distances that mark the upper end of the expected spatial distribution of observable supernovae.
As table~\ref{tab-ana-distance} shows, 300 events correspond to a distance of \SI{59}{kpc} or more for all supernova models considered here, which includes the whole Milky Way and many of its satellite galaxies, including the Large and Small Magellanic Clouds at distances of \SI{50}{kpc}~\cite{Pietrzynski2013} and \SI{61}{kpc}~\cite{Hilditch2005}, respectively.

\begin{table}[tbp]
\begin{center}
\begin{tabular}{lccc}
Model & $N_{\SI{10}{kpc}}$ & $d_{100}$ & $d_{300}$\\
\hline
Totani & \num{19716} & \SI{140}{kpc} & \SI{81}{kpc}\\
Nakazato & \num{17978} & \SI{134}{kpc} & \SI{77}{kpc}\\
Couch & \num{27539} & \SI{166}{kpc} & \SI{96}{kpc}\\
Vartanyan & \num{10372} & \SI{102}{kpc} & \SI{59}{kpc}\\
Tamborra & \num{25021} & \SI{158}{kpc} & \SI{91}{kpc}
\end{tabular}
\end{center}
\caption[Event numbers and corresponding distances for the supernova models considered in this work]{Number of events expected during the time interval of \SIrange{20}{520}{ms} for a supernova at the fiducial distance of \SI{10}{kpc} ($N_{\SI{10}{kpc}}$) and the distances at which 100 or 300 events are expected ($d_{100}$ and $d_{300}$, respectively) for the five supernova models considered in this work.}
\label{tab-ana-distance}
\end{table}%

A closer supernova---particularly one within the Milky Way, i.\,e. at distances of less than \SI{20}{kpc}---would of course result in a higher number of events and thus improve our ability to distinguish different models.
This will be particularly interesting for future, more targeted studies and I will give one example of such a study in section~\ref{ch-ana-siblings}.

Throughout this chapter, I consider only the time interval from \SIrange{20}{520}{ms} after the core bounce.
This time interval contains the shock stagnation and accretion phase (described in section~\ref{ch-intro-explosion-mechanism}), which contains clear signatures of the explosion mechanism and exhibits the largest differences between models.
The earlier~\cite{Kachelries2005} (\nue burst) and later\footnote{see e.\,g. reference~\cite{Suwa2019}, which also shows that the neutrino signal from the cooling phase can instead be used to determine properties of the resulting neutron star} (cooling) phases of neutrino emission are much better understood and exhibit only minor variations between models, making them less relevant for the analysis presented here.
Furthermore, due to the limited computing time available, many simulations---including the Couch, Vartanyan and Tamborra models used here---focus on the accretion phase and don’t include the full cooling phase.
Accordingly, by considering only this \SI{500}{ms} time interval I am able to include a wider range of models.

These events are then simulated in the detector simulation software WCSim (see section~\ref{ch-sim-wcsim}).
As discussed in section~\ref{ch-hk-id}, the exact photosensor configuration of the detector has not yet been decided upon. 
Here, I will initially consider two configurations.
The first is the reference configuration described in the design report~\cite{HKDR2018}. It consists of approximately \num{40000} box-and-line PMTs, resulting in a photocoverage of 40\,\%.
The other is a minimal configuration, which uses half the number of B\&L PMTs to achieve a 20\,\% photocoverage.%
\footnote{This configuration is \emph{not} being considered by the Hyper-Kamiokande collaboration. The alternative configurations that \emph{are} under consideration would augment this minimal configuration with several thousand multi-PMT modules. I use this minimal configuration only as the most conservative estimate since the exact number of mPMT modules is not yet known.}
Results for the smaller data set size (100 events), shown in section~\ref{ch-ana-modelfamilies}, demonstrate that there is no significant difference between both configurations in this analysis.
For the larger data set size (300 events), I have therefore focussed on the minimal configuration, which requires significantly less computing resources to simulate and reconstruct events.

The position, direction and energy of the outgoing lepton were then reconstructed using hk-\textsc{BONSAI} and the energy reconstruction script described in section~\ref{ch-sim-reco}.

\section{Data Reduction}\label{ch-ana-cuts}

After reconstruction, I apply two cuts to all reconstructed events: an energy cut, which removes all events with a reconstructed energy less than \SI{5}{MeV}, and a fiducial volume cut, which removes all events whose reconstructed vertex is less than \SI{1.5}{m} away from the top, bottom or side walls of the inner detector.

These cuts are intended to eliminate low-energy background from accidental coincidences of dark noise as well as radioactive decays in the detector.
Analogous cuts are also used for the solar neutrino analysis in Super-Kamiokande~\cite{Abe2016a}.
While several more advanced cuts used in that analysis cannot currently be applied to the analysis presented here,%
\footnote{Note that the Super-Kamiokande analysis includes events with energies as low as \SI{3.49}{MeV} within a reduced fiducial volume and instead uses several more advanced cuts to cut remaining background events. Since these cuts rely on comparison of Monte Carlo simulations with observed data, they cannot be applied to Hyper-Kamiokande at this time.}
the more stringent energy cut together with the much higher event rate (\SIrange{e2}{e3}{Hz} for the distant supernova bursts considered here, compared to about \SI{e-4}{Hz} for solar neutrinos in Super-Kamiokande) result in an effectively background-free data set.
Other backgrounds, including muon-induced spallation events or atmospheric neutrinos, occur at a much lower rate and are thus negligible during the single \SI{500}{ms} time interval considered here.

Once Hyper-Kamiokande is operating and the low-energy backgrounds are characterized in detail, it will likely be possible to develop more targeted cuts that allow us to include more low-energy events and extend the fiducial volume while remaining effectively background-free.
For the purposes of the current analysis, however, I have decided to err on the side of more conservative results, which are unaffected by variations in background levels, instead of presenting more optimistic results, which are dependent on uncertain assumptions regarding background levels.

Furthermore, I do not take into account the effects of other cuts that will occasionally occur during detector operations---e.\,g. due to calibration runs or hardware issues.

\begin{figure}[htbp]
	\centering
	\includegraphics[scale=0.5]{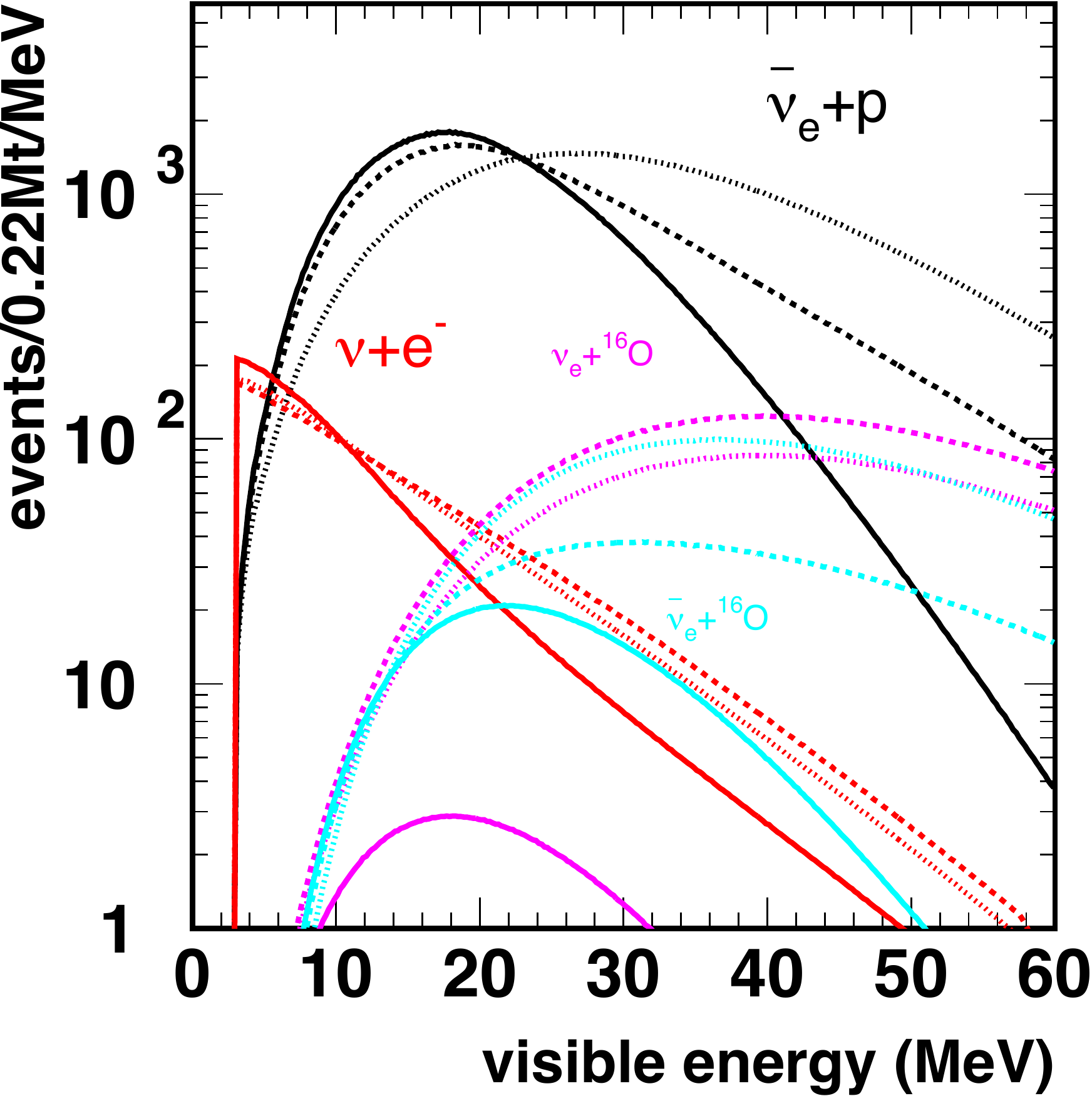}
	\caption[Energy spectra for different interaction channels in Hyper-Kamiokande]{Energy spectra of events from a supernova neutrino burst in Hyper-Kamiokande assuming the Totani model and a fiducial distance of \SI{10}{kpc}. Different colours stand for inverse beta decay (black), $\nu e$-scattering (red), \nue +$^{16}$O CC (purple) and \nuebar +$^{16}$O CC (light blue), while solid, dashed and dotted lines correspond to no oscillation, normal ordering and inverted ordering, respectively. Figure from reference~\cite{HKDR2018}.}
	\label{fig-ana-all_spectrum_hk_osc_resol}
\end{figure}

The fiducial volume cut described above removes about 13\,\% of all events in the inner detector.
The effect of the energy cut depends on the energy spectrum of the initial neutrino flux and therefore on the supernova model and the mass ordering.
As an example, figure~\ref{fig-ana-all_spectrum_hk_osc_resol} shows the energy spectra in different interaction channels for the Totani model.
Due to the strong energy-dependence of the cross sections, three of the four interaction channels produce almost no events at \SI{5}{MeV} or below.
Only elastic $\nu e$-scattering---a subdominant channel which contributes about 5\,\% of all events---has a significant contribution at energies below \SI{5}{MeV}.
Overall, out of the initial 100 or 300 events per data set more than 80\,\% typically remain after applying these cuts.

\section{Log-Likelihood Function}\label{ch-ana-ll}
After the cuts described above, I apply a log-likelihood function to the reconstructed times and energies of the remaining events in each data set to determine how well that data set matches each of the supernova models.

This log-likelihood function is similar to one that was originally derived for analysis of SN1987A taking into account only the main interaction channel, inverse beta decay~\cite{Loredo1989}.
However, the function used here includes all interaction channels.
It is derived in appendix~\ref{apx-likelihood} and given by
\begin{equation}
L = \ln \mathcal{L} = \sum_{i=1}^{N_\text{obs}} \ln N_i,
\end{equation}
where the index $i$ runs over the $N_\text{obs}$ events remaining in the data set and $N_i$ is the number of events predicted by a given supernova model in an infinitesimally small bin around the reconstructed time and energy of event $i$, summed over all interaction channels.

By using infinitesimally small bins in time and energy, this likelihood function makes optimal use of all available information.
In contrast, using a binned chi-squared test to compare observation with models requires a sufficiently large number of events per bin to be accurate. 
Especially in the case of a distant supernova, where only hundreds or thousands of events may be observed in Hyper-Kamiokande, two-dimensional binning in time and energy would only be possible in very coarse bins, which would lose a lot of the available information.

\begin{figure}[htbp]
	\centering
	\includegraphics[scale=0.65]{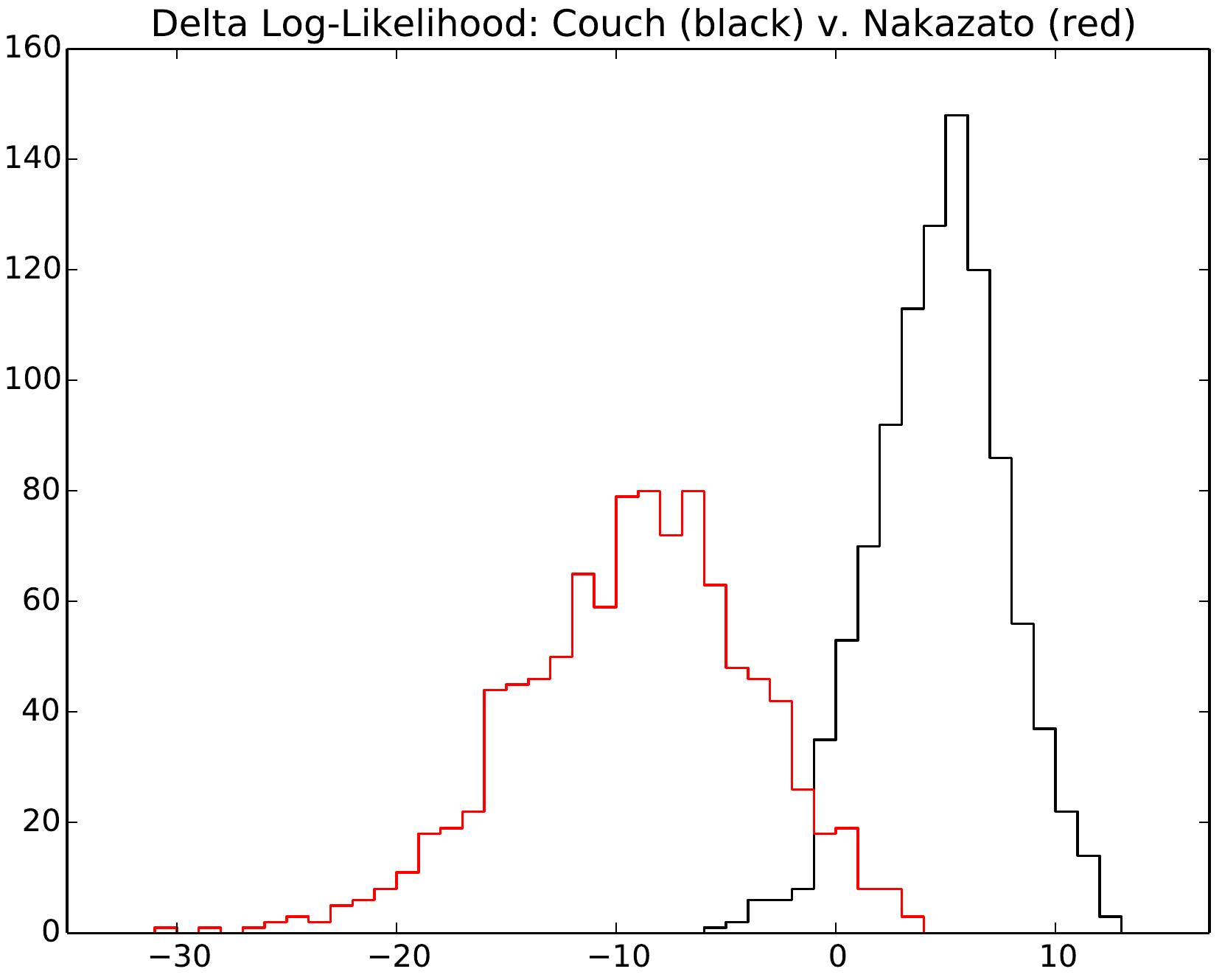}
	\caption[Distribution of $\Delta L$ for data sets generated from the Couch and Nakazato models]{Histograms showing the distribution of $\Delta L = L_\text{Couch} - L_\text{Nakazato}$ for the data sets generated from the Couch model (black) and from the Nakazato model (red), for 100 events per data set and normal mass ordering.}
	\label{fig-ana-deltall_explanation}
\end{figure}

The absolute numerical values of this likelihood function depend on the bin size chosen and are therefore not physically meaningful.
However, when calculating likelihood ratios for different models (i.\,e. differences in the log-likelihood, $\Delta L = L_A - L_B$), this dependence cancels out and the ratio describes whether model A or B is more likely to produce a given data set.
In the following sections, I will therefore exclusively use likelihood ratios to compare different models.

As an example, figure~\ref{fig-ana-deltall_explanation} shows a comparison of the Couch and Nakazato models.
For most data sets generated from the Couch (Nakazato) model, $\Delta L = L_\text{Couch} - L_\text{Nakazato}$ is positive (negative), indicating that this method is generally able to identify the true model.
However, the overlap of both histograms indicates that misidentification sometimes occurs because of random fluctuations in the data sets.
The accuracy of this method will be evaluated below.

\section{Distinguishing Different Models}\label{ch-ana-modelfamilies}
\subsection{N=100 Events Per Data Set}
\subsubsection{Normal Mass Ordering}
\begin{figure}[p]
	\centering
	\includegraphics[scale=0.28]{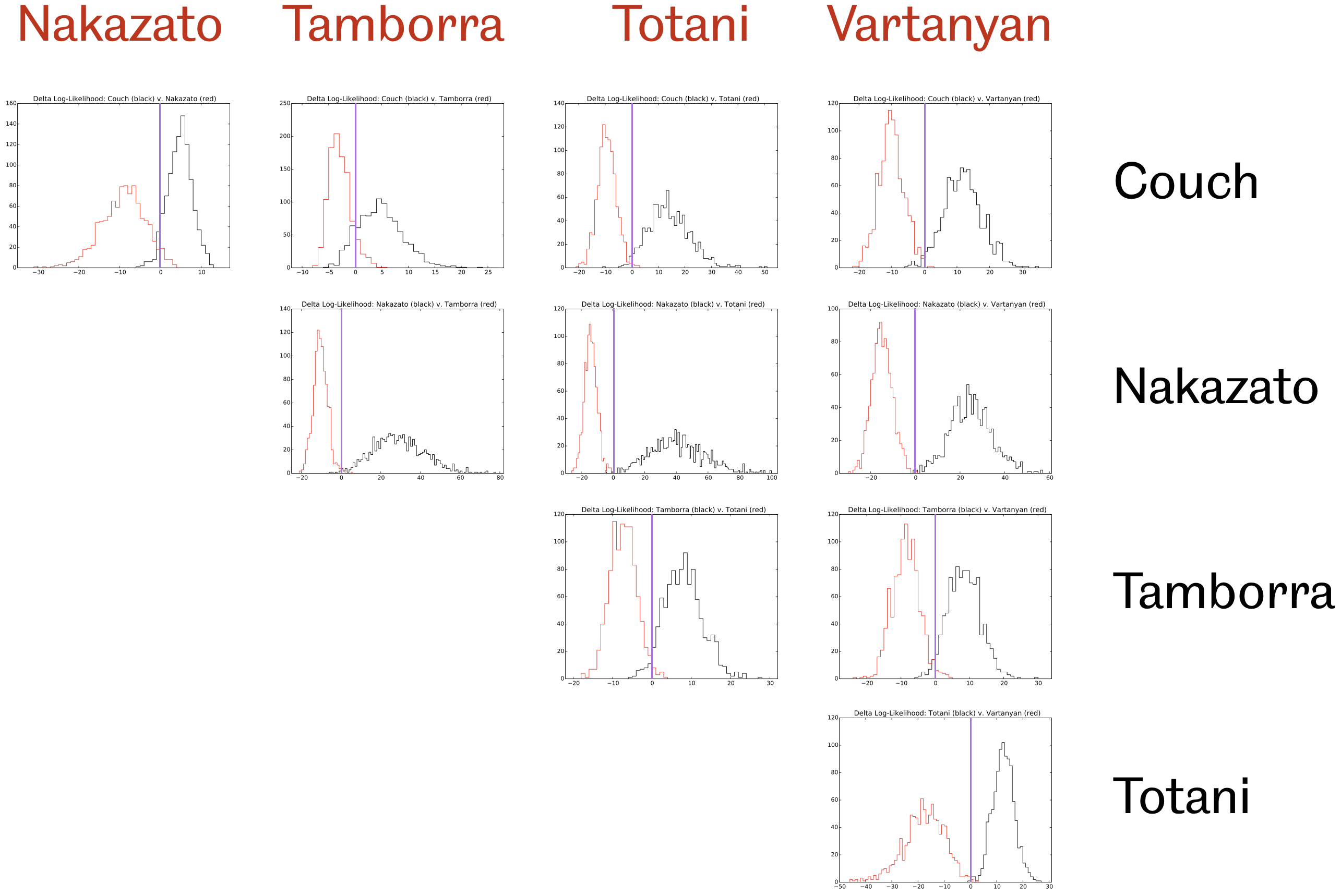}

	\vspace{2pc}

	\includegraphics[scale=0.28]{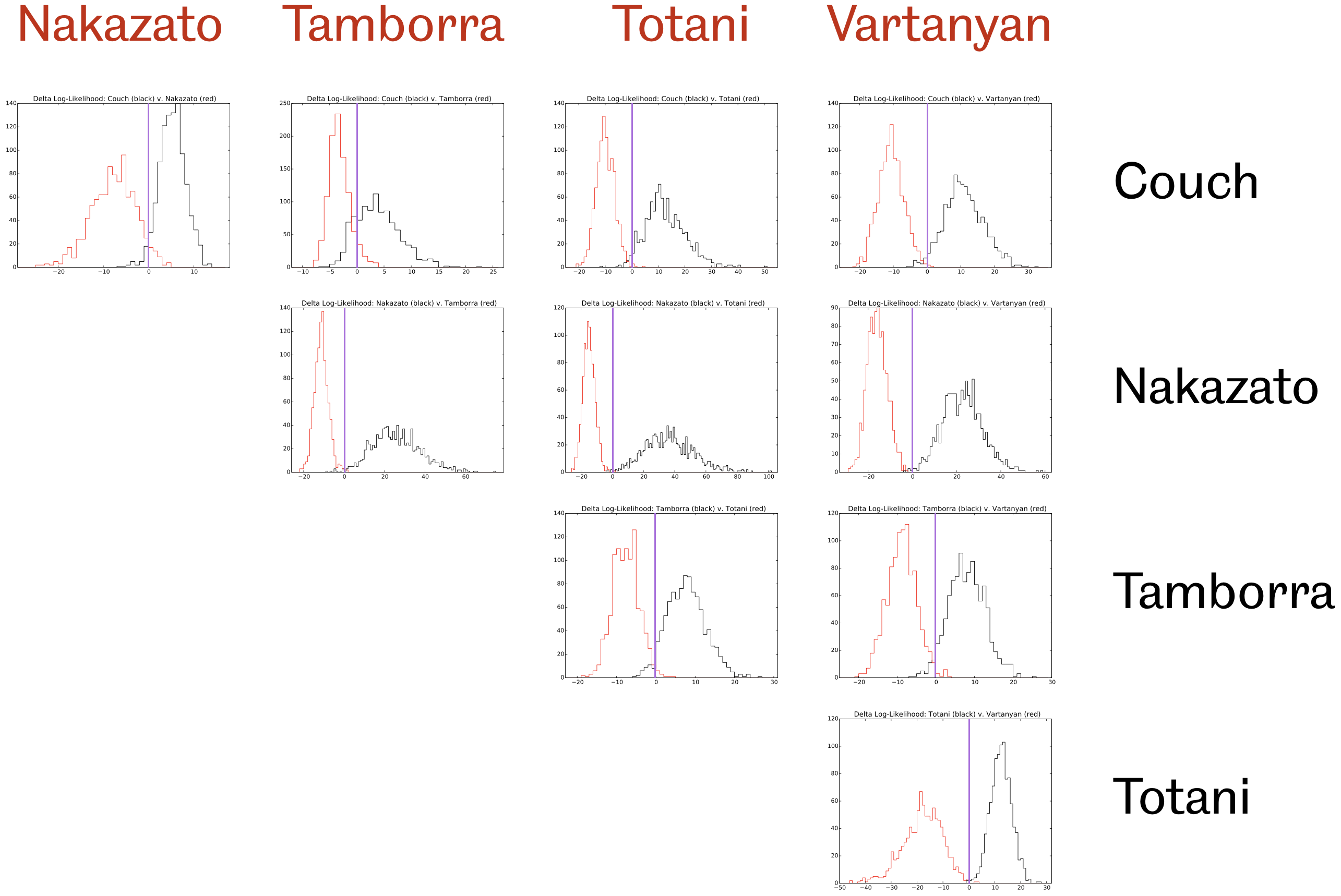}
	\caption[Distribution of $\Delta L$ for all pairs of supernova models: 100 events per data set and normal mass ordering, comparison of 20\,\% and 40\,\% photocoverage]{Top: Histograms showing the distribution of $\Delta L = L_\text{black} - L_\text{red}$ for all pairs of supernova models considered here, for 100 events per data set, normal mass ordering and 20\,\% photocoverage. The purple vertical line in each panel indicates $\Delta L = 0$. Bottom: As above, but for 40\,\% photocoverage.}
	\label{fig-ana-100normal}
\end{figure}
Figure~\ref{fig-ana-100normal} shows pairwise comparisons of the five different models described in section~\ref{ch-sim-models} for both 20 and 40\,\% photocoverage.
\enlargethispage{\baselineskip} 
Similar to the Couch/Nakazato comparison highlighted in figure~\ref{fig-ana-deltall_explanation}, the model pairs generally show clear separation with only minor overlap around $\Delta L = 0$.
The largest overlap is seen between the Couch and Tamborra models, indicating that these models are most similar and hardest to distinguish.

This can be seen more clearly in table~\ref{tab-ana-100normal}, which compares all five supernova models simultaneously by determining which of them produces the highest likelihood for a given data set.
For each model, the respective row indicates how many data sets (out of the 1000 that were generated) were identified as which model.
For example, in the 20\,\% (40\,\%) photocoverage scenario 853 (867) Tamborra data sets were identified correctly, while 84 (65) were misidentified as corresponding to the Couch model.
For the Couch model, almost 80\,\% of data sets were identified correctly in both photocoverage scenarios, with most of the remaining data sets being misidentified as the Tamborra model.
Finally, the three other models are identified correctly in over 95\,\% of all cases.

\begin{table}[tp]
\begin{center}
\begin{tabular}{lrrrrr}
True Model & Couch & Nakazato & Tamborra & Totani & Vartanyan\\
\hline
Couch & \textbf{795} & 57 & 122 & 12 & 14\\
Nakazato & 33 & \textbf{961} & 3 & 1 & 2\\
Tamborra & 84 & 0 & \textbf{853} & 33 & 30\\
Totani & 4 & 0 & 16 & \textbf{979} & 1\\
Vartanyan & 0 & 1 & 17 & 3 & \textbf{979}
\end{tabular}

\vspace{2pc}

\begin{tabular}{lrrrrr}
True Model & Couch & Nakazato & Tamborra & Totani & Vartanyan\\
\hline
Couch & \textbf{768} & 32 & 174 & 11 & 15\\
Nakazato & 41 & \textbf{951} & 5 & 0 & 3\\
Tamborra & 65 & 0 & \textbf{867} & 34 & 34\\
Totani & 2 & 0 & 11 & \textbf{985} & 2\\
Vartanyan & 1 & 0 & 10 & 2 & \textbf{987}
\end{tabular}
\end{center}
\caption[Accuracy of model identification: 100 events per data set and normal mass ordering, comparison of 20\,\% and 40\,\% photocoverage]{Top: Accuracy with which the true model can be identified, for 100 events per data set, normal mass ordering and 20\,\% photocoverage. Shows how many of the 1000 data sets generated for a given model (left column) were identified as each of the five models. Correctly identified models are \textbf{highlighted}. Bottom: As above, but for 40\,\% photocoverage.}
\label{tab-ana-100normal}
\end{table}%

Both figure~\ref{fig-ana-100normal} and table~\ref{tab-ana-100normal} indicate that differences between both photocoverage scenarios are within the range expected due to random fluctuations.
This is consistent with expectations:
Even for the 20\,\% photocoverage configuration, event times are reconstructed in Hyper-Kamiokande with an uncertainty of a few \si{ns}, while changes in the supernova neutrino fluxes take place on the scale of several \si{ms}.
Thus, any incremental improvements to time reconstruction offered by a higher photocoverage do not affect the likelihood.
The improved energy reconstruction could, in principle, have an impact.
However, the energy dependence of the event rate is relatively shallow across most of the energy range (see e.\,g. figure~\ref{fig-ana-all_spectrum_hk_osc_resol}).
Even for the 20\,\% photocoverage configuration, the energy reconstruction uncertainty is sufficiently small that it has only a minor impact on the likelihood.%
\footnote{In contrast, if a background component has a very steep energy cut-off, incrementally improved energy resolution can lead to a significant improvement, e.\,g. when distinguishing solar $^8$B neutrinos from \textit{hep} neutrinos~\cite{HKDR2018}.}
Additionally, this impact may partially cancel itself out as the reconstructed energy is equally likely to be too low or too high.

In the future, improvements to vertex and energy reconstruction offered by a higher photocoverage may allow us to relax the cuts described in section~\ref{ch-ana-cuts} beyond what may be possible with a lower photocoverage.
This would allow us to include more events which could improve the accuracy of model discrimination.
At this point, however, it is not possible to determine the quantitative effect this might have.
Since there is no significant difference between the two photocoverage scenarios in the current analysis, in the interest of clarity I will only show results for the 20\,\% photocoverage configuration in the following sections.

\subsubsection{Inverted Mass Ordering}

\begin{figure}[p]
	\centering
	\includegraphics[scale=0.28]{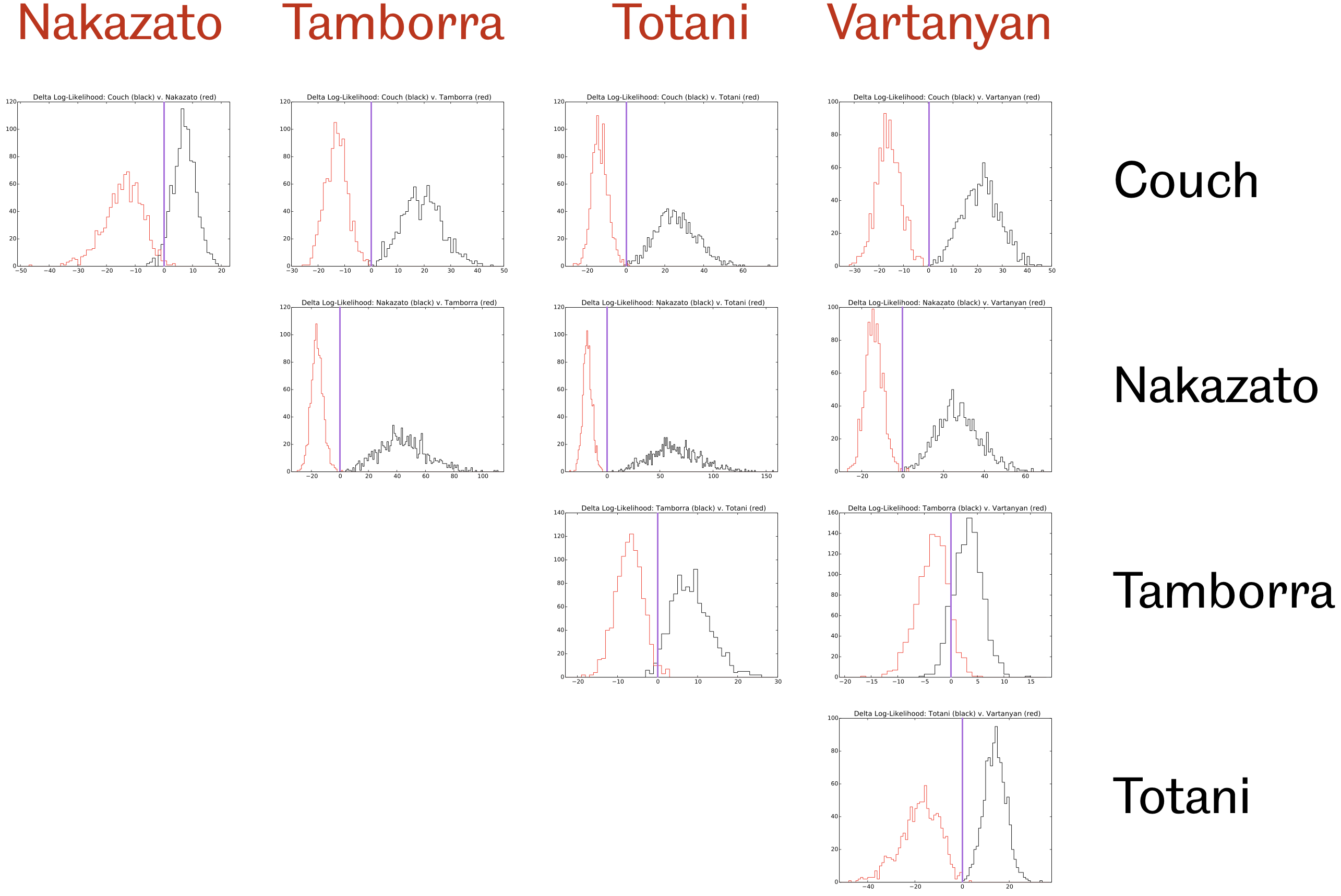}
	\caption[Distribution of $\Delta L$ for all pairs of supernova models: 100 events per data set and inverted mass ordering]{Histograms showing the distribution of $\Delta L = L_\text{black} - L_\text{red}$ for all pairs of supernova models considered here, for 100 events per data set, inverted mass ordering and 20\,\% photocoverage. The purple vertical line in each panel indicates $\Delta L = 0$.}
	\label{fig-ana-100inverted}
\end{figure}

\begin{table}[p]
\begin{center}
\begin{tabular}{lrrrrr}
True Model & Couch & Nakazato & Tamborra & Totani & Vartanyan\\
\hline
Couch & \textbf{960} & 35 & 4 & 1 & 0\\
Nakazato & 8 & \textbf{992} & 0 & 0 & 0\\
Tamborra & 0 & 1 & \textbf{858} & 21 & 120\\
Totani & 3 & 0 & 20 & \textbf{977} & 0\\
Vartanyan & 0 & 2 & 105 & 1 & \textbf{892}
\end{tabular}
\end{center}
\caption[Accuracy of model identification: 100 events per data set and inverted mass ordering]{Accuracy with which the true model can be identified, for 100 events per data set, inverted mass ordering and 20\,\% photocoverage. Shows how many of the 1000 data sets generated for a given model (left column) were identified as each of the five models. Correctly identified models are \textbf{highlighted}.}
\label{tab-ana-100inverted}
\end{table}%

Figure~\ref{fig-ana-100inverted} and table~\ref{tab-ana-100inverted} show results for the inverted mass ordering.
In this scenario, the largest overlap is observed between the Tamborra and Vartanyan models, with an 85--90\,\% chance of identifying those data sets correctly and a chance of just over 10\,\% of confusing these models for one another.
As for the normal mass ordering, the other three models are identified correctly in over 95\,\% of all cases.


\subsection{N=300 Events Per Data Set}

When considering larger data sets, the effect of random fluctuations between individual data sets will decrease.
As a result, the accuracy of model identification is expected to increase significantly.

\subsubsection{Normal Mass Ordering}

\begin{figure}[p]
	\centering
	\includegraphics[scale=0.28]{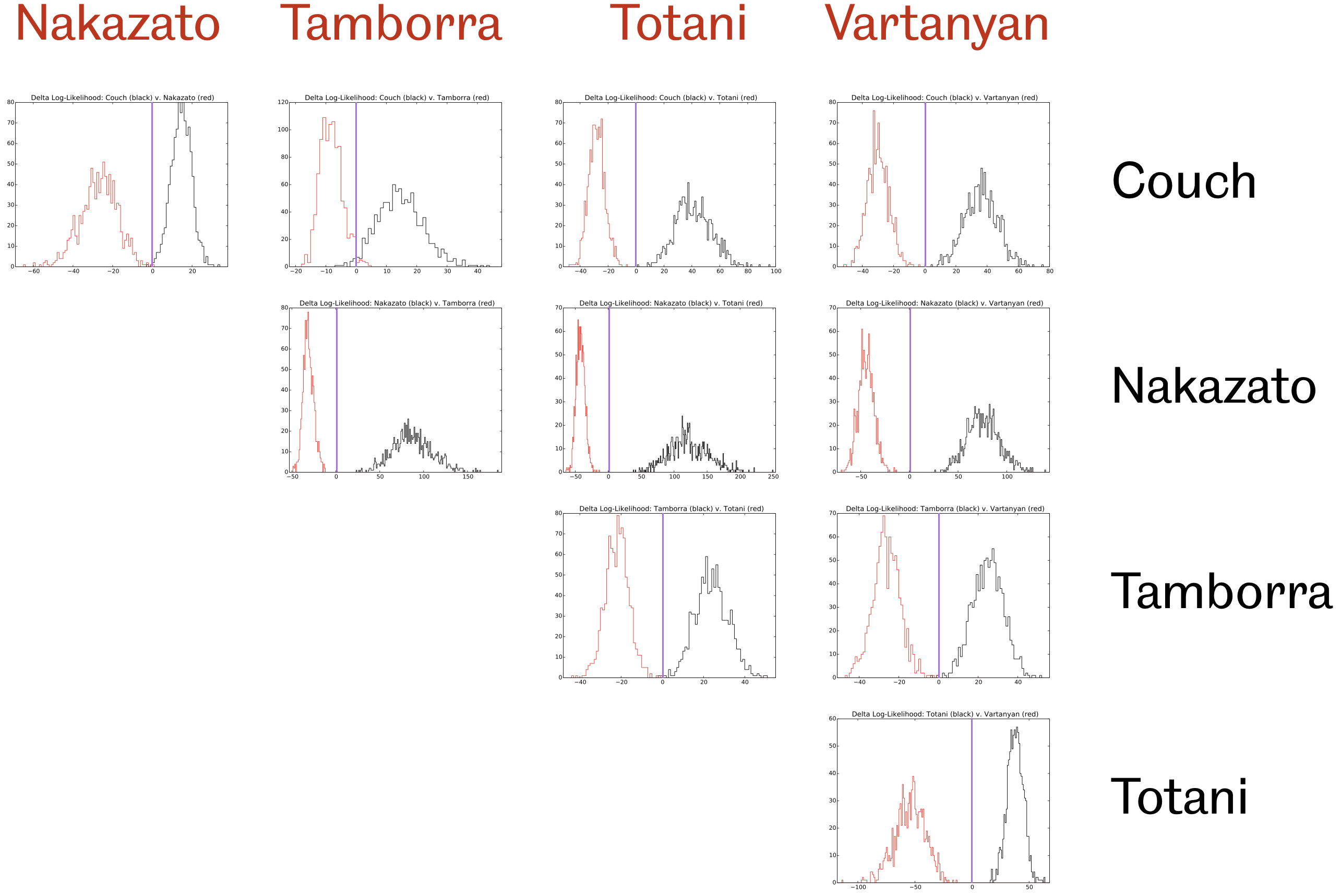}
	\caption[Distribution of $\Delta L$ for all pairs of supernova models: 300 events per data set and normal mass ordering]{Histograms showing the distribution of $\Delta L = L_\text{black} - L_\text{red}$ for all pairs of supernova models considered here, for 300 events per data set, normal mass ordering and 20\,\% photocoverage. The purple vertical line in each panel indicates $\Delta L = 0$.}
	\label{fig-ana-300normal}
\end{figure}

\begin{table}[p]
\begin{center}
\begin{tabular}{lrrrrr}
True Model & Couch & Nakazato & Tamborra & Totani & Vartanyan\\
\hline
Couch & \textbf{982} & 2 & 16 & 0 & 0\\
Nakazato & 1 & \textbf{999} & 0 & 0 & 0\\
Tamborra & 16 & 0 & \textbf{980} & 2 & 2\\
Totani & 0 & 0 & 0 & \textbf{1000} & 0\\
Vartanyan & 0 & 0 & 0 & 0 & \textbf{1000}
\end{tabular}
\end{center}
\caption[Accuracy of model identification: 300 events per data set and normal mass ordering]{Accuracy with which the true model can be identified, for 300 events per data set, normal mass ordering and 20\,\% photocoverage. Shows how many of the 1000 data sets generated for a given model (left column) were identified as each of the five models. Correctly identified models are \textbf{highlighted}.}
\label{tab-ana-300normal}
\end{table}%

Figure~\ref{fig-ana-300normal} shows the pairwise model comparisons for 300 events per data set and normal mass ordering.
Compared to figure~\ref{fig-ana-100normal}, the separation of models is clearly improved, consistent with expectations.
As a result, table~\ref{tab-ana-300normal} shows that the Couch and Tamborra models---which are most likely to be confused for each other in normal mass ordering---are now identified correctly with about 98\,\% accuracy and the probability of misidentifying one for the other is just 1.6\,\%.
The other three models are identified correctly with almost 100\,\% accuracy.

\subsubsection{Inverted Mass Ordering}

\begin{figure}[p]
	\centering
	\includegraphics[scale=0.28]{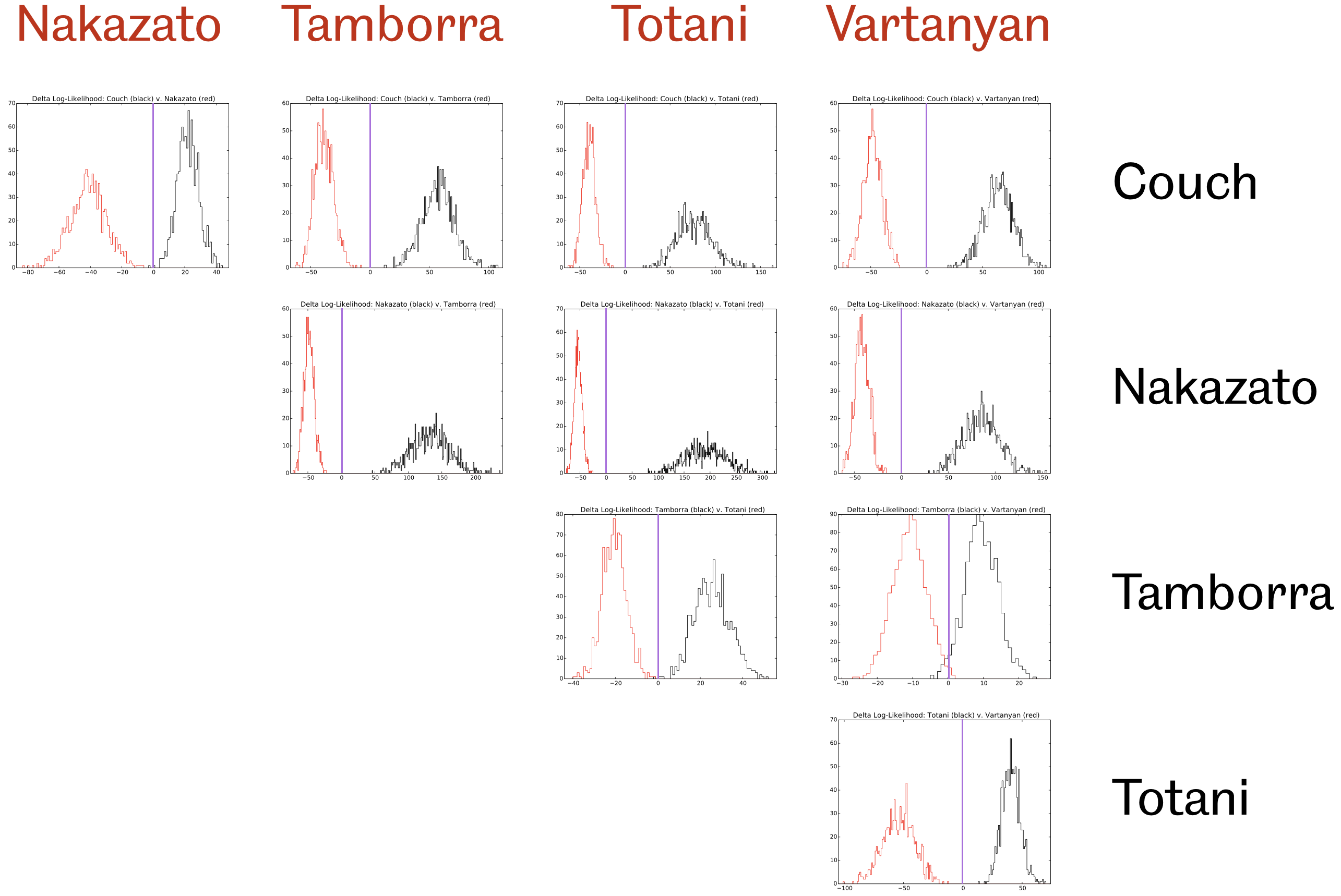}
	\caption[Distribution of $\Delta L$ for all pairs of supernova models: 300 events per data set and inverted mass ordering]{Histograms showing the distribution of $\Delta L = L_\text{black} - L_\text{red}$ for all pairs of supernova models considered here, for 300 events per data set, inverted mass ordering and 20\,\% photocoverage. The purple vertical line in each panel indicates $\Delta L = 0$.}
	\label{fig-ana-300inverted}
\end{figure}

\begin{table}[p]
\begin{center}
\begin{tabular}{lrrrrr}
True Model & Couch & Nakazato & Tamborra & Totani & Vartanyan\\
\hline
Couch & \textbf{999} & 1 & 0 & 0 & 0\\
Nakazato & 0 & \textbf{1000} & 0 & 0 & 0\\
Tamborra & 0 & 0 & \textbf{974} & 1 & 25\\
Totani & 0 & 0 & 0 & \textbf{1000} & 0\\
Vartanyan & 0 & 0 & 8 & 0 & \textbf{992}
\end{tabular}
\end{center}
\caption[Accuracy of model identification: 300 events per data set and inverted mass ordering]{Accuracy with which the true model can be identified, for 300 events per data set, inverted mass ordering and 20\,\% photocoverage. Shows how many of the 1000 data sets generated for a given model (left column) were identified as each of the five models. Correctly identified models are \textbf{highlighted}.}
\label{tab-ana-300inverted}
\end{table}%

Figure~\ref{fig-ana-300inverted} shows the pairwise model comparisons for 300 events per data set and inverted mass ordering.
Compared to figure~\ref{fig-ana-100inverted}, the separation of models is clearly improved, consistent with expectations.
As a result, table~\ref{tab-ana-300inverted} shows that the Tamborra and Vartanyan models---which are most likely to be confused for each other in inverted mass ordering---are now identified correctly with over 97\,\% accuracy.
The other three models are identified correctly with almost 100\,\% accuracy.

\section{Determining Progenitor Properties}\label{ch-ana-siblings}

In the previous section, I have compared supernova models simulated by different groups using different codes and employing different approximations that lead to quantitatively and sometimes qualitatively very different outcomes.
Such a comparison is important as long as details of the explosion mechanism are not understood, since it would help determine which model best reproduces the observed signatures of the explosion mechanism.

However, Hyper-Kamiokande is expected to start data taking in 2027 and, due to the low galactic supernova rate, it may be decades beyond that before it first detects a high-statistics supernova neutrino burst.
If computer models have started to converge by then---or if details of the explosion mechanism are confirmed as discussed above---it is of great interest to the field to determine what more detailed information can be extracted from the neutrino signal.
In particular, one important question is whether the neutrino signal can help us figure out details of the progenitor.

It was previously shown that some carefully chosen properties of the supernova neutrino signal may be used to determine the core compactness of the progenitor star~\cite{Horiuchi2017} or the mass and radius of the resulting proto-neutron star~\cite{Suwa2019}.
In contrast, the approach described in the previous sections is a fully general method that can, in principle, be used to determine any parameter that affects the neutrino fluxes emitted by a supernova.
As an example, in this section I will consider the mass and metallicity of progenitors.

I use fluxes provided by Nakazato \emph{et al.}~\cite{Nakazato2013}, who simulated a range of progenitors with different masses and metallicities using the same code.
Unlike above, all differences in the neutrino fluxes therefore correspond to differences in the progenitors.
Details of these simulations are provided in section~\ref{ch-sim-models-nakazato}.
In addition to the \SI{20}{\Msol} progenitor with solar metallicity (described as the “Nakazato” model in the previous section), I use \SI{13}{\Msol} and \SI{30}{\Msol} progenitors with solar metallicity and a \SI{20}{\Msol} progenitor with lower metallicity.
Below, progenitors are labelled “Nxxy0”, where xx corresponds to the mass in $M_\odot$ and y is either 0 (for solar metallicity, $Z=0.02$) or 1 (for the lower metallicity, $Z=0.004$).%
\footnote{This follows the convention adopted by the authors for the published flux files. See \href{http://asphwww.ph.noda.tus.ac.jp/snn/index.html}{\texttt{asphwww.ph.noda.tus.ac.jp/snn/index.html}}.}

\begin{figure}[tb]
	\centering
	\includegraphics[scale=0.65]{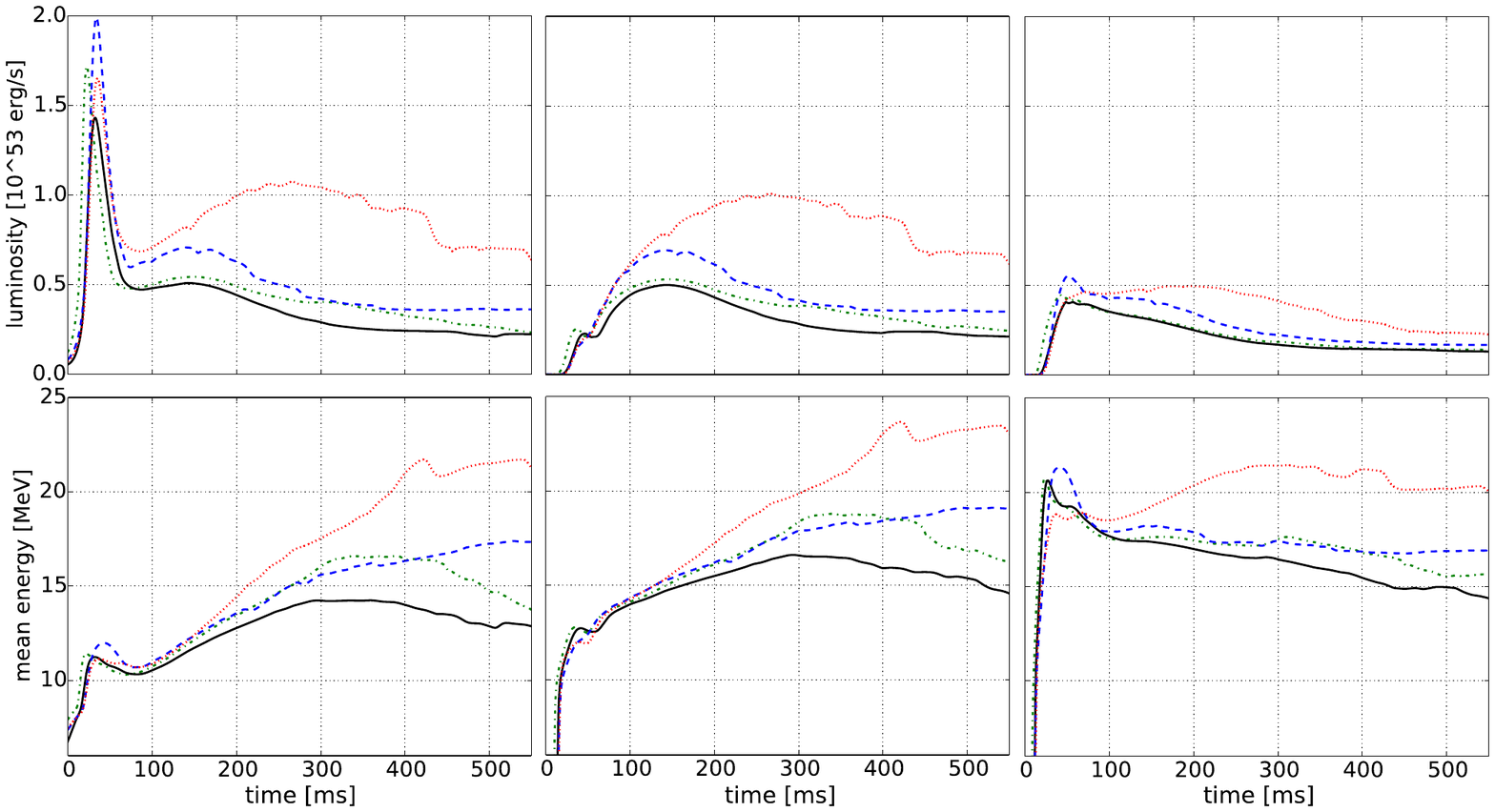}
	\caption[Luminosity and mean energy of \nue, \nuebar and \nux for progenitors with different initial masses and luminosities]{Luminosity (top) and mean energy (bottom) of \nue (left), \nuebar (centre) and \nux (right) as a function of time for the N1300 (dash-dotted green line), N2000 (solid black), N2010 (dashed blue) and N3000 (dotted red) progenitor.}
	\label{fig-ana-nakazato-comparison}
\end{figure}

Figure~\ref{fig-ana-nakazato-comparison} shows the luminosity and mean energy of neutrino fluxes predicted by these simulations.
Since the predicted fluxes for all progenitors are very similar, separation is difficult and I will only show results for the larger data set size of 300 events.
Apart from this, I have followed the same procedure described in the previous sections.

\begin{figure}[p]
	\centering
	\includegraphics[scale=0.35]{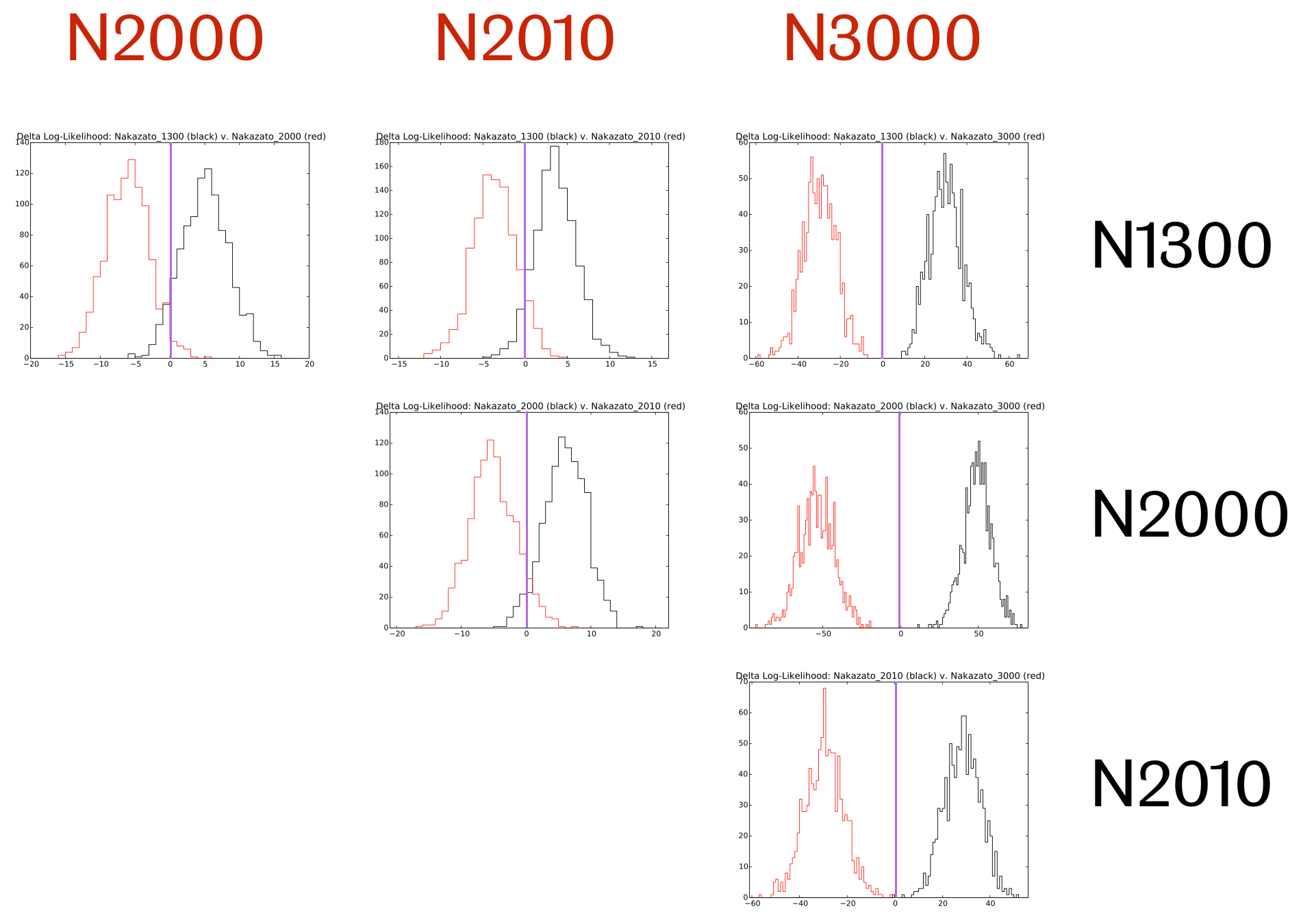}
	\caption[Distribution of $\Delta L$ for all pairs of progenitors: 300 events per data set and normal mass ordering]{Histograms showing the distribution of $\Delta L = L_\text{black} - L_\text{red}$ for all pairs of progenitors considered here, for 300 events per data set, normal mass ordering and 20\,\% photocoverage. The purple vertical line in each panel indicates $\Delta L = 0$.}
	\label{fig-ana-nakazato300normal}
\end{figure}

\begin{table}[p]
\begin{center}
\begin{tabular}{lrrrr}
True Progenitor & N1300 & N2000 & N2010 & N3000\\
\hline
N1300 & \textbf{878} & 61 & 61 & 0\\
N2000 & 17 & \textbf{944} & 39 & 0\\
N2010 & 74 & 75 & \textbf{850} & 1\\
N3000 & 0 & 0 & 0 & \textbf{1000}\\
\end{tabular}
\end{center}
\caption[Accuracy of progenitor identification: 300 events per data set and normal mass ordering]{Accuracy with which the true progenitor can be identified, for 300 events per data set, normal mass ordering and 20\,\% photocoverage. Shows how many of the 1000 data sets generated for a given progenitor (left column) were identified as each of the four progenitors. Correctly identified progenitors are \textbf{highlighted}.}
\label{tab-ana-nakazato300normal}
\end{table}%

Figure~\ref{fig-ana-nakazato300normal} shows a pairwise comparison of all four progenitors for normal mass ordering.
I find that the N3000 progenitor is clearly distinct from the other progenitors, whereas all other pairs show some overlap near $\Delta L = 0$, indicating a risk of misidentification.
Accordingly, table~\ref{tab-ana-nakazato300normal} shows that all data sets generated from the N3000 progenitor are identified correctly, while the accuracy is between 85 and 95\,\% for the three other progenitors.

\begin{figure}[p]
	\centering
	\includegraphics[scale=0.35]{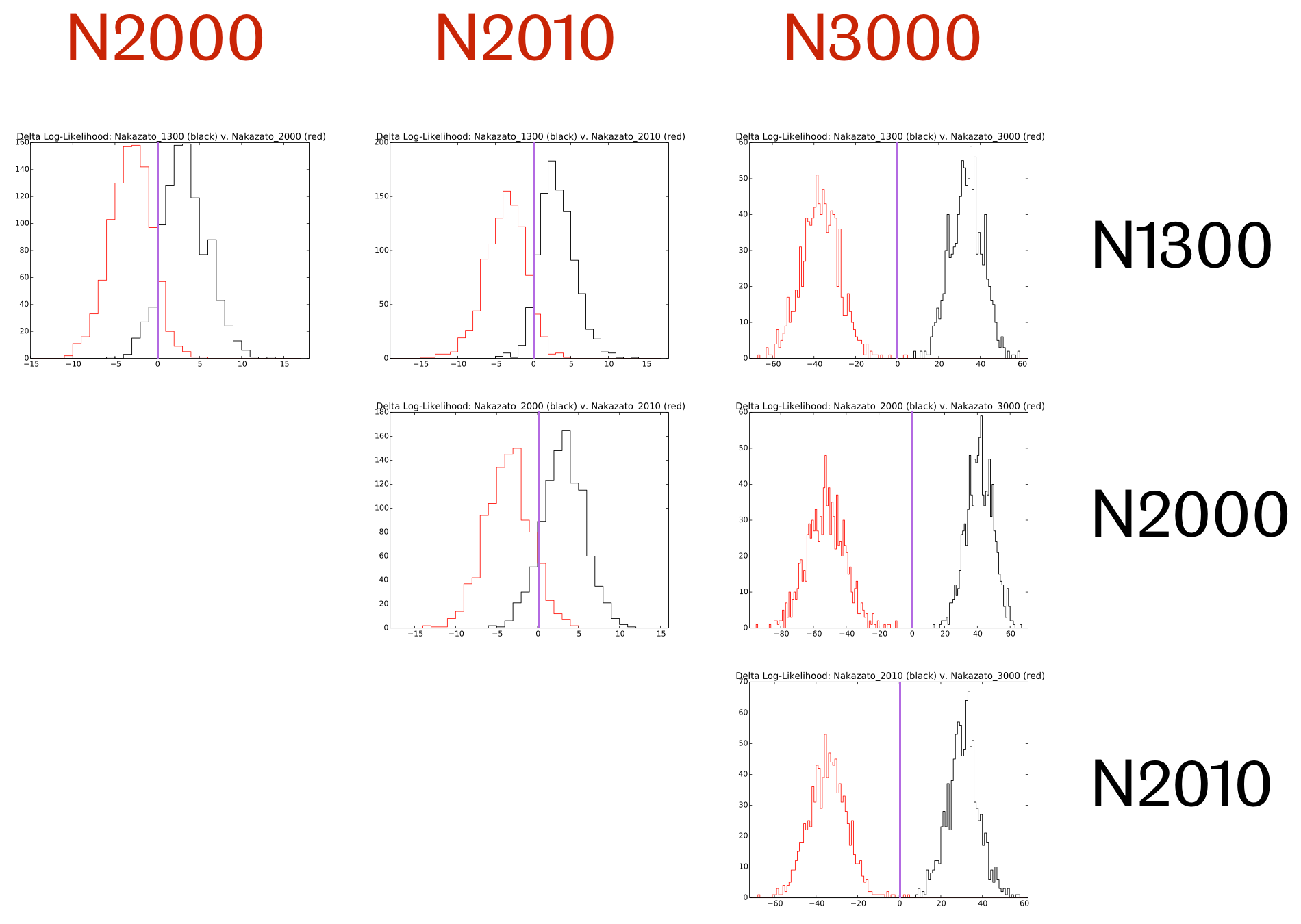}
	\caption[Distribution of $\Delta L$ for all pairs of progenitors: 300 events per data set and inverted mass ordering]{Histograms showing the distribution of $\Delta L = L_\text{black} - L_\text{red}$ for all pairs of progenitors considered here, for 300 events per data set, inverted mass ordering and 20\,\% photocoverage. The purple vertical line in each panel indicates $\Delta L = 0$.}
	\label{fig-ana-nakazato300inverted}
\end{figure}

\begin{table}[p]
\begin{center}
\begin{tabular}{lrrrr}
True Progenitor & N1300 & N2000 & N2010 & N3000\\
\hline
N1300 & \textbf{866} & 78 & 56 & 0\\
N2000 & 64 & \textbf{848} & 88 & 0\\
N2010 & 53 & 88 & \textbf{859} & 0\\
N3000 & 0 & 0 & 0 & \textbf{1000}\\
\end{tabular}
\end{center}
\caption[Accuracy of progenitor identification: 300 events per data set and inverted mass ordering]{Accuracy with which the true progenitor can be identified, for 300 events per data set, inverted mass ordering and 20\,\% photocoverage. Shows how many of the 1000 data sets generated for a given progenitor (left column) were identified as each of the four progenitors. Correctly identified progenitors are \textbf{highlighted}.}
\label{tab-ana-nakazato300inverted}
\end{table}%

Results for inverted mass ordering are shown in figure~\ref{fig-ana-nakazato300inverted} and table~\ref{tab-ana-nakazato300inverted}.
They are very similar, with N3000 clearly distinct while N1300, N2000 and N2010 are all at risk of being mistaken for each other with an identification accuracy of about 85\,\%.

Overall, while determining progenitor properties is more difficult than the model discrimination discussed in the previous section, these results show that it is clearly within the capabilities of Hyper-Kamiokande.
In fact, progenitor discrimination with 300 events per data set shows a similar accuracy to model discrimination with 100 events per data set.
This suggests that 3 to 10 times higher statistics may be sufficient to determine progenitor properties with a high accuracy.
Based on table~\ref{tab-ana-distance}, this would correspond to a supernova at distances larger than about \SI{20}{kpc}, which encompasses the whole Milky Way.

\section{Bayesian Interpretation}\label{ch-ana-bayes}

Throughout this chapter, I have answered the following question: Assuming model X describes the actual neutrino fluxes from a supernova, how likely are we to correctly identify X when comparing it with a range of other, different models?
This lets us identify which models are more or less similar to each other and assess Hyper-Kamiokande’s model discrimination capabilities.
However, the results are sensitive to adding other models to the comparison which are similar to the true model.
Furthermore, this does not reflect the scenario we will face in the future when we observe a single supernova neutrino burst and do not know the true model.

Thus, another question of interest is: Assuming that we observe a supernova neutrino burst that is best described by model X, how confident are we that we can exclude some alternative model Y?
To answer this, we need to consider the interpretation of the likelihood ratio.

In a Bayesian interpretation~\cite{Loredo2002}, the ratio of likelihoods for two models A and B is equal to the Bayes factor $B_\text{ij}$ and equivalently, the difference in log-likelihoods is $\Delta L = \ln B_\text{ij}$.
If there is no \emph{a priori} reason to prefer one model over the other, this can be used to exclude disfavoured models beyond a certain threshold.

\begin{table}[btp]
\begin{center}
\begin{tabular}{lll}
$\ln B_\text{AB}$ & $B_\text{AB}$ & Evidence for model A over model B\\
\hline
0 to 1 & 1 to 3 & Negligible\\
1 to 3 & 3 to 20 & Positive\\
3 to 5 & 20 to 150 & Strong\\
$>5$ & $>150$ & Very strong
\end{tabular}
\end{center}
\caption[Interpretation of Bayes factor when comparing two models A and B]{Interpretation of Bayes factor when comparing two models A and B. Adapted from~\cite{Kass1995}.}
\label{tab-ana-bayes}
\end{table}%

A suggested interpretation of Bayes factors is listed in table~\ref{tab-ana-bayes}.%
\footnote{Note that I show $B_\text{ij}$ here, whereas the original paper lists $2 B_\text{ij}$ due to its similarity with the more familiar $\Delta \chi^2$ values.~\cite{Kass1995,Ianni2009}}
Looking at the pairwise model comparison in figure~\ref{fig-ana-deltall_explanation}, 
we see that this interpretation matches our intuition:
The range from $\Delta L = -5$ to 5 contains almost the complete overlap between both histograms, where misidentification of data sets may occur, indicating that requiring $\Delta L \geq -5$ is unlikely to wrongly exclude the true model.
At the same time, a significant fraction of data sets based on the wrong model are excluded by this criterion.
Once we observe an actual supernova neutrino burst, this criterion would therefore allow us to narrow down the list of supernova models or progenitors that are compatible with the observed signal.


\chapter{Conclusions and Outlook}\label{ch-conclusions}

\setlength{\epigraphwidth}{.40\textwidth}
\epigraphhead[0]{\epigraph{A real Galactic supernova cannot, unfortunately, be guaranteed on the timescale of a PhD studentship~…}{\textit{Susan Cartwright}}}

\section{Conclusions}
The observation of neutrinos from SN1987A was a breakthrough for neutrino astronomy and dramatically improved our understanding of supernovae.
However, due to the small number of observed events, many questions still remain today and the neutrino community is desperate to prepare for the next galactic supernova.
As part of those preparations, in this thesis I have investigated the ability of the planned Hyper-Kamiokande detector to extract information from a supernova neutrino burst.
I showed that its large volume and ability to reconstruct individual events give it an excellent ability to discriminate between different supernova models and deduce details of the supernova explosion mechanism.

As part of this thesis, I have developed a complete toolchain for generating, simulating and reconstructing supernova neutrinos in Hyper-Kamiokande, which is described in chapter~\ref{ch-sim}.

In particular, I have developed sntools---a high-precision event generator for supernova neutrino burst observation with water Cherenkov detectors.
In addition to inverse beta decay---the main interaction channel, and often the only channel considered in the literature---sntools includes three subdominant interaction channels.
It implements precise, modern cross sections for all channels as well as a precise treatment of neutrino oscillations via the MSW effect.
sntools supports multiple input formats for neutrino fluxes and is designed to be modular and extensible to make it easy to add other supernova models.

I have also developed the first energy reconstruction for Hyper-Kamiokande.
It follows an approach pioneered by the Super-Kamiokande collaboration with some alterations in response to the modified detector design.
While this reconstruction is currently at an early stage and will require further adjustments once the detector design is finalized, it is already satisfactory for the analysis described here and I have demonstrated that further improvements would not affect the results.

To demonstrate the capabilities of Hyper-Kamiokande and this software toolchain, I have derived an improved likelihood function that supports multiple interaction channels.
It makes optimal use of event-by-event timing and energy information available in Hyper-Kamiokande to determine how well the neutrino fluxes predicted by a supernova model describe the observed events.
I have applied this likelihood function to a set of five supernova models representing the wide variety of models developed by different groups around the world.
I have found that by observing just 100 to 300 events, Hyper-Kamiokande will be able to distinguish these models with high accuracy.
This event count corresponds to supernovae at distances of at least \SI{59}{kpc}, which includes the whole Milky Way and many of its satellite galaxies, including the Large and Small Magellanic Clouds.
When an actual supernova neutrino burst is observed in the future, this model discrimination capability will allow us to determine which model best reproduces the supernova explosion mechanism realized in nature.

The method developed in this thesis is very general and can in principle be used to determine any parameter of a supernova that influences its neutrino emission.
As an example of a more targeted analysis, I have performed a comparison of different progenitors simulated with the same code.
The results demonstrate that by observing at least 300 events, Hyper-Kamiokande may be able to distinguish between progenitors with different masses or metallicities.


\section{Outlook}

Science is a team effort---and this is especially true in astroparticle physics.
Individual people or experiments may make important contributions but sustained progress is possible only through cooperation of groups from all different countries and subfields.
While this thesis focused on the contributions of Hyper-Kamiokande, I now want to take a step back and give an overview over other approaches that can be used in conjunction with Hyper-Kamiokande to study supernovae.

\subsection{Progress in Supernova Simulations}

Current supernova simulations are severely limited by the available computing power.
In response, two complementary approaches are possible:
On one hand, running large cohorts of simpler, one-dimensional simulations makes it possible to extensively study the effect of individual factors---like progenitor properties or simulation parameters---on the outcome of simulations.
On the other hand, performing a small number of increasingly realistic, multidimensional simulations enables the study of details of the supernova explosion mechanism.

In the coming years, both approaches are likely to see steady progress from a combination of increases in computing power, improvements to simulation codes and theoretical progress in input physics. 
The ultimate goal however---a high-resolution, fully three-dimensional simulation in General Relativity with exact treatment of neutrino transport and a complete set of precision microphysics
 ---remains elusive and convergence of different models is therefore unlikely. 

In the absence of such convergence, direct comparisons of different simulation codes are essential for estimating the uncertainty of simulation outcomes.
Recently, an extensive comparison of six simulation codes showed that, when using an identical set of basic input physics, the codes exhibit good qualitative and quantitative agreement~\cite{OConnor2018a}.
Future work extending such comparisons to more advanced setups---including multiple dimensions and more advanced input physics---would be particularly valuable.

Furthermore, it may be interesting to apply the analysis pipeline developed in this thesis to the outputs of such a model comparison and determine whether the inherent differences between simulation codes are large enough to be detectable.
This would be particularly useful if the progenitor of a galactic supernova is identified by telescopes, giving us independent information on the progenitor.

\subsection{Complementarity with Other Neutrino Detectors}

While Hyper-Kamiokande offers a unique combination of large detection volume and ability to precisely reconstruct individual events, other neutrino detectors may be more sensitive to different parts of the neutrino signal.
Below, important current or future neutrino detectors that offer such complementarity to Hyper-Kamiokande are briefly described.

\subsubsection{Super-Kamiokande}
Apart from its smaller size, Super-Kamiokande is very similar to Hyper-Kamiokande.
Since small differences in energy resolution have a negligible effect of the results presented in this thesis, they should approximately be equally valid for Super-Kamiokande.
Since Super-Kamiokande’s fiducial volume is 8.4 times smaller, 300 observed events corresponds to a distance that is $\sqrt{8.4} \approx 2.9$ smaller than listed in table~\ref{tab-ana-distance}, i.\,e. at least \SI{20}{kpc}, depending on the model.
Therefore, Super-Kamiokande is likely to have accurate model discrimination capabilities for a supernova anywhere within the Milky Way.
However, since backgrounds and energy reconstruction in Super-Kamiokande are well understood, the collaboration would be able to perform a more precise analysis than presented here, which may improve results further.

During the second half of 2018, extensive work took place to prepare Super-Kamiokande for the addition of gadolinium~\cite{Marti-Magro2018}.
Gadolinium has a high neutron-capture cross section and, upon capturing the neutron, emits a gamma cascade with about \SI{8}{MeV} energy.
This will enable Super-Kamiokande to efficiently tag neutrons, allowing it to distinguish between interaction channels that produce neutrons---such as inverse beta decay and some $^{16}$O charged-current interactions---and other interaction channels on an event-by-event basis.
This would help to identify an almost pure sample of elastic scattering events, which would enable a more accurate determination of the direction of the supernova and a more precise measurement of the initial \nue burst.
In the future, experiences gained from adding gadolinium to Super-Kamiokande may be used in an upgrade to Hyper-Kamiokande.

\subsubsection{Second Hyper-Kamiokande Detector in Korea}
Plans to build a second Hyper-Kamiokande detector in South Korea were published in a white paper in 2018~\cite{Abe2018a}.
While the primary motivation is significant improvements to neutrino oscillation measurements, supernova burst observations would of course benefit from the increased detection volume.
Furthermore, since a galactic supernova would likely be a once-in-a-lifetime event, the redundancy offered by a second detector would dramatically reduce the risk of missing the observation due to detector downtime e.\,g. during maintenance work.

Depending on position of the supernova, comparing the fluxes observed by both detectors may also allow for a direct observation of Earth matter effects~\cite{Mirizzi2006,Abe2018a}. 
In that case, the similar design of both detectors would cancel out some systematic uncertainties that would make such a comparison between other detectors much more difficult.

\subsubsection{DUNE}\label{ch-conclusions-dune}
DUNE~\cite{Acciarri2015} is a liquid argon time projection chamber that is currently under construction with a first \SI{10}{kt} detector module expected to begin operations shortly before Hyper-Kamiokande.
After construction of three additional modules expected over the following years, it may observe over \num{3000} events for a supernova at \SI{10}{kpc} distance.
Since its main interaction channel is \nue charged-current interaction on $^{40}$\!Ar, DUNE will observe a large sample of almost pure \nue interactions, which makes it highly complementary to Hyper-Kamiokande.

\subsubsection{JUNO}
JUNO~\cite{An2016} is a \SI{20}{kt} liquid scintillator detector that is currently under construction, with completion expected in 2021. 
In the case of a supernova at \SI{10}{kpc} distance, it is expected to observe about \num{5000} inverse beta decay events.
Its low energy threshold enabled by the liquid scintillator technology and a high photocoverage will allow it to observe elastic neutrino-proton scattering.
This channel is sensitive to neutrinos and antineutrinos of all flavours.
If proton quenching in the liquid scintillator is well understood, this channel would therefore enable a measurement of the full neutrino flux which will be important to understand flavour conversions and the total energy emitted by the supernova in neutrinos.

\subsubsection{THEIA}
THEIA~\cite{Askins2019}, previously known as the Advanced Scintillator Detector Concept~\cite{Alonso2014}, is a concept for a future neutrino detector employing water-based liquid scintillator as a detector material.
It would be comparable in size to or larger than Super-Kamiokande, 
combining the large volume of water Cherenkov detectors with the sub-Cherenkov threshold sensitivity and excellent energy resolution of liquid scintillator detectors.
In addition to observing a high-statistics sample of supernova neutrinos, THEIA could tag inverse beta decay events via neutron capture---similar to a gadolinium-loaded Super-Kamiokande, but with higher efficiency---which would give the benefits explained above.

\subsubsection{IceCube and KM3NeT}
Both IceCube~\cite{Abbasi2011} and KM3NeT~\cite{Adrian-Martinez2016} are water Cherenkov detectors whose large volume and sparse instrumentation are optimized for high-energy neutrinos at the GeV-scale and above.
At the energies typical for supernova burst neutrinos, both would likely detect just one photon per neutrino interaction, making any individual event indistinguishable from noise.
As a result, both detectors could detect a supernova neutrino burst because of a sudden and temporary increase of the “noise” rate across the whole detector, but unlike Hyper-Kamiokande and the other neutrino detectors described above, they would not be able to reconstruct individual events.
As a result, they will not deliver the event-by-event energy information used in this thesis, offering at best a measurement of the average anergy of all neutrinos in the burst by observing the rate of coincidences of multiple photosensors.

Furthermore, while Hyper-Kamiokande would make an effectively background-free detection of supernova burst neutrinos, IceCube and KM3NeT have a significant background coming from dark noise of photosensors and radioactive impurities in the detector material.
Due to their higher event rate, however, the signal-to-noise ratio in measuring the event rate may still be larger than that of Hyper-Kamiokande for a sufficiently close supernova.


\subsubsection{Others}
The HALO experiment~\cite{Zuber2015} detects neutrinos via neutral-current scattering on lead nuclei, while upcoming xenon-based dark matter detectors~\cite{Chakraborty2014a,Lang2016} could detect supernova neutrinos via coherent elastic neutrino-nucleus scattering (CE$\nu$NS).
Both interaction channels are sensitive to all neutrino flavours, making these detectors complementary to Hyper-Kamiokande and many of the detectors listed above.
However, due to their relatively small sizes, at most a few hundred events are expected in these detectors for a supernova at \SI{10}{kpc} distance, limiting their contribution.

Furthermore, a number of other, smaller neutrino detectors exist that use the same detector technologies as some of the future experiments described above.
The advantages and disadvantages of the respective detector technology discussed above apply equally to them. However, due to their smaller size their contribution may be statistics-limited and I have not highlighted them here individually.

\subsection{Beyond Neutrinos}\label{ch-conclusions-mma} 

Neutrinos are a particularly interesting channel for observing supernova, since they allow a direct observation of the stellar core at the moment of explosion.

Gravitational waves similarly allow for a direct observation of the explosion and are sensitive to the asymmetric component of the explosion.
After multiple successful detection runs of both LIGO detectors~\cite{Aasi2015} and the Virgo detector~\cite{Acernese2015} in the past years, the KAGRA detector~\cite{Somiya2012,Aso2013} is expected to join the currently ongoing detection run in late 2019~\cite{Abbott2018}. 
Like the neutrino signal, the gravitational wave signal from a supernova has large uncertainties and is highly model-dependent.
However, a galactic supernova burst would likely be within the detection range of these detectors~\cite{Abbott2018}.

Finally, observations across the electromagnetic spectrum become possible once the shock wave of the supernova reaches the surface of the star.
The initial signal is called the shock breakout and takes place after the initial explosion with a delay of about one minute for very compact Wolf-Rayet stars, one hour for blue supergiants or one day for red supergiants~\cite{Kistler2013}.
Afterwards, follow-up observations across multiple wavelengths may be possible for months.
While these observations are not directly connected to the explosion mechanism, they offer a wealth of information.
For example, the spectra give information on the abundance of light elements in the outer layers of the star which lets us identify the type of progenitor, while the light curves of some supernovae give information on the amount of $^{56}$Ni produced through nuclear burning in the wake of the shock wave~\cite{Smartt2009}.

To realize the full potential of multi-messenger astronomy, we need to draw all these different methods of observing the supernova together and combine them into a coherent picture.
While the phenomenology of supernovae is so broad that it would not be feasible to exhaustively discuss potential combinations of these observation channels here, a few remarks are in order.

If an optical counterpart to a galactic supernova is observed---which is not certain, due to either dust extinction in the galactic plane or the unknown fraction of “failed” supernovae which collapse into a black hole before the shock wave is revived---it would likely be possible to identify the progenitor.
This may make it possible to significantly narrow down the range of possible progenitor properties and thus the range of supernova simulations that would need to be compared using the method described in this thesis.

Furthermore, observing an optical counterpart would enable a distance measurement.
If the uncertainty in the measured distance is on the order of 10\,\% or smaller, this may further help distinguish different models, since different models often predict clear differences in the total number of events at a fixed distance.
In that case, the total likelihood would become
\begin{equation}
\mathcal{L}_\text{total} = P(N_\text{obs}, X) \cdot \mathcal{L}_\text{fixed N},
\end{equation}
where $\mathcal{L}_\text{fixed N}$ is the likelihood derived in section~\ref{ch-ana-ll} and $P(N_\text{obs}, X)$ is the Poisson probability of observing $N_\text{obs}$ events when $X$ events were predicted by a given model based on the observed distance.

Without an optical counterpart, it would still be possible to estimate the distance by using the size of the neutronization burst as a standard candle~\cite{Kachelries2005}.
However, even with Hyper-Kamiokande or DUNE, this method would be likely be able to determine the distance only to an uncertainty of a few tens of per cent.%
\footnote{While reference~\cite{Kachelries2005} claims an accuracy of 5--10\,\%, this assumes a megaton water Cherenkov detector with gadolinium loading enabling a 90\,\% neutron tagging efficiency to identify inverse beta decay events. Such a detector will, unfortunately, not exist in the foreseeable future.}
Considering the range of models shown in table~\ref{tab-ana-distance}, this would likely be of little help in trying to distinguish between models except in extreme cases.

Finally, coordination between all observers is essential to be able to exploit the full potential of the next galactic supernova.
For this purpose, the supernova early warning system (SNEWS) was formed in 1999~\cite{Scholberg2000,Antonioli2004}.
Its main goal is to alert astronomers if multiple neutrino detectors around the world observe a coincident neutrino burst, enabling them to observe the light curve of the supernova as quickly as possible.
After an initial test phase, SNEWS has been operating in automatic mode since 2005 but has not sent out a supernova alert.

Not \emph{yet.}



\appendix

\chapter{Neutrino Interaction Cross Sections}\label{apx-xs}

\section{Inverse Beta Decay}
In this thesis, I use the inverse beta decay cross section derived in reference~\cite{Strumia2003}.
Here, I will briefly summarize the equations necessary to calculate that cross section.

I start by defining the constants
\begin{align}
\Delta &= m_n - m_p \approx \SI{1.293}{MeV} \\
M &= \frac{m_p + m_n}{2} \approx \SI{938.9}{MeV} \\
M_V &= \SI{0.71}{GeV} \\
M_A &= \SI{1.03}{GeV}
\end{align}
and the quantities
\begin{align}
s &= m_p^2 + 2 m_p E_\nu \\
s - u &= 2 m_p \left(E_\nu + E_e\right) - m_e^2 \\
t &= m_n^2 - m_p^2 - 2 m_p \left(E_\nu - E_e\right) \\[0.2em]
f_1 &= \frac{1 - 4.706 \frac{t}{4 M^2}}{\left(1-\frac{t}{4 M^2_{}\!}\right)\left(1-\frac{t}{M_V^2}\right)^2} \\[0.4em]
f_2 &= \frac{3.706}{\left(1 - \frac{t}{4 M^2_{}\!}\right) \left(1-\frac{t}{M_V^2}\right)^2} \\[0.4em]
g_1 &= \frac{-1.27}{\left(1-\frac{t}{M_A^2}\right)^2} \\[0.2em]
g_2 &= \frac{2 M^2 g_1}{m_{\pi}^2-t}.
\end{align}

The differential cross section is then given by
\begin{equation} \label{eq-apx-ibd}
\tdiff{\sigma}{E_e} = \frac{G_\text{F}^2 \cos^2 \theta_\text{C}}{4 \pi m_p E_\nu^2} \left|\mathcal{M}^2\right|
\times
\left[ 1+\frac{\alpha}{\pi}\left(6 + \frac{3}{2} \log \frac{m_p}{2 E_e}+1.2\left(\frac{m_e}{E_e}\right)^{1.5}\right) \right],
\end{equation}
where the first part is the tree-level cross section and the term in square brackets corresponds to radiative corrections at one-loop level~\cite{Kurylov2003}.
The matrix element $\left|\mathcal{M}^2\right|$ is given by
\begin{equation}
\left|\mathcal{M}^{2}\right| = A - (s-u) B + (s-u)^{2} C,
\end{equation}
where $A$, $B$ and $C$ depend on the transferred 4-momentum $t = q^2 < 0$ as described by the expressions
\begin{align}
A &=
\begin{aligned}[t]
	& \frac{t - m_e^2}{16} \left[
		4 f_1^2 \left(4 M^2 + t + m_e^2\right)
		+ 4 g_1^2 \left(-4 M^2 + t + m_e^2\right)
		+ \frac{4 m_e^2 t g_2^2}{M^2} \right.\\
		&\hspace{5em}\left. + f_2^2 \left(\frac{t^2}{M^2} + 4 t + 4 m_e^2 \right)
		+ 8 f_1 f_2 \left(2 t + m_e^2\right)
		+ 16 m_e^2 g_1 g_2
	\right] \\
	& - \frac{\Delta^2}{16} \left[
		\left(4 f_1^2 + t \frac{f_2^2}{M^2} \right) \left(4 M^2 + t - m_e^2\right)
		+ 4 g_1^2 \left(4 M^2 - t + m_e^2\right) \right.\\
		&\hspace{5em}\left.  + \frac{4 m_e^2 g_2^2 \left(t - m_e^2\right)}{M^2}
		+ 8 f_1 f_2 \left(2 t - m_e^2\right)
		+ 16 m_e^2 g_1 g_2
		\right] \\
	& - 2 m_e^2 M \Delta g_1\left(f_1+f_2\right) \\
\end{aligned} \\[0.5em]
B &= t g_1 \left(f_1+f_2\right) + \frac{m_e^2 \Delta\left(f_2^2 + f_1 f_2 + 2 g_1 g_2\right)}{4 M} \\[0.5em]
C &= \frac{f_1^2 + g_1^2}{4} - \frac{t f_2^2}{16 M^2}.
\end{align}

The kinematically allowed range of positron energies $E_1 \le E_e \le E_2$ as a function of neutrino energy $E_\nu$ is given by 
\begin{equation}
E_{1,2} = E_\nu - \delta - \frac{1}{m_p} E_\nu^\text{CM} \left(E_e^\text{CM} \pm p_e^\text{CM}\right),
\end{equation}
where $\delta$ and the center-of-mass energies and momenta are given by
\begin{align}
\delta &= \frac{m_n^2 - m_p^2 - m_e^2}{2 m_p} \\
E_\nu^\text{CM} &= \frac{s - m_p^2}{2 \sqrt{s}} \\
E_{e}^{\mathrm{CM}} &= \frac{s-m_{n}^{2}+m_{e}^{2}}{2 \sqrt{s}} \\
p_{e}^{\mathrm{CM}} &= \frac{\sqrt{\left[s-\left(m_{n}-m_{e}\right)^{2}\right]\left[s-\left(m_{n}+m_{e}\right)^{2}\right]}}{2 \sqrt{s}}.
\end{align}

I have also implemented the angular distribution of outgoing positrons, $\tdiffx{\sigma}{\cos\theta}$ derived by reference~\cite{Strumia2003} in sntools.
However, since the angular distribution of events was not used in this analysis, I will not include that equation here.

\section{Neutrino-Electron Scattering}
In this thesis, I use the cross section derived in reference~\cite{Bahcall1995} for neutrino-electron scattering.
Here, I will briefly summarize the equations necessary to calculate that cross section.

I start by defining the quantities
\begin{align}
T &= E_e - m_e \\
l &= \sqrt{ E_e^2 - m_e^2} \\[0.2em]
\beta &= \frac{l}{E_e} \\[0.2em]
z &= \frac{T}{E_\nu} \\
\rho_\text{NC} &= 1.0126 \\
x &= \sqrt\frac{1+2 m_e}{T} \\[0.2em]
f_0 &= \frac{E_e}{l} \ln\left( \frac{E_e + l}{m_e} \right) - 1
\end{align}
and the functions
\begin{align}
I (T) &= \frac{1}{6} \left\{ \frac{1}{3} + (3 - x^2) \left[ \frac{1}{2} x \ln \left(\frac{x+1}{x-1}\right) - 1\right] \right\} \\[0.2em]
L(x) &= \int_0^x \frac{\ln |1-t|}{t} \d t.
\end{align}

For \nue-electron scattering,
\begin{align}
\kappa (T) &= 0.9791 + 0.0097\cdot I(T) \\
g_L (T) &= \rho_\text{NC} \left[ \frac{1}{2} - \kappa (T) \sin^2 \theta_W \right] - 1\\
g_R (T) &= - \rho_\text{NC} \kappa (T) \sin^2 \theta_W,
\end{align}
while for \nux-electron scattering,
\begin{align}
\kappa (T) &= 0.9970 + 0.00037\cdot I(T) \\
g_L (T) &= \rho_\text{NC} \left[ \frac{1}{2} - \kappa (T) \sin^2 \theta_W \right] \\
g_R (T) &= - \rho_\text{NC} \kappa (T) \sin^2 \theta_W.
\end{align}
For antineutrino-electron scattering, $\kappa$ remains unchanged while $g_L$ and $g_R$ are swapped.

The cross section, including QCD and electroweak loop corrections, is then given by
\begin{equation}
\begin{aligned}[t]
\tdiff{\sigma}{E_e} =& \frac{2 G_{F}^{2} m_e}{\pi} \bigg\{ g_{L}^{2}(T) \left[1+\frac{\alpha}{\pi} f_{-}(z)\right] \\
	&\hspace{4em}+g_{R}^{2}(T)(1-z)^{2}\left[1+\frac{\alpha}{\pi} f_{+}(z)\right] \\
	&\hspace{4em}-g_{R}(T) g_{L}(T) \frac{m_e}{E_\nu} z\left[1+\frac{\alpha}{\pi} f_\pm (z)\right] \bigg\}.
\end{aligned}
\end{equation}
The terms proportional to $\alpha / \pi$ describe QED corrections to the cross section due to virtual and real photons.
Instead of the exact expressions given in reference~\cite{Ram1967}, I use the following simplified expressions which are typically accurate to much better than 1\,\%~\cite{Bahcall1995}.
\begin{align}
f_- (z) &=
\begin{aligned}[t]
	&f_0 \left[ 2 \ln\left( 1 - z - \frac{m_e}{E_e+l}\right) - \ln(1-z) - \frac{\ln z}{2} - \frac{5}{12} \right]\\
	&+ \frac{1}{2} \left[ L(z) - L(\beta)\right] - \frac{\ln^2 (1-z)}{2} - \left(\frac{11}{12} + \frac{z}{2}\right) \ln(1-z) \\
	&+ z \left[ \ln z + \frac{1}{2} \ln\left(\frac{2 E_\nu}{m_e}\right) \right] - \left( \frac{31}{18} + \frac{\ln z}{12}\right) \beta - \frac{11 z}{12} + \frac{z^2}{24}
\end{aligned} \\[0.5em]
f_+ (z) &=
\begin{aligned}[t]
	&f_0 \left[ 2 \ln\left( 1 - z - \frac{m_e}{E_e+l}\right) - \ln(1-z) - \frac{\ln z}{2} - \frac{2}{3} - \frac{z^2 \ln z + 1-z}{2 (1-z)^2}\right] \\
	&- \frac{1}{2} \left\{ \ln^2(1-z) + \beta \left[ L(1-z) - \ln z \ln (1-z)\right] \right\} \\
	&+ \frac{\ln(1-z)}{(1-z)^2} \left[ \frac{z^2}{2} \ln z + \frac{1-z}{3} \left( 2z - \frac{1}{2}\right) \right] \\
	&- \frac{1}{(1-z)^2} \left[ \frac{z^2}{2} L(1-z) - \frac{z (1-2z)}{3} \ln z - \frac{z (1-z)}{6} \right] \\
	&- \frac{\beta}{12 (1-z)^2} \left[ \ln z + (1-z) \frac{115 - 109z}{6}\right]
\end{aligned} \\
f_\pm (z) &= 2 f_0 \ln\left( 1 - z - \frac{m_e}{E_e+l} \right)
\end{align}

The energy $E_e$ of the scattered electron depends on the neutrino energy $E_\nu$ and the angle $\theta$ between the initial neutrino direction and the final electron direction.
It is given by
\begin{equation}
E_e = m_e + \frac{2 m_e E_\nu^2 \cos^2 \theta}{(m_e + E_\nu)^2 - E_\nu^2 \cos^2 \theta}.
\end{equation}

\chapter{Derivation of Likelihood Function}\label{apx-likelihood}

In this appendix, I derive the likelihood function used for the analysis in chapter~\ref{ch-ana}.
It is based on the likelihood function derived by Loredo and Lamb to analyse events from SN1987A~\cite{Loredo1989}, but I extend it to account for multiple interaction channels.

I start by considering bins in time and observed energy, where the bin size $\Delta t \cdot \Delta E$ is arbitrary as long as the expected number of events per bin,
\begin{equation}
N_i = \sum_\alpha N_i^\alpha = \frac{\d^2\,N^\alpha (E_i, t_i)}{\d E \d t} \Delta E \Delta t,
\end{equation}
is much smaller than 1 for all bins $i$.
Here, $N^\alpha (E, t)$ is the observed event rate predicted by a supernova model in the interaction channel $\alpha$ as a function of time and energy.



Assuming a Poisson distribution, the probability of observing no events in a single interaction channel $\alpha$ in a bin around time $t_i$ and energy $E_i$ is $P^\alpha_{0, i} = \exp\left( - N^\alpha_{i} \right)$.
When considering multiple interaction channels, the probability of observing no events is simply the product of the probabilities of observing no events in every single interaction channel, i.\,e.
\begin{align}
P_{0,i} &= \prod_\alpha P^\alpha_{0, i} = \prod_\alpha \exp\left( -N^\alpha_i \right)\\
&= \exp\left( - \sum_\alpha N^\alpha_i \right) = \exp\left( -N_i \right).
\end{align}

The probability of observing exactly one event in the interaction channel $\alpha$ is $P^\alpha_{1, i} = N^\alpha_i \exp\left( -N^\alpha_i \right)$,
so the total probability of observing exactly one event is
{
\allowdisplaybreaks
\begin{align}
P_{1, i} &= \sum_\alpha \left( P^\alpha_{1, i} \cdot \prod_{\beta \neq \alpha} P^\beta_{0, i} \right) \\
&= \sum_\alpha \left[ N^\alpha_i \exp\left( -N^\alpha_i \right) \cdot \prod_{\beta \neq \alpha} \exp\left( -N^\beta_i \right) \right] \\
&= \sum_\alpha \left[ N^\alpha_i \cdot \prod_\beta \exp\left( -N^\beta_i \right) \right] \\
&= P_{0, i} \cdot \sum_\alpha N^\alpha_i = P_{0, i} \cdot N_i.
\end{align}
}

The bin size was chosen such that the probability of observing more than one event in a bin is negligible.

The likelihood of observing exactly $N_\text{obs}$ events in a certain set of bins---which I refer to as $\mathcal{B}$ here---is then given by
\begin{align}
\mathcal{L} &= \prod_{i \in \mathcal{B}} P_{1, i} \cdot \prod_{i \notin \mathcal{B}} P_{0, i} \\
&= \prod_{i \in \mathcal{B}} \left( P_{0, i} \cdot N_i \right) \cdot \prod_{i \notin \mathcal{B}} P_{0, i} \\
&= \prod_{i \in \mathcal{B}} N_i \cdot \prod_i P_{0, i}
\end{align}
where the products over $i \notin \mathcal{B}$ include all bins that do not contain an event.

For simplicity, I consider the log-likelihood $L = \ln \mathcal{L}$.
Using $\ln(a \cdot b) = \ln(a) + \ln(b)$, the log-likelihood function is
\begin{equation}
L = \sum_{i \in \mathcal{B}} \ln N_i + \sum_i \ln P_{0, i},
\end{equation}
where the second term simplifies to
\begin{equation}
\sum_i \ln P_{0, i} = \sum_i \ln \left[ \exp\left( -N_i \right) \right] = - \sum_i N_i,
\end{equation}
which is the number of events predicted by the model.
Since I assume in this thesis that the distance to the supernova is unknown, I have normalized all models so as to reproduce the observed number of events.
This term is therefore model-independent and since I only consider likelihood ratios\footnote{Note that the exact values of this log-likelihood function are not physically meaningful since they depend on the arbitrary choice of bin size. In the likelihood ratio, this dependence cancels out; the ratio thus is a meaningful measure of which model is more likely to produce the observed events.} of different models A and~B, $\Delta L = L_A - L_B$, it cancels out.
The final likelihood function I use is thus given by
\begin{equation}
L = \sum_{i \in \mathcal{B}} \ln N_i.
\end{equation}

\backmatter
\bibliography{thesis}

\begin{thebibliography}{100}
\providecommand{\url}[1]{\texttt{#1}}
\providecommand{\urlprefix}{URL }
\providecommand{\eprint}[2][]{\url{#2}}

\bibitem{Munroe2014}
\textsc{R.~Munroe}: \emph{What if? Serious Scientific Answers to Absurd
  Hypothetical Questions}, p. 175.
\newblock Houghton Mifflin Harcourt, Boston (2014).

\bibitem{Sanduleak1970}
\textsc{N.~{Sanduleak}}: {A deep objective-prism survey for Large Magellanic
  Cloud members}.
\newblock \emph{Contributions from the Cerro Tololo Inter-American Observatory}
  \textbf{89} (1970).

\bibitem{Buras2006}
\textsc{R.~Buras}, \textsc{M.~Rampp}, \textsc{H.~T. Janka},
  \textsc{K.~Kifonidis}: {Two-dimensional hydrodynamic core-collapse supernova
  simulations with spectral neutrino transport. 1. Numerical method and results
  for a 15 solar mass star}.
\newblock \emph{Astron. Astrophys.} \textbf{447} (2006), pp. 1049--1092.

\bibitem{Liebendorfer2009}
\textsc{M.~Liebendörfer}, \textsc{S.~C. Whitehouse}, \textsc{T.~Fischer}: The
  Isotropic Diffusion Source Approximation for supernova neutrino transport.
\newblock \emph{Astrophys. J.} \textbf{698} (2009), 2, p. 1174–1190.

\bibitem{Marek2006}
\textsc{A.~Marek} \emph{et~al.}: {Exploring the relativistic regime with
  Newtonian hydrodynamics: An Improved effective gravitational potential for
  supernova simulations}.
\newblock \emph{Astron. Astrophys.} \textbf{445} (2006), p. 273.

\bibitem{Skinner2015}
\textsc{M.~A. Skinner}, \textsc{A.~Burrows}, \textsc{J.~C. Dolence}: {Should
  One Use the Ray-by-Ray Approximation in Core-Collapse Supernova Simulations?}
\newblock \emph{Astrophys. J.} \textbf{831} (2015), 1, p.~81.

\bibitem{Ellis1927}
\textsc{C.~D. Ellis}, \textsc{W.~A. Wooster}: The average energy of
  disintegration of radium E.
\newblock \emph{Proc. Roy. Soc. Lond.} \textbf{A117} (1927), 776, pp. 109--123.

\bibitem{Meitner1930}
\textsc{L.~Meitner}, \textsc{W.~Orthmann}: Über eine absolute Bestimmung der
  Energie der primären $\beta$-Strahlen von Radium E.
\newblock \emph{Zeitschrift für Physik} \textbf{60} (1930), p. 143.

\bibitem{Bohr1932}
\textsc{N.~Bohr}: Faraday lecture: Chemistry and the quantum theory of atomic
  constitution.
\newblock \emph{Journal of the Chemical Society}  (1932), pp. 349--384.

\bibitem{Pauli1930}
\textsc{W.~Pauli}: {Pauli letter collection: Letter to Lise Meitner} (1930).
\newblock Typed copy,
  \href{http://cds.cern.ch/record/83282}{\texttt{cds.cern.ch/record/83282}}.

\bibitem{Chadwick1932}
\textsc{J.~Chadwick}: Possible Existence of a Neutron.
\newblock \emph{Nature} \textbf{129} (1932), 3252, p. 312.

\bibitem{Amaldi1998}
\textsc{U.~Amaldi}: \emph{20th Century Physics: Essays and Recollections. A
  Selection of Historical Writings by Edoardo Amaldi}, chap. Preface.
\newblock World Scientific, Singapore (1998).

\bibitem{Bonolis2005}
\textsc{L.~Bonolis}: Bruno Pontecorvo: From slow neutrons to oscillating
  neutrinos.
\newblock \emph{American J. Phys.} \textbf{73} (2005), 6, pp. 487--499.

\bibitem{Fermi1934d}
\textsc{E.~Fermi}: Versuch einer Theorie der $\beta$-Strahlen. I.
\newblock \emph{Zeitschrift für Physik} \textbf{88} (1934), 3-4, pp. 161--177.

\bibitem{Bethe1934}
\textsc{H.~Bethe}, \textsc{R.~Peierls}: The ``Neutrino''.
\newblock \emph{Nature} \textbf{133} (1934), p. 532.

\bibitem{Leipunski1936}
\textsc{A.~I. Leipunski}: Determination of the Energy Distribution of Recoil
  Atoms During $\beta$ Decay and the Existence of the Neutrino.
\newblock \emph{Mathematical Proceedings of the Cambridge Philosophical
  Society} \textbf{32} (1936), 2, pp. 301--303.

\bibitem{Crane1938}
\textsc{H.~R. Crane}, \textsc{J.~Halpern}: New Experimental Evidence for the
  Existence of a Neutrino.
\newblock \emph{Phys. Rev.} \textbf{53} (1938), p. 789.

\bibitem{Jacobsen1945}
\textsc{J.~C. Jacobsen}, \textsc{O.~Kofoed-Hansen}: On the Recoil of the
  Nucleus in Beta-Decay.
\newblock \emph{Matematisk-Fysiske Meddelelser} \textbf{23} (1945), 12.

\bibitem{Christy1947}
\textsc{R.~F. Christy} \emph{et~al.}: The Conservation of Momentum in the
  Disintegration of Li8.
\newblock \emph{Phys. Rev.} \textbf{72} (1947), pp. 698--714.

\bibitem{Rodeback1952}
\textsc{G.~W. Rodeback}, \textsc{J.~S. Allen}: Neutrino Recoils Following the
  Capture of Orbital Electrons in A37.
\newblock \emph{Phys. Rev.} \textbf{86} (1952), pp. 446--450.

\bibitem{LosAlamos1997}
{The Reines-Cowan experiments: Detecting the Poltergeist}.
\newblock \emph{Los Alamos Sci.} \textbf{25} (1997), pp. 4--27.

\bibitem{Reines1953a}
\textsc{F.~Reines}, \textsc{C.~L. Cowan}: A Proposed Experiment to Detect the
  Free Neutrino.
\newblock \emph{Phys. Rev.} \textbf{90} (1953), pp. 492--493.

\bibitem{Cowan1953}
\textsc{C.~L. Cowan} \emph{et~al.}: Large Liquid Scintillation Detectors.
\newblock \emph{Phys. Rev.} \textbf{90} (1953), pp. 493--494.

\bibitem{Reines1953b}
\textsc{F.~Reines}, \textsc{C.~L. Cowan}: Detection of the Free Neutrino.
\newblock \emph{Phys. Rev.} \textbf{92} (1953), pp. 830--831.

\bibitem{Reines1956}
\textsc{F.~Reines}, \textsc{C.~L. Cowan}: The Neutrino.
\newblock \emph{Nature} \textbf{178} (1956), 4531, pp. 446--449.

\bibitem{Cowan1956}
\textsc{C.~L. Cowan} \emph{et~al.}: Detection of the Free Neutrino: a
  Confirmation.
\newblock \emph{Science} \textbf{124} (1956), 3212, pp. 103--104.

\bibitem{Wu1957}
\textsc{C.~S. Wu} \emph{et~al.}: Experimental Test of Parity Conservation in
  Beta Decay.
\newblock \emph{Phys. Rev.} \textbf{105} (1957), pp. 1413--1415.

\bibitem{Goldhaber1958}
\textsc{M.~Goldhaber}, \textsc{L.~Grodzins}, \textsc{A.~W. Sunyar}: Helicity of
  Neutrinos.
\newblock \emph{Phys. Rev.} \textbf{109} (1958), pp. 1015--1017.

\bibitem{Konopinski1953}
\textsc{E.~J. Konopinski}, \textsc{L.~M. Langer}: The Experimental
  Clarification of the Theory of $\beta$-Decay.
\newblock \emph{Ann. Rev. Nucl. Sci.} \textbf{2} (1953), pp. 261--304.

\bibitem{Davis1955}
\textsc{R.~Davis}: Attempt to Detect the Antineutrinos from a Nuclear Reactor
  by the Cl$^{37}$($\bar{\nu}$, $e^-$)A$^{37}$ Reaction.
\newblock \emph{Phys. Rev.} \textbf{97} (1955), pp. 766--769.

\bibitem{Davis1959}
\textsc{R.~Davis}, \textsc{D.~S. Harmer}: Attempt to observe the
  Cl$^{37}$(\nubar $e^{−}$)Ar$^{37}$ reaction induced by reactor
  antineutrinos.
\newblock \emph{Bull. Am. Phys. Soc.} \textbf{4} (1959), p. 217.

\bibitem{Danby1962}
\textsc{G.~Danby} \emph{et~al.}: Observation of High-Energy Neutrino Reactions
  and the Existence of Two Kinds of Neutrinos.
\newblock \emph{Phys. Rev. Lett.} \textbf{9} (1962), pp. 36--44.

\bibitem{Perl1975}
\textsc{M.~L. Perl} \emph{et~al.}: Evidence for Anomalous Lepton Production in
  e$^+$-e$^-$ Annihilation.
\newblock \emph{Phys. Rev. Lett.} \textbf{35} (1975), pp. 1489--1492.

\bibitem{Kodama2001}
\textsc{K.~Kodama} \emph{et~al.}  (DONUT Collaboration): Observation of tau
  neutrino interactions.
\newblock \emph{Phys. Lett.} \textbf{B504} (2001), 3, pp. 218--224.

\bibitem{LEP2006}
\textsc{S.~Schael} \emph{et~al.}  (SLD Electroweak Group, DELPHI, ALEPH, SLD,
  SLD Heavy Flavour Group, OPAL, LEP Electroweak Working Group, L3): {Precision
  electroweak measurements on the $Z$ resonance}.
\newblock \emph{Phys. Rep.} \textbf{427} (2006), p. 257.

\bibitem{PDG2018}
\textsc{M.~Tanabashi} \emph{et~al.}  (Particle Data Group): Review of Particle
  Physics.
\newblock \emph{Phys. Rev.} \textbf{D98} (2018), 030001.

\bibitem{England2007}
\textsc{P.~England}, \textsc{P.~Molnar}, \textsc{F.~Richter}: John Perry's
  neglected critique of Kelvin's age for the Earth: A missed opportunity in
  geodynamics.
\newblock \emph{GSA Today} \textbf{17} (2007), 1, pp. 4--9.

\bibitem{Eddington1920}
\textsc{A.~S. Eddington}: The internal constitution of the stars.
\newblock \emph{The Observatory} \textbf{43} (1920), pp. 341--358.

\bibitem{Bethe1939}
\textsc{H.~Bethe}: Energy Production in Stars.
\newblock \emph{Phys. Rev.} \textbf{55} (1939), pp. 434--456.

\bibitem{Bilenky1999}
\textsc{S.~M. Bilenky}, \textsc{C.~Giunti}, \textsc{W.~Grimus}: Phenomenology
  of Neutrino Oscillations.
\newblock \emph{Prog. Part. Nucl. Phys.} \textbf{43} (1999), pp. 1--86.

\bibitem{Holmgren1959}
\textsc{H.~D. Holmgren}, \textsc{R.~L. Johnston}: H$^3 (\alpha,\gamma)$Li$^7$
  and He$^3 (\alpha, \gamma)$Be$^7$ Reactions.
\newblock \emph{Phys. Rev.} \textbf{113} (1959), 6, pp. 1556--1559.

\bibitem{Bahcall1964}
\textsc{J.~N. Bahcall}: Solar Neutrinos. I. Theoretical.
\newblock \emph{Phys. Rev. Lett.} \textbf{12} (1964), pp. 300--302.

\bibitem{Davis1964}
\textsc{R.~Davis}: Solar Neutrinos. II. Experimental.
\newblock \emph{Phys. Rev. Lett.} \textbf{12} (1964), pp. 303--305.

\bibitem{Davis1968}
\textsc{R.~Davis}, \textsc{D.~S. Harmer}, \textsc{K.~C. Hoffman}: Search for
  Neutrinos from the Sun.
\newblock \emph{Phys. Rev. Lett.} \textbf{20} (1968), pp. 1205--1209.

\bibitem{Bahcall1968}
\textsc{J.~N. Bahcall}, \textsc{N.~A. Bahcall}, \textsc{G.~Shaviv}: Present
  Status of the Theoretical Predictions for the $^{37}$Cl Solar-Neutrino
  Experiment.
\newblock \emph{Phys. Rev. Lett.} \textbf{20} (1968), pp. 1209--1212.

\bibitem{Bahcall2001}
\textsc{J.~N. Bahcall}, \textsc{M.~H. Pinsonneault}, \textsc{S.~Basu}: {Solar
  models: Current epoch and time dependences, neutrinos, and helioseismological
  properties}.
\newblock \emph{Astrophys. J.} \textbf{555} (2001), pp. 990--1012.

\bibitem{Cleveland1998}
\textsc{B.~T. Cleveland} \emph{et~al.}: {Measurement of the solar electron
  neutrino flux with the Homestake chlorine detector}.
\newblock \emph{Astrophys. J.} \textbf{496} (1998), pp. 505--526.

\bibitem{Suzuki2001}
\textsc{Y.~Suzuki}: {Solar Neutrino Results from Super-Kamiokande}.
\newblock \emph{Nucl. Phys. Proc. Suppl.} \textbf{91} (2001), pp. 29--35.

\bibitem{Hampel1999}
\textsc{W.~Hampel} \emph{et~al.}  (GALLEX Collaboration): GALLEX solar neutrino
  observations: results for GALLEX IV.
\newblock \emph{Phys. Lett.} \textbf{B447} (1999), pp. 127--133.

\bibitem{Altmann2000}
\textsc{M.~Altmann} \emph{et~al.}  (GNO Collaboration): GNO solar neutrino
  observations: results for GNO I.
\newblock \emph{Phys. Lett.} \textbf{B490} (2000), 1--2, pp. 16--26.

\bibitem{Gavrin2001}
\textsc{V.~N. Gavrin}  (SAGE): {Solar neutrino results from SAGE}.
\newblock \emph{Nucl. Phys. Proc. Suppl.} \textbf{91} (2001), pp. 36--43.

\bibitem{Fowler1968}
\textsc{W.~A. Fowler}: Solar Neutrino Astronomy.
\newblock In: \emph{{Proceedings, International Symposium on Contemporary
  Physics, Trieste, Italy, 7-28 June 1968}}, vol.~1. International Atomic
  Energy Agency (IAEA), Vienna (Austria) (1968) pp. 359--370.

\bibitem{Bahcall1972}
\textsc{J.~N. Bahcall}, \textsc{N.~Cabibbo}, \textsc{A.~Yahil}: Are Neutrinos
  Stable Particles?
\newblock \emph{Phys. Rev. Lett.} \textbf{28} (1972), pp. 316--318.

\bibitem{Demarque1973}
\textsc{P.~{Demarque}}, \textsc{J.~G. {Mengel}}, \textsc{A.~V. {Sweigart}}:
  {Rotating Solar Models with Low Neutrino Flux}.
\newblock \emph{Astrophys. J.} \textbf{183} (1973), pp. 997--1004.

\bibitem{Clayton1975}
\textsc{D.~D. {Clayton}}, \textsc{M.~J. {Newman}}, \textsc{R.~J. {Talbot},
  Jr.}: {Solar models of low neutrino-counting rate - The central black hole}.
\newblock \emph{Astrophys. J.} \textbf{201} (1975), pp. 489--493.

\bibitem{Davis2003}
\textsc{R.~Davis}: Nobel Lecture: A half-century with solar neutrinos.
\newblock \emph{Rev. Mod. Phys.} \textbf{75} (2003), pp. 985--994.

\bibitem{Hirata1988}
\textsc{K.~S. Hirata} \emph{et~al.}  (Kamiokande-II): {Experimental Study of
  the Atmospheric Neutrino Flux}.
\newblock \emph{Phys. Lett.} \textbf{B205} (1988), p. 416.

\bibitem{Boger2000}
\textsc{J.~Boger}  (SNO Collaboration): The Sudbury neutrino observatory.
\newblock \emph{Nucl. Instrum. Meth.} \textbf{A449} (2000), pp. 172--207.

\bibitem{Ahmad2002}
\textsc{Q.~R. Ahmad} \emph{et~al.}  (SNO Collaboration): Direct evidence for
  neutrino flavor transformation from neutral current interactions in the
  Sudbury Neutrino Observatory.
\newblock \emph{Phys. Rev. Lett.} \textbf{89} (2002), 011301.

\bibitem{Pontecorvo1958a}
\textsc{B.~Pontecorvo}: {Mesonium and anti-mesonium}.
\newblock \emph{Sov. Phys. JETP} \textbf{6} (1958), p. 429.
\newblock [Zh. Eksp. Teor. Fiz.33,549(1957)].

\bibitem{Pontecorvo1958b}
\textsc{B.~Pontecorvo}: {Inverse beta processes and nonconservation of lepton
  charge}.
\newblock \emph{Sov. Phys. JETP} \textbf{7} (1958), pp. 172--173.
\newblock [Zh. Eksp. Teor. Fiz.34,247(1957)].

\bibitem{Gell-Mann1955}
\textsc{M.~Gell-Mann}, \textsc{A.~Pais}: Behavior of Neutral Particles under
  Charge Conjugation.
\newblock \emph{Phys. Rev.} \textbf{97} (1955), pp. 1387--1389.

\bibitem{Maki1962}
\textsc{Z.~Maki}, \textsc{M.~Nakagawa}, \textsc{S.~Sakata}: Remarks on the
  Unified Model of Elementary Particles.
\newblock \emph{Prog. Theor. Phys.} \textbf{28} (1962), 5, pp. 870--880.

\bibitem{Cabibbo1963}
\textsc{N.~Cabibbo}: Unitary symmetry and leptonic decays.
\newblock \emph{Phys. Rev. Lett.} \textbf{10} (1963), 12, pp. 531--533.

\bibitem{Christenson1964}
\textsc{J.~H. Christenson}, \textsc{J.~W. Cronin}, \textsc{V.~L. Fitch},
  \textsc{R.~Turlay}: {Evidence for the $2\pi$ Decay of the $K_2^0$ Meson}.
\newblock \emph{Phys. Rev. Lett.} \textbf{13} (1964), pp. 138--140.

\bibitem{Kobayashi1973}
\textsc{M.~Kobayashi}, \textsc{T.~Maskawa}: CP-Violation in the Renormalizable
  Theory of Weak Interaction.
\newblock \emph{Prog. Theor. Phys.} \textbf{49} (1973), 2, pp. 652--657.

\bibitem{Bilenky1978}
\textsc{S.~M. Bilenky}, \textsc{B.~Pontecorvo}: {Lepton Mixing and Neutrino
  Oscillations}.
\newblock \emph{Phys. Rep.} \textbf{41} (1978), pp. 225--261.

\bibitem{Majorana1937}
\textsc{E.~Majorana}: {Teoria simmetrica dell’elettrone e del positrone}.
\newblock \emph{Nuovo Cim.} \textbf{14} (1937), pp. 171--184.

\bibitem{Fukuda1994}
\textsc{Y.~Fukuda} \emph{et~al.}  (Kamiokande): {Atmospheric muon-neutrino /
  electron-neutrino ratio in the multiGeV energy range}.
\newblock \emph{Phys. Lett.} \textbf{B335} (1994), pp. 237--245.

\bibitem{Fukuda1998}
\textsc{Y.~Fukuda} \emph{et~al.}  (Super-Kamiokande): {Evidence for oscillation
  of atmospheric neutrinos}.
\newblock \emph{Phys. Rev. Lett.} \textbf{81} (1998), pp. 1562--1567.

\bibitem{Aartsen2018}
\textsc{M.~G. Aartsen} \emph{et~al.}  (IceCube): {Measurement of Atmospheric
  Neutrino Oscillations at 6–56 GeV with IceCube DeepCore}.
\newblock \emph{Phys. Rev. Lett.} \textbf{120} (2018), 7, p. 071801.

\bibitem{Abe2018}
\textsc{K.~Abe} \emph{et~al.}  (T2K): {Search for CP violation in Neutrino and
  Antineutrino Oscillations by the T2K experiment with $2.2\times10^{21}$
  protons on target}.
\newblock \emph{Phys. Rev. Lett.} \textbf{121} (2018), 17, p. 171802.

\bibitem{Acero2018}
\textsc{M.~A. Acero} \emph{et~al.}  (NOvA): {New constraints on oscillation
  parameters from $\nu_e$ appearance and $\nu_\mu$ disappearance in the NOvA
  experiment}.
\newblock \emph{Phys. Rev.} \textbf{D98} (2018), p. 032012.

\bibitem{Ahmad2002a}
\textsc{Q.~R. Ahmad} \emph{et~al.}  (SNO Collaboration): Measurement of day and
  night neutrino energy spectra at SNO and constraints on neutrino mixing
  parameters.
\newblock \emph{Phys. Rev. Lett.} \textbf{89} (2002), 011302.

\bibitem{Eguchi2003}
\textsc{K.~Eguchi} \emph{et~al.}  (KamLAND Collaboration): First Results from
  KamLAND: Evidence for Reactor Antineutrino Disappearance.
\newblock \emph{Phys. Rev. Lett.} \textbf{90} (2003), 021802.

\bibitem{Abe2011}
\textsc{K.~Abe} \emph{et~al.}  (T2K Collaboration): Indication of Electron
  Neutrino Appearance from an Accelerator-Produced Off-Axis Muon Neutrino Beam.
\newblock \emph{Phys. Rev. Lett.} \textbf{107} (2011), p. 041801.

\bibitem{Adamson2011}
\textsc{P.~Adamson} \emph{et~al.}  (MINOS Collaboration): Improved Search for
  Muon-Neutrino to Electron-Neutrino Oscillations in MINOS.
\newblock \emph{Phys. Rev. Lett.} \textbf{107} (2011), p. 181802.

\bibitem{Abe2012}
\textsc{Y.~Abe} \emph{et~al.}  (Double Chooz Collaboration): Reactor
  $\overline{\nu}_e$ disappearance in the Double Chooz experiment.
\newblock \emph{Phys. Rev.} \textbf{D86} (2012), p. 052008.

\bibitem{An2012}
\textsc{F.~P. An} \emph{et~al.}: Observation of Electron-Antineutrino
  Disappearance at Daya Bay.
\newblock \emph{Phys. Rev. Lett.} \textbf{108} (2012), 171803.

\bibitem{Ahn2012}
\textsc{J.~K. Ahn} \emph{et~al.}  (RENO Collaboration): Observation of Reactor
  Electron Antineutrinos Disappearance in the RENO Experiment.
\newblock \emph{Phys. Rev. Lett.} \textbf{108} (2012), 191802.

\bibitem{Wolfenstein1978}
\textsc{L.~Wolfenstein}: Neutrino oscillations in matter.
\newblock \emph{Phys. Rev.} \textbf{D17} (1978), 9, pp. 2369--2374.

\bibitem{Mikheev1985}
\textsc{S.~P. Mikheev}, \textsc{A.~Y. Smirnov}: Resonance Amplification of
  Oscillations in Matter and Spectroscopy of Solar Neutrinos.
\newblock \emph{Sov. J. Nucl. Phys.} \textbf{42} (1985), pp. 913--917.

\bibitem{Bethe1986}
\textsc{H.~Bethe}: Possible Explanation of the Solar-Neutrino Puzzle.
\newblock \emph{Phys. Rev. Lett.} \textbf{56} (1986), 12, pp. 1305--1308.

\bibitem{Snowglobes}
SNOwGLoBES:
  \href{http://webhome.phy.duke.edu/~schol/snowglobes/}{\texttt{webhome.phy.duke.edu/\textasciitilde
  schol/snowglobes/}}.
\newblock  Last checked: 2019-11-18.

\bibitem{Abbasi2011}
\textsc{R.~Abbasi} \emph{et~al.}  (IceCube Collaboration): IceCube sensitivity
  for low-energy neutrinos from nearby supernovae.
\newblock \emph{Astron. Astrophys.} \textbf{535} (2011), A109.

\bibitem{Adrian-Martinez2016}
\textsc{S.~Adrian-Martinez} \emph{et~al.}  (KM3Net): {Letter of intent for
  KM3NeT 2.0}.
\newblock \emph{J. Phys.} \textbf{G43} (2016), 8, p. 084001.

\bibitem{Asakura2016}
\textsc{K.~Asakura} \emph{et~al.}  (KamLAND): {KamLAND Sensitivity to Neutrinos
  from Pre-Supernova Stars}.
\newblock \emph{Astrophys. J.} \textbf{818} (2016), 1, p.~91.

\bibitem{Andringa2016}
\textsc{S.~Andringa} \emph{et~al.}  (SNO+): {Current Status and Future
  Prospects of the SNO+ Experiment}.
\newblock \emph{Adv. High Energy Phys.} \textbf{2016} (2016), p. 6194250.

\bibitem{An2016}
\textsc{F.~An} \emph{et~al.}  (JUNO): {Neutrino Physics with JUNO}.
\newblock \emph{J. Phys.} \textbf{G43} (2016), 3, p. 030401.

\bibitem{Gruszko2019}
\textsc{J.~Gruszko} \emph{et~al.}: {Detecting Cherenkov light from 1–2 MeV
  electrons in linear alkylbenzene}.
\newblock \emph{JINST} \textbf{14} (2019), 02, p. P02005.

\bibitem{Yeh2011}
\textsc{M.~Yeh} \emph{et~al.}: {A new water-based liquid scintillator and
  potential applications}.
\newblock \emph{Nucl. Instrum. Meth.} \textbf{A660} (2011), pp. 51--56.

\bibitem{Alonso2014}
\textsc{J.~R. Alonso} \emph{et~al.}: {Advanced Scintillator Detector Concept
  (ASDC): A Concept Paper on the Physics Potential of Water-Based Liquid
  Scintillator}  (2014).
\newblock \href{https://arxiv.org/abs/1409.5864}{\texttt{arXiv:1409.5864}}.

\bibitem{Askins2019}
\textsc{M.~Askins} \emph{et~al.}: {Theia: An advanced optical neutrino
  detector}  (2019).
\newblock \href{https://arxiv.org/abs/1911.03501}{\texttt{arXiv:1911.03501}}.

\bibitem{Amerio2004}
\textsc{S.~Amerio} \emph{et~al.}  (ICARUS): {Design, construction and tests of
  the ICARUS T600 detector}.
\newblock \emph{Nucl. Instrum. Meth.} \textbf{A527} (2004), pp. 329--410.

\bibitem{Acciarri2017}
\textsc{R.~Acciarri} \emph{et~al.}  (MicroBooNE): {Design and Construction of
  the MicroBooNE Detector}.
\newblock \emph{JINST} \textbf{12} (2017), 02, p. P02017.

\bibitem{Acciarri2015}
\textsc{R.~Acciarri} \emph{et~al.}  (DUNE): {Long-Baseline Neutrino Facility
  (LBNF) and Deep Underground Neutrino Experiment (DUNE) Conceptual Design
  Report, Vol. 2: The Physics Program for DUNE at LBNF}  (2015).
\newblock \href{https://arxiv.org/abs/1512.06148}{\texttt{arXiv:1512.06148}}.

\bibitem{Zhao2006}
\textsc{F.-Y. Zhao}, \textsc{R.~G. Strom}, \textsc{S.-Y. Jiang}: The Guest Star
  of AD185 must have been a Supernova.
\newblock \emph{Chinese J. Astron. Astrophys.} \textbf{6} (2006), 5, p.
  635–640.

\bibitem{Baade1934a}
\textsc{W.~Baade}, \textsc{F.~Zwicky}: On Super-Novae.
\newblock \emph{Proceedings of the National Academy of Sciences} \textbf{20}
  (1934), 5, pp. 254--259.

\bibitem{Baade1934b}
\textsc{W.~Baade}, \textsc{F.~Zwicky}: Cosmic Rays from Super-Novae.
\newblock \emph{Proceedings of the National Academy of Sciences} \textbf{20}
  (1934), 5, pp. 259--263.

\bibitem{Baade1934}
\textsc{W.~Baade}, \textsc{F.~Zwicky}: Remarks on Super-Novae and Cosmic Rays.
\newblock \emph{Phys. Rev.} \textbf{46} (1934), pp. 76--77.

\bibitem{Zwicky1964}
\textsc{F.~Zwicky}: Basic results of the international search for Supernovae.
\newblock \emph{Annales d'Astrophysique} \textbf{27} (1964), p. 300.

\bibitem{Minkowski1941}
\textsc{R.~Minkowski}: Spectra of Supernovae.
\newblock \emph{Publications of the Astronomical Society of the Pacific}
  \textbf{53} (1941), 314, p. 224.

\bibitem{Milisavljevic2018}
\textsc{D.~Milisavljevic}, \textsc{R.~Margutti}: {Peculiar Supernovae}.
\newblock \emph{Space Sci. Rev.} \textbf{214} (2018), 4, p.~68.

\bibitem{Heger2003}
\textsc{A.~Heger} \emph{et~al.}: How massive single stars end their life.
\newblock \emph{Astrophys. J.} \textbf{591} (2003), pp. 288--300.

\bibitem{Hoyle1960}
\textsc{F.~Hoyle}, \textsc{W.~A. Fowler}: Nucleosynthesis in Supernovae.
\newblock \emph{Astrophys. J.} \textbf{132} (1960), p. 565.

\bibitem{Whelan1973}
\textsc{J.~Whelan}, \textsc{I.~Iben}: Binaries and Supernovae of Type I.
\newblock \emph{Astrophys. J.} \textbf{186} (1973), pp. 1007--1014.

\bibitem{Maoz2014}
\textsc{D.~Maoz}, \textsc{F.~Mannucci}, \textsc{G.~Nelemans}: {Observational
  clues to the progenitors of Type-Ia supernovae}.
\newblock \emph{Ann. Rev. Astron. Astrophys.} \textbf{52} (2014), pp. 107--170.

\bibitem{Hillebrandt2013}
\textsc{W.~Hillebrandt}, \textsc{M.~Kromer}, \textsc{F.~K. Röpke},
  \textsc{A.~J. Ruiter}: {Towards an understanding of Type Ia supernovae from a
  synthesis of theory and observations}.
\newblock \emph{Front. Phys.} \textbf{8} (2013), pp. 116--143.

\bibitem{Wright2016}
\textsc{W.~P. Wright} \emph{et~al.}: {Neutrinos from type Ia supernovae: The
  deflagration-to-detonation transition scenario}.
\newblock \emph{Phys. Rev.} \textbf{D94} (2016), 2, p. 025026.

\bibitem{Wright2017}
\textsc{W.~P. Wright} \emph{et~al.}: {Neutrinos from type Ia supernovae: The
  gravitationally confined detonation scenario}.
\newblock \emph{Phys. Rev.} \textbf{D95} (2017), 4, p. 043006.

\bibitem{Guillochon2017}
\textsc{J.~Guillochon}, \textsc{J.~Parrent}, \textsc{L.~Z. Kelley},
  \textsc{R.~Margutti}: {An Open Catalog for Supernova Data}.
\newblock \emph{Astrophys. J.} \textbf{835} (2017), 1, p.~64.

\bibitem{TOSC}
The Open Supernova Catalog: \href{https://sne.space}{\texttt{sne.space}}.
\newblock  Last checked: 2019-11-18.

\bibitem{Lien2009}
\textsc{A.~Lien}, \textsc{B.~D. Fields}: {Cosmic Core-Collapse Supernovae from
  Upcoming Sky Surveys}.
\newblock \emph{JCAP} \textbf{0901} (2009), p. 047.

\bibitem{Abell2009}
\textsc{P.~A. Abell} \emph{et~al.}  (LSST Science, LSST Project): {LSST Science
  Book, Version 2.0}  (2009).
\newblock \href{https://arxiv.org/abs/0912.0201}{\texttt{arXiv:0912.0201}}.

\bibitem{Tammann1994}
\textsc{G.~A. Tammann}, \textsc{W.~Loeffler}, \textsc{A.~Schroder}: {The
  Galactic supernova rate}.
\newblock \emph{Astrophys. J. Suppl.} \textbf{92} (1994), pp. 487--493.

\bibitem{Colgate1961}
\textsc{S.~A. {Colgate}}, \textsc{W.~H. {Grasberger}}, \textsc{R.~H. {White}}:
  {The Dynamics of Supernova Explosions}.
\newblock \emph{Astron. J.} \textbf{70} (1961), p. 280.

\bibitem{Colgate1966}
\textsc{S.~A. Colgate}, \textsc{R.~H. White}: The Hydrodynamic Behavior of
  Supernovae Explosions.
\newblock \emph{Astrophys. J.} \textbf{143} (1966), pp. 626--681.

\bibitem{Arnett1967}
\textsc{W.~D. Arnett}: Mass Dependence in Gravitational Collapse of Stellar
  Cores.
\newblock \emph{Canadian J. Phys.} \textbf{45} (1967), 5, pp. 1621--1641.

\bibitem{Wilson1971}
\textsc{J.~R. Wilson}: A Numerical Study of Gravitational Stellar Collapse.
\newblock \emph{Astrophys. J.} \textbf{163} (1971), pp. 209--219.

\bibitem{Wilson1982}
\textsc{J.~R. Wilson}: Supernovae and post-collapse behavior.
\newblock In: \textsc{J.~M. Centrella}, \textsc{J.~M. LeBlanc}, \textsc{R.~L.
  Bowers}, eds.: \emph{Numerical Astrophysics. Proceedings of the Symposium in
  honour of James R. Wilson, held at the University of Illinois Urbana
  Champaign, October, 1982}. Jones and Bartlett Publ., Boston (1985) p. 422.

\bibitem{Hillebrandt1989}
\textsc{W.~Hillebrandt}, \textsc{P.~Hoflich}: The supernova 1987A in the Large
  Magellanic Cloud.
\newblock \emph{Rep. Prog. Phys.} \textbf{52} (1989), 11, pp. 1421--1473.

\bibitem{Janka2012}
\textsc{H.-T. Janka}: Explosion Mechanisms of Core-Collapse Supernovae.
\newblock \emph{Ann. Rev. Nucl. Part. Sci.} \textbf{62} (2012), pp. 407--451.

\bibitem{Hanke2013}
\textsc{F.~Hanke} \emph{et~al.}: SASI Activity in Three-Dimensional
  Neutrino-Hydro\-dynamics Simulations of Supernova Cores.
\newblock \emph{Astrophys. J.} \textbf{770} (2013), 1, p.~66.

\bibitem{Muller2012}
\textsc{B.~Müller}, \textsc{H.-T. Janka}, \textsc{A.~Marek}: A new
  multi-dimensional general relativistic neutrino hydrodynamics code for
  core-collapse supernovae: II. Relativistic explosion models of core-collapse
  supernovae.
\newblock \emph{Astrophys. J.} \textbf{756} (2012), pp. 84--105.

\bibitem{Burbidge1957}
\textsc{E.~M. Burbidge}, \textsc{G.~R. Burbidge}, \textsc{W.~A. Fowler},
  \textsc{F.~Hoyle}: Synthesis of the Elements in Stars.
\newblock \emph{Rev. Mod. Phys.} \textbf{29} (1957), pp. 547--650.

\bibitem{Bethe1985}
\textsc{H.~A. Bethe}, \textsc{J.~R. Wilson}: Revival of a stalled supernova
  shock by neutrino heating.
\newblock \emph{Astrophys. J.} \textbf{295} (1985), pp. 14--23.

\bibitem{Janka2007}
\textsc{H.-T. Janka} \emph{et~al.}: Theory of core-collapse supernovae.
\newblock \emph{Phys. Rep.} \textbf{442} (2007), pp. 38--74.

\bibitem{Nakazato2013}
\textsc{K.~Nakazato} \emph{et~al.}: {Supernova Neutrino Light Curves and
  Spectra for Various Progenitor Stars: From Core Collapse to Proto-neutron
  Star Cooling}.
\newblock \emph{Astrophys. J. Suppl.} \textbf{205} (2013), p.~2.

\bibitem{Kachelries2005}
\textsc{M.~Kachelrieß} \emph{et~al.}: Exploiting the neutronization burst of a
  galactic supernova.
\newblock \emph{Phys. Rev.} \textbf{D71} (2005), 063003.

\bibitem{Blondin2003}
\textsc{J.~M. Blondin}, \textsc{A.~Mezzacappa}, \textsc{C.~DeMarino}: Stability
  of Standing Accretion Shocks, with an Eye toward Core-Collapse Supernovae.
\newblock \emph{Astrophys. J.} \textbf{584} (2003), 2, pp. 971--980.

\bibitem{Lund2010}
\textsc{T.~Lund} \emph{et~al.}: Fast time variations of supernova neutrino
  fluxes and their detectability.
\newblock \emph{Phys. Rev.} \textbf{D82} (2010), p. 063007.

\bibitem{Tamborra2013}
\textsc{I.~Tamborra} \emph{et~al.}: {Neutrino signature of supernova
  hydrodynamical instabilities in three dimensions}.
\newblock \emph{Phys. Rev. Lett.} \textbf{111} (2013), 12, p. 121104.

\bibitem{Migenda2015}
\textsc{J.~Migenda}: \emph{{Detecting Fast Time Variations in the Supernova
  Neutrino Flux with Hyper-Kamiokande}}.
\newblock Master's thesis, Munich, Max Planck Inst. (2015).
\newblock \href{https://arxiv.org/abs/1609.04286}{\texttt{arXiv:1609.04286}}.

\bibitem{Hannestad1998}
\textsc{S.~Hannestad}, \textsc{G.~Raffelt}: Supernova Neutrino Opacity from
  Nucleon-Nucleon Bremsstrahlung and Related Processes.
\newblock \emph{Astrophys. J.} \textbf{507} (1998), pp. 339--352.

\bibitem{Thompson2000}
\textsc{T.~A. Thompson}, \textsc{A.~Burrows}, \textsc{J.~E. Horvath}: $\mu$ and
  $\tau$ neutrino thermalization and production in supernovae: Processes and
  time scales.
\newblock \emph{Phys. Rev.} \textbf{C62} (2000), 035802.

\bibitem{Buras2003}
\textsc{R.~Buras} \emph{et~al.}: Electron Neutrino Pair Annihilation: A New
  Source for Muon and Tau Neutrinos in Supernovae.
\newblock \emph{Astrophys. J.} \textbf{587} (2003), pp. 320--326.

\bibitem{Keil2003}
\textsc{M.~T. Keil}, \textsc{G.~G. Raffelt}, \textsc{H.-T. Janka}: Monte Carlo
  Study of Supernova Neutrino Spectra Formation.
\newblock \emph{Astrophys. J.} \textbf{590} (2003), pp. 971--991.

\bibitem{Hirata1987}
\textsc{K.~Hirata} \emph{et~al.}: Observation of a neutrino burst from the
  supernova SN1987A.
\newblock \emph{Phys. Rev. Lett.} \textbf{58} (1987), 14, pp. 1490--1493.

\bibitem{Hirata1988a}
\textsc{K.~S. Hirata} \emph{et~al.}: Observation in the Kamiokande-II detector
  of the neutrino burst from supernova SN1987A.
\newblock \emph{Phys. Rev.} \textbf{D38} (1988), 2, pp. 448--458.

\bibitem{Bionta1987}
\textsc{R.~M. Bionta} \emph{et~al.}  (IMB Collaboration): Observation of a
  neutrino burst in coincidence with supernova 1987A in the Large Magellanic
  Cloud.
\newblock \emph{Phys. Rev. Lett.} \textbf{58} (1987), 14, pp. 1494--1496.

\bibitem{Alekseev1987}
\textsc{E.~Alekseev}, \textsc{L.~Alekseeva}, \textsc{V.~Volchenko},
  \textsc{I.~Krivosheina}: Possible Detection of a Neutrino Signal on 23
  February 1987 at the Baksan Underground Scintillation Telescope of the
  Institute of Nuclear Research.
\newblock \emph{JETP Letters} \textbf{45} (1987), pp. 589--592.

\bibitem{Raffelt1999}
\textsc{G.~G. Raffelt}: {Particle physics from stars}.
\newblock \emph{Ann. Rev. Nucl. Part. Sci.} \textbf{49} (1999), pp. 163--216.

\bibitem{Longo1987}
\textsc{M.~J. Longo}: {Tests of Relativity From SN1987A}.
\newblock \emph{Phys. Rev.} \textbf{D36} (1987), p. 3276.

\bibitem{Arnett1989}
\textsc{W.~D. Arnett}, \textsc{J.~N. Bahcall}, \textsc{R.~P. Kirshner},
  \textsc{S.~E. Woosley}: {Supernova SN1987A}.
\newblock \emph{Ann. Rev. Astron. Astrophys.} \textbf{27} (1989), pp. 629--700.

\bibitem{Cooperstein1988}
\textsc{J.~Cooperstein}: Neutron stars and the equation of state.
\newblock \emph{Phys. Rev.} \textbf{C37} (1988), pp. 786--796.

\bibitem{Dadykin1989}
\textsc{V.~L. Dadykin}, \textsc{G.~T. Zatsepin}, \textsc{O.~G. Ryazhskaya}:
  {Events detected by underground detectors on February 23, 1987}.
\newblock \emph{Sov. Phys. Usp.} \textbf{32} (1989), pp. 459--468.
\newblock [Usp. Fiz. Nauk158,139(1989)].

\bibitem{Aglietta1987}
\textsc{M.~Aglietta} \emph{et~al.}: {On the event observed in the Mont Blanc
  Underground Neutrino observatory during the occurrence of Supernova 1987a}.
\newblock \emph{Europhys. Lett.} \textbf{3} (1987), pp. 1315--1320.

\bibitem{De-Rujula1987}
\textsc{A.~De~Rujula}: {May a Supernova Bang Twice?}
\newblock \emph{Phys. Lett.} \textbf{B193} (1987), pp. 514--524.

\bibitem{Galeotti2016}
\textsc{P.~Galeotti}, \textsc{G.~Pizzella}: {New analysis for the correlation
  between gravitational wave and neutrino detectors during SN1987A}.
\newblock \emph{Eur. Phys. J.} \textbf{C76} (2016), 8, p. 426.

\bibitem{Ryazhskaya2018}
\textsc{O.~G. Ryazhskaya}: {Problems of Neutrino Radiation from SN 1987A: 30
  Years Later}.
\newblock \emph{Phys. Atom. Nucl.} \textbf{81} (2018), 1, pp. 113--119.
\newblock [Yad. Fiz.81,no.1,98(2018)].

\bibitem{Aglietta1987a}
\textsc{M.~Aglietta} \emph{et~al.}: {Comments on the two events observed in
  neutrino detectors during the supernova 1987a outburst.}
\newblock \emph{Europhys. Lett.} \textbf{3} (1987), pp. 1321--1324.

\bibitem{Berezinsky1988}
\textsc{V.~S. Berezinsky}, \textsc{C.~Castagnoli}, \textsc{V.~I. Dokuchaev},
  \textsc{P.~Galeotti}: {On the Possibility of a Two-Bang Supernova Collapse}.
\newblock \emph{Nuovo Cim.} \textbf{C11} (1988), pp. 287--303.

\bibitem{Imshennik2004}
\textsc{V.~S. Imshennik}, \textsc{O.~G. Ryazhskaya}: {A rotating collapsar and
  possible interpretation of the LSD neutrino signal from SN 1987A}.
\newblock \emph{Astron. Lett.} \textbf{30} (2004), pp. 14--31.

\bibitem{Ehrlich2018}
\textsc{R.~Ehrlich}: {The Mont Blanc neutrinos from SN 1987A: Could they have
  been monochromatic (8 MeV) tachyons with $m^2=-0.38$ keV$^2$?}
\newblock \emph{Astropart. Phys.} \textbf{99} (2018), pp. 21--29.

\bibitem{Burrows2012}
\textsc{A.~Burrows}, \textsc{J.~C. Dolence}, \textsc{J.~W. Murphy}: An
  investigation into the character of pre-explosion core-collapse supernova
  shock motion.
\newblock \emph{Astrophys. J.} \textbf{759} (2012), pp. 5--15.

\bibitem{Kuroda2016}
\textsc{T.~Kuroda}, \textsc{K.~Kotake}, \textsc{T.~Takiwaki}: {A new
  Gravitational-wave Signature From Standing Accretion Shock Instability in
  Supernovae}.
\newblock \emph{Astrophys. J.} \textbf{829} (2016), 1, p. L14.

\bibitem{Kuroda2017}
\textsc{T.~Kuroda}, \textsc{K.~Kotake}, \textsc{K.~Hayama},
  \textsc{T.~Takiwaki}: {Correlated Signatures of Gravitational-Wave and
  Neutrino Emission in Three-Dimensional General-Relativistic Core-Collapse
  Supernova Simulations}.
\newblock \emph{Astrophys. J.} \textbf{851} (2017), 1, p.~62.

\bibitem{OConnor2018}
\textsc{E.~P. O'Connor}, \textsc{S.~M. Couch}: {Exploring Fundamentally
  Three-dimensional Phenomena in High-fidelity Simulations of Core-collapse
  Supernovae}.
\newblock \emph{Astrophys. J.} \textbf{865} (2018), 2, p.~81.

\bibitem{Summa2018}
\textsc{A.~Summa}, \textsc{H.~T. Janka}, \textsc{T.~Melson}, \textsc{A.~Marek}:
  {Rotation-supported Neutrino-driven Supernova Explosions in Three Dimensions
  and the Critical Luminosity Condition}.
\newblock \emph{Astrophys. J.} \textbf{852} (2018), 1, p.~28.

\bibitem{Walk2018}
\textsc{L.~Walk}, \textsc{I.~Tamborra}, \textsc{H.-T. Janka},
  \textsc{A.~Summa}: {Identifying rotation in SASI-dominated core-collapse
  supernovae with a neutrino gyroscope}.
\newblock \emph{Phys. Rev.} \textbf{D98} (2018), 12, p. 123001.

\bibitem{Vartanyan2019}
\textsc{D.~Vartanyan} \emph{et~al.}: {A Successful 3D Core-Collapse Supernova
  Explosion Model}.
\newblock \emph{Mon. Not. Roy. Astron. Soc.} \textbf{482} (2019), 1, pp.
  351--369.

\bibitem{Melson2015}
\textsc{T.~Melson}, \textsc{H.-T. Janka}, \textsc{A.~Marek}: Neutrino-driven
  supernova of a low-mass iron-core progenitor boosted by three-dimensional
  turbulent convection.
\newblock \emph{Astrophys. J.} \textbf{801} (2015), L24.

\bibitem{Turner1988}
\textsc{M.~S. Turner}: Axions from SN1987A.
\newblock \emph{Phys. Rev. Lett.} \textbf{60} (1988), pp. 1797--1800.

\bibitem{Janka1996}
\textsc{H.-T. Janka}, \textsc{W.~Keil}, \textsc{G.~G. Raffelt},
  \textsc{D.~Seckel}: Nucleon Spin Fluctuations and the Supernova Emission of
  Neutrinos and Axions.
\newblock \emph{Phys. Rev. Lett.} \textbf{76} (1996), pp. 2621--2624.

\bibitem{Dreiner2014}
\textsc{H.~K. Dreiner}, \textsc{J.-F. Fortin}, \textsc{C.~Hanhart},
  \textsc{L.~Ubaldi}: Supernova constraints on MeV dark sectors from $e^+ e^-$
  annihilations.
\newblock \emph{Phys. Rev.} \textbf{D89} (2014), 105015.

\bibitem{Kazanas2015}
\textsc{D.~Kazanas} \emph{et~al.}: Supernova bounds on the dark photon using
  its electromagnetic decay.
\newblock \emph{Nucl. Phys.} \textbf{B890} (2015), pp. 17--29.

\bibitem{Alberini1986}
\textsc{C.~Alberini} \emph{et~al.}  (LVD): {The Large Volume Detector (LVD) of
  the Gran Sasso Laboratory}.
\newblock \emph{Nuovo Cim.} \textbf{C9} (1986), pp. 237--261.

\bibitem{Suzuki1987}
\textsc{A.~Suzuki}: {(Super-Kamiokande): Next Generation Underground Facility
  at Kamioka}.
\newblock In: \emph{{Proceedings: Workshop on Elementary Particle Picture of
  the Universe, Tsukuba, Japan, Feb 6-7, 1987}} (1987) .

\bibitem{HKDR2018}
\textsc{K.~Abe} \emph{et~al.}  (Hyper-Kamiokande): {Hyper-Kamiokande Design
  Report}  (2018).
\newblock \href{https://arxiv.org/abs/1805.04163}{\texttt{arXiv:1805.04163}}.

\bibitem{Ikeda2007}
\textsc{M.~Ikeda} \emph{et~al.}  (Super-Kamiokande): {Search for Supernova
  Neutrino Bursts at Super-Kamiokande}.
\newblock \emph{Astrophys. J.} \textbf{669} (2007), pp. 519--524.

\bibitem{Vigorito2018}
\textsc{C.~F. Vigorito}, \textsc{G.~Bruno}, \textsc{W.~Fulgione},
  \textsc{A.~Molinario}  (LVD): {Search for Supernova Neutrinos with the LVD
  experiment: the 2017 update}.
\newblock \emph{PoS} \textbf{ICRC2017} (2018), p. 1017.

\bibitem{Abe2016}
\textsc{K.~Abe} \emph{et~al.}  (Super-Kamiokande): {Real-Time Supernova
  Neutrino Burst Monitor at Super-Kamiokande}.
\newblock \emph{Astropart. Phys.} \textbf{81} (2016), pp. 39--48.

\bibitem{Dighe2003a}
\textsc{A.~S. Dighe}, \textsc{M.~T. Keil}, \textsc{G.~G. Raffelt}: {Detecting
  the neutrino mass hierarchy with a supernova at IceCube}.
\newblock \emph{JCAP} \textbf{0306} (2003), p. 005.

\bibitem{Kato2017}
\textsc{C.~Kato} \emph{et~al.}: {Neutrino emissions in all flavors up to the
  pre-bounce of massive stars and the possibility of their detections}.
\newblock \emph{Astrophys. J.} \textbf{848} (2017), p.~48.

\bibitem{Capozzi2018}
\textsc{F.~Capozzi}, \textsc{B.~Dasgupta}, \textsc{A.~Mirizzi}:
  {Model-independent diagnostic of self-induced spectral equalization versus
  ordinary matter effects in supernova neutrinos}.
\newblock \emph{Phys. Rev.} \textbf{D98} (2018), 6, p. 063013.

\bibitem{Scholberg2018}
\textsc{K.~Scholberg}: {Supernova Signatures of Neutrino Mass Ordering}.
\newblock \emph{J. Phys.} \textbf{G45} (2018), 1, p. 014002.

\bibitem{Tamborra2012}
\textsc{I.~Tamborra} \emph{et~al.}: High-resolution supernova neutrino spectra
  represented by a simple fit.
\newblock \emph{Phys. Rev.} \textbf{D86} (2012), 125031.

\bibitem{Gallo-Rosso2018}
\textsc{A.~Gallo~Rosso}, \textsc{F.~Vissani}, \textsc{M.~C. Volpe}: {What can
  we learn on supernova neutrino spectra with water Cherenkov detectors?}
\newblock \emph{JCAP}  (2018), 04, p. 040.

\bibitem{Nikrant2018}
\textsc{A.~Nikrant}, \textsc{R.~Laha}, \textsc{S.~Horiuchi}: {Robust
  measurement of supernova $\nu_e$ spectra with future neutrino detectors}.
\newblock \emph{Phys. Rev.} \textbf{D97} (2018), 2, p. 023019.

\bibitem{Totani1998}
\textsc{T.~Totani}, \textsc{K.~Sato}, \textsc{H.~E. Dalhed}, \textsc{J.~R.
  Wilson}: Future detection of supernova neutrino burst and explosion
  mechanism.
\newblock \emph{Astrophys. J.} \textbf{496} (1998), pp. 216--225.

\bibitem{Horiuchi2018}
\textsc{S.~Horiuchi}, \textsc{J.~P. Kneller}: {What can be learned from a
  future supernova neutrino detection?}
\newblock \emph{J. Phys.} \textbf{G45} (2018), 4, p. 043002.

\bibitem{Friedland2004}
\textsc{A.~Friedland}, \textsc{C.~Lunardini}, \textsc{C.~Pena-Garay}: {Solar
  neutrinos as probes of neutrino matter interactions}.
\newblock \emph{Phys. Lett.} \textbf{B594} (2004), p. 347.

\bibitem{Barger2005}
\textsc{V.~Barger}, \textsc{P.~Huber}, \textsc{D.~Marfatia}: {Solar
  mass-varying neutrino oscillations}.
\newblock \emph{Phys. Rev. Lett.} \textbf{95} (2005), p. 211802.

\bibitem{Holanda2004}
\textsc{P.~C. de~Holanda}, \textsc{A.~{\relax Yu}. Smirnov}: {Homestake result,
  sterile neutrinos and low-energy solar neutrino experiments}.
\newblock \emph{Phys. Rev.} \textbf{D69} (2004), p. 113002.

\bibitem{Abbott2017}
\textsc{B.~P. Abbott} \emph{et~al.}  (LIGO Scientific, Virgo, Fermi GBM,
  INTEGRAL, IceCube, AstroSat Cadmium Zinc Telluride Imager Team, IPN,
  Insight-Hxmt, ANTARES, Swift, AGILE Team, 1M2H Team, Dark Energy Camera
  GW-EM, DES, DLT40, GRAWITA, Fermi-LAT, ATCA, ASKAP, Las Cumbres Observatory
  Group, OzGrav, DWF (Deeper Wider Faster Program), AST3, CAASTRO, VINROUGE,
  MASTER, J-GEM, GROWTH, JAGWAR, CaltechNRAO, TTU-NRAO, NuSTAR, Pan-STARRS,
  MAXI Team, TZAC Consortium, KU, Nordic Optical Telescope, ePESSTO, GROND,
  Texas Tech University, SALT Group, TOROS, BOOTES, MWA, CALET, IKI-GW
  Follow-up, H.E.S.S., LOFAR, LWA, HAWC, Pierre Auger, ALMA, Euro VLBI Team, Pi
  of Sky, Chandra Team at McGill University, DFN, ATLAS Telescopes, High Time
  Resolution Universe Survey, RIMAS, RATIR, SKA South Africa/MeerKAT):
  {Multi-messenger Observations of a Binary Neutron Star Merger}.
\newblock \emph{Astrophys. J.} \textbf{848} (2017), 2, p. L12.

\bibitem{Abe2018a}
\textsc{K.~Abe} \emph{et~al.}  (Hyper-Kamiokande): {Physics potentials with the
  second Hyper-Kamiokande detector in Korea}.
\newblock \emph{PTEP} \textbf{2018} (2018), 6, p. 063C01.

\bibitem{Glashow1961}
\textsc{S.~L. Glashow}: {Partial Symmetries of Weak Interactions}.
\newblock \emph{Nucl. Phys.} \textbf{22} (1961), pp. 579--588.

\bibitem{Weinberg1967}
\textsc{S.~Weinberg}: {A Model of Leptons}.
\newblock \emph{Phys. Rev. Lett.} \textbf{19} (1967), pp. 1264--1266.

\bibitem{Salam1968}
\textsc{A.~Salam}: {Weak and Electromagnetic Interactions}.
\newblock \emph{Conf. Proc.} \textbf{C680519} (1968), pp. 367--377.

\bibitem{Gell-Mann1964}
\textsc{M.~Gell-Mann}: {A Schematic Model of Baryons and Mesons}.
\newblock \emph{Phys. Lett.} \textbf{8} (1964), pp. 214--215.

\bibitem{Zweig1964}
\textsc{G.~Zweig}: {An SU(3) model for strong interaction symmetry and its
  breaking}  (1964).
\newblock CERN-TH-401.

\bibitem{Gross1973}
\textsc{D.~J. Gross}, \textsc{F.~Wilczek}: {Ultraviolet Behavior of Nonabelian
  Gauge Theories}.
\newblock \emph{Phys. Rev. Lett.} \textbf{30} (1973), pp. 1343--1346.

\bibitem{Politzer1973}
\textsc{H.~D. Politzer}: {Reliable Perturbative Results for Strong
  Interactions?}
\newblock \emph{Phys. Rev. Lett.} \textbf{30} (1973), pp. 1346--1349.

\bibitem{Fritzsch1973}
\textsc{H.~Fritzsch}, \textsc{M.~Gell-Mann}, \textsc{H.~Leutwyler}: {Advantages
  of the Color Octet Gluon Picture}.
\newblock \emph{Phys. Lett.} \textbf{47B} (1973), pp. 365--368.

\bibitem{Georgi1974}
\textsc{H.~Georgi}, \textsc{S.~L. Glashow}: {Unity of All Elementary Particle
  Forces}.
\newblock \emph{Phys. Rev. Lett.} \textbf{32} (1974), pp. 438--441.

\bibitem{Langacker1981}
\textsc{P.~Langacker}: {Grand Unified Theories and Proton Decay}.
\newblock \emph{Phys. Rep.} \textbf{72} (1981), p. 185.

\bibitem{Nakamura1989}
\textsc{K.~Nakamura}: Present status and future of Kamiokande.
\newblock In: \textsc{K.~Hidaka}, \textsc{C.~S. Lim}, eds.: \emph{Proceedings
  of the Third Meeting on Physics at the TeV Energy Scale}. National Laboratory
  for High-Energy Physics, Tsukuba, Japan (1989) pp. 297--312.

\bibitem{Suzuki1988}
\textsc{A.~Suzuki}  (Kamiokande-II): {Kamiokande-II Low Background Detector}
  (1988).
\newblock KEK-Preprint-88-65.

\bibitem{NobelPrize2002}
The Nobel Prize in Physics 2002:
  \href{https://www.nobelprize.org/prizes/physics/2002/summary/}{\texttt{www.nobelprize.org/prizes/physics/2002/summary/}}.
\newblock  Last checked: 2019-11-18.

\bibitem{Kasuga1996}
\textsc{S.~Kasuga} \emph{et~al.}: {A Study on the e / mu identification
  capability of a water Cherenkov detector and the atmospheric neutrino
  problem}.
\newblock \emph{Phys. Lett.} \textbf{B374} (1996), pp. 238--242.

\bibitem{Fukuda2003}
\textsc{S.~Fukuda} \emph{et~al.}: The Super-Kamiokande detector.
\newblock \emph{Nucl. Instrum. Meth.} \textbf{A501} (2003), pp. 418--462.

\bibitem{Abe2016a}
\textsc{K.~Abe} \emph{et~al.}  (Super-Kamiokande): {Solar Neutrino Measurements
  in Super-Kamiokande-IV}.
\newblock \emph{Phys. Rev.} \textbf{D94} (2016), 5, p. 052010.

\bibitem{NobelPrize2015}
The Nobel Prize in Physics 2015:
  \href{https://www.nobelprize.org/prizes/physics/2015/summary/}{\texttt{www.nobelprize.org/prizes/physics/2015/summary/}}.
\newblock  Last checked: 2019-11-18.

\bibitem{Abe2011a}
\textsc{K.~Abe} \emph{et~al.}  (T2K): {The T2K Experiment}.
\newblock \emph{Nucl. Instrum. Meth.} \textbf{A659} (2011), pp. 106--135.

\bibitem{Abe2017}
\textsc{K.~Abe} \emph{et~al.}  (Super-Kamiokande): {Search for proton decay via
  $p \to e^+\pi^0$ and $p \to \mu^+\pi^0$ in 0.31  megaton·years exposure
  of the Super-Kamiokande water Cherenkov detector}.
\newblock \emph{Phys. Rev.} \textbf{D95} (2017), 1, p. 012004.

\bibitem{Suzuki2000}
\textsc{Y.~Suzuki}: {Comments on future proton decay experiments}.
\newblock \emph{AIP Conf. Proc.} \textbf{533} (2000), 1, p.~25.

\bibitem{Shiozawa2000}
\textsc{M.~Shiozawa}: {Study of 1-Megaton water Cherenkov detectors for the
  future proton decay search}.
\newblock \emph{AIP Conf. Proc.} \textbf{533} (2000), 1, pp. 21--24.

\bibitem{Abe2011b}
\textsc{K.~Abe} \emph{et~al.}  (Hyper-Kamiokande Working Group): Letter of
  Intent: The Hyper-Kamiokande Experiment---Detector Design and Physics
  Potential  (2011).
\newblock \href{https://arxiv.org/abs/1109.3262}{\texttt{arXiv:1109.3262}}.

\bibitem{HKDR2016}
\textsc{K.~Abe} \emph{et~al.}  (Hyper-Kamiokande): {Hyper-Kamiokande Design
  Report}  (2016).
\newblock KEK-PREPRINT-2016-21.

\bibitem{Gonokami2018}
Concerning the Start of Hyper-Kamiokande:
  \href{http://www.hyper-k.org/en/news/news-20180912.html}{\texttt{www.hyper-k.org/en/news/news-20180912.html}}.
\newblock  Last checked: 2019-11-18.

\bibitem{Beacom2004}
\textsc{J.~F. Beacom}, \textsc{M.~R. Vagins}: {GADZOOKS! Anti-neutrino
  spectroscopy with large water Cherenkov detectors}.
\newblock \emph{Phys. Rev. Lett.} \textbf{93} (2004), p. 171101.

\bibitem{Marti-Magro2018}
\textsc{L.~Marti-Magro}  (Super-Kamiokande): {SuperK-Gd}.
\newblock \emph{PoS} \textbf{ICRC2017} (2018), p. 1043.

\bibitem{R12860}
Large Photocathode Area Photomultiplier Tubes:
  \href{https://www.hamamatsu.com/resources/pdf/etd/LARGE_AREA_PMT_TPMH1376E.pdf}{\texttt{www.hamamatsu.com/resources/pdf/etd/LARGE\_AREA\_PMT\_TPMH1376E.pdf}}.
\newblock  Last checked: 2019-11-18.

\bibitem{HKTR2018}
Hyper-Kamiokande Technical Report (2018).
\newblock Unpublished.

\bibitem{Adrian-Martinez2014}
\textsc{S.~Adrián-Martínez} \emph{et~al.}  (KM3NeT): {Deep sea tests of a
  prototype of the KM3NeT digital optical module}.
\newblock \emph{Eur. Phys. J.} \textbf{C74} (2014), 9, p. 3056.

\bibitem{ET9302}
9302B series data sheet:
  \href{http://et-enterprises.com/images/data_sheets/9302B.pdf}{\texttt{et-enterprises.com/images/data\_sheets/9302B.pdf}}.
\newblock  Last checked: 2019-11-18.

\bibitem{ET9320KFLB}
9320KFLB data sheet (provisional):
  \href{http://et-enterprises.com/images/data_sheets/9320KFLB.pdf}{\texttt{et-enterprises.com/images/data\_sheets/9320KFLB.pdf}}.
\newblock  Last checked: 2019-11-18.

\bibitem{Richards2018}
\textsc{B.~Richards}: ToolDAQ Framework v2.2.1 (2019).
\newblock
  \href{http://doi.org/10.5281/zenodo.3229251}{\texttt{DOI:10.5281/zenodo.3229251}}.

\bibitem{Scholberg2000}
\textsc{K.~Scholberg}: SNEWS: The SuperNova Early Warning System.
\newblock \emph{AIP Conf. Proc.} \textbf{523} (2000), pp. 355--361.

\bibitem{Pronost2018}
\textsc{G.~Pronost} \emph{et~al.}: {Development of new radon monitoring systems
  in the Kamioka mine}.
\newblock \emph{PTEP} \textbf{2018} (2018), 9, p. 093H01.

\bibitem{Antonioli1997}
\textsc{P.~Antonioli} \emph{et~al.}: {A Three-dimensional code for muon
  propagation through the rock: Music}.
\newblock \emph{Astropart. Phys.} \textbf{7} (1997), pp. 357--368.

\bibitem{Ferrari2005}
\textsc{A.~Ferrari}, \textsc{P.~R. Sala}, \textsc{A.~Fasso}, \textsc{J.~Ranft}:
  {FLUKA: A multi-particle transport code}  (2005).
\newblock CERN-2005-010.

\bibitem{Hosaka2006}
\textsc{J.~Hosaka} \emph{et~al.}  (Super-Kamiokande): {Solar neutrino
  measurements in Super-Kamiokande-I}.
\newblock \emph{Phys. Rev.} \textbf{D73} (2006), p. 112001.

\bibitem{Li2014}
\textsc{S.~W. Li}, \textsc{J.~F. Beacom}: {First calculation of cosmic-ray muon
  spallation backgrounds for MeV astrophysical neutrino signals in
  Super-Kamiokande}.
\newblock \emph{Phys. Rev.} \textbf{C89} (2014), p. 045801.

\bibitem{Malek2003}
\textsc{M.~S. Malek}: \emph{{A search for supernova relic neutrinos}}.
\newblock Ph.D. thesis, SUNY, Stony Brook (2003).

\bibitem{Malek2003a}
\textsc{M.~Malek} \emph{et~al.}  (Super-Kamiokande): {Search for supernova
  relic neutrinos at SUPER-KAMIOKANDE}.
\newblock \emph{Phys. Rev. Lett.} \textbf{90} (2003), p. 061101.

\bibitem{Bays2012}
\textsc{K.~Bays} \emph{et~al.}  (Super-Kamiokande): {Supernova Relic Neutrino
  Search at Super-Kamiokande}.
\newblock \emph{Phys. Rev.} \textbf{D85} (2012), p. 052007.

\bibitem{Li2015}
\textsc{S.~W. Li}, \textsc{J.~F. Beacom}: {Spallation Backgrounds in
  Super-Kamiokande Are Made in Muon-Induced Showers}.
\newblock \emph{Phys. Rev.} \textbf{D91} (2015), 10, p. 105005.

\bibitem{Li2015a}
\textsc{S.~W. Li}, \textsc{J.~F. Beacom}: {Tagging Spallation Backgrounds with
  Showers in Water-Cherenkov Detectors}.
\newblock \emph{Phys. Rev.} \textbf{D92} (2015), 10, p. 105033.

\bibitem{Bollig2017}
\textsc{R.~Bollig} \emph{et~al.}: {Muon Creation in Supernova Matter
  Facilitates Neutrino-driven Explosions}.
\newblock \emph{Phys. Rev. Lett.} \textbf{119} (2017), 24, p. 242702.

\bibitem{Sukhbold2016}
\textsc{T.~Sukhbold} \emph{et~al.}: {Core-Collapse Supernovae from 9 to 120
  Solar Masses Based on Neutrino-powered Explosions}.
\newblock \emph{Astrophys. J.} \textbf{821} (2016), 1, p.~38.

\bibitem{Couch2019}
\textsc{S.~M. Couch}, \textsc{M.~L. Warren}, \textsc{E.~P. O'Connor}:
  {Simulating Turbulence-aided Neutrino-driven Core-collapse Supernova
  Explosions in One Dimension}  (2019).
\newblock \href{https://arxiv.org/abs/1902.01340}{\texttt{arXiv:1902.01340}}.

\bibitem{Ibeling2013}
\textsc{D.~Ibeling}, \textsc{A.~Heger}: {The Metallicity Dependence of the
  Minimum Mass for Core-Collapse Supernovae}.
\newblock \emph{Astrophys. J.} \textbf{765} (2013), p. L43.

\bibitem{Takiwaki2016}
\textsc{T.~Takiwaki}, \textsc{K.~Kotake}, \textsc{Y.~Suwa}: {Three-dimensional
  simulations of rapidly rotating core-collapse supernovae: finding a
  neutrino-powered explosion aided by non-axisymmetric flows}.
\newblock \emph{Mon. Not. Roy. Astron. Soc.} \textbf{461} (2016), 1, pp.
  L112--L116.

\bibitem{Andresen2019}
\textsc{H.~Andresen} \emph{et~al.}: {Gravitational waves from 3D core-collapse
  supernova models: The impact of moderate progenitor rotation}.
\newblock \emph{Mon. Not. Roy. Astron. Soc.} \textbf{486} (2019), 2, pp.
  2238--2253.

\bibitem{Obergaulinger2018}
\textsc{M.~Obergaulinger}, \textsc{O.~Just}, \textsc{M.~Aloy}: {Core collapse
  with magnetic fields and rotation}.
\newblock \emph{J. Phys.} \textbf{G45} (2018), 8, p. 084001.

\bibitem{Podsiadlowski2004}
\textsc{P.~Podsiadlowski} \emph{et~al.}: {The effects of binary evolution on
  the dynamics of core collapse and neutron-star kicks}.
\newblock \emph{Astrophys. J.} \textbf{612} (2004), pp. 1044--1051.

\bibitem{Aufderheide1991}
\textsc{M.~B. {Aufderheide}}, \textsc{E.~{Baron}}, \textsc{F.~K. {Thielemann}}:
  {Shock Waves and Nucleosynthesis in Type II Supernovae}.
\newblock \emph{Astrophys. J.} \textbf{370} (1991), p. 630.

\bibitem{Ugliano2012}
\textsc{M.~Ugliano}, \textsc{H.~T. Janka}, \textsc{A.~Marek},
  \textsc{A.~Arcones}: {Progenitor-Explosion Connection and Remnant Birth
  Masses for Neutrino-Driven Supernovae of Iron-Core Progenitors}.
\newblock \emph{Astrophys. J.} \textbf{757} (2012), p.~69.

\bibitem{Perego2015}
\textsc{A.~Perego} \emph{et~al.}: {Pushing Core-Collapse Supernovae to
  Explosions in Spherical Symmetry I: The Model and the Case of SN1987A}.
\newblock \emph{Astrophys. J.} \textbf{806} (2015), 2, p. 275.

\bibitem{OConnor2018a}
\textsc{E.~O'Connor} \emph{et~al.}: {Global Comparison of Core-Collapse
  Supernova Simulations in Spherical Symmetry}.
\newblock \emph{J. Phys.} \textbf{G45} (2018), 10, p. 104001.

\bibitem{Wilson1986}
\textsc{J.~R. Wilson}, \textsc{R.~Mayle}, \textsc{S.~E. Woosley},
  \textsc{T.~Weaver}: {Stellar core collapse and supernova}.
\newblock \emph{Annals N. Y. Acad. Sci.} \textbf{470} (1986), pp. 267--293.

\bibitem{Mayle1987}
\textsc{R.~Mayle}, \textsc{J.~R. Wilson}, \textsc{D.~N. Schramm}: {Neutrinos
  from gravitational collapse}.
\newblock \emph{Astrophys. J.} \textbf{318} (1987), pp. 288--306.

\bibitem{Geha1998}
\textsc{M.~C. Geha} \emph{et~al.}  (WFPC2 ID Team): {Stellar populations in
  three outer fields of the lmc}.
\newblock \emph{Astron. J.} \textbf{115} (1998), pp. 1045--1056.

\bibitem{Mottini2006}
\textsc{M.~Mottini} \emph{et~al.}: {The chemical composition of cepheids in the
  milky way and the magellanic clouds}.
\newblock \emph{Mem. Soc. Ast. It.} \textbf{77} (2006), pp. 156--159.

\bibitem{Shen1998}
\textsc{H.~Shen}, \textsc{H.~Toki}, \textsc{K.~Oyamatsu},
  \textsc{K.~Sumiyoshi}: {Relativistic equation of state of nuclear matter for
  supernova explosion}.
\newblock \emph{Prog. Theor. Phys.} \textbf{100} (1998), p. 1013.

\bibitem{Sumiyoshi2005}
\textsc{K.~Sumiyoshi} \emph{et~al.}: {Postbounce evolution of core-collapse
  supernovae: Long-term effects of equation of state}.
\newblock \emph{Astrophys. J.} \textbf{629} (2005), pp. 922--932.

\bibitem{Warren2019}
\textsc{M.~L. Warren}: {Test upload of neutrino emission from \SI{20}{\Msol}
  progenitor} (2019).
\newblock
  \href{https://sandbox.zenodo.org/record/257869}{\texttt{https://sandbox.zenodo.org/record/257869}}.

\bibitem{Sukhbold2014}
\textsc{T.~Sukhbold}, \textsc{S.~Woosley}: {The Compactness of Presupernova
  Stellar Cores}.
\newblock \emph{Astrophys. J.} \textbf{783} (2014), p.~10.

\bibitem{Fryxell2000}
\textsc{B.~Fryxell} \emph{et~al.}: {FLASH: An Adaptive Mesh Hydrodynamics Code
  for Modeling Astrophysical Thermonuclear Flashes}.
\newblock \emph{Astrophys. J. Suppl.} \textbf{131} (2000), pp. 273--334.

\bibitem{Dubey2009}
\textsc{A.~Dubey} \emph{et~al.}: Extensible component-based architecture for
  FLASH, a massively parallel, multiphysics simulation code.
\newblock \emph{Parallel Computing} \textbf{35} (2009), 10, pp. 512 -- 522.

\bibitem{OConnor2018b}
\textsc{E.~P. O'Connor}, \textsc{S.~M. Couch}: {Two Dimensional Core-Collapse
  Supernova Explosions Aided by General Relativity with Multidimensional
  Neutrino Transport}.
\newblock \emph{Astrophys. J.} \textbf{854} (2018), 1, p.~63.

\bibitem{Steiner2013}
\textsc{A.~W. Steiner}, \textsc{M.~Hempel}, \textsc{T.~Fischer}: {Core-collapse
  supernova equations of state based on neutron star observations}.
\newblock \emph{Astrophys. J.} \textbf{774} (2013), p.~17.

\bibitem{OConnor2015}
\textsc{E.~O'Connor}: {An Open-Source Neutrino Radiation Hydrodynamics Code for
  Core-Collapse Supernovae}.
\newblock \emph{Astrophys. J. Suppl.} \textbf{219} (2015), 2, p.~24.

\bibitem{Tamborra2014}
\textsc{I.~Tamborra} \emph{et~al.}: {Neutrino emission characteristics and
  detection opportunities based on three-dimensional supernova simulations}.
\newblock \emph{Phys. Rev.} \textbf{D90} (2014), p. 045032.

\bibitem{Woosley2002}
\textsc{S.~E. Woosley}, \textsc{A.~Heger}, \textsc{T.~A. Weaver}: The evolution
  and explosion of massive stars.
\newblock \emph{Rev. Mod. Phys.} \textbf{74} (2002), p. 1015.

\bibitem{Fryxell1991}
\textsc{B.~Fryxell}, \textsc{E.~Müller}, \textsc{W.~D. Arnett}: {Instabilities
  and clumping in SN 1987A. I - Early evolution in two dimensions}.
\newblock \emph{Astrophys. J.} \textbf{367} (1991), pp. 619--634.

\bibitem{Colella1984}
\textsc{P.~Colella}, \textsc{P.~R. Woodward}: {The Piecewise Parabolic Method
  (PPM) for Gas Dynamical Simulations}.
\newblock \emph{J. Comput. Phys.} \textbf{54} (1984), pp. 174--201.

\bibitem{Rampp2002}
\textsc{M.~Rampp}, \textsc{H.-T. Janka}: Radiation hydrodynamics with
  neutrinos.
\newblock \emph{Astron. Astrophys.} \textbf{396} (2002), 1, pp. 361--392.

\bibitem{Lattimer1991}
\textsc{J.~M. Lattimer}, \textsc{F.~D. Swesty}: {A Generalized equation of
  state for hot, dense matter}.
\newblock \emph{Nucl. Phys.} \textbf{A535} (1991), pp. 331--376.

\bibitem{Radice2017}
\textsc{D.~Radice} \emph{et~al.}: {Electron-Capture and Low-Mass
  Iron-Core-Collapse Supernovae: New Neutrino-Radiation-Hydrodynamics
  Simulations}.
\newblock \emph{Astrophys. J.} \textbf{850} (2017), 1, p.~43.

\bibitem{Seadrow2018}
\textsc{S.~Seadrow} \emph{et~al.}: {Neutrino Signals of Core-Collapse
  Supernovae in Underground Detectors}.
\newblock \emph{Mon. Not. Roy. Astron. Soc.} \textbf{480} (2018), 4, pp.
  4710--4731.

\bibitem{Skinner2019}
\textsc{M.~A. Skinner} \emph{et~al.}: {Fornax: a Flexible Code for Multiphysics
  Astrophysical Simulations}.
\newblock \emph{Astrophys. J. Suppl.} \textbf{241} (2019), 1, p.~7.

\bibitem{Shibata2011}
\textsc{M.~Shibata}, \textsc{K.~Kiuchi}, \textsc{Y.-i. Sekiguchi},
  \textsc{Y.~Suwa}: {Truncated Moment Formalism for Radiation Hydrodynamics in
  Numerical Relativity}.
\newblock \emph{Prog. Theor. Phys.} \textbf{125} (2011), pp. 1255--1287.

\bibitem{Murchikova2017}
\textsc{L.~M. Murchikova}, \textsc{E.~Abdikamalov}, \textsc{T.~Urbatsch}:
  {Analytic Closures for M1 Neutrino Transport}.
\newblock \emph{Mon. Not. Roy. Astron. Soc.} \textbf{469} (2017), 2, pp.
  1725--1737.

\bibitem{Burrows2006}
\textsc{A.~Burrows}, \textsc{S.~Reddy}, \textsc{T.~A. Thompson}: {Neutrino
  opacities in nuclear matter}.
\newblock \emph{Nucl. Phys.} \textbf{A777} (2006), pp. 356--394.

\bibitem{sntools}
sntools: Event generator for supernova burst neutrinos in Hyper-Kamiokande:
  \href{https://github.com/JostMigenda/sntools}{\texttt{github.com/JostMigenda/sntools}}.
\newblock  Last checked: 2019-11-18.

\bibitem{Hayato2009}
\textsc{Y.~Hayato}: {A neutrino interaction simulation program library NEUT}.
\newblock \emph{Acta Phys. Polon.} \textbf{B40} (2009), pp. 2477--2489.

\bibitem{Andreopoulos2015}
\textsc{C.~Andreopoulos} \emph{et~al.}: {The GENIE Neutrino Monte Carlo
  Generator: Physics and User Manual}  (2015).
\newblock \href{https://arxiv.org/abs/1510.05494}{\texttt{arXiv:1510.05494}}.

\bibitem{Virtanen2019}
\textsc{P.~Virtanen} \emph{et~al.}: {SciPy 1.0--Fundamental Algorithms for
  Scientific Computing in Python}  (2019).
\newblock \href{https://arxiv.org/abs/1907.10121}{\texttt{arXiv:1907.10121}}.

\bibitem{Walt2011}
\textsc{S.~van~der Walt}, \textsc{S.~C. Colbert}, \textsc{G.~Varoquaux}: {The
  NumPy Array: A Structure for Efficient Numerical Computation}.
\newblock \emph{Comput. Sci. Eng.} \textbf{13} (2011), 2, pp. 22--30.

\bibitem{Dighe2000}
\textsc{A.~S. Dighe}, \textsc{A.~{\relax Yu}. Smirnov}: {Identifying the
  neutrino mass spectrum from the neutrino burst from a supernova}.
\newblock \emph{Phys. Rev.} \textbf{D62} (2000), p. 033007.

\bibitem{Fogli2005}
\textsc{G.~L. Fogli}, \textsc{E.~Lisi}, \textsc{A.~Mirizzi},
  \textsc{D.~Montanino}: {Probing supernova shock waves and neutrino flavor
  transitions in next-generation water-Cerenkov detectors}.
\newblock \emph{JCAP} \textbf{0504} (2005), p. 002.

\bibitem{Schirato2002}
\textsc{R.~C. Schirato}, \textsc{G.~M. Fuller}: {Connection between supernova
  shocks, flavor transformation, and the neutrino signal}  (2002).
\newblock
  \href{https://arxiv.org/abs/astro-ph/0205390}{\texttt{arXiv:astro-ph/0205390}}.

\bibitem{Duan2006a}
\textsc{H.~Duan}, \textsc{G.~M. Fuller}, \textsc{J.~Carlson}, \textsc{Y.-Z.
  Qian}: {Coherent Development of Neutrino Flavor in the Supernova
  Environment}.
\newblock \emph{Phys. Rev. Lett.} \textbf{97} (2006), p. 241101.

\bibitem{Duan2006}
\textsc{H.~Duan}, \textsc{G.~M. Fuller}, \textsc{J.~Carlson}, \textsc{Y.-Z.
  Qian}: {Simulation of Coherent Non-Linear Neutrino Flavor Transformation in
  the Supernova Environment. 1. Correlated Neutrino Trajectories}.
\newblock \emph{Phys. Rev.} \textbf{D74} (2006), p. 105014.

\bibitem{Raffelt2007}
\textsc{G.~G. Raffelt}, \textsc{A.~{\relax Yu}. Smirnov}: {Self-induced
  spectral splits in supernova neutrino fluxes}.
\newblock \emph{Phys. Rev.} \textbf{D76} (2007), p. 081301.

\bibitem{Dasgupta2009}
\textsc{B.~Dasgupta}, \textsc{A.~Dighe}, \textsc{G.~G. Raffelt},
  \textsc{A.~{\relax Yu}. Smirnov}: {Multiple Spectral Splits of Supernova
  Neutrinos}.
\newblock \emph{Phys. Rev. Lett.} \textbf{103} (2009), p. 051105.

\bibitem{Friedland2010}
\textsc{A.~Friedland}: {Self-refraction of supernova neutrinos: mixed spectra
  and three-flavor instabilities}.
\newblock \emph{Phys. Rev. Lett.} \textbf{104} (2010), p. 191102.

\bibitem{Duan2011}
\textsc{H.~Duan}, \textsc{A.~Friedland}: {Self-induced suppression of
  collective neutrino oscillations in a supernova}.
\newblock \emph{Phys. Rev. Lett.} \textbf{106} (2011), p. 091101.

\bibitem{Izaguirre2017}
\textsc{I.~Izaguirre}, \textsc{G.~Raffelt}, \textsc{I.~Tamborra}: {Fast
  Pairwise Conversion of Supernova Neutrinos: A Dispersion-Relation Approach}.
\newblock \emph{Phys. Rev. Lett.} \textbf{118} (2017), 2, p. 021101.

\bibitem{Chakraborty2016}
\textsc{S.~Chakraborty}, \textsc{R.~Hansen}, \textsc{I.~Izaguirre},
  \textsc{G.~Raffelt}: {Collective neutrino flavor conversion: Recent
  developments}.
\newblock \emph{Nucl. Phys.} \textbf{B908} (2016), pp. 366--381.

\bibitem{Esteban-Pretel2008}
\textsc{A.~Esteban-Pretel} \emph{et~al.}: {Role of dense matter in collective
  supernova neutrino transformations}.
\newblock \emph{Phys. Rev.} \textbf{D78} (2008), p. 085012.

\bibitem{Zaizen2018}
\textsc{M.~Zaizen}, \textsc{T.~Yoshida}, \textsc{K.~Sumiyoshi},
  \textsc{H.~Umeda}: {Collective neutrino oscillations and detectabilities in
  failed supernovae}.
\newblock \emph{Phys. Rev.} \textbf{D98} (2018), 10, p. 103020.

\bibitem{Chakraborty2014}
\textsc{S.~Chakraborty}, \textsc{A.~Mirizzi}, \textsc{N.~Saviano},
  \textsc{D.~d.~S. Seixas}: {Suppression of the multi-azimuthal-angle
  instability in dense neutrino gas during supernova accretion phase}.
\newblock \emph{Phys. Rev.} \textbf{D89} (2014), 9, p. 093001.

\bibitem{Mirizzi2006}
\textsc{A.~Mirizzi}, \textsc{G.~G. Raffelt}, \textsc{P.~D. Serpico}: Earth
  matter effects in supernova neutrinos: optimal detector locations.
\newblock \emph{JCAP} \textbf{05} (2006), 012.

\bibitem{Dighe2003}
\textsc{A.~S. Dighe}, \textsc{M.~T. Keil}, \textsc{G.~G. Raffelt}: Identifying
  Earth matter effects on supernova neutrinos at a single detector.
\newblock \emph{JCAP}  (2003), 06, p. 006.

\bibitem{Beacom2002}
\textsc{J.~F. Beacom}, \textsc{W.~M. Farr}, \textsc{P.~Vogel}: {Detection of
  supernova neutrinos by neutrino proton elastic scattering}.
\newblock \emph{Phys. Rev.} \textbf{D66} (2002), p. 033001.

\bibitem{Strumia2003}
\textsc{A.~Strumia}, \textsc{F.~Vissani}: Precise quasielastic neutrino/nucleon
  cross-section.
\newblock \emph{Phys. Lett.} \textbf{B564} (2003), pp. 42--54.

\bibitem{Kurylov2003}
\textsc{A.~Kurylov}, \textsc{M.~J. Ramsey-Musolf}, \textsc{P.~Vogel}:
  {Radiative corrections to low-energy neutrino reactions}.
\newblock \emph{Phys. Rev.} \textbf{C67} (2003), p. 035502.

\bibitem{t-Hooft1971}
\textsc{G.~'t~Hooft}: {Predictions for neutrino - electron cross-sections in
  Weinberg's model of weak interactions}.
\newblock \emph{Phys. Lett.} \textbf{37B} (1971), pp. 195--196.

\bibitem{Bahcall1995}
\textsc{J.~N. Bahcall}, \textsc{M.~Kamionkowski}, \textsc{A.~Sirlin}: {Solar
  neutrinos: Radiative corrections in neutrino - electron scattering
  experiments}.
\newblock \emph{Phys. Rev.} \textbf{D51} (1995), pp. 6146--6158.

\bibitem{Nakazato2018}
\textsc{K.~Nakazato}, \textsc{T.~Suzuki}, \textsc{M.~Sakuda}: {Charged-current
  scattering off $^{16}$O nucleus as a detection channel for supernova
  neutrinos}.
\newblock \emph{PTEP} \textbf{2018} (2018), 12, p. 123E02.

\bibitem{Suzuki2018}
\textsc{T.~Suzuki} \emph{et~al.}: {Neutrino-nucleus reactions on $^{16}$O based
  on new shell-model Hamiltonians}.
\newblock \emph{Phys. Rev.} \textbf{C98} (2018), 3, p. 034613.

\bibitem{Kolbe2002}
\textsc{E.~Kolbe}, \textsc{K.~Langanke}, \textsc{P.~Vogel}: {Estimates of weak
  and electromagnetic nuclear decay signatures for neutrino reactions in
  Super-Kamiokande}.
\newblock \emph{Phys. Rev.} \textbf{D66} (2002), p. 013007.

\bibitem{Tomas2003}
\textsc{R.~Tomas} \emph{et~al.}: {Supernova pointing with low-energy and
  high-energy neutrino detectors}.
\newblock \emph{Phys. Rev.} \textbf{D68} (2003), p. 093013.

\bibitem{WCSim}
WCSim: The WCSim GEANT4 application:
  \href{https://github.com/WCSim/WCSim}{\texttt{github.com/WCSim/WCSim}}.
\newblock  Last checked: 2019-11-18.

\bibitem{Agostinelli2003}
\textsc{S.~Agostinelli} \emph{et~al.}  (\textsc{Geant4}): {\textsc{Geant4}: A
  Simulation toolkit}.
\newblock \emph{Nucl. Instrum. Meth.} \textbf{A506} (2003), pp. 250--303.

\bibitem{Brun1997}
\textsc{R.~Brun}, \textsc{F.~Rademakers}: {ROOT: An object oriented data
  analysis framework}.
\newblock \emph{Nucl. Instrum. Meth.} \textbf{A389} (1997), pp. 81--86.

\bibitem{Goon2012}
\textsc{J.~Goon} \emph{et~al.}  (LBNE): {The Long Baseline Neutrino Experiment
  (LBNE) Water Cherenkov Detector (WCD) Conceptual Design Report (CDR)}
  (2012).
\newblock \href{https://arxiv.org/abs/1204.2295}{\texttt{arXiv:1204.2295}}.

\bibitem{Casper2002}
\textsc{D.~Casper}: {The Nuance neutrino physics simulation, and the future}.
\newblock \emph{Nucl. Phys. Proc. Suppl.} \textbf{112} (2002), pp. 161--170.

\bibitem{nuance-format}
\textsc{Nuance} format:
  \href{http://neutrino.phy.duke.edu/nuance-format/}{\texttt{neutrino.phy.duke.edu/nuance-format/}}.
\newblock  Last checked: 2019-11-18.

\bibitem{Smy2007}
\textsc{M.~Smy}: {Low Energy Event Reconstruction and Selection in
  Super-Kamiokande-III}.
\newblock In: \emph{{Proceedings, 30th International Cosmic Ray Conference
  (ICRC 2007): Merida, Yucatan, Mexico, July 3-11, 2007}}, vol.~5 (2007) pp.
  1279--1282.

\bibitem{Cravens2008}
\textsc{J.~P. Cravens} \emph{et~al.}  (Super-Kamiokande): {Solar neutrino
  measurements in Super-Kamiokande-II}.
\newblock \emph{Phys. Rev.} \textbf{D78} (2008), p. 032002.

\bibitem{Nakano2016}
\textsc{Y.~Nakano}: \emph{{8B solar neutrino spectrum measurement using
  Super-Kamiokande IV}}.
\newblock Ph.D. thesis, U. Tokyo (main) (2016).

\bibitem{Abe2011c}
\textsc{K.~Abe} \emph{et~al.}  (Super-Kamiokande): {Solar neutrino results in
  Super-Kamiokande-III}.
\newblock \emph{Phys. Rev.} \textbf{D83} (2011), p. 052010.

\bibitem{energetic-bonsai}
energetic-bonsai: Energy reconstruction for (hk-)BONSAI:
  \href{https://github.com/JostMigenda/energetic-bonsai/}{\texttt{github.com/JostMigenda/energetic-bonsai/}}.
\newblock  Last checked: 2019-11-18.

\bibitem{Ikeda2009}
\textsc{M.~Ikeda}: \emph{{Precise Measurement of Solar Neutrinos with
  Super-Kamiokande III}}.
\newblock Ph.D. thesis, Tokyo U. (2009).

\bibitem{Pietrzynski2013}
\textsc{G.~Pietrzyński} \emph{et~al.}: {An eclipsing binary distance to the
  Large Magellanic Cloud accurate to 2 per cent}.
\newblock \emph{Nature} \textbf{495} (2013), pp. 76--79.

\bibitem{Hilditch2005}
\textsc{R.~W. Hilditch}, \textsc{I.~D. Howarth}, \textsc{T.~J. Harries}: {Forty
  eclipsing binaries in the Small Magellanic Cloud: Fundamental parameters and
  cloud distance}.
\newblock \emph{Mon. Not. Roy. Astron. Soc.} \textbf{357} (2005), pp. 304--324.

\bibitem{Suwa2019}
\textsc{Y.~Suwa} \emph{et~al.}: {Observing Supernova Neutrino Light Curves with
  Super-Kamiokande: Expected Event Number over 10s}.
\newblock \emph{Astrophys. J.} \textbf{881} (2019), p. 139.

\bibitem{Loredo1989}
\textsc{T.~J. Loredo}, \textsc{D.~Q. Lamb}: Neutrino from SN1987A: Implications
  for cooling of the nascent neutron star and the mass of the electron
  anti-neutrino.
\newblock \emph{Annals N. Y. Acad. Sci.} \textbf{571} (1989), pp. 601--630.

\bibitem{Horiuchi2017}
\textsc{S.~Horiuchi}, \textsc{K.~Nakamura}, \textsc{T.~Takiwaki},
  \textsc{K.~Kotake}: {Estimating the core compactness of massive stars with
  Galactic supernova neutrinos}.
\newblock \emph{J. Phys.} \textbf{G44} (2017), 11, p. 114001.

\bibitem{Loredo2002}
\textsc{T.~J. Loredo}, \textsc{D.~Q. Lamb}: Bayesian analysis of neutrinos
  observed from supernova SN-1987A.
\newblock \emph{Phys. Rev.} \textbf{D65} (2002), p. 063002.

\bibitem{Kass1995}
\textsc{R.~E. Kass}, \textsc{A.~E. Raftery}: {Bayes Factors}.
\newblock \emph{J. Am. Statist. Assoc.} \textbf{90} (1995), 430, pp. 773--795.

\bibitem{Ianni2009}
\textsc{A.~Ianni} \emph{et~al.}: {The Likelihood for supernova neutrino
  analyses}.
\newblock \emph{Phys. Rev.} \textbf{D80} (2009), p. 043007.

\bibitem{Zuber2015}
\textsc{K.~Zuber}  (HALO): {HALO, a supernova neutrino observatory}.
\newblock \emph{Nucl. Part. Phys. Proc.} \textbf{265-266} (2015), pp. 233--235.

\bibitem{Chakraborty2014a}
\textsc{S.~Chakraborty}, \textsc{P.~Bhattacharjee}, \textsc{K.~Kar}: {Observing
  supernova neutrino light curve in future dark matter detectors}.
\newblock \emph{Phys. Rev.} \textbf{D89} (2014), 1, p. 013011.

\bibitem{Lang2016}
\textsc{R.~F. Lang} \emph{et~al.}: {Supernova neutrino physics with xenon dark
  matter detectors: A timely perspective}.
\newblock \emph{Phys. Rev.} \textbf{D94} (2016), 10, p. 103009.

\bibitem{Aasi2015}
\textsc{J.~Aasi} \emph{et~al.}  (LIGO Scientific): {Advanced LIGO}.
\newblock \emph{Class. Quant. Grav.} \textbf{32} (2015), p. 074001.

\bibitem{Acernese2015}
\textsc{F.~Acernese} \emph{et~al.}  (VIRGO): {Advanced Virgo: a
  second-generation interferometric gravitational wave detector}.
\newblock \emph{Class. Quant. Grav.} \textbf{32} (2015), 2, p. 024001.

\bibitem{Somiya2012}
\textsc{K.~Somiya}  (KAGRA): {Detector configuration of KAGRA: The Japanese
  cryogenic gravitational-wave detector}.
\newblock \emph{Class. Quant. Grav.} \textbf{29} (2012), p. 124007.

\bibitem{Aso2013}
\textsc{Y.~Aso} \emph{et~al.}  (KAGRA): {Interferometer design of the KAGRA
  gravitational wave detector}.
\newblock \emph{Phys. Rev.} \textbf{D88} (2013), 4, p. 043007.

\bibitem{Abbott2018}
\textsc{B.~P. Abbott} \emph{et~al.}  (KAGRA, LIGO Scientific, VIRGO):
  {Prospects for Observing and Localizing Gravitational-Wave Transients with
  Advanced LIGO, Advanced Virgo and KAGRA}.
\newblock \emph{Living Rev. Rel.} \textbf{21} (2018), 1, p.~3.

\bibitem{Kistler2013}
\textsc{M.~D. Kistler}, \textsc{W.~C. Haxton}, \textsc{H.~Yüksel}: {Tomography
  of Massive Stars from Core Collapse to Supernova Shock Breakout}.
\newblock \emph{Astrophys. J.} \textbf{778} (2013), p.~81.

\bibitem{Smartt2009}
\textsc{S.~J. Smartt}: {Progenitors of core-collapse supernovae}.
\newblock \emph{Ann. Rev. Astron. Astrophys.} \textbf{47} (2009), pp. 63--106.

\bibitem{Antonioli2004}
\textsc{P.~Antonioli} \emph{et~al.}  (SNEWS Collaboration): SNEWS: The
  SuperNova Early Warning System.
\newblock \emph{New Journal of Physics} \textbf{6} (2004), 114.

\bibitem{Ram1967}
\textsc{M.~Ram}: {Inner Bremsstrahlung in Low-Energy Electron-Neutrino
  (Antineutrino) Scattering}.
\newblock \emph{Phys. Rev.} \textbf{155} (1967), pp. 1539--1553.

\end{thebibliography}

\end{document}